\documentclass[useAMS,usenatbib]{mn2e}
\usepackage{url}
\usepackage{graphicx}
\usepackage{lscape}
\usepackage{amsmath,amssymb}

\newcommand{\HI}{\hbox{\rmfamily H\,{\textsc i}}}

\title[Deep Near-Infrared Surface Photometry and Properties of Local Volume Dwarf Irregular Galaxies]{Deep Near-Infrared Surface Photometry and Properties of Local Volume Dwarf Irregular Galaxies}

\author[T. Young et al.]{T. Young$^{1,2,4}$\thanks{E-mail: tyoung@mso.anu.edu.au}, H. Jerjen$^{1}$, \'{A}. R. L\'{o}pez-S\'{a}nchez$^{2,3}$, B. S. Koribalski$^{4}$\\
$^{1}$ Research School of Astronomy and Astrophysics, The Australian National University, Mt Stromlo Observatory, via Cotter Rd,\\ Weston, ACT 2611, Australia\\
$^{2}$ Australian Astronomical Observatory, 105 Delhi Road, North Ryde, NSW 2113, PO Box 915, North Ryde, NSW 1670, Australia \\
$^{3}$ Department of Physics and Astronomy, Macquarie University, NSW 2109, Australia \\
$^{4}$ CSIRO Astronomy and Space Science, Australia Telescope National Facility, PO Box 76, Epping, NSW 1710, Australia}

\begin{document}

\date{}

\pagerange{\pageref{firstpage}--\pageref{lastpage}} \pubyear{}

\maketitle

\label{firstpage}

\begin{abstract}
We present deep \textit{H}-band surface photometry and analysis of 40 Local Volume galaxies, a sample primarily composed of dwarf irregulars in the Cen A group, obtained using the IRIS2 detector at the 3.9m Anglo-Australian Telescope. We probe to a surface brightness of \mbox{$\sim$25 mag arcsec$^{-2}$}, reaching a 40 times lower stellar density than the Two Micron All Sky Survey (2MASS). Employing extremely careful and rigorous cleaning techniques to remove contaminating sources, we perform surface photometry on 33 detected galaxies deriving the observed total magnitude, effective surface brightness and best fitting S\'{e}rsic parameters. We make image quality and surface photometry comparisons to 2MASS and VISTA Hemispheric Survey (VHS) demonstrating that deep targeted surveys are still the most reliable means of obtaining accurate surface photometry. We investigate the \textit{B}-\textit{H} colours with respect to mass for Local Volume galaxies, finding that the colours of dwarf irregulars are significantly varied, eliminating the possibility of using optical-NIR colour transformations to facilitate comparison to the more widely available optical data sets. The structure-luminosity relationships are investigated for our `clean' sample of dwarf irregulars. We demonstrate that a significant fraction of the Local Volume dwarf irregular population have underlying structural properties similar to both Local Volume and Virgo Cluster dwarf ellipticals. Linear regressions to structure-luminosity relationships for the Local Volume galaxies and Virgo Cluster dwarf ellipticals show significant differences in both slope and scatter around the established trend lines, suggesting that environment might regulate the structural scaling relationships of dwarf galaxies in comparison to their more isolated counterparts.

\end{abstract}


\section{Introduction} \label{intro}
In hierarchical structure formation scenarios \citep{White1978,Blumenthal1984,Bullock2001}, numerous low mass dark matter halos in the distant past were important building blocks for the high mass galaxies observed today. Dwarf galaxies in the Local Volume \mbox{(LV, D $\lesssim$ 10 Mpc)} represent the optical manifestation and closest analogue to these dark matter halos. Furthermore, dwarfs are the most common galaxy system \citep{Marzke1997,Ellis1997} and span a wide range of physical characteristics and environments \citep{Karachentsev2013}. As such dwarf galaxies through their physical properties and star formation histories represent excellent laboratories to study the initial conditions of the LV.

Galaxies evolving with little external feedback processes and in low density environments will contain significant levels of dust which can significantly attenuate and distort the optical flux \citep[e.g.,][]{Driver2007}. Furthermore, the stellar mass of most galaxies is dominated by the quiescent old stellar component whose energy output peaks at near infrared (NIR) wavelengths in contrast to the ultra-violet dominated spectrum of young massive stars. Even in Blue Compact Dwarf galaxies (BCDs), the evolutionary synthesis models by \citet{Krueger1995} demonstrate that the NIR flux contribution of young massive stars are only significant in moderately strong star bursts where the gas continuum can contribute $\sim$20-40\% of the NIR flux for at least 100 Myrs \citep{Krueger1995,Vanzi2000,Vanzi2002}. In such galaxies, multi-band NIR analysis can help to further disentangle old ($>$500 Myr) and young ($<$100Myr) stellar populations \cite[e.g.][and references therein]{Lopez-Sanchez2008}. As such NIR observations at the 1.65 $\mu$m (\textit{H}-band) will invariably trace a more accurate distribution of the underlying old stellar population and therefore the mass potential \citep{Gavazzi1996b,Noeske2003}, whilst significantly reducing attenuation due to dust.

Whilst the 2MASS sample \citep{Jarrett2000,Jarrett2003} covers the entire sky in the \textit{J H K}$_s$ NIR photometric bands and down to 14.7, 13.9 and 13.1 mag respectively, \citet[][hereafter KJRD08]{Kirby2008}, showed that due to the shallowness of the photometry, the survey missed many low surface brightness dwarf systems and underestimates the fluxes of those it did detect by as much as 70\%. This is perhaps reflected best by \citet[hereafter KMK13]{Karachentsev2013}, a volume-limited catalogue of galaxies within the local spherical volume of 10 Mpc radius or by the condition \mbox{V$_{LG}$ $\leq$ 500 km s$^{âˆ’1}$}, where \mbox{$\sim$60\%} of the 869 galaxies contained do not have a direct \textit{K}-band (2.15 $\mu$m) measurement. Quoting directly, of those that do, most are measured from 2MASS data, supplemented with photometric measurements from \citet{Fingerhut2010}, \citet{Vaduvescu2005,Vaduvescu2006}, although we note other literature sources present in the full on-line catalogue. Even including the deep \textit{H}-band photometric measurements of KJRD08, (which are not catalogued in KMK13), the availability of accurate total stellar masses fall significantly short of the availability of accurate distances in KMK13 (311 total which are measured using, the Tully-Fisher relation, tip of the red giant branch
magnitude, Cepheid luminosity, supernova, horizontal branch or RR lyrae; by surface brightness fluctuation and planetary nebula luminosity function and the remainder through membership and Hubble flow estimates).

This is in stark contrast to the availability of data for investigating galaxy properties related to other wavelengths. The neutral hydrogen gas masses of nearby southern hemispheric galaxies are well documented courtesy of the H\textsc{i} Parkes All Sky Survey \citep[HIPASS,][]{Barnes2001,Meyer2004,Koribalski2004}. Higher resolution H\textsc{i} studies are more limited in coverage but are more extensive than current infrared surveys, such as the Local Volume H\textsc{i} Survey \citep[e.g.][]{Koribalski2008,VanEymeren2009,Koribalski2009,VanEymeren2010,Lopez-Sanchez2012,Kirby2012}, and other H\textsc{i} studies \citep[e.g][]{Begum2008,Walter2008a,Hunter2012a,Ott2012}, although these may have limited utility in the ultra faint dwarf regime due to limited spatial and kinematic resolution \citep[see][]{Kirby2012}. Star formation properties are readily available for many LV galaxies in large part due to the 11 Mpc H$\alpha$ and Ultraviolet galaxy survey \citep{Kennicutt2008,Lee2009,Lee2009a}, the Spitzer Local Volume legacy survey \citep{Dale2009} and other targeted H$\alpha$ studies of dwarf galaxies such as those in the Cen A and Sculptor groups by \cite{Bouchard2009}. Clearly, deeper infrared observations and analysis are needed to complement the current wavelength coverage.

Constraining the evolutionary relationship between dwarf irregular (dIrr) galaxies and dwarf ellipticals (dE) is one such example of how more infrared data may contribute. Whether a direct evolutionary sequence between dIrr and dE galaxies exists remains an open question despite a connection being suggested in studies conducted over 30 years ago \citep{Lin1983,Kormendy1985,Binggeli1986,Davies1988}. Although gas removal mechanisms were devised allowing a morphological transformation from dIrrs to dEs \citep[e.g.]{Mayer2006}, gas stripping of present day dIrrs are unable to account for the extremely luminous dEs found in the Virgo cluster (VC) \citep{Bothun1986,James1991} and differences in the chemical abundances \citep{Grebel2003} suggested that dIrrs and dEs form a distinct parallel morphological sequence of dwarf galaxies. The rarer dwarf transition galaxies \citep[dTrans,][]{Sandage1991,Skillman1995,Mateo1998,Dellenbusch2008} may however imply an evolutionary link from dIrrs to the less luminous dEs. More recent studies \citep[e.g.,][]{Orban2008,Kazantzidis2011, Weisz2011b,Kenney2014} have increasingly illuminated on the possibility and mechanisms for dwarf irregulars evolving into dEs, but a definitive conclusion remains elusive.

Photometric studies of dIrrs \citep[e.g.][]{Binggeli1991,Patterson1996,Bremnes1998,Bremnes1999,Bremnes2000,Barazza2001,Parodi2002} demonstrated that the derived surface brightness profiles in the optical passbands were well described by an exponential curve in first approximation. However, due to the recent or ongoing star formation in dIrrs their optical structural properties are disproportionately affected by the luminous young stellar population invalidating direct comparison to the dEs whose optical properties are instead reflective of the underlying distribution. As more NIR observations of nearby dIrrs became available \citep[e.g.,][]{Noeske2003,Vaduvescu2005,Vaduvescu2006,Vaduvescu2008,Kirby2008,Fingerhut2010,DeSwardt2010,McCall2012} it became increasingly clear however that analytical approximations to the underlying stellar distribution were improved by fitting a function with a free shape parameter, such as via a S\'{e}rsic function or by a hyperbolic-secant (sech) either of which can account for the flattened structural profiles. Given the variation in the structural properties of dIrrs, does a near-infrared luminosity-structural scaling relationship exist? Furthermore, given that the near-infrared allows us to sample the underlying stellar mass distribution despite recent star formation, what fraction of the relationship might overlap with the optically derived equivalent for dEs?

This paper is dedicated to presenting our deep \textit{H}-band surface photometry and analysis of 40 LV galaxies. We probe to a surface brightness of \mbox{$\sim$25 mag arcsec$^{-2}$} ($\approx$ 1 $L_{\odot}$ pc$^{-1}$), reaching a 40 times lower stellar density than 2MASS. We perform photometry on all detected sources deriving the observed total magnitude, effective surface brightness and best fitting S\'{e}rsic parameters. Physical parameters are inferred from these quantities using the best available distances in the literature \cite[e.g.][and references therein]{Karachentsev2002, Karachentsev2004, Karachentsev2007, Karachentsev2013}. Image quality comparisons are made with respect to the 2MASS and Vista Hemispheric Survey (VHS)\footnote{\url{http://horus.roe.ac.uk/vsa/index.html}}. Finally we investigate the optical-NIR and structural scaling relationships, particularly comparing the latter to the scaling relationships for dwarf ellipticals.

The paper is organised as follows, in Sect. \ref{sample} we discuss the sample selection. The observations and data reduction methods are outlined in Sect. \ref{obs}. The photometric calibration and procedures are discussed in Sect. \ref{phot}. Further sections are dedicated to the outlining and discussion of results. 

\section{Sample Selection} \label{sample}

\begin{figure}
\centering
\includegraphics[scale=0.45]{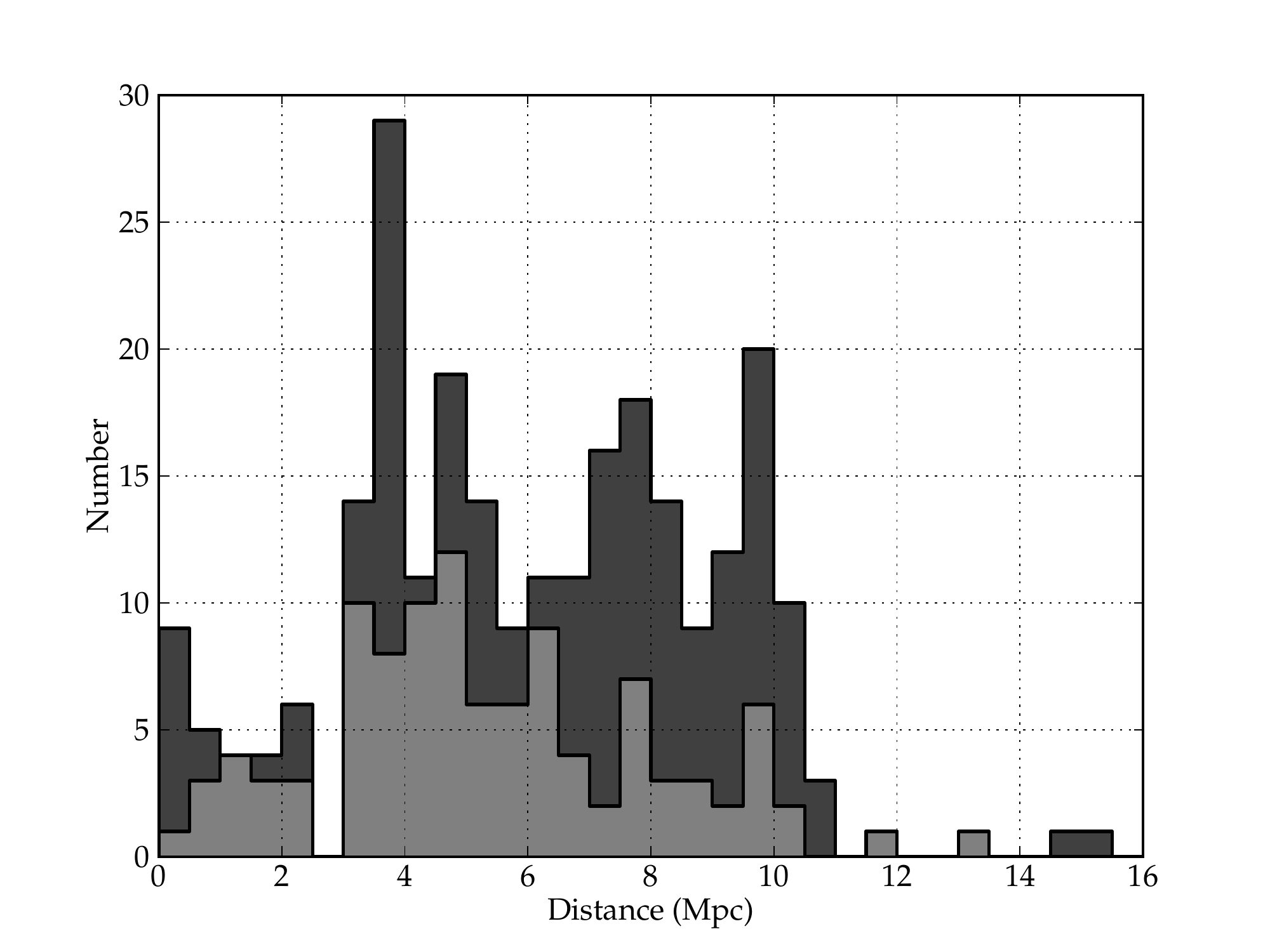}
\includegraphics[scale=0.45]{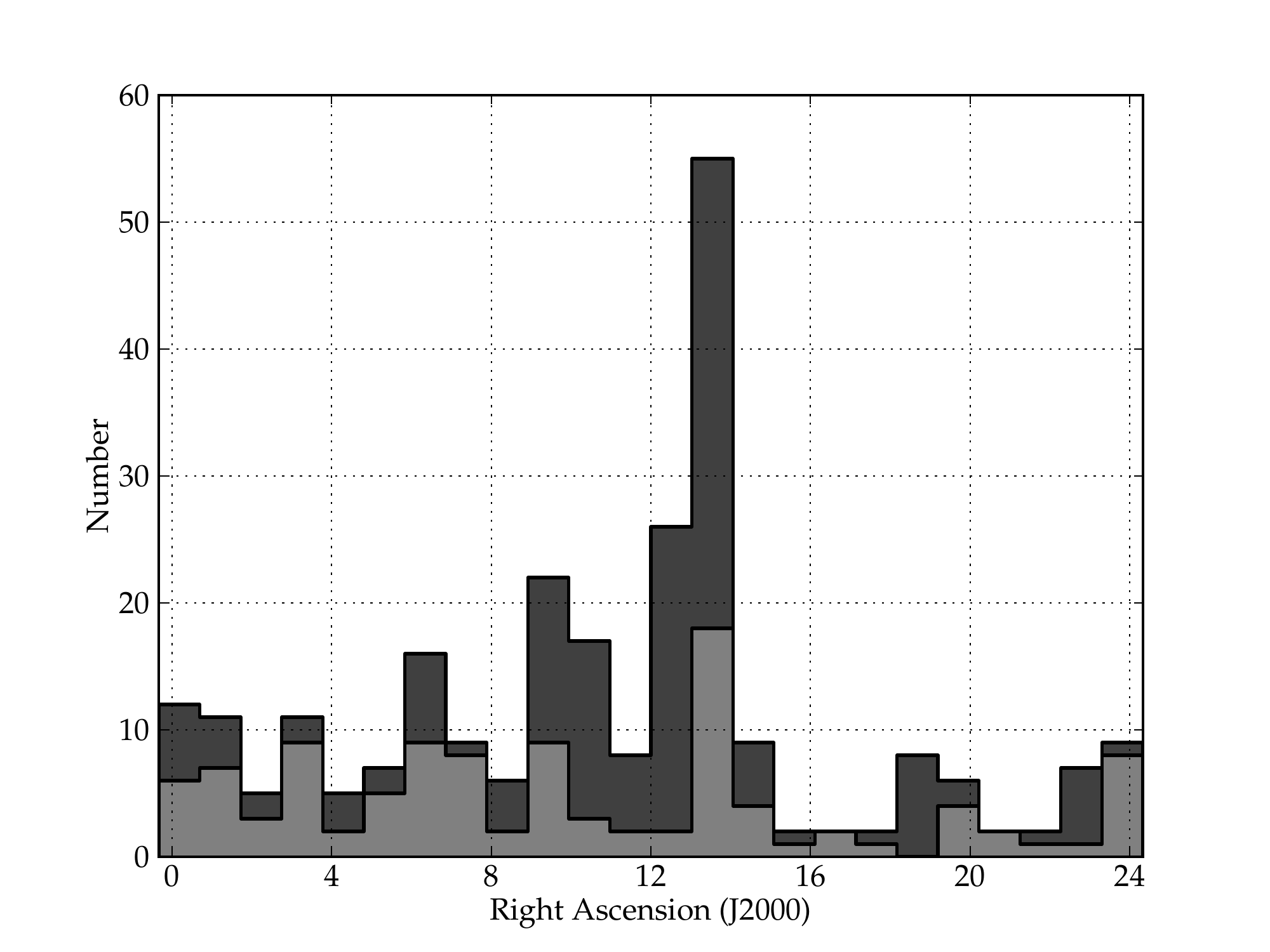}
\includegraphics[scale=0.45]{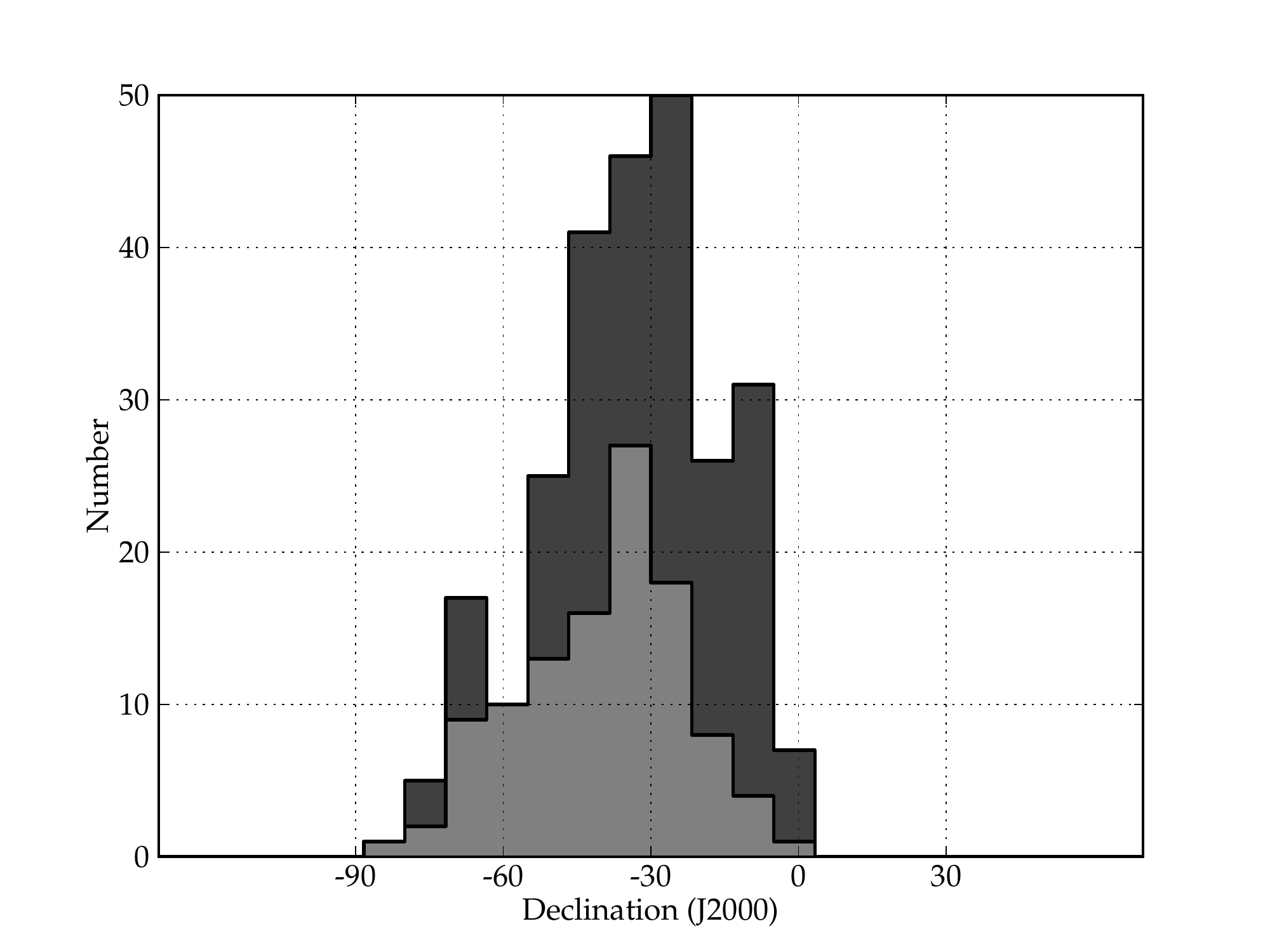}
\caption[Sample coverage histograms]{Sample coverage histograms, limited to the southern hemisphere, with respect to Right Ascension, Declination and distance for the: (\textit{dark grey}) KMK13 catalog; (\textit{light grey}) and the combined study sample and KJDR08}
\label{samplehist}
\end{figure}

KJRD08 selected 68 program galaxies of mixed morphology (Hubble types E3 through to Sc, including many irregular and dwarf galaxies) within the LV, preferentially selecting from the field and the nearby Sculptor group. Surface photometry was performed on 57 galaxies where sufficient signal to noise was obtained, whilst the remaining eleven where only marginally detected or not at all. At the time of investigation the total sample consisted of 80\% of all known field galaxies and 17\% of all known group members within the Local Volume (group members are defined to be those with positive tidal indices sourced from KMK13). In the following, we refer to these galaxies as the KJRD08 sample.

Another significant survey is the Local Volume H\textsc{i} Survey\footnote{\url{http://www.atnf.csiro.au/research/LVHIS/}} (LVHIS) \citep[][Koribalski et al. 2014 in prep]{Koribalski2008}, a deep 20cm radio-continuum and H\textsc{i} line survey of nearby gas rich galaxies performed with the Australia Telescope Compact Array (ATCA). LVHIS consists of a complete sample of galaxies selected to be within v$_{\text{LG}}$ $<$ 550 km s$^{-1}$ and DEC $<$ -30$^{\circ}$ and detected in HIPASS. The sample consists of about 80 galaxies all of which were observed with the ATCA in the 21-cm spectral line (Koribalski et al. 2014 in prep).

The selection criteria for our study's sample was designed to best complement the KJRD08 and LVHIS samples. We therefore targeted the remaining LVHIS galaxies not covered in the KJRD08 sample and those predominately located in the nearby Centaurus A group (Cen A) \citep{Rejkuba2004}. Conveniently, the Cen A group has been the subject of detailed distance studies \citep[e.g.,][]{Jerjen2000a,Karachentsev2007} and so accurate galaxy distance are available for computing physical quantities; as well as \textit{B}-band optical magnitudes \citep[e.g.][]{Lauberts1989,DeVaucouleurs1991}. We also observed an additional 7 galaxies which did not qualify for the LVHIS sample, 5 dIrrs from \cite{Doyle2005}, 1 from \cite{Wong2006} and HIPASS J1919-68 \citep{Kilborn2002,Karachentsev2004}. We outline the morphological type, distance and distance indicator, \textit{B}-band optical magnitude and error as well corresponding references for our sample in Table \ref{sampleprops}.

Since we target galaxies with LVHIS H\textsc{i} observations, the study sample is almost exclusively composed of nearby dIrrs and irregular magellanic (Im) type galaxies (for convenience we refer to both as dIrrs hereafter). The main differences between our sample and the KJRD08 sample are the mass range and the environments they trace. The galaxies within this study have an estimated median stellar mass of $\log_{10}(\mathcal{M}_{*}/\mathcal{M}_{\odot})$ = 8.1 and maximum of 9.3, whereas the KJRD08 sample contains galaxies with a median mass of 8.5 and maximum 11.1 $\log_{10}(\mathcal{M}_{*}/\mathcal{M}_{\odot})$ assuming a stellar mass to light ratio of one (see the end of Sect. \ref{phot} for further details). The differences in the median masses are due to the morphological differences between the two samples. The KJRD08 sample contains more massive spirals and lenticular galaxies, in comparison to the dIrrs predominant in the study sample. In terms of environment, the KJRD08 sample and our sample conveniently trace the main cosmic structures of the southern hemisphere out to 10 Mpc (The Sculptor and Cen A groups). The Cen A group is by far the richest group in the LV \citep[see Fig 7 in][]{Jerjen2000} and has a similar bi-modal structure to the Local Group \citep{Karachentsev2002}. Contrastingly the Sculptor group is a loose filament of galaxies \citep{Jerjen1998} expanding with the Hubble flow \citep{Karachentsev2003}. \cite{Bouchard2009} quantified the environmental differences of the Sculptor and Cen A groups using a \textit{B}-band luminosity density. Although galaxies in both environments have equally low luminosity densities, the mean for the Sculptor group is lower than that of the Cen A group \cite[see Fig 9 in][]{Bouchard2009}.

\begin{figure*}
\includegraphics[trim = 10mm 10mm 10mm 30mm, clip=true, scale=0.95]{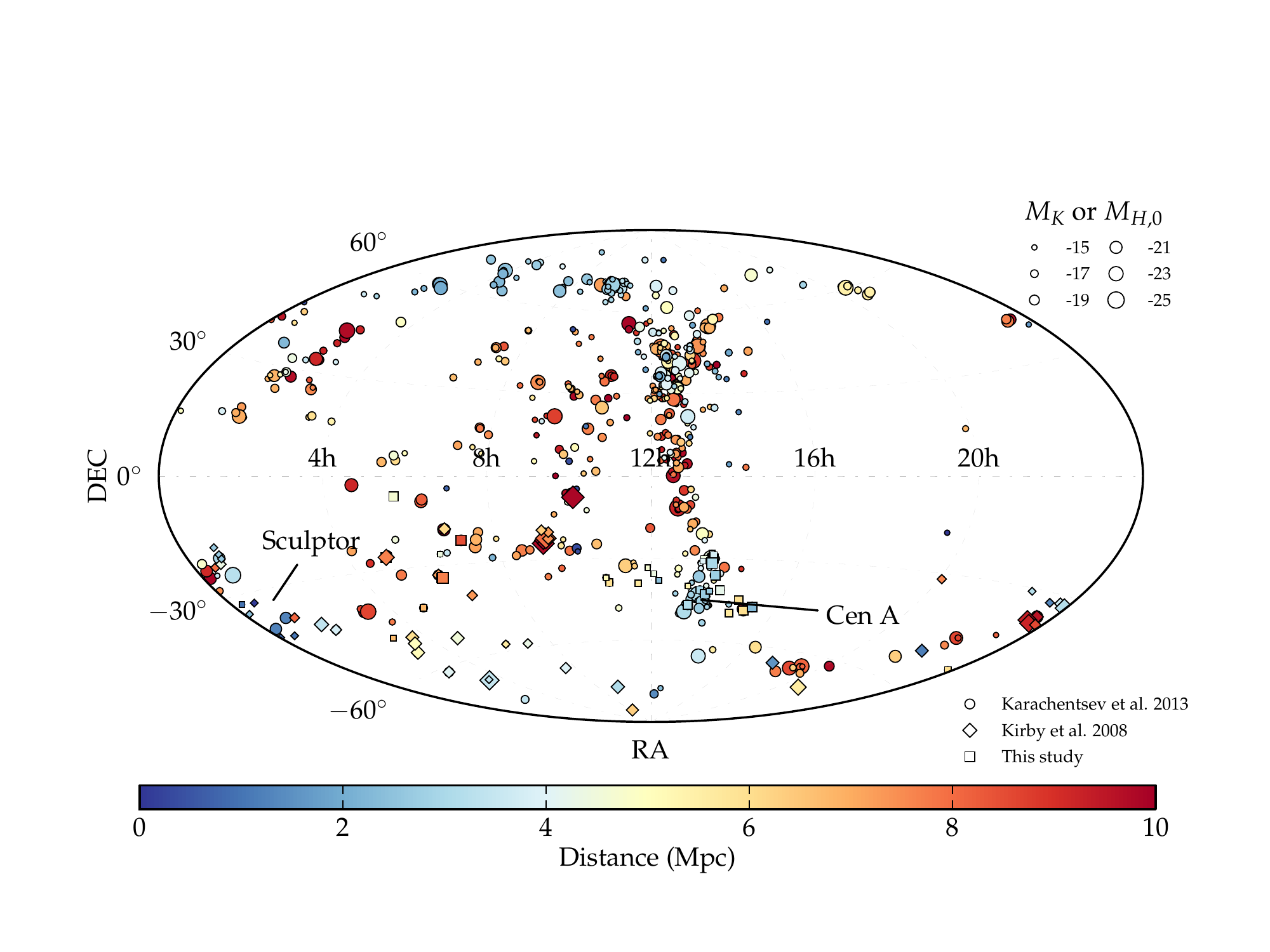}
\caption{Aitoff projection of all known galaxies within the local Volume (1 $<$ D $<$ 10 Mpc). The circles represent the galaxies listed in the updated nearby galaxy catalog \protect\citep{Karachentsev2013}. The squares represent represent the galaxies studied within this report. The diamonds represent the KJRD08 sample. The distance is indicated by the colour bar on the bottom of the image. Similarly, the near infrared luminosity, (\textit{K}-band luminosities for the updated nearby galaxy catalogue and extinction correct \textit{H}-band luminosities for the KJRD08 and study samples), is indicated by the marker size in the upper right hand corner of the image.}
\label{atlas}
\end{figure*}

The KMK13 updated nearby galaxy catalog contains 259 galaxies in the southern hemisphere ($\delta$ $<$ 0). Combining our sample with the KJRD08 sample, we cover $\sim$ 42\% (106/259) of their cataloged galaxies in the southern hemisphere. Figure \ref{samplehist} presents histograms of the study sample and the KMK13 catalog in RA, DEC and distance, restricted to the southern hemisphere. Figure \ref{atlas} illustrates the spatial and luminosity distribution of the KJRD08 sample, study sample and the KMK13 catalog within the volume 1 $<$ D $<$ 10 Mpc.

\section{Observation and Reduction} \label{obs}
Near Infrared \textit{H}-band images were obtained for 40 galaxies on two observing runs using the Infrared Imager and Spectrograph 2 \citep[IRIS2]{2004SPIE.5492..998T} on the 3.9m Anglo-Australian Telescope (AAT) located at Siding Spring Observatory. The IRIS2 detector is a 1024x1024 Rockwell HgCdTe Astronomical Wide Area Infrared Imager-1 (HAWAII) array with a pixel scale of 0.4486$''$ pixel$^{-1}$, resulting in an instantaneous field of view (FOV) of 7.7$'$ $\times$ 7.7$'$ 

The first observing run took place on the 6th through 10th of June 2011. The second took place on the 27th-28th October 2012. A one off observation was conducted during the 4th of October 2012 during Director's service time for ESO 252-IG001. Weather conditions varied over the course of each observation and thus much care was taken in screening individual object frames to ensure they had sufficient photometric depth before reduction. The seeing ranged from 0.8$''$ to 3.1$''$ with a mean seeing of 1.7$''$.

We employed two observing strategies depending on the angular size of the target galaxy source and the FOV of IRIS2.

\begin{itemize}
\item \textbf{Jitter Self Flat} (JSF). Most target galaxies had an optical diameter less than 4 arcmin and so these objects are expected to fill less than 20\% of the detector array. These targets were observed in a 3 $\times$ 3 grid pattern with a spacing of 90$''$ resulting in a 4.7$'$ $\times$ 4.7$'$ arcmin region common to all pointings which includes the target galaxy and a significant portion of the sky. A maximum of 60 sec was spent on any pointing, consisting of 10 $\times$ 6 sec integrations (in order to keep the sky + object counts within the linear regime) which were averaged before being written to a file. This nine point jitter pattern was repeated four times leading to a total integration time of 36 mins on source.

\item \textbf{Chop Sky Jitter} (CSJ) Using the recommendations of \cite{Vaduvescu2004} objects filling $\gtrsim 40$\% of the array FOV require a matching observation of blank sky to measure changes in the background level and illumination pattern (since a significant fraction of the sky cannot be sampled in the source frame). Five jittered observations (10$''$ offsets) of the target galaxy were bracketed and interleaved with six jittered observations of the sky, 10$'$ north or south. At each object or sky jitter position, 3 $\times$ 10 sec or 6 $\times$ 5 sec integrations were averaged. This pattern was repeated between five and 12 times, for a total on-source exposure time of up to half an hour per galaxy.

\end{itemize}
We follow the reduction techniques outlined in Sect. 3 of KJRD08 with some modification. We perform the data reduction using the \textsc{orac-dr}, \textsc{pyfits}\footnote{\url{http://www.stsci.edu/institute/software_hardware/pyfits}}, the \textsc{iraf} and the \textsc{imsurfit} and custom scripts utilizing the \textsc{ccdpack} routines, \textsc{pairndf}, \textsc{register}, \textsc{tranndf} and \textsc{makemos} as part of the Starlink Software Collection\footnote{\url{http://starlink.jach.hawaii.edu/starlink}}. Observations made with the JSF method employ a slightly modified \textsc{jitter\_self\_flat} recipe, while those observed with CSJ use a similarly modified \textsc{chop\_sky\_jitter} recipe. Table \ref{sources} outlines the observation log of the sample which is organised in to the following columns: Column (1) - Galaxy name; Column (2) - HIPASS designation; Columns (3) and (4) are the source's respective equatorial coordinates, Right Ascension and Declination, in the J2000 epoch; Column (5) - Observation Date; Column (6) - Observation strategy; Column (7) - Total exposure time (mins); Column (8) - Mean seeing during the observation. For coordinate references, please refer to Table \ref{sampleprops}.

\begin{table*}
\centering
\begin{tabular}{llcccccc}
\hline\hline
Galaxy name & HIPASS & R.A. & Dec & Obs date & Obs mode & t$_{tot}$ & Seeing \\
 & & (J2000) & (J2000) & (YYYY-MM-DD) & & (min) & (arcsec) \\
(1) & (2) & (3) & (4) & (5) & (6) & (7) & (8) \\
\hline
ESO 410-G005 & J0015--32 & 00:15:32 & -32:10:47 & 2011-06-08 & JSF & 43 & 3.1 \\
ESO 199-G007 & J0258--49 & 02:58:04 & -49:22:56 & 2011-06-10 & JSF & 57 & 1.9 \\
ESO 252-IG001 & J0457--42 & 04:56:59 & -42:48:14 & 2012-10-04 & JSF & 9 & 1.8 \\
KK 49 & J0541+06 & 05:41:42 & 06:40:51 & 2012-10-28 & JSF & 22 & 2.1 \\
AM 0605-341 & J0607--34 & 06:07:20 & -34:12:14 & 2012-10-27 & JSF & 18 & 1.9 \\
NGC 2188 & J0610--34 & 06:10:09 & -34:06:22 & 2012-10-27 & CSJ & 15 & 1.5 \\
ESO 489-G?056 & J0626--26 & 06:26:17 & -26:15:55 & 2012-10-28 & JSF & 36 & 2.0 \\
AM 0704-582 & J0705--58 & 07:05:18 & -58:31:13 & 2012-10-27 & JSF & 35 & 1.6 \\
ESO 558-G011 & J0706--22 & 07:06:56 & -22:02:26 & 2012-10-28 & JSF & 18 & 2.0 \\
ESO 376-G016 & J1043--37 & 10:43:27 & -37:02:37 & 2012-10-27 & JSF & 29 & 2.6 \\
ESO 318-G013 & J1047--38 & 10:47:42 & -38:51:13 & 2012-10-28 & JSF & 12 & 1.6 \\
ESO 320-G014 & J1137--39 & 11:37:53 & -39:13:13 & 2011-06-06 & JSF & 36 & 0.9 \\
ESO 379-G007 & J1154--33 & 11:54:43 & -33:33:36 & 2011-06-08 & JSF & 44 & 1.5 \\
ESO 379-G024 & J1204--35 & 12:04:57 & -35:44:30 & 2011-06-06 & JSF & 34 & 1.0 \\
ESO 321-G014 & J1214--38 & 12:13:50 & -38:13:51 & 2011-06-08 & JSF & 27 & 1.0 \\
CEN 6 & J1305--40 & 13:05:02 & -40:05:02 & 2011-06-06 & JSF & 31 & 0.9 \\
ESO 269-G058 & J1310--46A & 13:10:33 & -46:59:29 & 2011-06-07 & CSJ & 22 & 1.3 \\
KK 195 & J1321--31 & 13:21:08 & -31:31:45 & 2011-06-07 & JSF & 39 & 1.1 \\
AM 1321-304 & J1324--30 & 13:24:36 & -30:58:19 & 2011-06-06 & JSF & 36 & 0.8 \\
IC 4247 & J1326--30A & 13:26:44 & -30:21:45 & 2011-06-07 & JSF & 18 & 1.2 \\
ESO 324-G024 & J1327--41 & 13:27:37 & -41:29:00 & 2011-06-08 & JSF & 54 & 2.0 \\
UGCA 365 & J1336--29 & 13:36:31 & -29:14:05 & 2011-06-06 & JSF & 43 & 2.1 \\
ESO 444-G084 & J1337--28 & 13:37:20 & -28:02:43 & 2011-06-06 & JSF & 22 & 1.4 \\
LEDA 592761 & J1337--39 & 13:37:25 & -39:53:48 & 2011-06-08 & JSF & 45 & 2.1 \\
NGC 5237 & J1337--42 & 13:37:39 & -42:50:52 & 2011-06-10 & JSF & 22 & 1.4 \\
NGC 5253 & J1339--31A & 13:39:55 & -31:38:24 & 2011-06-10 & JSF & 9 & 1.2 \\
IC 4316 & J1340--28 & 13:40:19 & -28:53:29 & 2011-06-08 & JSF & 27 & 2.5 \\
NGC 5264 & J1341--29 & 13:41:36 & -29:54:47 & 2011-06-10 & JSF & 15 & 1.7 \\
ESO 325-G011 & J1345--41 & 13:45:00 & -41:51:40 & 2011-06-10 & JSF & 30 & 1.8 \\
- & J1348-37 & 13:48:33 & -37:58:03 & 2011-06-10 & JSF & 44 & 1.4 \\
ESO 383-G087 & J1349--36 & 13:49:18 & -36:03:41 & 2011-06-10 & CSJ & 30 & 1.5 \\
LEDA 3097113 & J1351--47 & 13:51:22 & -47:00:00 & 2011-06-10 & JSF & 22 & 1.7 \\
NGC 5408 & J1403--41 & 14:03:20 & -41:22:39 & 2011-06-08 & JSF & 35 & 3.0 \\
UKS 1424-460 & J1428--46 & 14:28:03 & -46:18:06 & 2011-06-10 & JSF & 30 & 1.5 \\
ESO 222-G010 & J1434--49 & 14:35:02 & -49:25:14 & 2011-06-06 & JSF & 21 & 1.4 \\
ESO 272-G025 & J1443--44 & 14:43:25 & -44:42:18 & 2011-06-06 & JSF & 44 & 1.9 \\
ESO 223-G009 & J1501--48 & 15:01:09 & -48:17:31 & 2011-06-08 & JSF & 38 & 3.1 \\
ESO 274-G001* & J1514--46 & 15:14:14 & -46:48:17 & 2011-06-10 & CSJ* & 35* & 2.4 \\
- & J1919--68 & 19:19:58 & -68:39:13 & 2012-10-28 & JSF & 27 & 1.5 \\
ESO 149-G003 & J2352--52 & 23:52:02 & -52:34:40 & 2011-06-10 & JSF & 32 & 1.2 \\
\hline
\end{tabular}
\caption{Sample observation log. *Due to its large angular size, this galaxy required several pointings and the total exposure time refers to a 7' $\times$ 7' region centred on the galaxy. Note that this does not cover the full angular extent of the galaxy but sufficiently sampled the surface brightness profile.}
\label{sources}
\end{table*}

We first carefully examine each frame and remove those impaired by atmospheric conditions such as the presence of cirrus clouds. For JSF observations the preprocessing of all raw frames included the subtraction
of a matching dark frame, linearity and interquadrant crosstalk, correction, and bad pixel masking is handled by the \textsc{jitter\_self\_flat} recipe. Considerable care is taken to ensure accurate flat-fielding over the entire field of the array. An interim flat field is created by taking the median at each pixel of the nine normalized object frames, then divide each of the nine images by this interim flat field. Extended sources within these flat-fielded object frames are automatically detected and masked, and an improved flat field is created from the masked versions of the nine normalized object frames. We apply a correction for astrometric distortion internal to IRIS2 by resampling the proper flat-fielded image. For CSJ observations, the six sky frames are first offset in intensity to a common modal value, then a flat field is formed from the median value at each pixel. All six sky frames and five object frames are flat-fielded, then the modal pixel values of the two sky frames bracketing each object frame are averaged and subtracted from that object frame. Ordinarily in both cases, spatial additive offsets would be computed for the resulting flat-fielded/sky subtracted object frames and the mosaic formed within the recipe.

We find that several improvements to the background subtraction and corrections to bright star channel effects can be made by further processing before mosaicing. Thus our reduction procedure diverges from KJRD08 in the following way. Firstly, by passing the flat-fielded object frames through \textsc{iraf} task \textsc{imsurfit} we were able to obtain a much improved background linearity. Custom scripts utilizing \textsc{pyfits} repair the bad pixel mask and re-scale dark rows in the array (as caused by saturated stars), by measuring the sigma-clipped median of each row and correcting their offset to a nominal value. The files are then passed through custom scripts utlizing \textsc{ccdpack}, where each object frame is registered. The coordinate and sky offsets are then computed before the frames are finally mosaiced.

\begin{figure}
\includegraphics[scale=0.6]{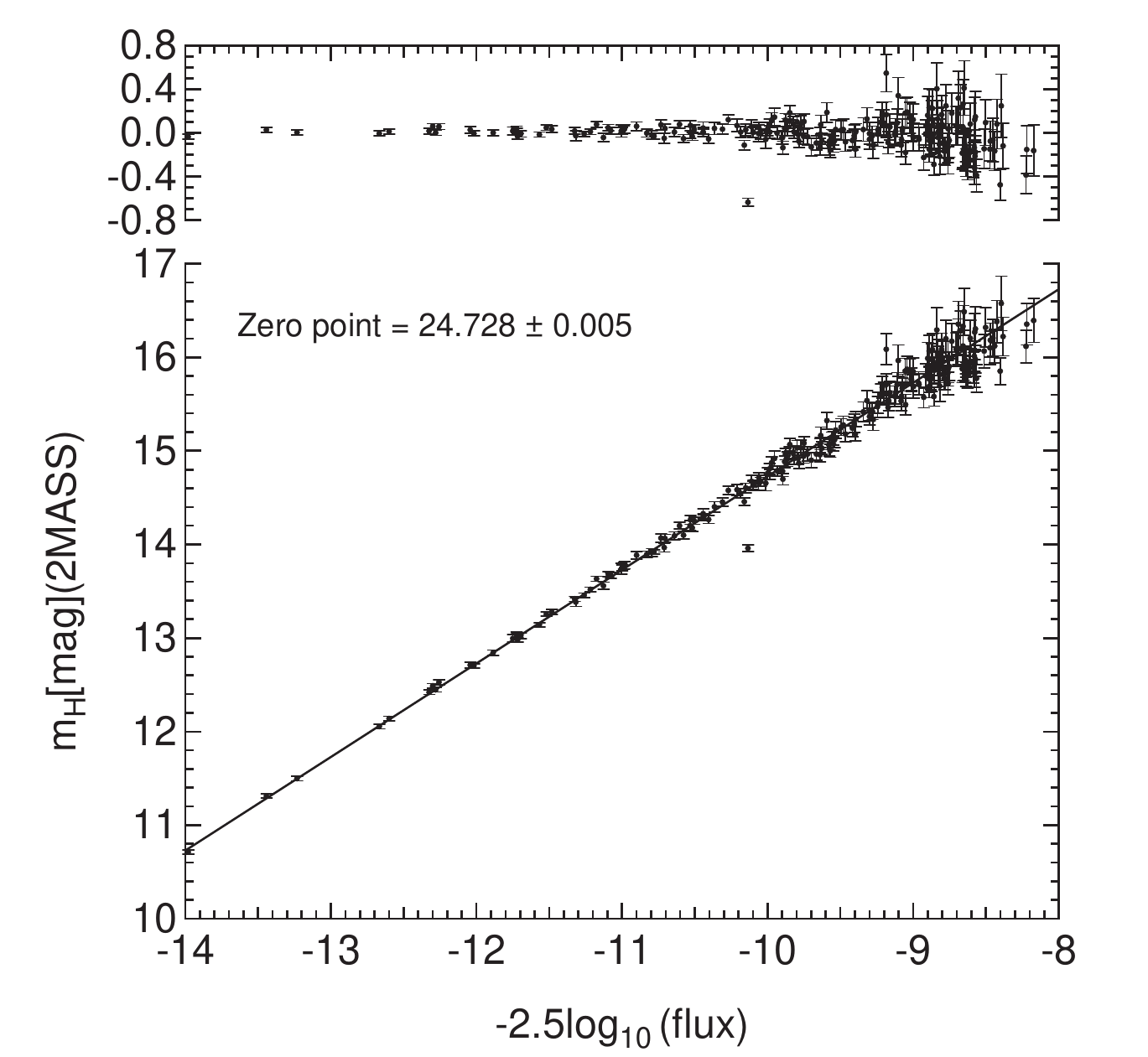} \includegraphics[scale=0.6]{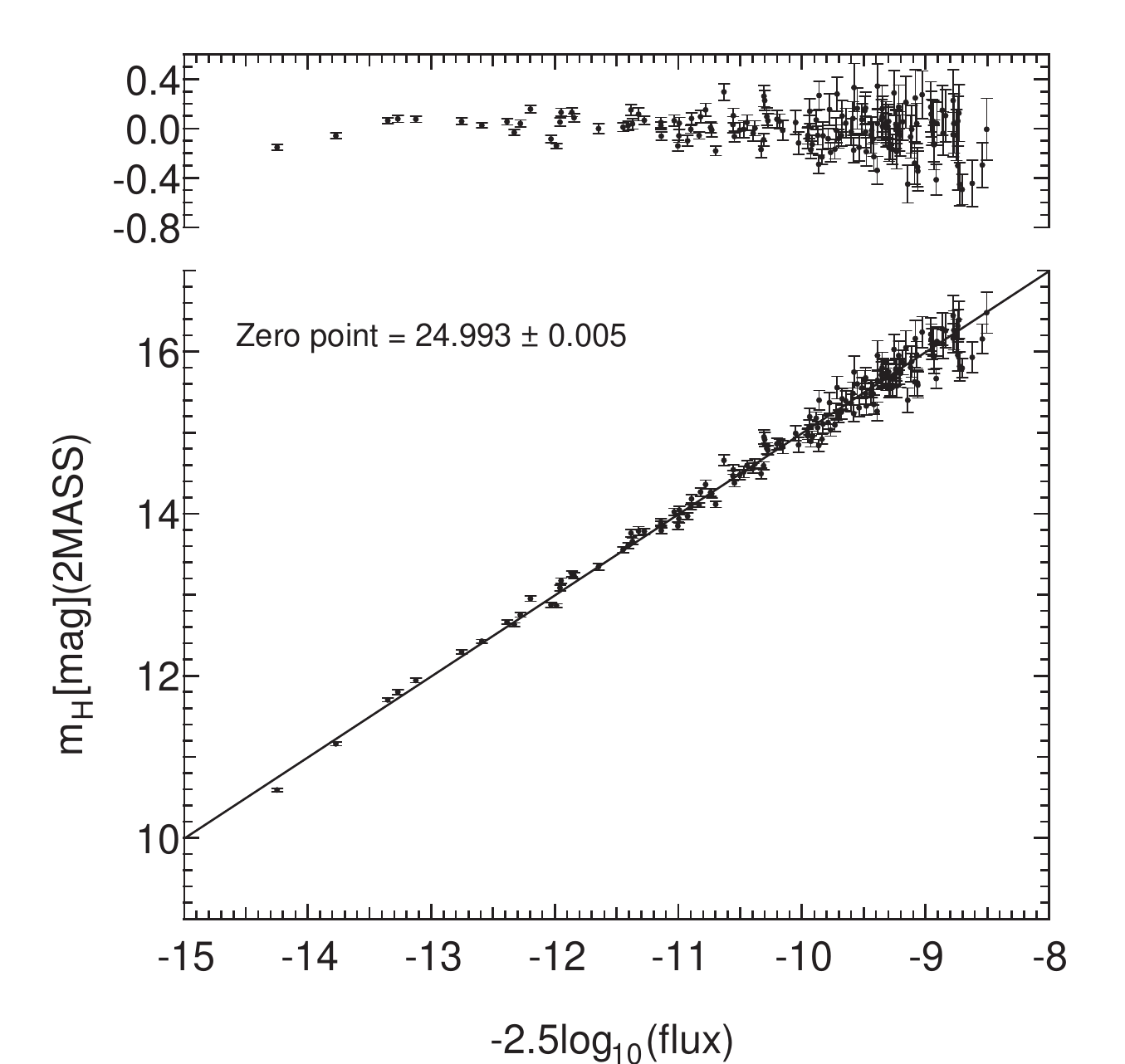}
\caption{Correlation between instrumental and 2MASS \textit{H}-band magnitudes for field stars around the galaxies ESO 269-G058 (\textit{top}) and NGC 5237 (\textit{bottom}). The spherical points indicate the correlated instrumental magnitude versus the 2MASS \textit{H}-band magnitudes, where error bars correspond to the uncertainty in the 2MASS \textit{H}-band magnitude for a given star. The sub-plot is the residual to a linear fit of the data. The y-intercept (`zero-point') is indicated on the plot itself with the corresponding 1$\sigma$ uncertainty. Total exposure times were equal for both sources and the difference in zero points are due to the presence of atmospheric cirrus clouds during observations of ESO
269-G058. }
\label{cal}
\end{figure}

The instrumental magnitudes of field stars were measured using the \textsc{iraf} task \textsc{phot}. Photometric calibration of each field was performed through cross-correlation of the stellar positions with those in the 2MASS catalog (Fig. \ref{cal}) and adopting a linear fit to catalog and instrumental magnitudes to determine the instrumental zero point for a given image. Typically, greater than 95\% of stars were matched in the cross correlation. The 1$\sigma$ uncertainty in the zero point was no greater than 0.01 mag for all our sources.

\begin{figure}
\includegraphics[scale=0.44]{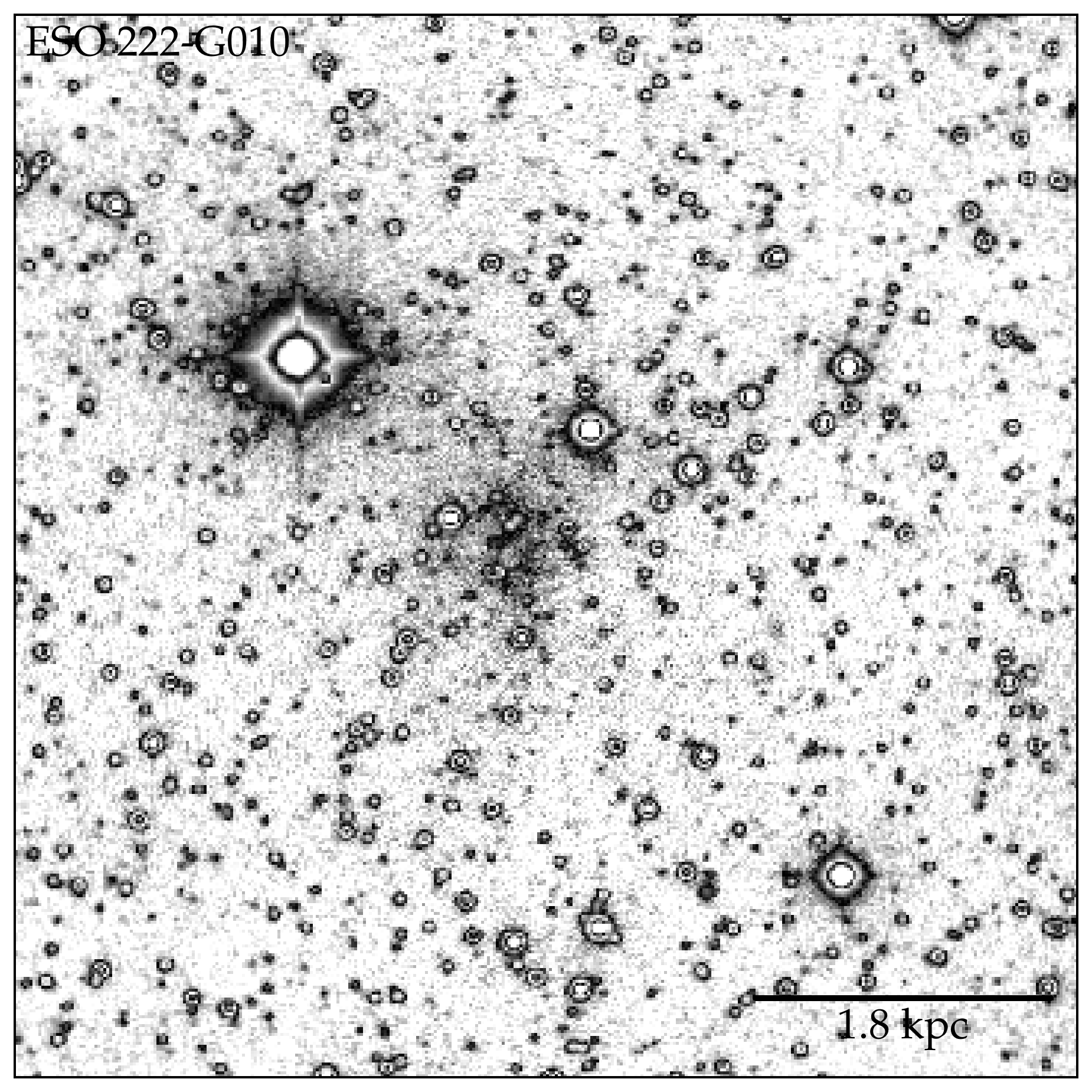} \includegraphics[scale=0.44]{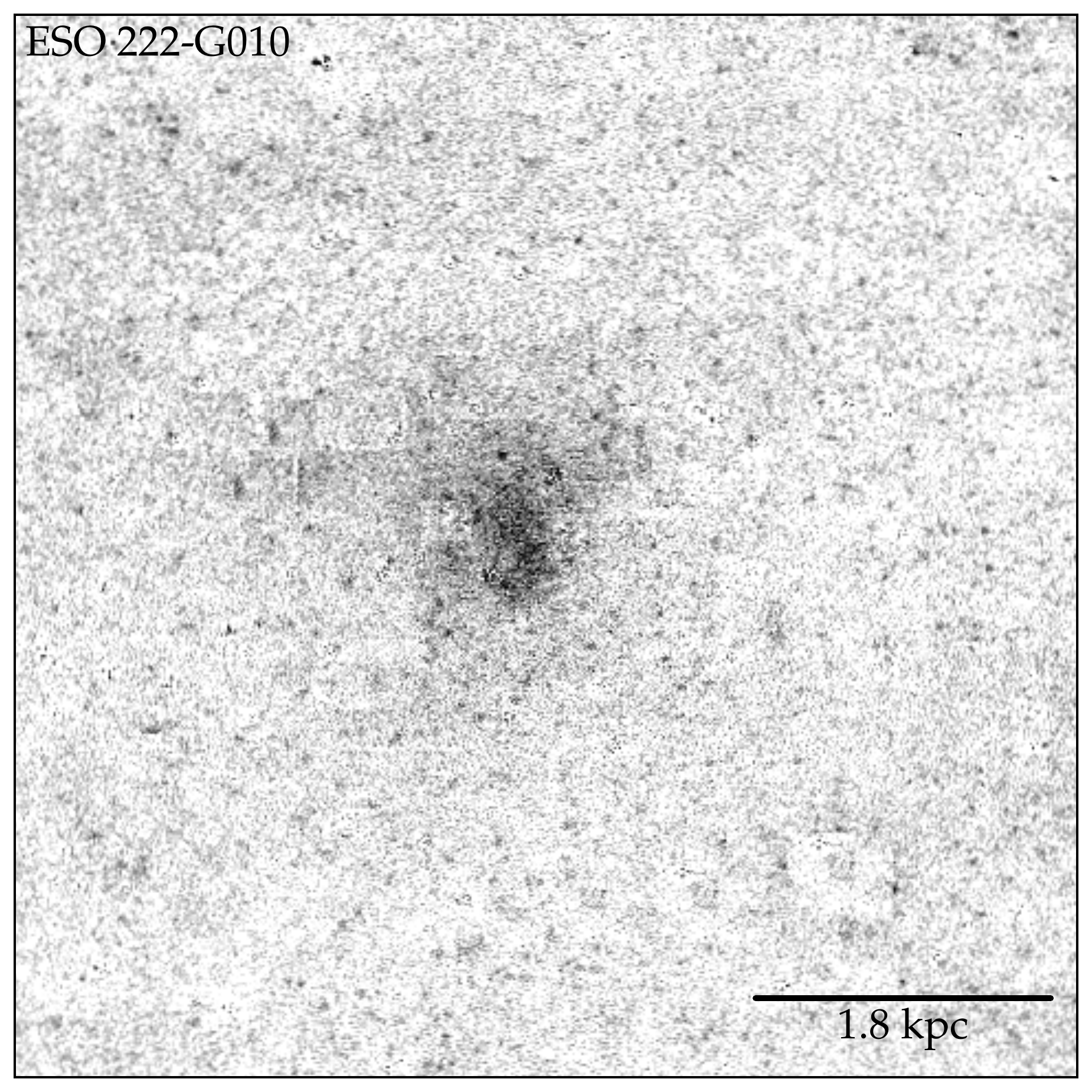}
\caption{Demonstrating the effectiveness of the foreground star subtraction procedures. The top image prior to cleaning and the bottom image is the result after cleaning for the galaxy ESO 222-G010.}
\label{sub} 
\end{figure}

In order to achieve accurate surface photometry extremely careful cleaning was conducted for each image in which a galaxy was detected. Foreground stars were removed using the custom \textsc{iraf} script \textsc{killall} \citep{KILLALL}. This proved more effective than standard \textsc{daophot} tasks in removing the medium and low intensity stars. Both tasks failed in removing very bright or saturated stars and background galaxies however. These unwanted sources were masked from the automated procedures and removed manually by selecting a nearby patch of sky and replacing the star with this patch. For a star superposed on the galaxy, its light was replaced with a patch within the same isophote if the size of the star was small relative to the galaxy. In the case of a large relative size, the contaminated area was replaced with its mirror image with respect to the galactic center. Care was taken to ensure no significant galaxy structures were removed or artifacts introduced within the measured galaxy isophotes. Figure \ref{sub} demonstrates the effectiveness of the removal process on a small, low surface brightness galaxy.


\section{Photometry} \label{phot}
An intensity growth curve as a function of the azimuthally averaged radius were produced for the cleaned \textit{H}-band images using synthetic aperture tasks within \textsc{iraf} and accounting for the ellipticity. The ellipticity and position angle of the galaxy is measured by generating isophotes down to within ten percent of the maximum optical radius and averaging the generated ellipticities and position angles, or in the case of irregular galaxies by selecting the isophote which robustly describes the disk component of the galaxy. In cases of either extremely low surface brightness or small size, the ellipticity and the position angle were estimated by eye.  The asymptotic intensity of the growth curve corresponded to the total apparent magnitude, $m_H$ that could be measured down the sky level of the image. The largest source of uncertainty is the variation of the sky level. When producing the growth curve, by systematically varying the assumed sky level we were able to determine the best convergence beyond the visually identifiable radii of the galaxy. Similarly, by varying the assumed sky level above and below that which produced the best growth curve; and by noting the ranges of values which produced no noticeable difference in the growth curve, we were able to estimate the uncertainties in the derived photometric parameters

The half light geometric radius $r_{\text{eff}}$ was measured at half the total intensity and the mean effective surface brightness $\langle \mu_{H} \rangle_{\text{eff}}$ was calculated within this radius. The uncertainties in the total apparent magnitude, $m_H$ varied mostly between 0.03 and 0.15 mag excluding two extreme cases. In one case, the galaxy ESO223-G009 is hidden behind a significant amount of foreground stars. Unfortunately, this galaxy was observed in poor seeing (3.1 arcsec) conditions, which made replacing stars with a blank patch of sky or galaxy within a given isophote extremely challenging. Consequently the variation in the sky is significantly higher than what is typical for the rest of the sample and so the uncertainty is correspondingly higher at 0.4 mag. In the other case, the galaxy ESO222-G010 also contains a significant number of foreground stars as well as a very bright star near to the galaxy in addition to having very low surface brightness. The uncertainty in the magnitude for this galaxy is 0.23 mag.

We produced the surface brightness profile for each galaxy by differentiating the growth curve with respect to radius. Depending on exposure times and atmospheric conditions the image photometric zero point ranged between 23.6 $<$ $\mu_{\text{lim}}$ $<$ 25.1 mag arcsec$^{-2}$. For a given galaxy, the surface brightness profile could be recovered down to the surface brightness limit (note this is not the same as the photometric zero point but is derived from it and the sky variance, see below for further details), except in the case of the edge on spiral ESO274-G001 whose angular extent was larger than the \textsc{IRIS2} field of view. The surface brightness profiles are presented in Fig. \ref{SBresults1}. The error bars are the computed rms scatter of the intensity within each isophote bin. Applying the \textsc{iraf} \textsc{ellipse} task we computed the ellipticity and position angle, which we hold fixed when generating the surface brightness profiles. The sample of KJRD08 however, had a wider morphological range and finds that it was necessary to leave ellipticity and position angle unfixed for each isophote, especially for the inner isophotes. Since our sample consists of mostly dwarf irregulars, holding the geometric parameters fixed averages out irregularities (which are less significant in the NIR relative to the optical) and better samples the underlying structural distribution.

Table \ref{mpars} lists the measured properties of our sources obtained using the photometric procedures described above. The table is arranged in the following manner: Column (1) Galaxy name; Column (2) - HIPASS designation; Column (3) - The total apparent \textit{H}-band magnitude $m_H$; Column (4) - The effective radius $r_{\text{eff}}$; Column (5) mean effective surface brightness $\langle \mu_{H} \rangle_{\text{eff}}$, Column (6) - mean ellipticity; Column (7) position angle.

Through extrapolating the generated surface brightness profiles to infinity, we are able to recover the flux below the background noise level. We achieve this analytically through the use of a S\'{e}rsic function \citep{Sersic1963},

\begin{align}
\mu(r) = \mu_0 + 1.086(r/\alpha)^n, \label{sersic}
\end{align}
where $\mu_0$ is the central surface brightness, $r$ is the geometric mean radius, $\alpha$ is a scale length parameter and $n$ is the shape parameter (S\'{e}rsic index). S\'{e}rsic profiles have historically successfully fit the radial profiles of dwarf ellipticals galaxies \citep[e.g.][]{Binggeli1998}. Similarly the photometric study of KJRD08 demonstrated that for the dwarf irregulars in their sample, S\'{e}rsic profiles are equally as successful. The studies of \cite{Vaduvescu2005} and \cite{Fingerhut2010} find that the surface brightness profiles of dwarf irregulars with flattened inner profiles follow a hyperbolic-secant (sech) function, $\mu = \mu_0 - 2.5 \log (\text{sech}(r/r_0))$, and where necessary adding a Gaussian component to describe a nucleated inner region. Although either a single component S\'{e}rsic or sech function is adequate for the majority of our sample, we prefer the S\'{e}rsic function as it is successfully used for early type dwarf galaxies. In some cases where a single component fit is inaccurate for sources with a nucleated or truncated inner component we find that fitting a double S\'{e}rsic function adequately describes the overall profile. We extend the procedures for performing photometry with a single S\'{e}rsic function (KJRD08) by defining a double S\'{e}rsic function:

\begin{align}
I(r_{i},r_{o}) = I_{0,i} \exp\left(-(r_i/\alpha_i)^{n_i}\right) + I_{0,o} \exp\left(-(r_o/\alpha_o)^{n_o}\right)
\end{align}

is the double surface brightness profile in intensity units with $i$ and $o$ representing the inner and outer profile parameters as defined in Equation \ref{sersic}. By extrapolating the 
function to infinity we are able to compute the missing flux ($\Delta m$) for a given set of S\'{e}rsic parameters $I_{0,i}$, $\alpha_i$, $n_i$, $I_{0,o}$, $\alpha_o$, $n_o$.

\begin{align}
\Delta m &= -2.5 \log_{10}(1 - I_{\text{missing}}/I_{\text{tot}}) \nonumber\\
&= -2.5 \log_{10}\left(1 - \frac{\sum\limits_{k = i,o} w_k.\Gamma[2/n_k, (r_{\text{max},k}/\alpha_k)]}{\sum\limits_{k = i,o} w_k.\Gamma[2/n_k]}\right)
\end{align}
where
\begin{align}
I_{\text{missing}} &= 2\pi\int^{\infty}_{\text{rmax}} \sum\limits_{k = i,o} I_{0,k} \exp\left(-(r_k/\alpha_k)^{n_k}\right) rdr \nonumber\\
&= 2\pi\left(\sum\limits_{k = i,o} w_k.\Gamma[2/n_k, (r_{\text{max},k}/\alpha_k)]\right) \\\nonumber\\
I_{\text{total}} &= 2\pi\int^{\infty}_{0} \sum\limits_{k = i,o} I_{0,k} \exp\left(-(r_k/\alpha_k)^{n_k}\right) rdr \nonumber\\
&= 2\pi \left( \sum\limits_{k = i,o} w_k.\Gamma[2/n_k] \right)
\end{align}
and
\begin{align}
w_k = I_{0,k} \alpha^{2}_k / n_k. \nonumber
\end{align}
$\Gamma(a,x)$ is the upper incomplete gamma function and $r_{\text{max}}$ is the radius where the growth curve reaches the asymptotic intensity (to within a few percent). We subtract this quantity from the observed magnitudes when deriving the absolute magnitudes,
\begin{align}
M_{H,0} = m_H + \Delta m - 5\log_{10} D_{\text{Mpc}} - 25 - A_H, 
\end{align}
where $A_H$ is the Galactic extinction correction factor of \cite{Schlafly2011}, which is a re-calibration of the \cite{Schlegel1998} extinction map; and luminosity
\begin{align}
L = 10^{0.4(M_{H,\odot} - M_{H,0})}, \nonumber
\end{align}
where $M_{H,\odot}$ = 3.35 is the \textit{H}-band luminosity of the sun \citep{Colina1996}.

The effective radius, $r_{\text{eff}}$, is systematically underestimated unless a correction is applied to account for the missing flux. The corrected effective radius, $r_{\text{eff,0}}$ can be numerically solved for multiple profiles using the implicit equation:
\begin{align}
\frac{I_{tot}}{2} = 2\pi \sum\limits_{k = i,o} w_{k}\Gamma \left[ \frac{2}{n_{k}},\frac{r_{\text{eff,0}}}{\alpha_{k}} \right].
\end{align}
Employing the colour dependent stellar mass-to-light ratio relation from \cite{Bell2003} for the B-H colours of our galaxy sample (Tables \ref{sampleprops} and \ref{dpars}) we find the stellar mass-to-light ratios to fall in the range, 0.4 $<$ $\Upsilon^{H,0}_{*}$ $<$ 1.3. This is consistent with a unison mass-to-light ratio ($\Upsilon^{H,0}_{*}$ = 1.0) yielded with the \cite{DeJong1996} model, when considering a typical 12 Gyr, solar metallicity stellar population with constant star formation rate and Salpeter initial mass function. Our empirically derived mass-to-light ratios are within the ranges of those determined by KJRD08, $\Upsilon^{H,0}_{*}$ = 0.9 $\pm$ 0.6, and the study of \citet{Bell2003}, 0.7 $<$ $\Upsilon^{H,0}_{*}$ $<$ 1.3. We adopt an error-weighted mean, $\Upsilon^{H,0}_{*}$ = 1.0 $\pm$ 0.2, of these mass-to-light ratios when estimating the stellar masses of the galaxies in our sample.

\begin{table*}
\centering
\begin{tabular}{llccccc}
\hline\hline
Galaxy Name & HIPASS &  $m_{H,\text{obs}}$  & r$_{\text{eff}}$  & $\left\langle \mu_{H} \right\rangle _{\text{eff}}$ & e & PA \\
&  & (mag)     & (arcsec)   & (mag arcsec$^{-2}$) &    &   ($^{\circ}$)   \\
(1)  & (2)      & (3)    & (4)     & (5)   & (6) & (7)     \\
\hline
ESO	410-G005	&	J0015--32	&	12.33	$\pm$	0.07	&	24.6	$\pm$	1.0	&	21.28	$\pm$	0.02	&	0.32	$\pm$	0.02	&	49	$\pm$	3\\
ESO	199-G007	&	J0258--49	&	14.27	$\pm$	0.15	&	20.0	$\pm$	3.0	&	22.77	$\pm$	0.13	&	0.63	$\pm$	0.01	&	14	$\pm$	1\\
ESO	252-IG001	&	J0457--42	&	13.18	$\pm$	0.04	&	15.2	$\pm$	0.5	&	21.09	$\pm$	0.03	&	0.45	$\pm$	0.06	&	65	$\pm$	5\\
KK	49	&	J0541+06	&	10.54	$\pm$	0.04	&	17.2	$\pm$	0.6	&	18.71	$\pm$	0.04	&	0.24	$\pm$	0.03	&	-7	$\pm$	2\\
AM	0605-341	&	J0607--34	&	11.98	$\pm$	0.07	&	11.7	$\pm$	0.7	&	19.32	$\pm$	0.05	&	0.37	$\pm$	0.04	&	86	$\pm$	1\\
NGC	2188	&	J0610--34	&	9.62	$\pm$	0.05	&	34.0	$\pm$	1.3	&	19.27	$\pm$	0.03	&	0.76	$\pm$	0.03	&	-3	$\pm$	1\\
ESO	489-G?056	&	J0626--26	&	13.69	$\pm$	0.09	&	13.0	$\pm$	0.7	&	21.25	$\pm$	0.02	&	0.31	$\pm$	0.06	&	20	$\pm$	2\\
ESO	558-G011	&	J0706--22	&	10.95	$\pm$	0.10	&	26.9	$\pm$	0.6	&	20.14	$\pm$	0.03	&	0.50	$\pm$	0.05	&	47	$\pm$	3\\
ESO	376-G016	&	J1043--37	&	14.46	$\pm$	0.08	&	10.6	$\pm$	0.5	&	21.59	$\pm$	0.03	&	0.50	&	40	\\			
ESO	318-G013	&	J1047--38	&	12.56	$\pm$	0.10	&	19.7	$\pm$	0.5	&	21.02	$\pm$	0.06	&	0.67	$\pm$	0.06	&	70	$\pm$	3\\
ESO	320-G014	&	J1137--39	&	13.20	$\pm$	0.05	&	16.3	$\pm$	0.7	&	21.26	$\pm$	0.03	&	0.37	$\pm$	0.01	&	77	$\pm$	\\
ESO	379-G007	&	J1154--33	&	14.15	$\pm$	0.20	&	22.7	$\pm$	1.5	&	22.93	$\pm$	0.02	&	0.45	$\pm$	0.04	&	-81	$\pm$	4\\
ESO	379-G024	&	J1204--35	&	13.71	$\pm$	0.20	&	17.5	$\pm$	2.5	&	21.92	$\pm$	0.14	&	0.50	$\pm$	0.03	&	27	$\pm$	1\\
ESO	321-G014	&	J1214--38	&	12.74	$\pm$	0.12	&	21.2	$\pm$	1.2	&	21.36	$\pm$	0.01	&	0.44	$\pm$	0.08	&	12	$\pm$	1\\
CEN	6	&	J1305--40	&	14.44	$\pm$	0.10	&	12.2	$\pm$	0.6	&	21.88	$\pm$	0.01	&	0.35	$\pm$	0.08	&	-80	$\pm$	8\\
ESO	269-G058	&	J1310--46A	&	9.77	$\pm$	0.04	&	27.9	$\pm$	0.5	&	18.99	$\pm$	0.02	&	0.32	$\pm$	0.04	&	66	$\pm$	5\\
AM	1321-304	&	J1324--30	&	13.28	$\pm$	0.10	&	18.6	$\pm$	1.5	&	21.62	$\pm$	0.07	&	0.43	$\pm$	0.01	&	-71	$\pm$	4\\
IC	4247	&	J1326--30A	&	11.81	$\pm$	0.07	&	15.7	$\pm$	1.0	&	19.84	$\pm$	0.06	&	0.57	$\pm$	0.02	&	-27	$\pm$	1\\
ESO	324-G024	&	J1327--41	&	11.22	$\pm$	0.08	&	44.6	$\pm$	1.8	&	21.46	$\pm$	0.01	&	0.32	$\pm$	0.04	&	49	$\pm$	1\\
UGCA	365	&	J1336--29	&	12.99	$\pm$	0.12	&	20.4	$\pm$	1.1	&	21.54	$\pm$	0.09	&	0.54	$\pm$	0.01	&	32	$\pm$	2\\
ESO	444-G084	&	J1337--28	&	13.20	$\pm$	0.07	&	21.0	$\pm$	0.6	&	21.80	$\pm$	0.01	&	0.09	$\pm$	0.06	&	-38	$\pm$	\\
NGC	5237	&	J1337--42	&	9.86	$\pm$	0.04	&	20.9	$\pm$	0.2	&	18.46	$\pm$	0.03	&	0.20	$\pm$	0.04	&	-71	$\pm$	5\\
NGC	5253	&	J1339--31A	&	8.44	$\pm$	0.04	&	24.3	$\pm$	0.8	&	17.23	$\pm$	0.05	&	0.58	$\pm$	0.08	&	41	$\pm$	1\\
IC	4316	&	J1340--28	&	11.67	$\pm$	0.06	&	25.9	$\pm$	1.0	&	20.73	$\pm$	0.03	&	0.39	$\pm$	0.02	&	50	$\pm$	5\\
NGC	5264	&	J1341--29	&	10.31	$\pm$	0.10	&	29.1	$\pm$	1.7	&	19.66	$\pm$	0.04	&	0.35	$\pm$	0.04	&	58	$\pm$	4\\
ESO	325-G011	&	J1345--41	&	12.31	$\pm$	0.10	&	32.9	$\pm$	1.2	&	21.90	$\pm$	0.02	&	0.50	$\pm$	0.08	&	-50	$\pm$	9\\
ESO	383-G087	&	J1349--36	&	9.55	$\pm$	0.10	&	49.5	$\pm$	0.6	&	19.93	$\pm$	0.03	&	0.12	$\pm$	0.04	&	84	$\pm$	3\\
NGC	5408	&	J1403--41	&	10.76	$\pm$	0.07	&	24.6	$\pm$	1.4	&	19.71	$\pm$	0.04	&	0.52	$\pm$	0.03	&	72	$\pm$	5\\
ESO	222-G010	&	J1434--49	&	12.41	$\pm$	0.23	&	22.5	$\pm$	2.6	&	21.16	$\pm$	0.08	&	0.25	&	10	\\			
ESO	272-G025	&	J1443--44	&	11.72	$\pm$	0.05	&	19.1	$\pm$	0.5	&	20.04	$\pm$	0.03	&	0.38	$\pm$	0.05	&	63	$\pm$	2\\
ESO	223-G009	&	J1501--48	&	11.07	$\pm$	0.40	&	26.4	$\pm$	8.0	&	20.18	$\pm$	0.10	&	0.28	$\pm$	0.09	&	-38	$\pm$	9\\
ESO	274-G001	&	J1514--46	&	9.63	$\pm$	0.03	&	31.6	$\pm$	0.7	&	19.12	$\pm$	0.02	&	0.77	$\pm$	0.08	&	36	$\pm$	1\\
ESO	149-G003	&	J2352--52	&	13.14	$\pm$	0.05	&	12.7	$\pm$	0.6	&	20.56	$\pm$	0.05	&	0.72	$\pm$	0.07	&	-32	$\pm$	2\\

\hline
\end{tabular}
\caption{Measured sample properties.}
\label{mpars}
\end{table*}

\section{Results} \label{results}

\subsection{Derived parameters}
We compute physical quantities for our detected sample using the best available distances and applying the photometric procedures described in Sect. \ref{phot}. These parameters are presented in Table \ref{dpars} which is arranged as follows: Column(1) - Galaxy name; Column (2) - HIPASS designation; Column (3-5) - S\'{e}rsic parameters $\mu_0$, $\alpha$, $n$ respectively; Column (6) - missing flux correction $\Delta m$; Columns (7) and (8) - the corrected effective radius in arcsec and kpc unites respectively; Column (9) The extinction corrected \textit{H}-band absolute magnitude $M_{H,0}$; and Column (10) - the total stellar mass adopting a mass-to-light ratio, $\Upsilon^{H,0}_{*}$ = 1.0.

\begin{table*}
   \tabcolsep 5.4pt
\begin{tabular}{@{}llrrrrrrrrrr@{}}
\hline\hline
Galaxy name  & HIPASS & $\mu_{0}$  & $\alpha$  & $n$   & $\Delta m$     & $r_{\text{eff,0}}$  & $r_\text{eff,0}$   & $M_{H,0}$  & $\log_{10}(\mathcal{M}_{*})$ \\  
 & & [mag arcsec$^{-2}$] & [arcsec]  &    & [mag]      & [arcsec]  & [kpc]   & [mag]   & $\log_{10}(\mathcal{M}_{\odot})$\\
(1)  & (2)   & (3)   & (4)   & (5)      & (6)   & (7)   & (8)   & (9) &  (10)\\
\hline
S\'{e}rsic fits*\\
\hline
ESO 410-G005 & J0015--32 & 20.67 $\pm$ 0.01 & 29.8 $\pm$ 0.3 & 1.40 $\pm$ 0.02 & 0.05 & 31 $\pm$ 1 & 0.29 $\pm$ 0.04 & -14.1 $\pm$ 0.2 & 7.0 $\pm$ 0.1 \\
ESO 199-G007 & J0258--49 & 21.42 $\pm$ 0.08 & 18.2 $\pm$ 1.4 & 1.10 $\pm$ 0.08 & 0.19 & 27 $\pm$ 6 & 0.8 $\pm$ 0.4 & -15.0 $\pm$ 0.6 & 7.3 $\pm$ 0.2 \\
ESO 252-IG001 & J0457--42 & 20.41 $\pm$ 0.04 & 19.4 $\pm$ 0.6 & 1.50 $\pm$ 0.06 & 0.03 & 20 $\pm$ 1 & 0.7 $\pm$ 0.2 & -16.1 $\pm$ 0.3 & 7.8 $\pm$ 0.1 \\
NGC 2188 & J0610--34 & 18.31 $\pm$ 0.01 & 60.4 $\pm$ 0.3 & 1.50 $\pm$ 0.01 & 0.23 & 61 $\pm$ 1 & 2.2 $\pm$ 0.4 & -20.0 $\pm$ 0.4 & 9.3 $\pm$ 0.2 \\
ESO 489-G?056 & J0626--26 & 20.88 $\pm$ 0.04 & 17.8 $\pm$ 0.4 & 2.00 $\pm$ 0.10 & 0.02 & 15 $\pm$ 1 & 0.4 $\pm$ 0.1 & -14.8 $\pm$ 0.2 & 7.3 $\pm$ 0.1 \\
ESO 376-G016 & J1043--37 & 20.88 $\pm$ 0.02 & 13.7 $\pm$ 2.2 & 1.10 $\pm$ 0.20 & 0.70 & 19 $\pm$ 11 & 0.7 $\pm$ 0.5 & -15.5 $\pm$ 2.8 & 7.5 $\pm$ 1.1 \\
ESO 318-G013 & J1047--38 & 20.10 $\pm$ 0.05 & 28.5 $\pm$ 1.1 & 1.40 $\pm$ 0.08 & 0.12 & 30 $\pm$ 3 & 0.9 $\pm$ 0.3 & -16.7 $\pm$ 0.4 & 8.0 $\pm$ 0.2 \\
ESO 320-G014 & J1137--39 & 20.40 $\pm$ 0.05 & 16.0 $\pm$ 0.6 & 1.20 $\pm$ 0.04 & 0.12 & 21 $\pm$ 2 & 0.6 $\pm$ 0.1 & -15.9 $\pm$ 0.3 & 7.7 $\pm$ 0.1 \\
ESO 379-G007 & J1154--33 & 22.00 $\pm$ 0.10 & 21.7 $\pm$ 2.6 & 1.00 $\pm$ 0.09 & 0.48 & 35 $\pm$ 11 & 0.9 $\pm$ 0.4 & -15.0 $\pm$ 1.1 & 7.3 $\pm$ 0.4 \\
ESO 379-G024 & J1204--35 & 19.80 $\pm$ 0.20 & 4.2 $\pm$ 1.3 & 0.60 $\pm$ 0.05 & 0.39 & $<$80   & $<$2.2   & -15.2 $\pm$ 0.5 & 7.4 $\pm$ 0.2 \\
ESO 321-G014 & J1214--38 & 20.81 $\pm$ 0.03 & 31.2 $\pm$ 0.7 & 1.80 $\pm$ 0.06 & 0.22 & 28 $\pm$ 1 & 0.4 $\pm$ 0.1 & -15.0 $\pm$ 0.4 & 7.4 $\pm$ 0.1 \\
CEN 6 & J1305-40 & 21.46 $\pm$ 0.04 & 19.1 $\pm$ 0.6 & 1.90 $\pm$ 0.10 & 0.24 & 16 $\pm$ 1 & 0.5 $\pm$ 0.1 & -14.6 $\pm$ 0.4 & 7.2 $\pm$ 0.2 \\
ESO 269-G058 & J1310--46A & 18.17 $\pm$ 0.01 & 28.3 $\pm$ 0.2 & 1.30 $\pm$ 0.01 & 0.05 & 35 $\pm$ 1 & 0.6 $\pm$ 0.1 & -18.2 $\pm$ 0.2 & 8.6 $\pm$ 0.1 \\
AM 1321-304 & J1324--30 & 20.68 $\pm$ 0.05 & 17.9 $\pm$ 0.8 & 1.10 $\pm$ 0.05 & 0.12 & 25 $\pm$ 3 & 0.6 $\pm$ 0.1 & -15.2 $\pm$ 0.3 & 7.4 $\pm$ 0.1 \\
IC 4247 & J1326--30A & 18.72 $\pm$ 0.03 & 16.4 $\pm$ 0.5 & 1.10 $\pm$ 0.03 & 0.14 & 24 $\pm$ 2 & 0.6 $\pm$ 0.1 & -16.8 $\pm$ 0.3 & 8.1 $\pm$ 0.1 \\
UGCA 365 & J1336--29 & 20.58 $\pm$ 0.03 & 23.9 $\pm$ 0.6 & 1.30 $\pm$ 0.03 & 0.11 & 28 $\pm$ 2 & 0.7 $\pm$ 0.1 & -15.7 $\pm$ 0.3 & 7.6 $\pm$ 0.1 \\
ESO 444-G084 & J1337--28 & 21.34 $\pm$ 0.05 & 24.9 $\pm$ 0.8 & 1.80 $\pm$ 0.10 & 0.07 & 22 $\pm$ 2 & 0.5 $\pm$ 0.1 & -15.2 $\pm$ 0.3 & 7.4 $\pm$ 0.1 \\
IC 4316 & J1340--28 & 19.59 $\pm$ 0.02 & 20.5 $\pm$ 0.5 & 1.00 $\pm$ 0.02 & 0.23 & 36 $\pm$ 2 & 0.8 $\pm$ 0.1 & -16.8 $\pm$ 0.4 & 8.1 $\pm$ 0.1 \\
NGC 5264 & J1341--29 & 18.96 $\pm$ 0.01 & 34.2 $\pm$ 0.2 & 1.50 $\pm$ 0.01 & 0.03 & 35 $\pm$ 0 & 0.8 $\pm$ 0.1 & -18.0 $\pm$ 0.2 & 8.5 $\pm$ 0.1 \\
ESO 325-G011 & J1345--41 & 21.54 $\pm$ 0.05 & 60.9 $\pm$ 1.7 & 2.60 $\pm$ 0.02 & 0.07 & 46 $\pm$ 1 & 0.8 $\pm$ 0.1 & -15.5 $\pm$ 0.3 & 7.5 $\pm$ 0.1 \\
NGC 5408 & J1403--41 & 18.89 $\pm$ 0.01 & 29.8 $\pm$ 0.3 & 1.30 $\pm$ 0.01 & 0.13 & 35 $\pm$ 1 & 0.8 $\pm$ 0.1 & -17.8 $\pm$ 0.3 & 8.5 $\pm$ 0.1 \\
ESO 222-G010 & J1434--49 & 20.14 $\pm$ 0.05 & 21.0 $\pm$ 1.1 & 1.00 $\pm$ 0.04 & 0.27 & 35 $\pm$ 5 & 1.0 $\pm$ 0.3 & -16.8 $\pm$ 0.6 & 8.1 $\pm$ 0.2 \\
ESO 272-G025 & J1443--44 & 19.07 $\pm$ 0.01 & 16.7 $\pm$ 0.2 & 1.10 $\pm$ 0.01 & 0.15 & 24 $\pm$ 1 & 0.7 $\pm$ 0.2 & -17.3 $\pm$ 0.5 & 8.3 $\pm$ 0.2 \\
ESO 223-G009 & J1501--48 & 19.41 $\pm$ 0.01 & 27.2 $\pm$ 0.4 & 1.20 $\pm$ 0.02 & 0.09 & 35 $\pm$ 1 & 1.1 $\pm$ 0.2 & -18.2 $\pm$ 0.5 & 8.6 $\pm$ 0.2 \\
ESO 274-G001 & J1514--46 & 18.12 $\pm$ 0.01 & 58.4 $\pm$ 0.6 & 1.00 $\pm$ 0.01 & 0.59 & 95 $\pm$ 3 & 1.4 $\pm$ 0.2 & -18.5 $\pm$ 0.7 & 8.7 $\pm$ 0.3 \\
ESO 149-G003 & J2352--52 & 19.65 $\pm$ 0.03 & 18.2 $\pm$ 0.6 & 1.30 $\pm$ 0.03 & 0.13 & 22 $\pm$ 1 & 0.6 $\pm$ 0.1 & -15.9 $\pm$ 0.4 & 7.7 $\pm$ 0.1 \\

 \hline                                 
 Double S\'{e}rsic fits* \\                              
 \hline                                 
KK 49 & J0541+06 & 18.40 $\pm$ 0.04 & 15.0 $\pm$ 0.2 & 2.20 $\pm$ 0.09 & 0.01 & 16 $\pm$ 1 & 0.4 $\pm$ 0.1 & -18.3 $\pm$ 0.4 & 8.7 $\pm$ 0.2 \\
   & & 19.70 $\pm$ 0.13 & 32.7 $\pm$ 2.0 & 1.60 $\pm$ 0.08 & & & & & \\             
AM 0605-341 & J0607--34 & 19.40 $\pm$ 0.20 & 14.0 $\pm$ 0.5 & 2.40 $\pm$ 0.20 & 0.07 & 15 $\pm$ 1 & 0.5 $\pm$ 0.2 & -17.4 $\pm$ 0.4 & 8.3 $\pm$ 0.2 \\
   & & 19.20 $\pm$ 0.23 & 11.3 $\pm$ 0.8 & 1.00 $\pm$ 0.01 & & & & & \\             
ESO 558-G011 & J0706--22 & 21.00 $\pm$ 0.25 & 28.1 $\pm$ 0.9 & 5.00 $\pm$ 2.00 & 0.07 & 36 $\pm$ 6 & 1.5 $\pm$ 0.5 & -18.9 $\pm$ 0.4 & 8.9 $\pm$ 0.1 \\
   & & 19.70 $\pm$ 0.11 & 38.2 $\pm$ 2.4 & 1.40 $\pm$ 0.08 & & & & & \\             
ESO 324-G024 & J1327--41 & 21.20 $\pm$ 0.14 & 13.2 $\pm$ 1.1 & 1.30 $\pm$ 0.20 & 0.16 & 48 $\pm$ 2 & 0.9 $\pm$ 0.1 & -16.8 $\pm$ 0.3 & 8.1 $\pm$ 0.1 \\
   & & 21.30 $\pm$ 0.04 & 73.9 $\pm$ 1.2 & 2.30 $\pm$ 0.10 & & & & & \\             
NGC 5237 & J1337--42 & 17.70 $\pm$ 0.02 & 18.0 $\pm$ 0.2 & 1.40 $\pm$ 0.03 & 0.01 & 21 $\pm$ 1 & 0.3 $\pm$ 0.1 & -17.9 $\pm$ 0.2 & 8.5 $\pm$ 0.1 \\
   & & 21.50 $\pm$ 0.37 & 63.1 $\pm$ 5.6 & 3.00 $\pm$ 0.50 & & & & & \\             
NGC 5253 & J1339--31A & 15.60 $\pm$ 0.14 & 10.4 $\pm$ 0.9 & 0.90 $\pm$ 0.08 & 0.05 & 26 $\pm$ 9 & 0.5 $\pm$ 0.2 & -19.4 $\pm$ 0.3 & 9.1 $\pm$ 0.1 \\
   & & 18.50 $\pm$ 0.48 & 63.8 $\pm$ 9.5 & 1.60 $\pm$ 0.20 & & & & & \\             
ESO 383-G087 & J1349--36 & 21.20 $\pm$ 0.10 & 21.1 $\pm$ 0.4 & 4.90 $\pm$ 0.70 & 0.07 & 49 $\pm$ 2 & 0.8 $\pm$ 0.1 & -18.2 $\pm$ 0.3 & 8.6 $\pm$ 0.1 \\
   & & 19.50 $\pm$ 0.02 & 54.6 $\pm$ 0.7 & 1.60 $\pm$ 0.03 & & & & & \\             

\hline
\end{tabular}
\caption{Derived sample properties. * See section \ref{phot} for further details on surface brightness profile fits.}
\label{dpars}
\end{table*}

\subsection{Comparisons to 2MASS}
\begin{figure}
\includegraphics[scale=0.6]{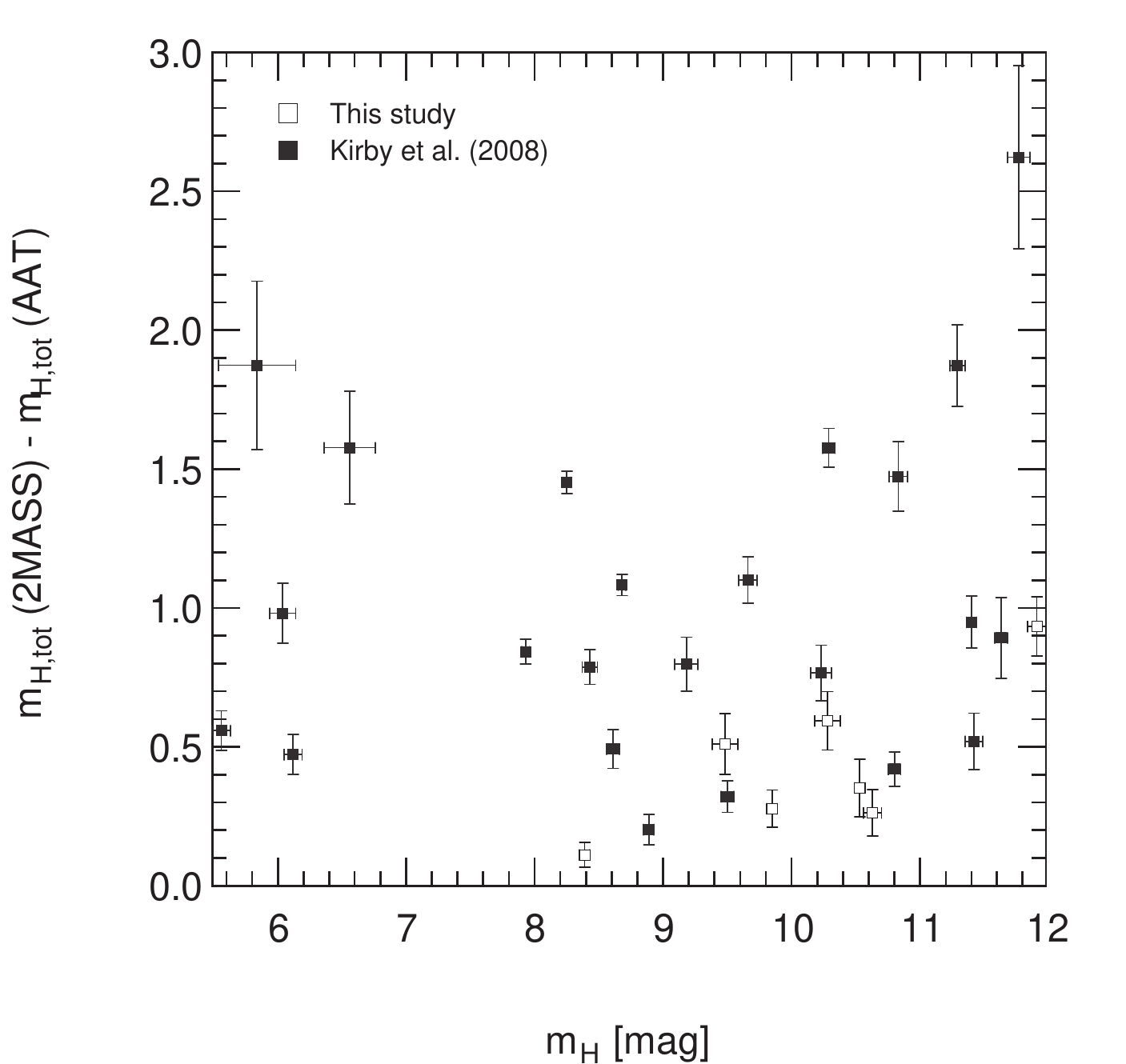}\\ \includegraphics[scale=0.6]{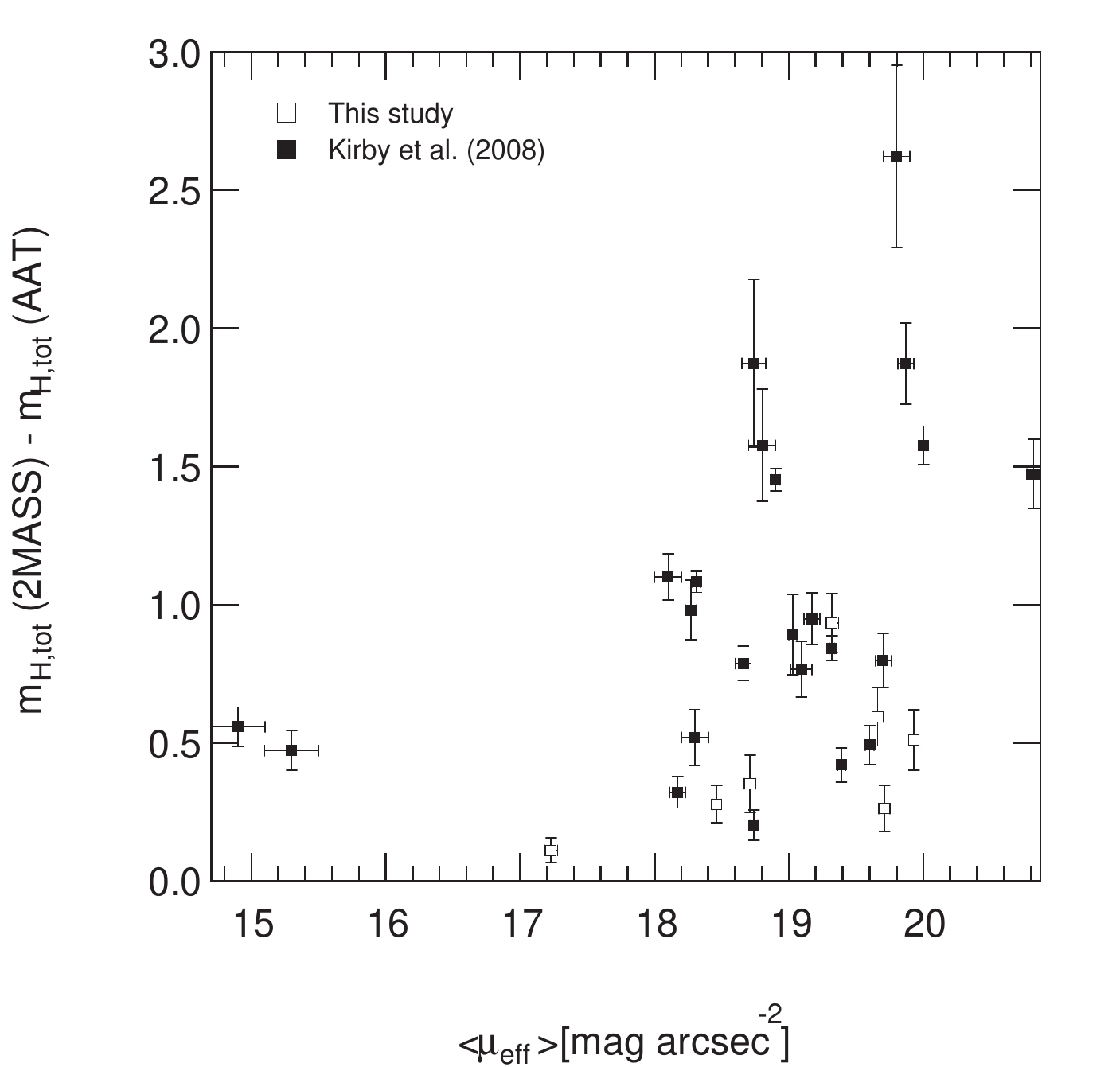}
\caption{2MASS photometry compared with the KJRD08 and our sample for \textit{H}-band luminosity (\textit{top}) and mean effective surface brightness (\textit{bottom}). These plots demonstrate the systematic underestimation of galaxy fluxes in the 2MASS catalog.}
\label{2masscomp}
\end{figure}

Following on from the extensive discussion of 2MASS magnitudes in KJRD08 and \cite{Andreon2002} we provide an updated comparison of the total observed magnitudes to include where possible the 33 galaxies analysed in this study. Out of the 90 galaxies with photometric analysis in the KJRD08 sample and this study, only 30 are detected within the 2MASS All Sky Extended Source Catalog \citep[][ESC]{Jarrett2000}. Figure \ref{2masscomp} plots the difference between the IRIS2 total \textit{H}-band magnitudes 
and the total magnitudes for galaxies common to the ESC. The plot illustrates the systematic underestimation of luminosity especially for galaxies with mean effective surface brightness lower than 18 mag arcsec$^{-2}$. KJRD08 re-analyzed 2MASS images with photmetric procedures employed for the KJRD08 sample. They found that the re-derived magnitudes corresponded to those in the ESC and thus differences in photometric procedure can be ruled out except in cases where artifacts (1-5\% of the ESC) have contaminated the photometery \citep{Jarrett2000}. \cite{Jarrett2000} states these artifacts consist of false sources generated in the vicinity of bright stars, transient phenomena and infrared `airglow', and the applied removal procedures not being 100\% effective. These cases may account for at least some of the scatter in Figure \ref{2masscomp}. 

We emphasize that these comparisons have been made using 2MASS magnitudes which recover flux lost beneath the background noise by extrapolating the surface profile \cite[see,][]{Jarrett2000}. Since the surface brightness profiles of galaxies sampled by the 2MASS imagery are largely confined to the inner, high surface brightness regions due to the shallowness of the photometry, the derived S\'{e}rsic index is particularly susceptible to variations in slope, particularly for galaxies with a nucleated core and disk component. To elaborate further, one could imagine a scenario in which missing flux corrections are underestimated for a galaxy with a nucleated inner surface brightness profile, and an outer exponential disk component in cases where the image depth only allows the nucleated component to be sufficiently sampled. This could lead to a derived shape parameter(s) with a slope well below the exponential disk component and the resulting corrections for the flux missing below the background level will completely miss the flux contributed by the disk. It is therefore crucial when attempting to perform photometery of dwarf galaxies (and all galaxies in general) that the surface brightness profile is sufficiently sampled to allow proper correction for the missing flux. Deep targeted, photometric surveys hold the advantage over all sky surveys regarding this issue, as the former is able to tailor the total integration time to sufficiently sample the surface brightness profile. 

The large scatter in Fig \ref{2masscomp} shows mild trends in mean effective surface brightness and observed total luminosities as expected. However, the intrinsic scatter in these plots demonstrate that applying a simple correction function to the 2MASS catalog is not possible for obtaining accurate NIR magnitudes for (low-surface brightness) galaxies.

\subsection{Comparisons to the VISTA Hemispheric Survey}

The VISTA Hemispheric survey (VHS) is a wide-field infrared southern sky survey conducted on the Visible and Infrared Survey Telescope for Astronomy (VISTA) at ESO. The VHS collaboration in conjunction with other VISTA Public Surveys aim to cover the whole southern celestial hemisphere to a depth 30 times fainter than 2MASS in at least two photometric wavebands (\textit{J} and \textit{K$_s$}), with an exposure time of 60 seconds per waveband to produce median 5$\sigma$ point source (Vega) limits of \textit{J} = 20.2 and \textit{K$_s$} = 18.1.

\begin{figure*}
\includegraphics[scale=0.3]{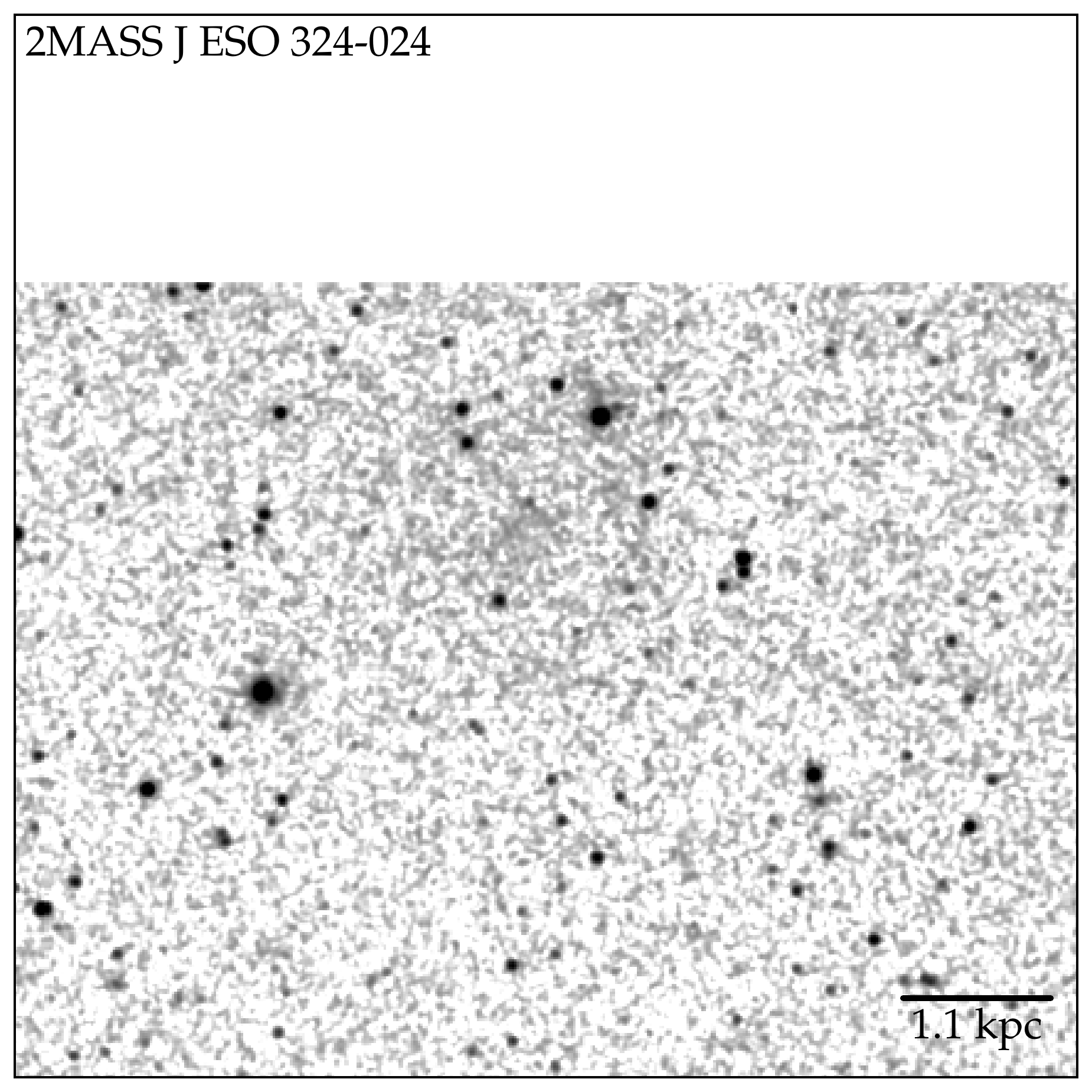}
\includegraphics[scale=0.3]{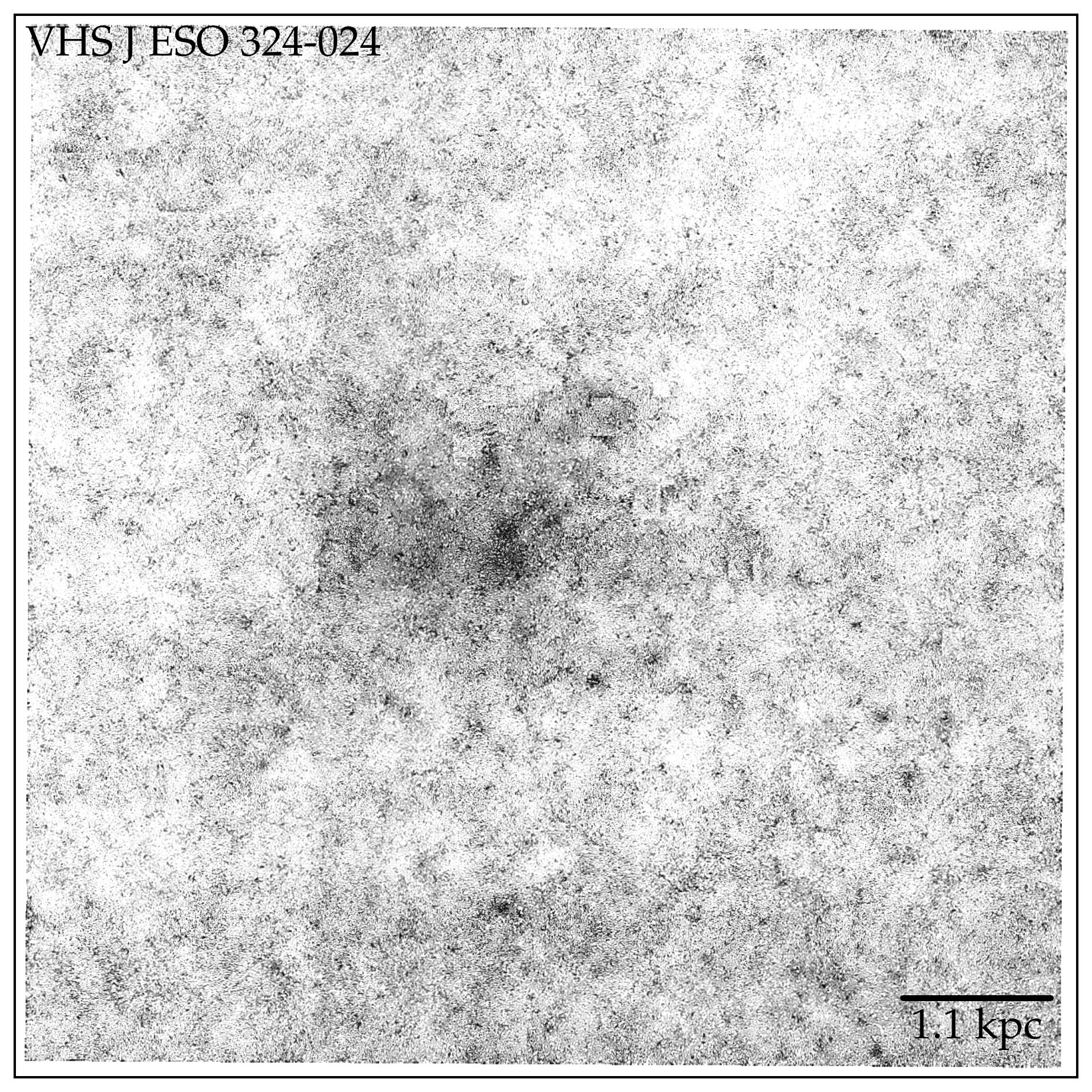}
\includegraphics[scale=0.3]{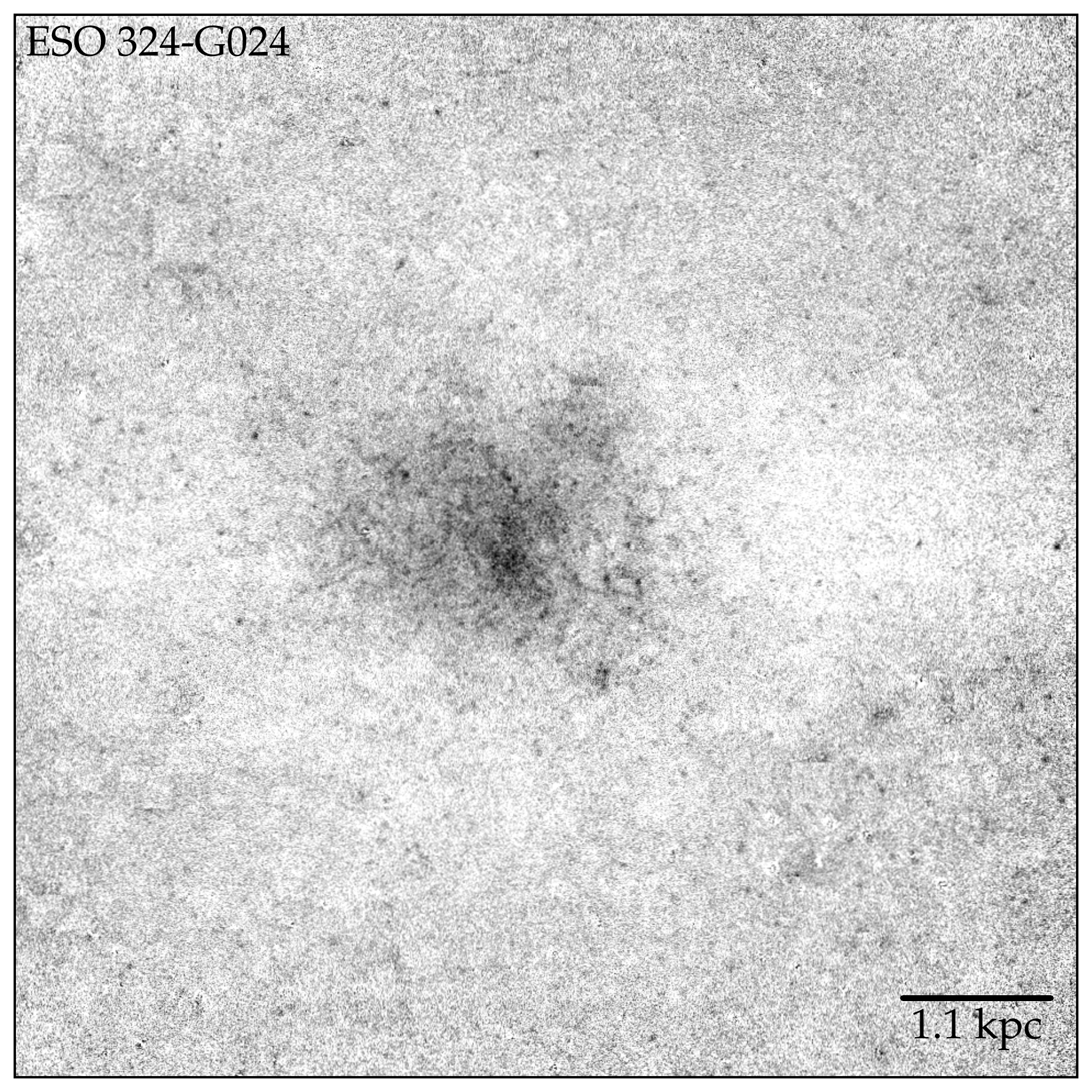}\\
\includegraphics[scale=0.3]{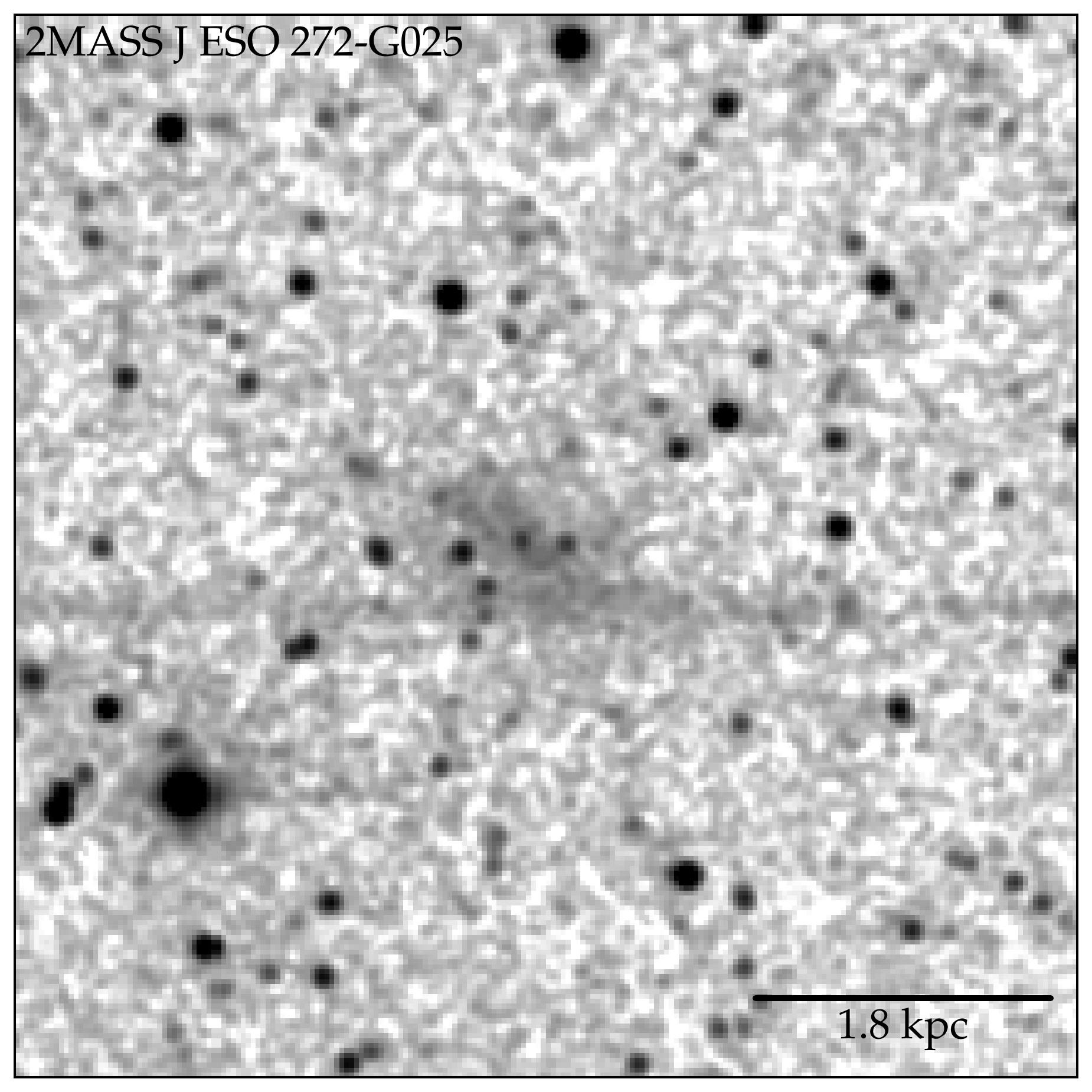}
\includegraphics[scale=0.3]{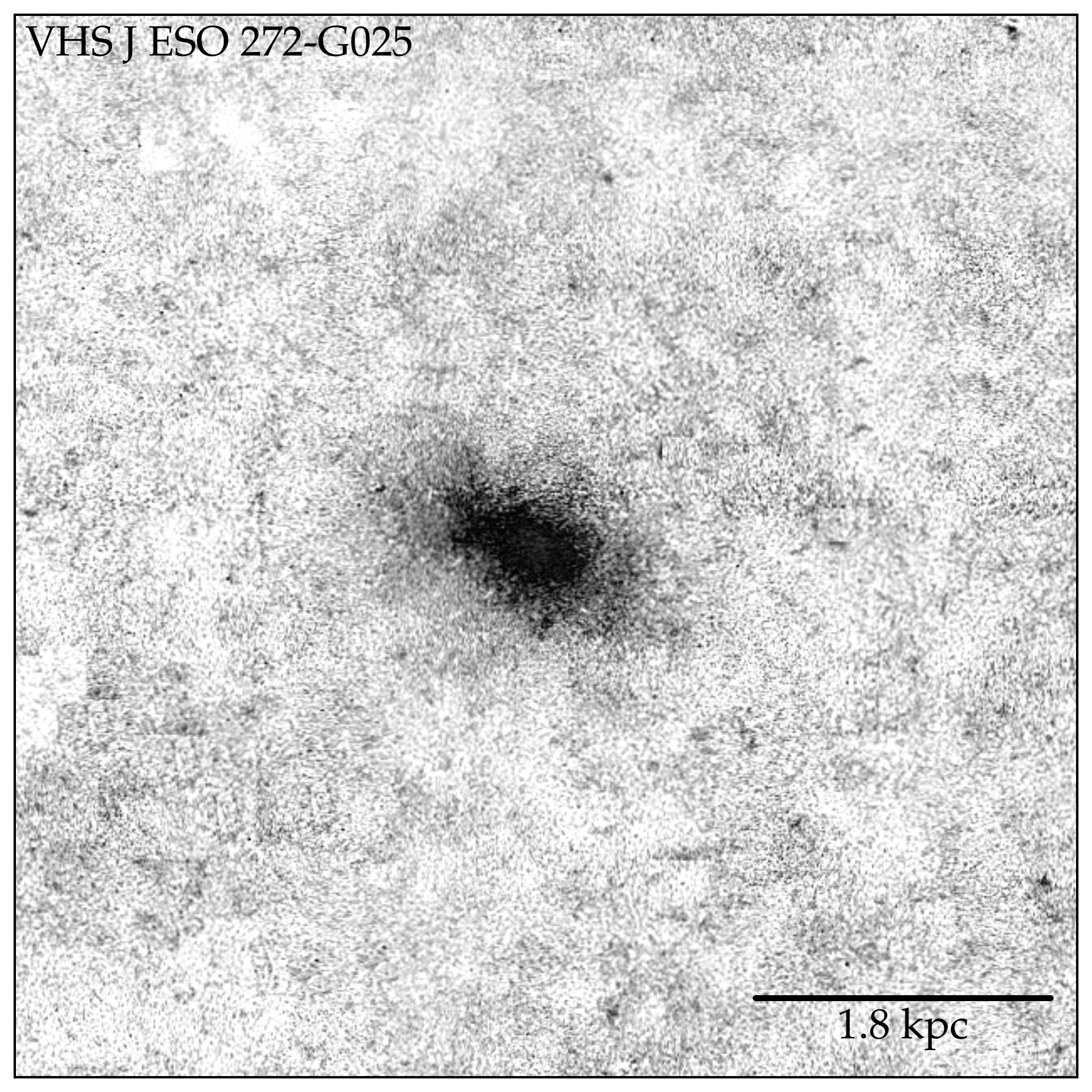}
\includegraphics[scale=0.3]{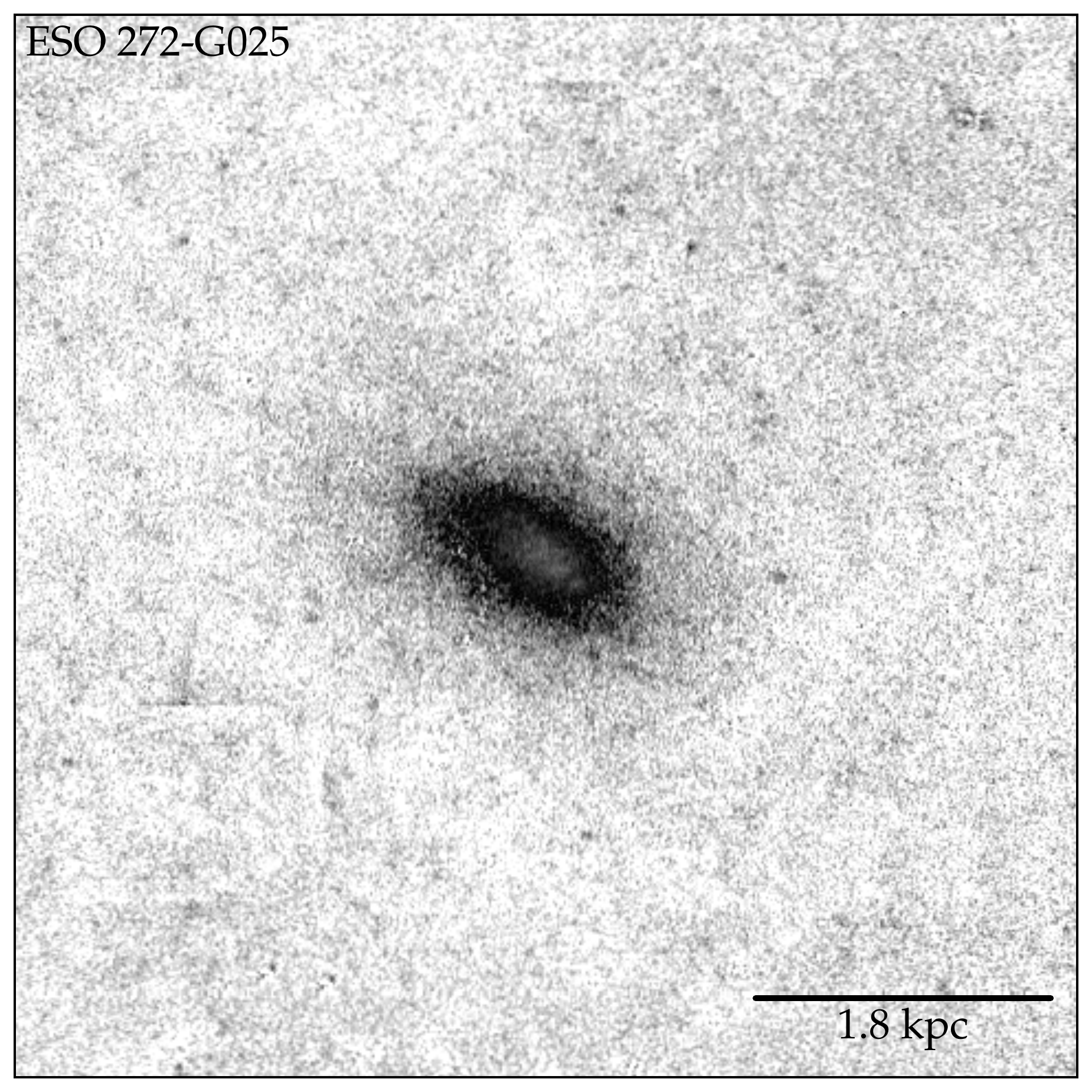}

\caption{(\textit{Left}) 2MASS all sky survey images. (\textit{Middle}) The cleaned Vista VHS survey images. (\textit{Right}) Our cleaned \textit{H}-band images. The galaxies used in the comparison are ESO 324-G024 (\textit{Top}) and ESO 272-G025 (\textit{Bottom}).}
\label{VHScomp}
\end{figure*}

\begin{figure*}
\includegraphics[scale=0.6]{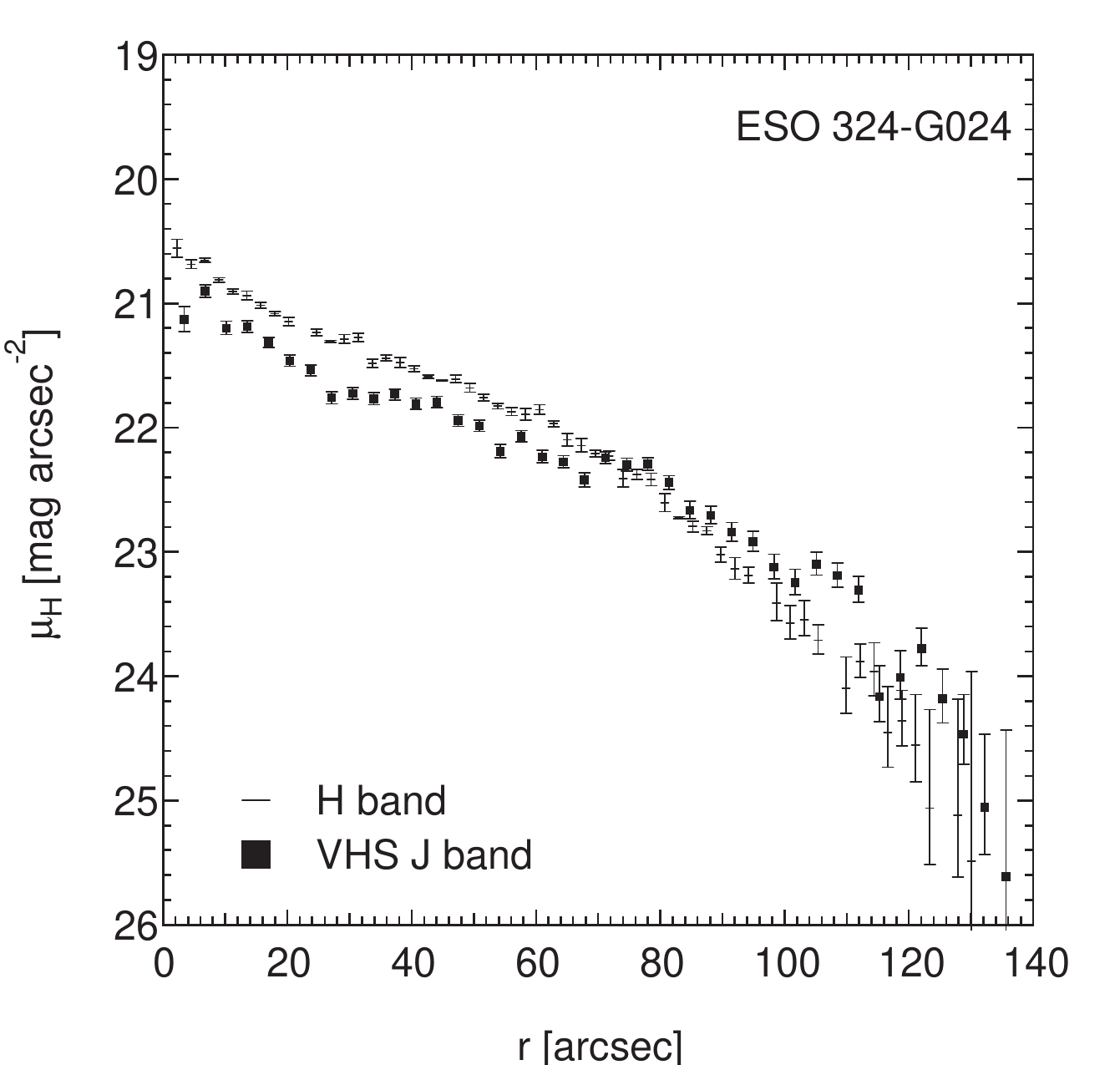}
\includegraphics[scale=0.6]{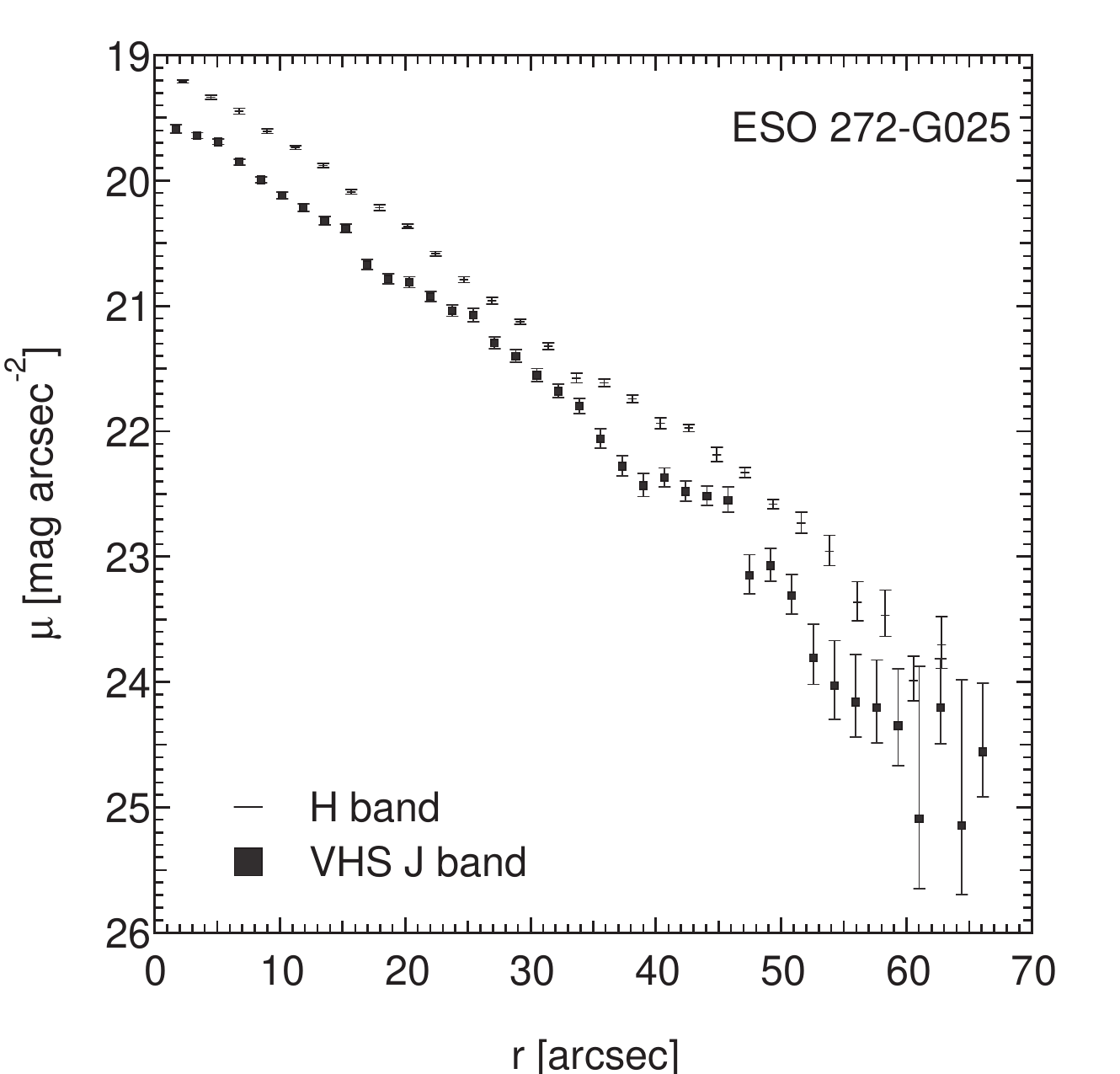}
\caption{Corresponding \textit{H}-band and VHS \textit{J}-band surface brightness profiles produced by adopting the center, ellipticity and PA as derived from the IRIS2 data. The galaxies used in the comparison are ESO 324-G024 (\textit{Left}) and ESO 272-G025 (\textit{Right}).}
\label{VHScompSB}
\end{figure*}

Although the science goals\footnote{\url{http://www.vista-vhs.org/science}} do not specifically include the photometry of LV dwarf galaxies, there may be an opportunity to perform NIR photometry on dwarf galaxies as has been conducted in this research. To test the suitability, we obtain images for a moderately high surface brightness (relative to our sample) galaxy ESO 272-G025 and a low surface brightness galaxy ESO 324-G024. These pre-reduced, VHS DR2, \textit{J}-band images were obtained using the `multiget' tool located on the VISTA Science Archive (VSA) website\footnote{\url{http://horus.roe.ac.uk/vsa/dboverview.html}}. The strategies employed for the VISTA NIR reductions are identical to the ones employed in this study\footnote{\url{http://apm49.ast.cam.ac.uk/surveys-projects/vista/technical/data-processing}\\\url{http://apm49.ast.cam.ac.uk/surveys-projects/vista/technical/sky-subtraction}}. We clean the VHS \textit{J}-band images using the same procedures described in Sect. \ref{obs} and employ the same analysis techniques described in Sect. \ref{phot} for deriving the surface brightness profile. The cleaned \textit{J}-band galaxy images and associated surface brightness profiles are qualitatively compared to those in our sample in Fig \ref{VHScomp} and \ref{VHScompSB} as well as the 2MASS images for these galaxies obtained with 2MASS image services\footnote{\url{http://irsa.ipac.caltech.edu/applications/2MASS/IM/}}.

Immediately clear from the comparison in Figs \ref{VHScomp} and \ref{VHScompSB} is the significant increase in depth the VHS images provide over the 2MASS images. For galaxies with central surface brightness higher than $\mu_{0,J}$ $\sim$ 20-21 mag arcsec$^{-2}$ VHS should achieve a sufficient surface brightness limit to adequately sample the S\'{e}rsic index and correctly account for the flux lost below the background level. In contrast, 2MASS barely detects galaxies with central surface brightnesses brighter than $\mu_H$ $=$ 20 mag arcsec$^{-2}$. The VHS may therefore provide a promising data set to extract galaxy photometry down to $\sim$23-24 mag arcsec$^{-2}$. In terms of low surface brightness galaxy photometry, the VHS is not without its limitations, especially when compared to deep, targeted surveys such as the one conducted in this study. Due to the shorter exposure times in the VHS images, the variance in the sky background is significantly higher relative to our \textit{H}-band images obtained at the AAT. In the case of the two galaxies compared in Fig \ref{VHScomp} and \ref{VHScompSB}, this variance was of order 3-4 times higher. A stable sky background is the critical factor in determining accurate galaxy photometry. Under or overestimating the background level will lead to changes in slope of the surface brightness profile and thus the derived total magnitude. Therefore beyond the maximum visual angular extent of the galaxy, one requires that the background level relative to the signal of the galaxy remains stable. In practical terms, the azimuthally averaged growth curve must remain asymptotically flat for large enough radii in order to adopt a sky background level. In cases where the variance in the sky background level becomes comparable to the surface brightness of the galaxy, it becomes difficult to determine for which sky background value will the growth curve converge. 

\begin{figure}
\includegraphics[scale=0.6]{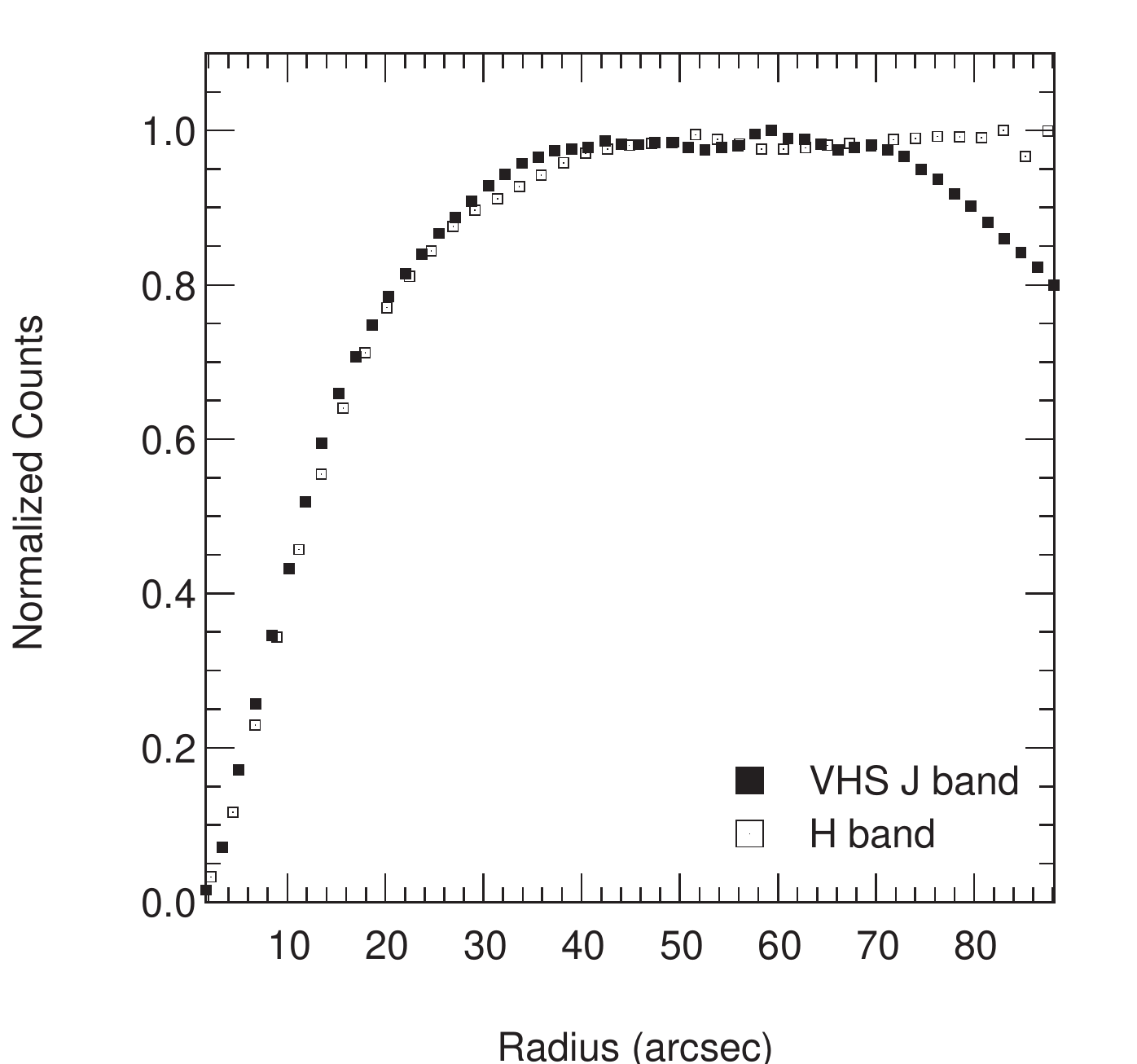}
\caption{Comparing the background stability of VHS \textit{J}-band image and the study \textit{H}-band image for the galaxy ESO 149-G003. The plot shows the azimutally averaged counts against radius above the adopted sky level. The counts are normalized to the maximum above the sky level within the plotted radius to provide an easier comparison between the two curves.}
\label{growthcomp}
\end{figure}

To investigate this issue we obtain a third VHS \textit{J}-band image, ESO149-G003. Figure \ref{growthcomp} compares the \textit{H}-band and VHS \textit{J}-band growth curve convergences for our sample galaxy ESO 149-G003 (r$_{\text{eff}}$ = 12.7 arcsec, $\langle\mu_H\rangle$ = 20.56). Fluctuations in the sky background level for the VHS \textit{J}-band image on the 40-50 arcsec scale cause the growth curve to diverge at large radii whilst the \textit{H}-band image remains convergent. In the case of the low surface brightness galaxy, ESO 324-G024, determining the sky background level for the VHS \textit{J}-band image was also difficult. This is a possible explanation for the J-H colour variation observed in the surface brightness profiles in Figure \ref{VHScomp} left panel. Conversely, a colour gradient is not observed in ESO 272-G025 whose mean surface brightness is 1.5 magnitudes higher than ESO 324-G024.

Taking all into account, VHS images of moderately high surface brightness galaxies (by our sample standards) should provide accurate photometric magnitudes and will be a major improvement over 2MASS photometry. However for low surface brightness galaxies such as ESO 324-G024 deep targeted observations are still highly desirable to sufficiently stabilize the background and accurately sample the surface brightness profile. It is also important to mention that the center of an irregular galaxy is notoriously difficult to find in shallow images due to the dominance of irregularly distributed, bright and prominent H\textsc{ii} regions. Only a sufficiently deep image reveals the geometric center as measured from the low surface brightness, more regular isophotes.

\subsection{Null Detections}
\begin{figure}
\includegraphics[scale=0.45]{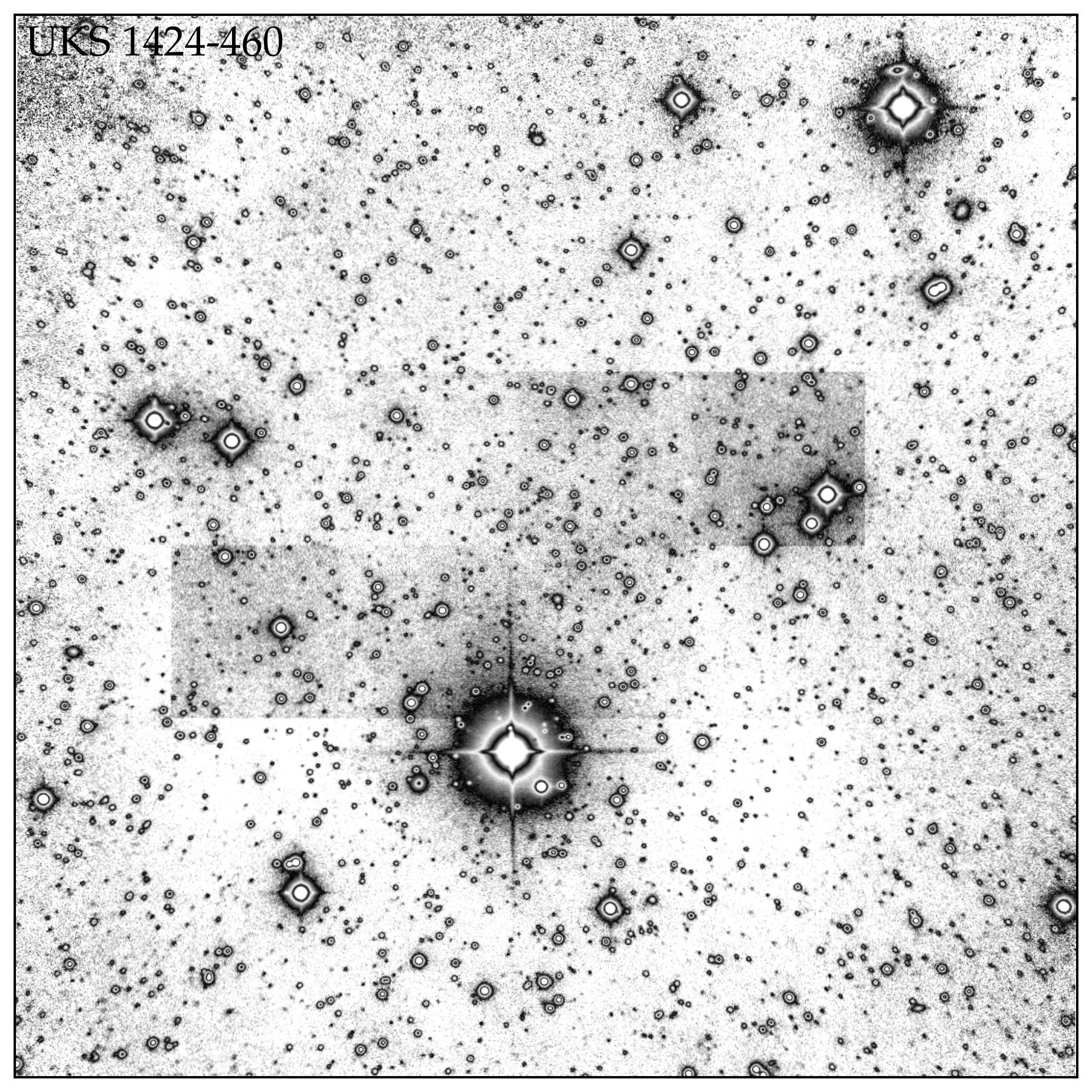}
\caption{Determining the detection limit by artificially adding counts to an image of UKS 1424-460. The top row of adjacent square regions ($\sim$50 arcsec wide) have added artificially from left to right (0.25, 0.5, 0.75, 1)$\sigma_{\text{sky}}$ and the reverse is true for the bottom row. The detection limit is equivalent to the minimum detectable counts per pixel visible above the background noise which in the case of the image for UKS 1424-460 corresponds to 0.5$\sigma_{\text{sky}}$}
\label{addnoise}
\end{figure}

The galaxies listed in Table \ref{nulls} were those observed but not included in the photometric analysis due to an inadequate image depth, or significant foreground stars. We provide an estimate on the lower bound of the total apparent magnitude of these galaxies. Galaxies not detected or only marginally detected were determined through visual inspection. Although we recover the surface brightness profiles of galaxies down to $\sim$25 mag arcsec$^{-2}$, this strategy requires a priori knowledge of the geometric parameters, such as the center and ellipticity. In a marginal detection, the parameters are uncertain at best or non-existent in the case of a complete null detection. Since we are interested in upper bound estimates, we instead require the surface brightness limit that represents the faintest visually detectable galaxy. We estimate this quantity by artificially adding counts proportional to $\sigma_{\text{sky}}$ to all pixels in a square region on the image of one of our non-detections, the galaxy UKS 1424-460. We then repeat this process for different fractions of sigma, (0.25, 0.5, 0.75, 1)$\sigma_{\text{sky}}$ and arrange the square regions to be adjacent. Figure \ref{addnoise} illustrates the result of this process. Importantly, we apply this process to the uncleaned images since the presence of bright stars, particularly in poor seeing conditions, increase the difficulty in visually identifying a galaxy. In the image of UKS 1424-460 we were only barely able to detect the 0.25$\sigma_{\text{sky}}$ level. Typically we adopted $\sigma_{\text{lim}}$ = 0.5$\sigma_{\text{sky}}$. This result can be converted into a surface brightness limit for a given photometric zero point and pixel scale in arcsec ($w$) using the equation,

\begin{align}
\mu_{lim} = -2.5\log_{10}\left(\frac{\sigma_{\text{lim}}}{w^2} \right) + \text{zero point},
\end{align}

The brightest apparent magnitude that a galaxy can have corresponds to the product of the surface intensity and the area defined by the outer isophote $r_{\text{max}}$,
\begin{align}
m_{\text{tot}} > \langle \mu_{\text{eff},lim} \rangle - 2.5\log_{10}(\pi r^2_{\text{max}}). \label{sblimit}
\end{align}

In cases where no optical detection of the galaxy exists, we adopt an $r_{\text{max}}$ of 60 arcsec which is representative of the median half light radius of 20 arcsec. In the following subsections we discuss each non detected galaxy individually comparing them to known parameters in the literature where possible, particular the H\textsc{i} mass-to-light ratios.
\begin{table*}

\begin{tabular}{llcccc}

\hline\hline
Galaxy Name & HIPASS	&	Reason		&	$m_H$	&	$M_{H,0}$&	$\log_{10}(\mathcal{M})$\\
(1)	&	(2)		&	(3)	&	(4)	&	(5) & (6)\\
\hline
AM 0704-582 & J0705--58&	Not Detected&	$>$10.9&	$>$-17.6&	$<$8.4\\
KK 195 & J1321--31&	Not detected&	$>$12.3&	$>$-16.3&	$<$7.9\\
LEDA 592761 & J1337--39&	Marginal detection&	$>$12.7&	$>$-15.8&	$<$7.7\\
- & J1348--37 &	Not detected&	$>$13.0&	$>$-15.9&	$<$7.7\\
LEDA 3097113 & J1351--47&	Not detected&	$>$12.5&	$>$-16.3&	$<$7.9\\
UKS 1424-460 & J1428--46&	Not detected&	$>$11.8&	$>$-16.0&	$<$7.7\\
- & J1919--68&	Source confusion? \\
\hline

\end{tabular}
\caption{Derived upper limits on null detected sources.}
\label{nulls}
\end{table*}

\subsubsection{Argo Dwarf Irregular (AM 0704-582)}
This galaxy was included in this study to serve as a consistency check on reduction and analysis procedures as it was analyzed in KJRD08. Despite 35 minutes of total on source integration (2.6 times longer than KJRD08), we estimate the surface brightness limit at 22.4 mag arcsec$^{-2}$ from equation \ref{sblimit}, which is slightly brighter than the mean effective surface brightness measure (22.57 $\pm$ 0.2) in KJRD08.  Seeing conditions (1.6 arcsec) were poorer in our observations than KJRD08 (1.2 arcsec) and the image zero point was an order of magnitude fainter than our sample median. It is likely the null detection was the result of poor atmospheric conditions, e.g. cirrus, during the observation.

\subsubsection{KK 195 (HIPASS J1321-31)}
Subject to prior study with the Hubble space telescope, this anomalous dwarf galaxy has a unique star formation history and a large \HI\, mass-to-light ratio $\mathcal{M}_{HI,\odot}/L_{\odot,B}$ = 5 \citep{Pritzl2003,Grossi2007}. Additionally, the absence of H$\alpha$ in emission \cite{Bouchard2009} shows that this galaxy is not experiencing any current star formation. HST imaging in the \textit{V} and \textit{I}-bands \citep{Grossi2007} show an extremely diffuse resolved stellar distribution with a semi-major angular extent of $\sim$40 arc seconds.

\cite{Fingerhut2010} obtains a marginal detection in the \textit{K$_s$}-band and derives a total apparent magnitude of $m_{Ks}$ = 15.87 $\pm$ 0.5. The authors cautioned the use of this value other than a faint limit to the total magnitude, citing concerns over the inherent tendency for dIrr to have flattened inner profiles and the fact that the study did not measure the faintest isophote to a depth of greater than 2.5 mag arcsec$^{-2}$ below the central surface brightness. Nevertheless, this value can serve as a first order input for deriving the stellar mass. Adopting a distance of 5.2 Mpc \citep{Grossi2007} we infer an extinction corrected magnitude of $M_{K_s,0} = -12.7$ and a corresponding lower stellar mass limit of $\log_{10}(\mathcal{M}_{*}/\mathcal{M}_{\odot}) =$ 6.4, nearly an order of magnitude less massive than any of the detected dwarfs in our sample. Adopting a HIPASS \HI\, mass of 7.4 ($\log_{10} (\mathcal{M}_{*}/\mathcal{M}_{\odot})$) \citep{Koribalski2004}, the corresponding upper limit for the \textit{K}-band \HI\, mass-to-light-ratio is 12.3 ($\mathcal{M}_{HI,\odot}/L_{\odot,Ks}$), which is higher than the corresponding H\textsc{i} mass-to-light ratio for the \textit{B}-band of 4.9 ($\mathcal{M}_{HI,\odot}/L_{\odot,B}$) \citep{Pritzl2003,Grossi2007}. Indeed this galaxy ranks as one of the least efficient star forming systems in the LV \citep{Warren2007}.
 
\subsubsection{HIPASS J1337-39 (LEDA 592761)}
HIPASS J1337-39 is a gas rich dIrr and has active ongoing star formation \citep{Grossi2007}. We are only able to get a marginal detection of the galaxy with our detection limit of 22.8 mag arcsec$^{-2}$. Employing the same analysis as in the previous section using the total apparent magnitude of K$_s$ 16.61 from 
\cite{Fingerhut2010}, we derive a \textit{K$_s$}-band \HI\, mass-to-light ratio of 17.4 ($\mathcal{M}_{HI,\odot}/L_{\odot,Ks}$), which is significantly higher than the corresponding mass-to-light ratio for the \textit{B}-band, 2.1 ($\mathcal{M}_{HI,\odot}/L_{\odot,B}$) \citep{Grossi2007}. Since the NIR flux contribution of young massive stars and associated gas continuum can typically contribute only $\sim$20-40\%, the large difference in the \textit{B}-band and \textit{K$_s$}-band \HI\, mass-to-light ratio must be at least partially attributable to the active ongoing star formation. According to figure 4 in \cite{Grossi2007}, the galaxy has a rather high surface brightness in the \textit{B}-band and is morphological dominated by H\textsc{ii} emission knots. These knots are barely discernible in our \textit{H}-band Figure \ref{results2}, however insufficient depth was achieved to adequately determine the geometric parameters, or sufficiently sample the surface brightness profile.

\subsubsection{HIPASS J1348-37}
Despite our image limiting surface brightness of 23.1 mag arcsec$^{-2}$, we were just barely able to detect this galaxy that was found in the blind H\textsc{i} survey of the southern sky with the Parkes Telescope \citep{Banks1999}. Furthermore, the coordinates used in our observation were slightly incorrect (note they are correct in Table \ref{sample}), and so the location of the galaxy is off center (see Figure \ref{results3}). \cite{Karachentsev2007} presented a HST image of the galaxy (in their Figure 3) and derived a TRGB-based distance of $5.75 \pm 0.66$ Mpc. The HST image shows a morphologically diffuse, low surface brightness dIrr with a major axis greater than 30 arcsec in length. This extremely low surface brightness dwarf galaxy appears as a relatively large object in the field of view compared to a point-source-like high surface brightness background galaxy located approximately 30 arcsec to the northwest. The photometry performed by \cite{Fingerhut2010} reported a scale length of 3.5 arcsec and a central surface brightness of 20.55 mag arcsec$^{-2}$, which suggests that this photometry was accidentally performed on the background galaxy rather than HIPASS J1348-37.

\subsubsection{HIPASS J1351-47 (LEDA 3097113)}
The estimated limiting surface brightness of 23.5 mag arcsec$^{-2}$ was insufficient to unambiguously detect this galaxy, mainly due to crowding from a significant number of galactic foreground stars (See Fig. \ref{results3}). Using the total apparent \textit{K}-band magnitude K$_s$ = 15.8 from 
\cite{Fingerhut2010} and the HIPASS H\textsc{i} mass of 7.4 ($\log_{10} (\mathcal{M}_{\odot})$) \citep{Koribalski2004}, we derive a \textit{K$_{s}$}-band H\textsc{i} mass-to-light-ratio of $\mathcal{M}_{HI,\odot}/L_{\odot,Ks}$ = 4.0. Although the H\textsc{i} mass-to-light ratio is not as extreme as for KK195 or HIPASS J1337-39 these galaxies do share a low stellar mass estimate, $\log_{10}(\mathcal{M}_{*}/\mathcal{M}_{\odot})$ = 6.0-6.5 , as determined through the K$_s$ surface brightness photometery of \cite{Fingerhut2010}. These mass estimates are over an order of magnitude lower than what our surface brightness limits could detect. Indeed, ESO 410-G005 is the least massive galaxy successfully detected in our sample with a mass estimate of $\log_{10}(\mathcal{M}_{*}/\mathcal{M}_{\odot})$ = 7.0. Given that this galaxy had the advantage of being very nearby, it is very unlikely that galaxies less massive or luminous ($M_H$ $<$ 14.0) than ESO 410-G005 can be detected with $\sim$1800 sec total on-source integration time on a 4m class telescope. The photometry performed in \cite{Kirby2008} is also in agreement with this statement. In order to penetrate down to even lower masses and luminosities, significantly longer exposure times are needed or larger telescopes.

\subsubsection{UKS 1424-460}
The estimated limiting surface brightness of 22.8 mag arcsec$^{-2}$ was insufficient to detect this galaxy. A crowded field compounded with the shallow image depth precluded even a marginal detection.

\subsubsection{HIPASS J1919-68}
The low-velocity ($v_{hel}$ = 242 km s$^{-1}$) of the H\textsc{i} source, HIPASS J1919-68 detected in \citep{Banks1999}, and inspection of its H\textsc{i} environment suggests that is most likely an isolated high velocity cloud complex. Coincidental to HIPASS J1919-68 our \textit{H}-band image shows a pair of faint smudges, potentially a pair of dIrrs which are less than one arcmin in size. Given the small optical diameter of this source relative to the H\textsc{i} heliocentric velocity, it is an unlikely optical counterpart.

\section{Discussion} \label{discussion}

\subsection{Classifying the luminosity profiles of dIrrs} \label{SBP}

The surface brightness profiles (SBPs) of spiral galaxies have long been known to deviate from simple exponentials \citep{Freeman1970}. Modern deep imaging studies of local spirals have further revealed the extent of variation in their structural profiles \citep[][and references therein]{Erwin2005,Pohlen2006,Erwin2008}. In addition to classical Freeman types, modern classification of spirals are threefold. Two are extensions to the the classical `Freeman Types': The first (Type I) describes profiles which exhibit no `break' in slope and are well described by an exponential fit; and second (Type II), describes a `truncated' profile where the slope of the outer component is steeper than the inner component after the break \citep{Pohlen2006}. A third type (Type III) describes an `anti-truncated' profile and could be considered the inverse of Type II \citep{Erwin2005,Pohlen2006}.

In Sect. \ref{intro} we described how surface photometry of dIrrs were typically performed with a function of fixed shape, such as modified exponential function, sech function or exponential. Yet significant variation exists in the shapes of the SBPs of dIrrs. Indeed many SBPs of dIrrs in this study (see, Fig. \ref{SBresults1} and Table \ref{dpars}) are not well described by simple exponential fits with only $\sim$20\% of dIrrs in this study and KJRD08 considered Type I. Rather a single or double component function with a free shape parameter such as the S\'{e}rsic is robust enough to describe the SBPs of the dIrrs contained within this study and KJDR08 \citep[see,][for an alternative discussion on the S\'{e}rsic function in comparison to the modified exponential distribution function]{Noeske2003,Meyer2014}. 

The majority of galaxies in this study and others \citep[e.g.][]{Vaduvescu2005,Fingerhut2010} exhibit, `flattened' or `downbending' profiles like the classical Freeman Type II profiles for spiral galaxies. Type II breaks in the structural profile of irregular galaxies have been observed in other studies \citep{Herrmann2013,Hunter2006}, however the radius bins of their SBPs are much coarser than those presented in this study, KJRD08, and others \citep[][references therein]{McCall2012}; and so these might otherwise look like classical Type IIs. 

Type III profiles are characteristic of galaxies undergoing a starburst or star formation event \citep{Noeske2003,Hunter2006}. Type III dwarf irregulars are not readily identifiable in KJRD08, however this study can find two examples, NGC5253 and ESO324-G024. As expected, NGC5253 is a starbursting dIrr \citep[][references]{Lopez-Sanchez2012}, whilst ESO324-G024 is also undergoing a star formation event \citep{Bouchard2009}. Additional structural types of dwarf irregulars exist in addition to the three SBP types commonplace in spiral galaxies. Figure 1 in \cite{Herrmann2013} illustrates these different types. 

Yet a fourth type of dIrr SBP is identified in \cite{Herrmann2013}. Type Flat Increasing (FI) SBPs are characterized by a flat or even increasing central SBP followed by a decreasing profile after some break. No Type FI SBPs are identified in the dIrrs analysed in KJDR08, or this study. We suspect that Type FI optical SBPs are representative of dwarf galaxies with bright star forming regions dispersed in the disk outside the geometric centre. NIR observations of dIrrs are significantly less sensitive to irregularities that would otherwise contribute more flux in an isophote bin outside the geometric centre. The resulting SBF might otherwise look more like a Type II than a Type FI when viewed in the NIR. This statement is supported by Figure 18 in \citep{Herrmann2013} which plots the averaged profiles of Type II and III galaxies for the NUV and \textit{V}-bands. The change in slope at the break in the \textit{V}-band is less severe than in the \textit{NUV}-band as expected.

The type III galaxies, NGC5253 and ESO324-G024 were fit with a double S\'{e}rsic to account for the `upbending' central component. The residuals of these galaxies in Fig. \ref{SBresults1} demonstrate the success of a double S\'{e}rsic function in fitting a component in addition to the underlying stellar disk. The double S\'{e}rsic function was also successfully used to fit the profiles of dwarf galaxies which include a nucleated, Type II component in addition to the underlying exponential or Type II disk. Example galaxies in this study include the BCD galaxies AM 0605-341 and KK 49. \cite{Vaduvescu2006} also found success fitting BCD galaxies in the LV and VC using a sech function and an additional Gaussian to describe the central starburst.

\subsection{Colour-Stellar mass scaling relationships} \label{ONIR}

\begin{figure}
\includegraphics[scale=0.6]{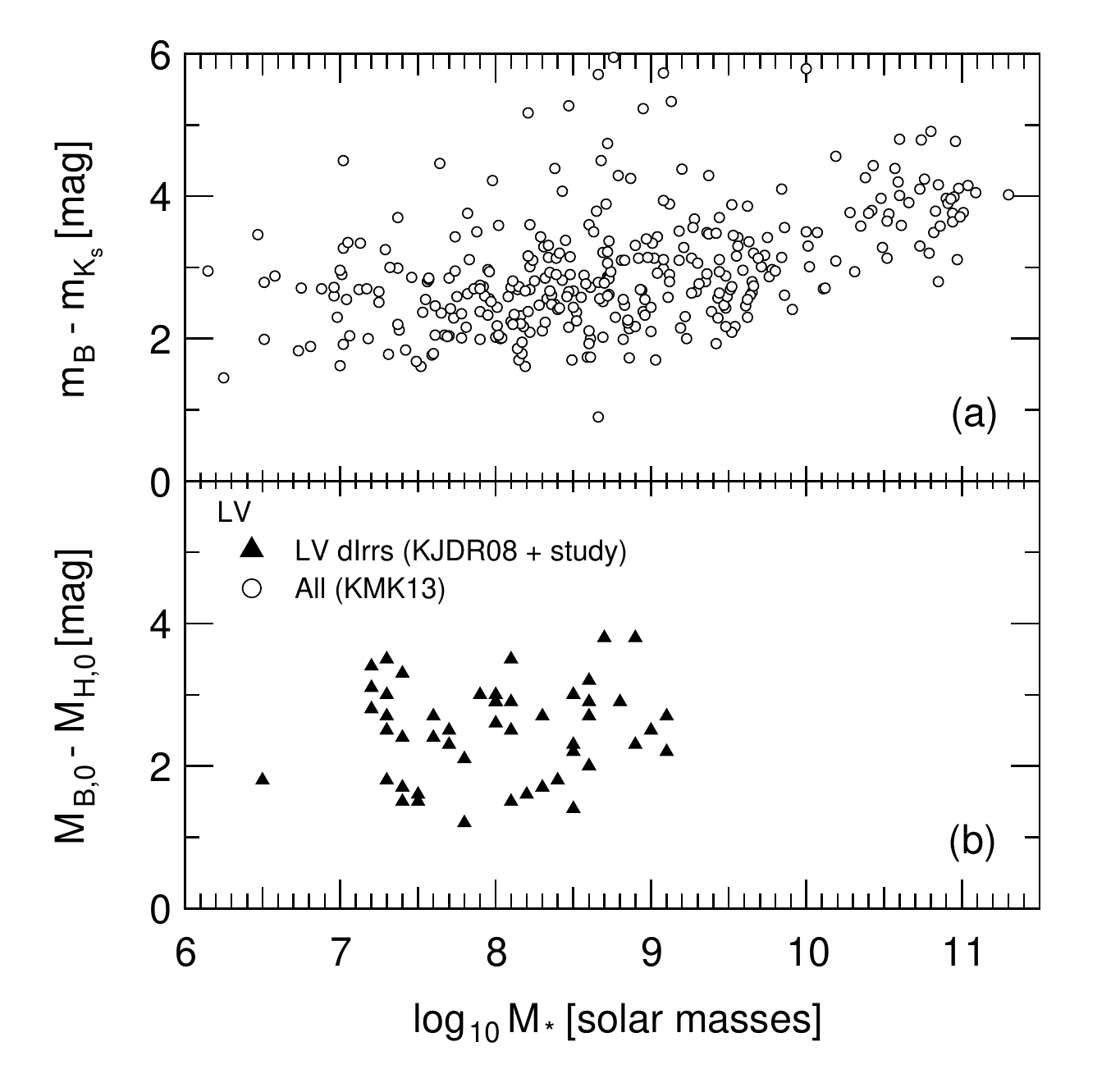}
\caption{The color-stellar mass relationship for various data sets: \textbf{(\textit{a})} for the various observed \textit{B}-\textit{K} colours of LV galaxies in KMK13; \textbf{(\textit{b})} for extinction corrected \textit{B}-\textit{H} colours of the LV dIrrs in this study and KJDR08.}
\label{BvH}
\end{figure}

We investigate the trends in optical-NIR colour for the entire KMK13 sample and the LV dIrrs in this study and KJDR08 in Fig. \ref{BvH} in order to determine whether colour-scaling relationships will allows us to compare our H-band parameters to those in the optical. Panel (\textit{a}) in Fig. \ref{BvH} contains the KMK13 LV galaxies. The subplot has been deliberately restricted to 6 mags to facilitate comparison to the other data sets. As a result, a handful of extreme outliers lie above the plotted region. We confirm trends in NIR-optical colour stellar mass relationships such that more massive galaxies are also more red for both the LV galaxies and the VC galaxies. Close to the completeness limit of the VC data set, a breakdown in the relationship occurs at $M_{*}/M_{\odot}$ $\sim$ 10$^{9}$ for the KMK13 LV galaxies. Galaxies less massive than the break mass 10$^{9}$ $M_{*}/M_{\odot}$ are predominantly dwarf galaxies with 21\% of those dE/dSph and 66\% dIrrs. 

The study sample LV dIrrs exhibit high variance in the colours and no discernible trend with respect to mass. We note that the uncertainties in the \textit{B}-band plate photometery may account for at least some of the scatter where used in deriving the colours of our data and the KJRD08 sample, however the variance in the colour is systematic (the calculated Pearson-correlation coefficient, $\rho$ = 0.15) across luminosity/stellar mass. Regardless, colour scaling relationships do not provide us a method to easily compare our dIrr properties to the more numerous optical samples.

\subsection{Structural scaling relations - connecting dIrrs and dEs}
The alleged structural dichotomy between the dwarf elliptical (dE) and the regular/luminous elliptical populations in the structural parameter-luminosity space have long since been rectified, particularly when accounting for `core' ellipticals either through exclusion \citep{Gavazzi2005} or via inward extrapolation of the measured surface brightness profile \citep{Jerjen1997}. The current paradigm regarding dEs is that structurally they are the low mass end of a continuous population which include the luminous and regular ellipticals \cite[see][and references therin]{Graham2013}. In Sect. \ref{SBP} we elucidated on the variation of dIrr SBPs. How might the structural properties of dIrrs compare to current understanding? Are the structural properties of dIrrs and dEs dichotomous, or are they instead compatible?

Despite the paradigm that late type galaxies might morphologically evolve into early types through environmental processing \cite[e.g.][]{Kormendy1985,Boselli2006} via gas stripping \citep[e.g.][]{Mayer2006} and/or tidal stirring \citep[e.g.][]{Kazantzidis2011} a structural comparison of dIrrs and dEs has been performed only recently for VC dwarfs \citep{Meyer2014}. The truncation of the star forming and H\textsc{i} disks independently of the underlying stellar disk \cite[][references therein]{Koopmann2004,Gavazzi2013} suggests that the underlying stellar structural properties of late type galaxies are retained through evolution. Encouragingly, \cite{Meyer2014} found that the structural properties of VC BCDs overlap with typically more compact dEs and VC dIrrs overlap with more diffuse dEs. Recently \cite{Weisz2011b} found for their nearby sample of 60 dwarf galaxies (D $<$ 4Mpc) that irrespective of morphological type the majority formed the bulk of their mass prior to z $\sim$ 1. Furthermore, the mean SFHs of each morphological type are similar and only diverge within the last Gyr. The evidence is largely in favour of a similarity between the structural properties of dIrrs and dEs, but to what extent is this the case for the LV dIrrs contained within this study?

\begin{figure}
\includegraphics[trim = 2mm 2mm 2mm 2mm,scale=0.6]{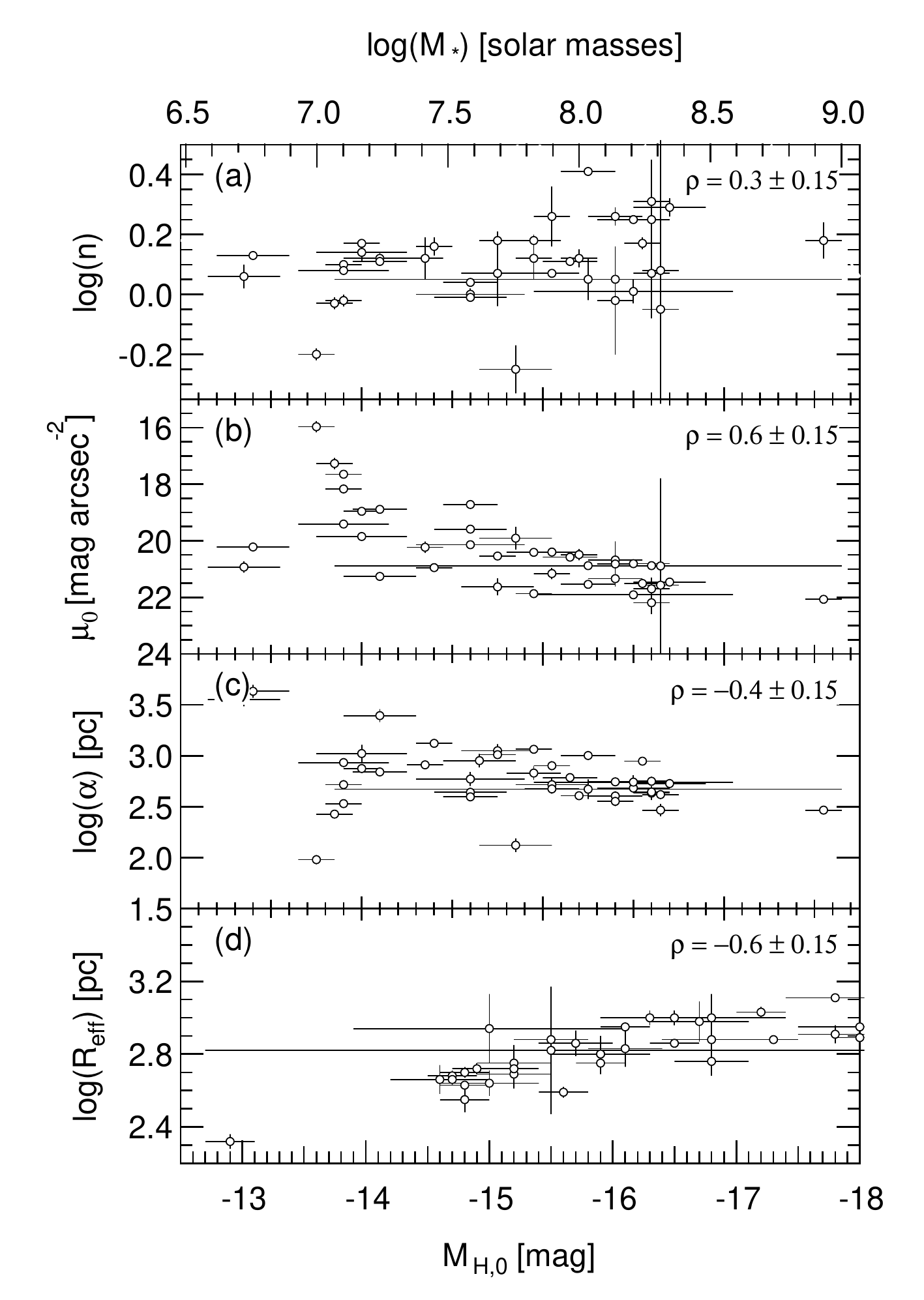}
\caption{The structural scaling relationships and correlations for the dIrrs in this study and KJDR08 with respect to luminosity for: \textit{\textbf{(a)}} the logarithm of the S\'{e}rsic index; \textit{\textbf{(b)}} the central surface brightness; \textit{\textbf{(c)}} the logarithm of the scale length; and \textit{\textbf{(d)}} the logarithm of the effective radius.}
\label{scalerel}
\end{figure}

Previous studies \citep{Binggeli1998,Jerjen2000a} have found correlations between the structural properties of Cen A, Scl and VC dE with respect to their stellar luminosity. We explore these correlations in the four plots of Fig. \ref{scalerel} for our morphologically `clean' sample of dIrrs. This sample contains the 42 single component dIrrs that were successfully detected and analysed in KJDRO8 and this study with redshift independent distance measurements. This selection naturally excludes the two component BCDs. \cite{Vaduvescu2006} showed that for their sample of 16 VC BCDs the diffuse component represents the overwhelming majority of the NIR light for most BCDs, the star-burst enhancing the flux by less than about 0.3 mag. However, a relatively constant J-Ks = 0.7-0.9 mag at all radii suggests that the enhanced star formation could also lower the stellar mass to light ratio for the entire galaxy. We exclude analysis of the BCDs in this study but see \cite{Amorin2009}. The structural properties of dIrrs are generally correlated with the underlying stellar mass. We find the weakest correlation with the shape of the luminosity profile log(n), but find relatively strong correlations between their central surface brightness, effective radius, and scale lengths. 

\begin{figure*}
\includegraphics[trim = 5mm 5mm 5mm 5mm,scale=0.6]{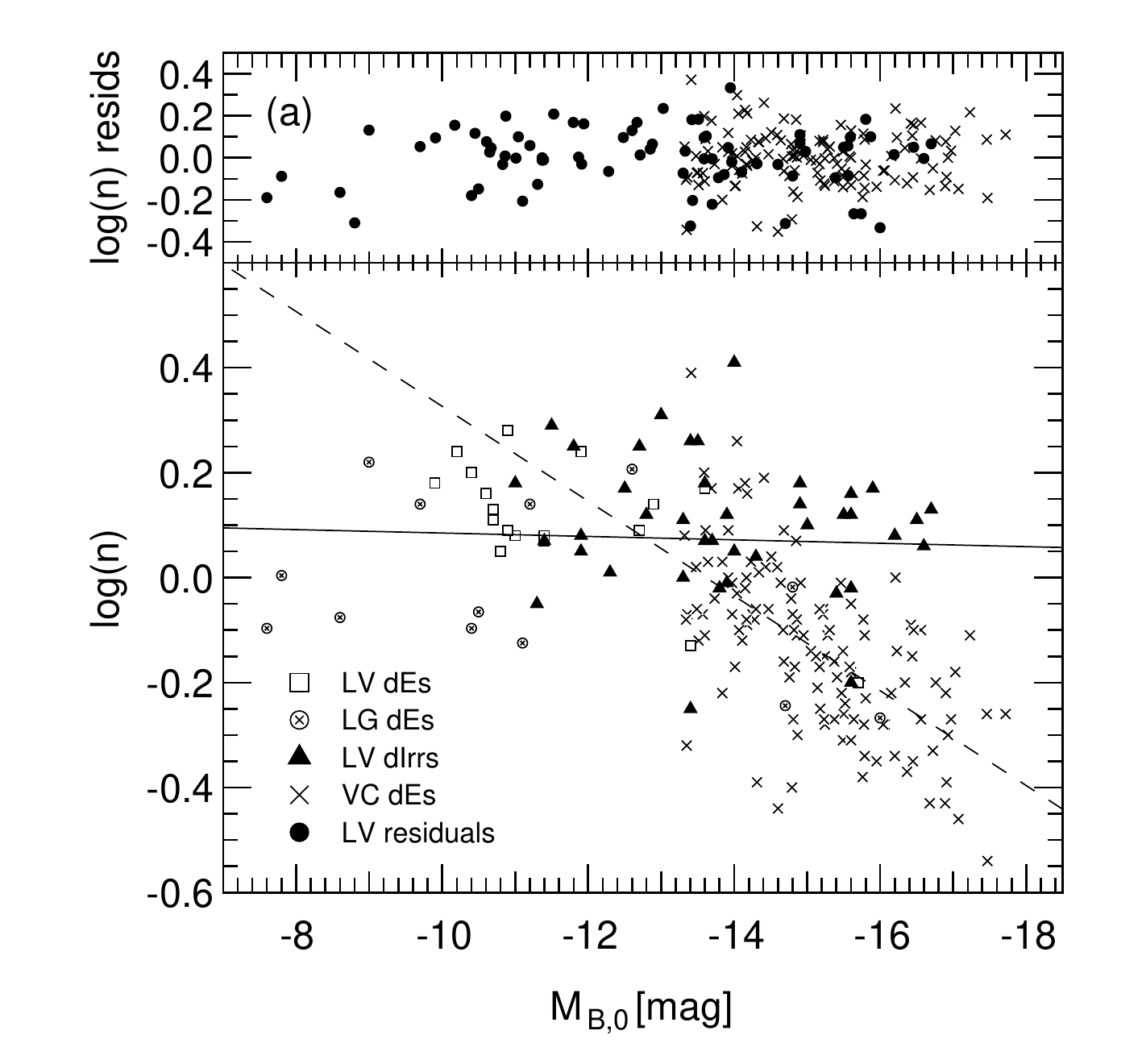} \includegraphics[trim = 5mm 5mm 5mm 5mm,scale=0.6]{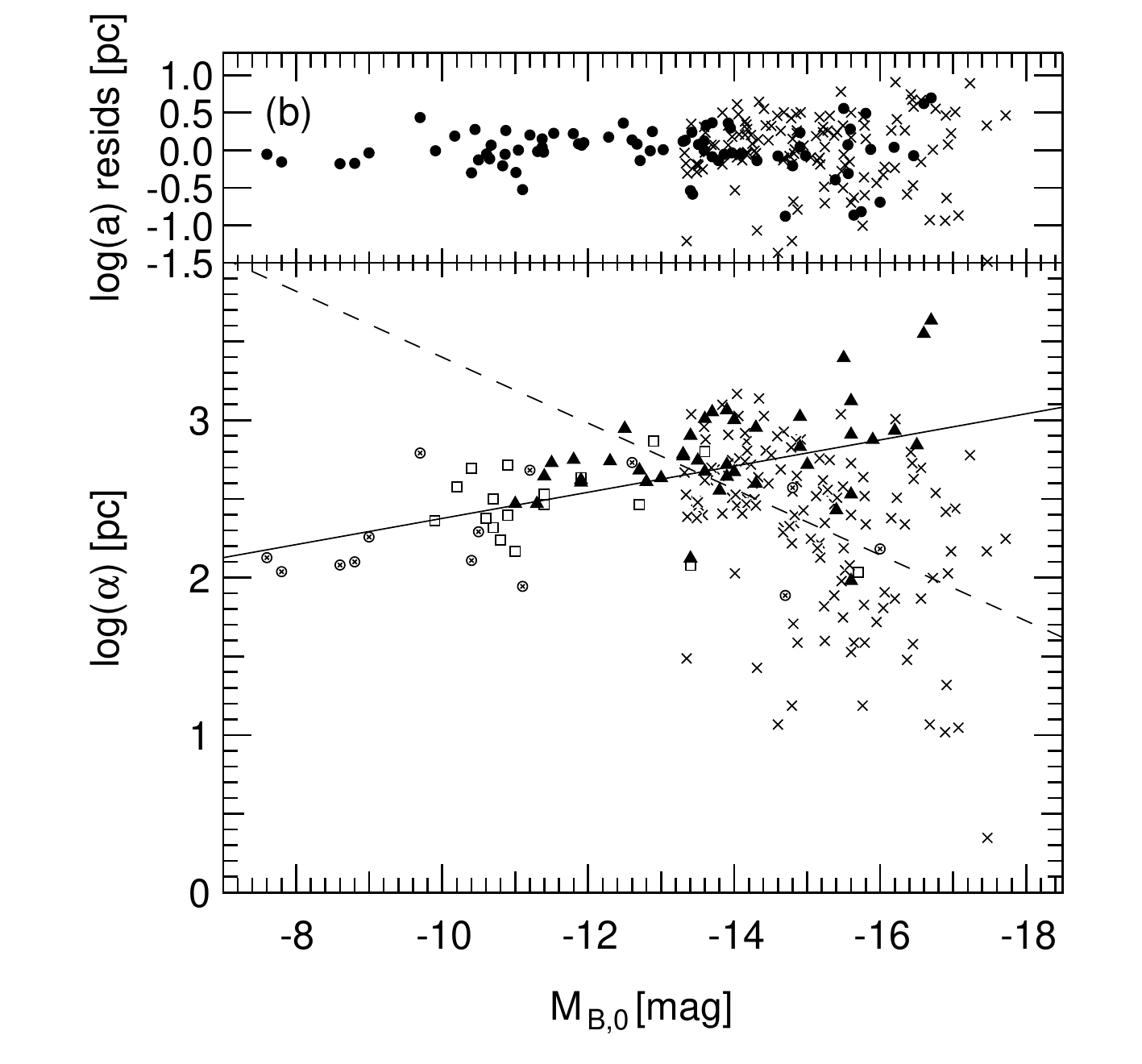} \includegraphics[trim = 5mm 5mm 5mm 5mm,scale=0.6]{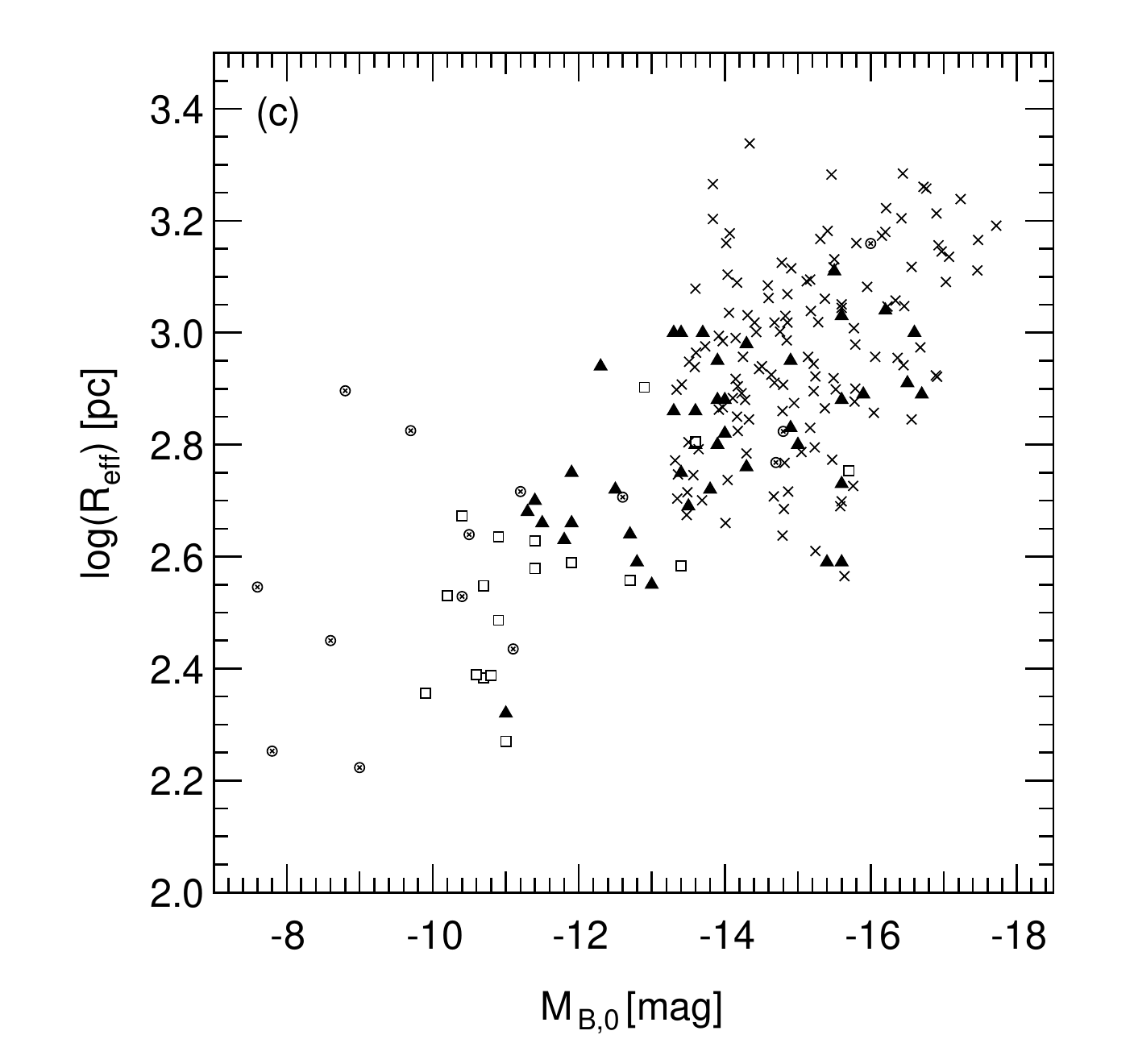}
\caption{Selected structure-luminosity relationships for dIrrs and dEs in the LV and VC. \textbf{(\textit{a})} Comparison of the logarithmic S\'{e}rsic Index and absolute \textit{B}-band magnitude for LV dIrrs in this study and KJDR08 (\textit{black triangles}), LV and LG dEs \protect\citep{Jerjen2000a} (\textit{open squares} and \textit{plussed circles} respectively) and VC dEs \protect\citep{Binggeli1998} (\textit{crosses}). Residuals in the linear regressions are included in the upper subplot. \textbf{(\textit{b})} As for panel (\textit{a}) but instead comparing the logarithmic scale length. \textbf{(\textit{c})} As for panel (\textit{a}) but instead comparing the logarithmic effective radius. Panels \textit{(a)} and \textit{(b)} include fits and their associated residuals such that the dotted line is the linear regression to the VC galaxies and the solid line is the linear regression to the LV galaxies.}
\label{scalecomp}
\end{figure*}

We compare the trend in structural properties with respect to the \textit{B}-band luminosity (Fig. \ref{scalecomp}), in a similar fashion to Fig. 6 in \cite{Amorin2009} with two important differences. Firstly, we compare the \textit{H}-band derived structural parameters to the prominent optically derived properties of dEs in the literature. In doing so we reduce the flux contribution of the young stellar component of dIrrs, sampling the underlying stellar disk and thus providing the fairest comparison to the typically quiescent dEs. Additionally, we include the Scl and Cen A dwarf ellipticals contained within \cite{Jerjen2000a} and the VC dEs \citep{Binggeli1998} to attempt to address the influence of environment. Figure \ref{scalecomp} shows that a significant fraction of the LV dIrr population exhibit underlying structural properties similar to both LV and VC dEs. Many dIrrs are therefore structurally eligible for the various hypothetical morphological transformation scenarios. \cite{Meyer2014} found for their sample of VC dIrrs and BCDs ``...after termination of starburst activity, the BCDs will presumably fade into galaxies that are structurally
similar to ordinary early-type dwarfs. In contrast, the irregulars are more diffuse than the BCDs and are structurally similar to the more diffuse half of the Virgo early-type dwarfs.'' Indeed, the structural relationships explored in their Fig. 6 indicates that the luminosity-central surface brightness, and luminosity-effective radius relationships are largely compatible for the VC dIrr and dE populations. Indeed, we also find a common relationship between the effective radius and luminosity for the LV dwarfs and VC dEs (see Fig \ref{scalecomp} panel \textit{(c)}), and a significant fraction of the LV dIrrs overlap with both the LV and VC dEs. However, is it very surprising that larger galaxies are more luminous irrespective of dwarf morphological sub-type or environment?

One advantage of fitting SBPs profiles of dIrrs with the S\'{e}rsic function over a sech or modified exponential is that we have access to shape and size information through the S\'{e}rsic parameters, n and $\alpha$. Whilst concentration indexes are model independent we have demonstrated in Sect. \ref{SBP} how successfully the S\'{e}rsic function fits the NIR stellar profiles of dIrrs. Therefore the scale length is an excellent indicator of the `compactness' irrespective of the shape of the luminosity profile, unlike the effective radius or Concentration index. We therefore explore the shape-luminosity and scale length-luminosity relationships in panels \textit{(a)} and \textit{(b)} of Fig. \ref{scalecomp}. 

Linear regressions to structure-luminosity relationships for the LV galaxies and VC cluster galaxies show significant differences in both slope and scatter around the established trend lines. In panel (\textit{a}), the logarithm of the S\'{e}rsic index is plotted against the \textit{B}-band luminosity along with separate fits to the LV galaxies and VC dEs. Remember that the structural parameters for the dIrrs in this study and KJDR08 are obtained with the \textit{H}-band data, but somewhat unorthodoxly plotted against their \textit{B}-band magnitudes (obtained from the literature, see Tab. \ref{sampleprops} so as to avoid resorting to scaling relationships which we showed were uncertain for low mass galaxies in Sect. \ref{ONIR}). The LV galaxies exhibit similar scatter around the established trends in the log(n)-luminosity scale relationship with respect to the VC dEs. Additionally, the overall trend is flat whilst there is a negative trend between the VC dE. Conversely in panel (\textit{b}), which plots the log of the scale parameter against the \textit{B}-band luminosity, the variance around the trends is much higher for the VC dEs when compared to the LV galaxies. The evidence therefore suggests that the VC environment strongly regulates the underlying structural size and shape of dwarf galaxies with respect to what is typical in the LV.

There are however some caveats to this conclusion. Notably, the scatter in the scale length relationship for dIrrs increases to similar levels to the VC dEs at high luminosities. We have not included VC dIrrs in our analysis due to a lack of photometric coverage for LV galaxies in the southern hemisphere.  Modern surveys \citep[e.g.]{Ferrarese2012} tend to use ugriz whilst the optical coverage of many LV galaxies in the southern hemisphere are restricted to Landolt photometric systems. As we have discussed previously, we prefer not to use colour transformation relations. Therefore it remains to be seen what deep NIR observations of low mass VC dIrrs (and dEs) would reveal with respect to the shape-luminosity and scale-luminosity relationships. The second consideration is sample completeness limits for the VC dEs. How might the properties of significantly less massive VC dEs change the picture established in this study?

\section{Summary}
We have presented deep \textit{H}-band surface photometry and analysis of 40 LV galaxies obtained using the IRIS2 at the 3.9m AAT. We probed to a typical surface brightness of $\sim$25 mag arcsec$^{-2}$, reaching a 40 times lower stellar density than 2MASS. By employing extremely careful and rigorous cleaning techniques to remove contaminating sources, we performed surface photometry on 33 detected sources deriving the observed total magnitude, effective surface brightness and best fitting S\'{e}rsic parameters. We discuss the null detected galaxies in context with what is known from the literature and derive upper limits to their stellar luminosity. For the sufficiently detected galaxies we infer physical parameters from these measured quantities using the best available distances in the literature. 

We make images quality and surface photometry comparisons to 2MASS and VHS. The comparisons demonstrate that deep targeted surveys are still the most reliable means of obtaining accurate surface photometry for low surface brightness galaxies, although VHS represents a significant increase in image depth over 2MASS and may provide a promising data set to extract galaxy photometry down to $\sim$23-24 mag arcsec$^{-2}$.  

We confirm the correlation between \textit{B} and \textit{H}-band luminosities demonstrated in \cite{Kirby2008}, however LV dwarf galaxies below the break mass $M_{*}/M_{\odot}$ $\sim$ 10$^{9}$ do not obey this relationship. Local Volume dIrrs are significantly varied with respect to colour, eliminating the possibility of using optical-NIR colour transformations to facilitate comparison to the more widely available optical data sets.

Significant variation exists in the shapes of the SBPs of dIrrs. We have successfully characterized these profiles using a pure S\'{e}rsic or a double S\'{e}rsic function. The majority of galaxies in this study exhibit `flattened' or `downbending' profiles like the classical Freeman Type II profiles for spiral galaxies, whilst only $\sim$20\% are well described by a pure exponential profile.

The structure-luminosity relationships are investigated for our `clean' sample of dIrrs. {We recover the expected correlations between the luminosity, effective radius, scale length and central surface brightness but find no obvious correlation with respect to the shape of the luminosity profile.  We have also demonstrated that a significant fraction of the LV dIrr population have underlying structural properties similar to both LV dEs and diffuse VC dEs. Linear regressions to structure-luminosity relationships for the LV galaxies and VC cluster galaxies show significant differences in both slope and scatter around the established trend lines, suggesting that environment regulates the structural scaling relationships of dwarf galaxies in comparison to their more isolated counterparts.

\section{Acknowledgements}

TY would like to thank the Research School of Astronomy and Astrophysics, CSIRO Astronomy and Space Science division and the Australian Astronomical Observatory through their support with the Research Training Scheme, ATNF Graduate program and the Trevor Burgess Scholarship respectively. TY would also like to thank Marshall McCall, Ovidiu Vaduvescu and Robin Fingerhut for kindly providing the KILLALL routine and further assistance through private correspondence. The authors would like to extend their gratitude to Stuart Ryder for his assistance with the 2011 observation runs. We thank Igor Karachentsev, whose revision comments have improved and streamlined the discussions in this paper.

This paper is based on data obtained with the AAT. The study made use of data products from the 2MASS, which is a joint project of the University of Massachusetts and the Infrared Processing and Analysis Center at the California Institute of Technology, funded by the National Aeronautics and Space Administration and the National Science Foundation. Support for IRIS2 data reduction within ORAC-DR is provided by the Joint Astronomy Centre. This research has made use of the GOLD Mine Database. This research has made use of the NASA/IPAC Extragalactic Database (NED) which is operated by the Jet Propulsion Laboratory, California Institute of Technology, under contract with the National Aeronautics and Space Administration.

\footnotesize
\bibliographystyle{mn2e}
\bibliography{bibclean}
\label{lastpage}

\appendix
\section{Tables and Figures}
\begin{table*}
\centering
\begin{tabular}{llcccccccc}
\hline\hline
Galaxy Name & HIPASS & Type & D & Method & B$_T$ & B$_T$ err & D Ref & B$_T$ Ref \\
& & & [Mpc] & & [mag] & [mag] &\\
(1) & (2) & (3) & (4) & (5) & (6) & (7) & (8) & (9)\\
\hline
ESO 410-G005 & J0015--32 & dTrans & 1.92 & TRGB & 14.9 & 0.1 & 1 & 9  \\
ESO 199-G007 & J0258--49 & dIrr & 6.56 & H & 16.4 & 0.5 &  & 10 \\
ESO 252-IG001 & J0457--42 & dIrr & 7.2 & TF & 14.4 & 0.2 & 2 & 11 \\
KK 49 & J0541+06 & dIrr/BCDG & 5.15 & H & 16.1 & 0.4 &  & 12 \\
AM 0605-341 & J0607--34 & dIrr/BCDG & 7.4 & MEM & 14.1 & 0.5 & 2 & 2 \\
NGC 2188 & J0610--34 & Sm & 7.4 & TF & 12.1 & 0.2 & 2 & 11 \\
ESO 489-G?056 & J0626--26 & dIrr & 4.99 & TRGB & 15.7 & 0.1 & 3 & 9 \\
AM 0704-582 & J0705--58 & Sm & 4.9 & TRGB & 15.0 & 0.1 & 3 & 9 \\
ESO 558-G011 & J0706--22 & dIrr & 8.4 & TF & 14.4 & 0.1 & 2 & 9 \\
ESO 376-G016 & J1043--37 & dIrr & 7.1 & TF & 15.5 & 0.1 & 2 & 10 \\
ESO 318-G013 & J1047--38 & dIrr & 6.5 & TF & 15.0 & 0.1 & 2 & 13 \\
ESO 320-G014 & J1137--39 & dIrr & 6.08 & TRGB & 15.9 & 0.1 & 4 & 10 \\
ESO 379-G007 & J1154--33 & dIrr & 5.22 & TRGB & 16.6 & 0.1 & 5 & 10 \\
ESO 379-G024 & J1204--35 & dIrr & 4.88 & H & 16.6 & 0.1 &  & 10 \\
ESO 321-G014 & J1214--38 & dIrr & 3.18 & TRGB & 15.2 & 0.1 & 5 & 10 \\
CEN 6 & J1305--40 & dIrr & 5.78 & TRGB & 17.65 & 0.08 & 4 & 14 \\
ESO 269-G058 & J1310--46A & dIrr & 3.8 & TRGB & 13.3 & 0.2 & 4 & 11 \\
KK 195 & J1321--31 & dSph/dIrr & 5.22 & TRGB & 17.1 &  & 6 & 13 \\
AM 1321-304 & J1324--30 & dIrr & 4.63 & TRGB & 16.67 &  & 5 & 15 \\
IC 4247 & J1326--30A & dIrr & 4.97 & TRGB & 14.4 & 0.1 & 4 & 10 \\
ESO 324-G024 & J1327--41 & dIrr & 3.73 & TRGB & 12.9 & 0.1 & 5 & 16 \\
UGCA 365 & J1336--29 & dIrr & 5.25 & TRGB & 15.5 & 0.1 & 4 & 10 \\
ESO 444-G084 & J1337--28 & dIrr & 4.61 & TRGB & 15.06 &  & 5 & 13 \\
LEDA 592761 & J1337--39 & dIrr & 4.83 & TRGB & 16.5 &  & 7 & 13 \\
NGC 5237 & J1337--42 & dIrr/BCDG & 3.4 & TRGB & 13.2 & 0.2 & 4 & 11 \\
NGC 5253 & J1339--31A & dIrr & 3.56 & TRGB & 10.8 & 0.2 & 8 & 11 \\
IC 4316 & J1340--28 & dIrr & 4.41 & TRGB & 14.56 &  & 5 & 13 \\
NGC 5264 & J1341--29 & dIrr & 4.53 & TRGB & 12.6 & 0.2 & 5 & 11 \\
ESO 325-G011 & J1345--41 & dIrr & 3.4 & TRGB & 14.0 & 0.2 & 5 & 10 \\
- & J1348--37 & new & 5.75 & TRGB & 17.6 &  & 4 & 2 \\
ESO 383-G087 & J1349--36 & Sm & 3.45 & TRGB & 11.0 & 0.1 & 4 & 10 \\
LEDA 3097113 & J1351--47 & new & 5.73 & TRGB & 17.5 &  & 4 & 13 \\
NGC 5408 & J1403--41 & dIrr/BCDG & 4.81 & TRGB & 12.2 & 0.2 & 5 & 11 \\
UKS 1424-460 & J1428--46 & dIrr & 3.58 & TRGB & 15.2 &  & 5 & 2 \\
ESO 222-G010 & J1434--49 & dIrr & 5.8 & TF & 16.33 &  & 2 & 13 \\
ESO 272-G025 & J1443--44 & dIrr & 5.88 & H & 14.8 &  &  & 10 \\
ESO 223-G009 & J1501--48 & dIrr & 6.49 & TRGB & 13.82 &  &  4 & 13 \\
ESO 274-G001 & J1514--46 & Sd & 3.09 & TRGB & 11.7 & 0.2 &  4 & 11 \\
ESO 149-G003 & J2352--52 & dIrr & 5.9 & TF & 15.05 & 0.1 & 2 & 10 \\
\hline
\end{tabular}
\caption{Sample properties obtained from external sources. References: (1) \protect\citet{Karachentsev2000}, (2) \protect\citet{Karachentsev2013}, (3) \protect\citet{Karachentsev2003}, (4) \protect\citet{Karachentsev2007}, (5) \protect\citet{Karachentsev2002}, (6) \protect\cite{Pritzl2003}, (7) \protect\cite{Grossi2007}, (8) \protect\cite{Mould2008}, (9) \protect\citet{Parodi2002}, (10) \protect\citet{Lauberts1989}, (11) \protect\citet{DeVaucouleurs1991}, (12) \protect\cite{Huchtmeier2000}, (13) \protect\citep{Karachentsev2004}, (14) \protect\citet{Jerjen2000}, (15) \protect\citet{Metcalfe1994}, (16) \protect\cite{Lee2011}.}
\label{sampleprops}
\end{table*}
\clearpage

\begin{figure*}
\setlength{\arraycolsep}{1pt}
\setlength{\parskip}{0pt}
$
\begin{array}{ccc}

\includegraphics[trim = 2mm 2mm 2mm 2mm, clip = true, scale = 0.3]{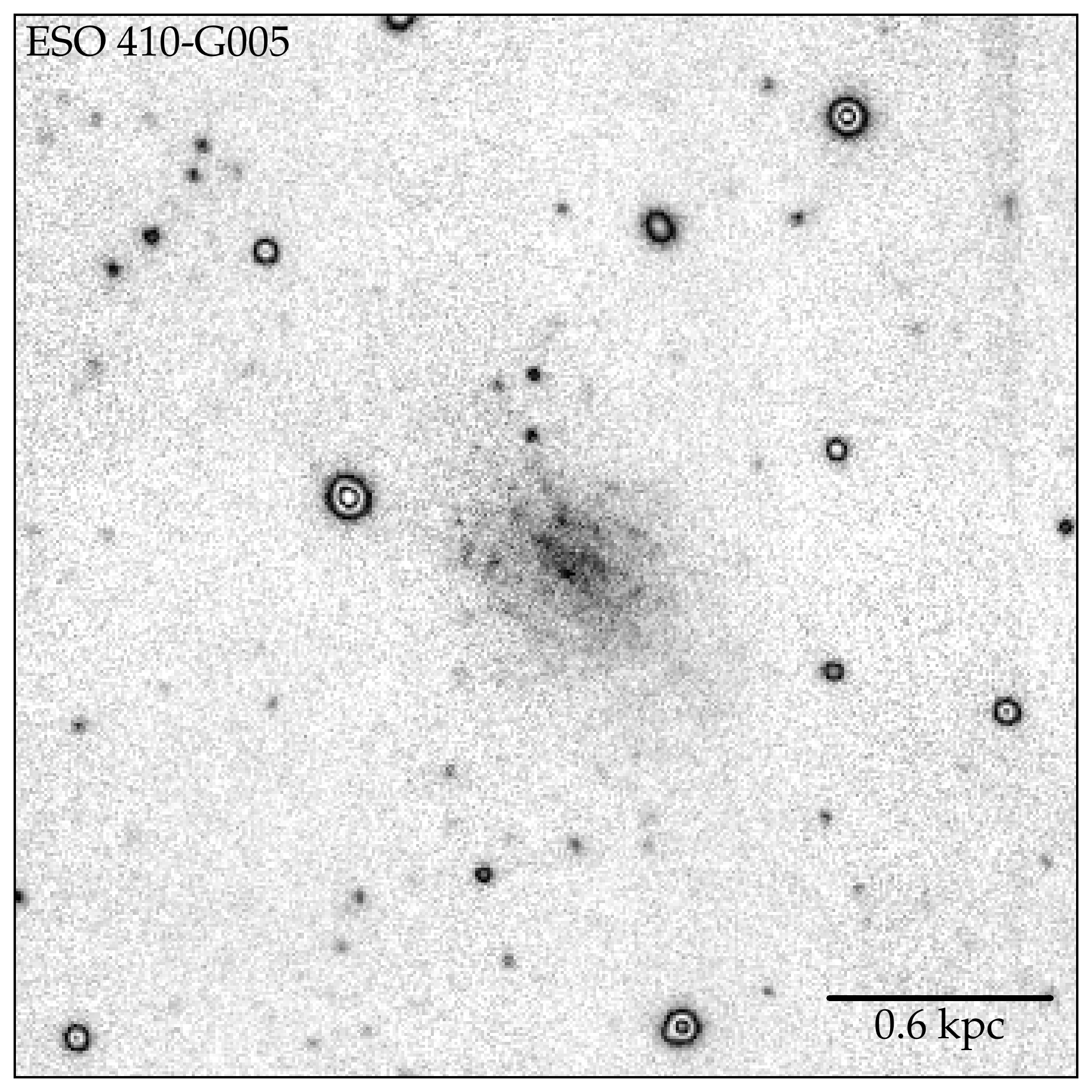}&
\includegraphics[trim = 2mm 2mm 2mm 2mm, clip = true, scale = 0.3]{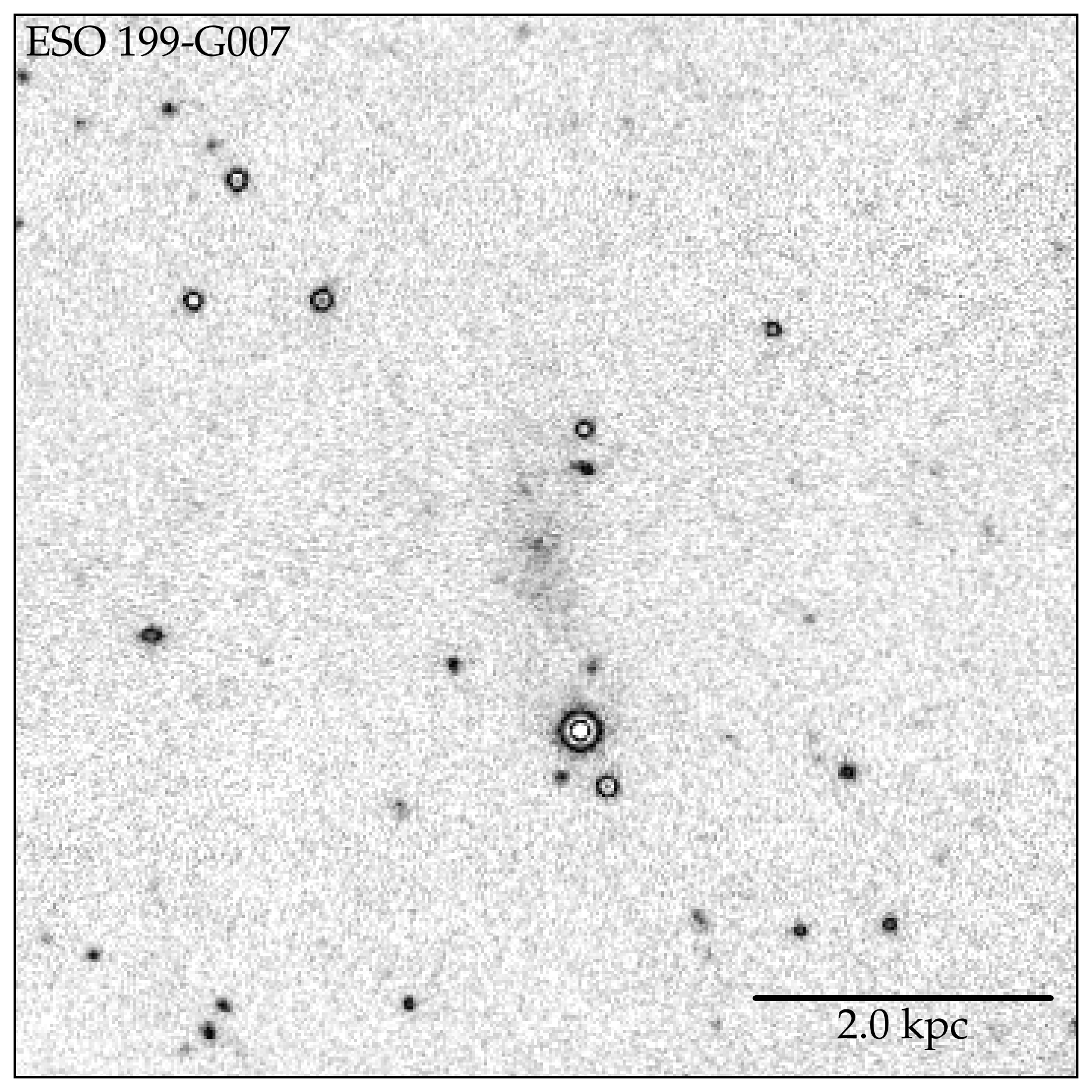}&
\includegraphics[trim = 2mm 2mm 2mm 2mm, clip = true, scale = 0.3]{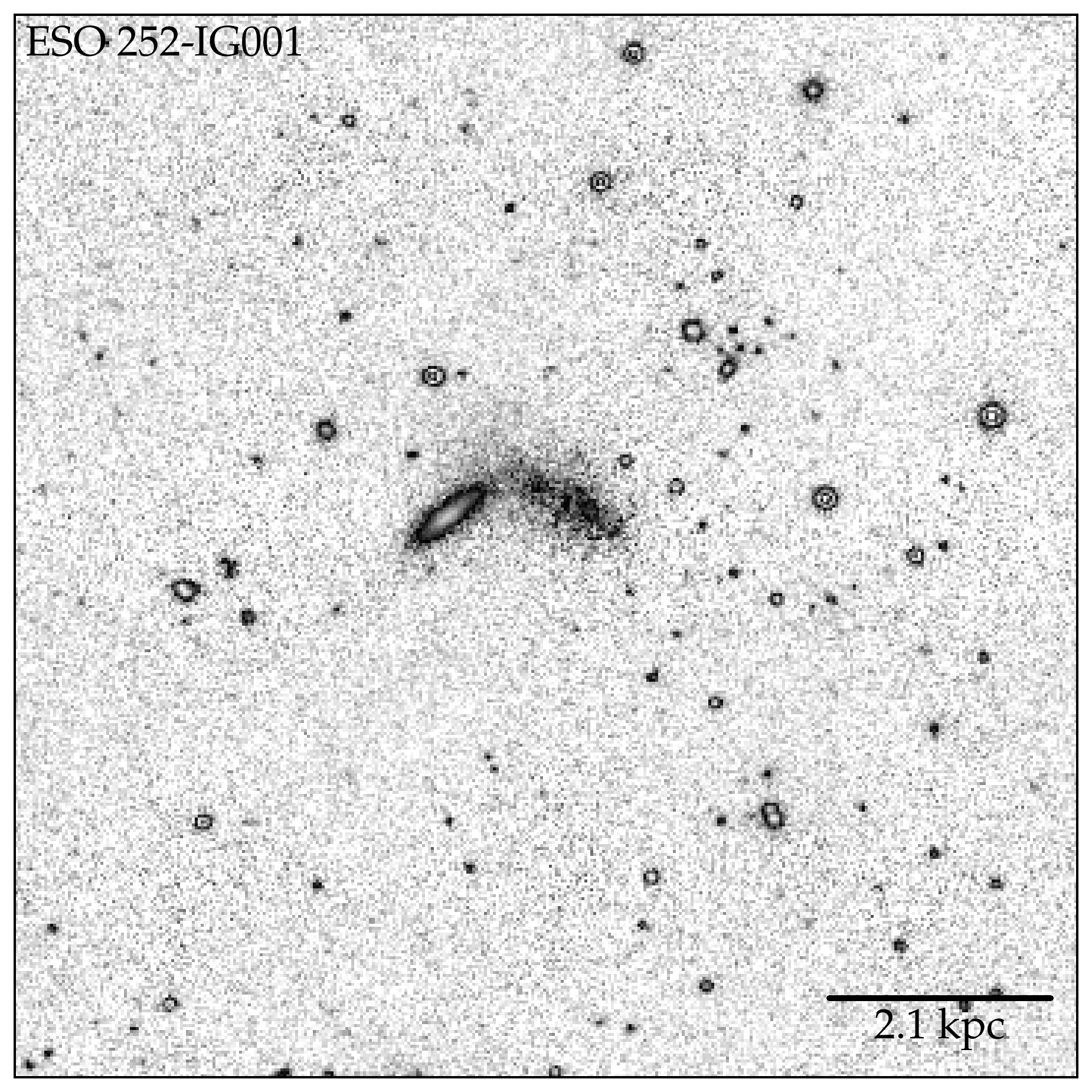}\\

\includegraphics[trim = 2mm 2mm 2mm 2mm, clip = true, scale = 0.3]{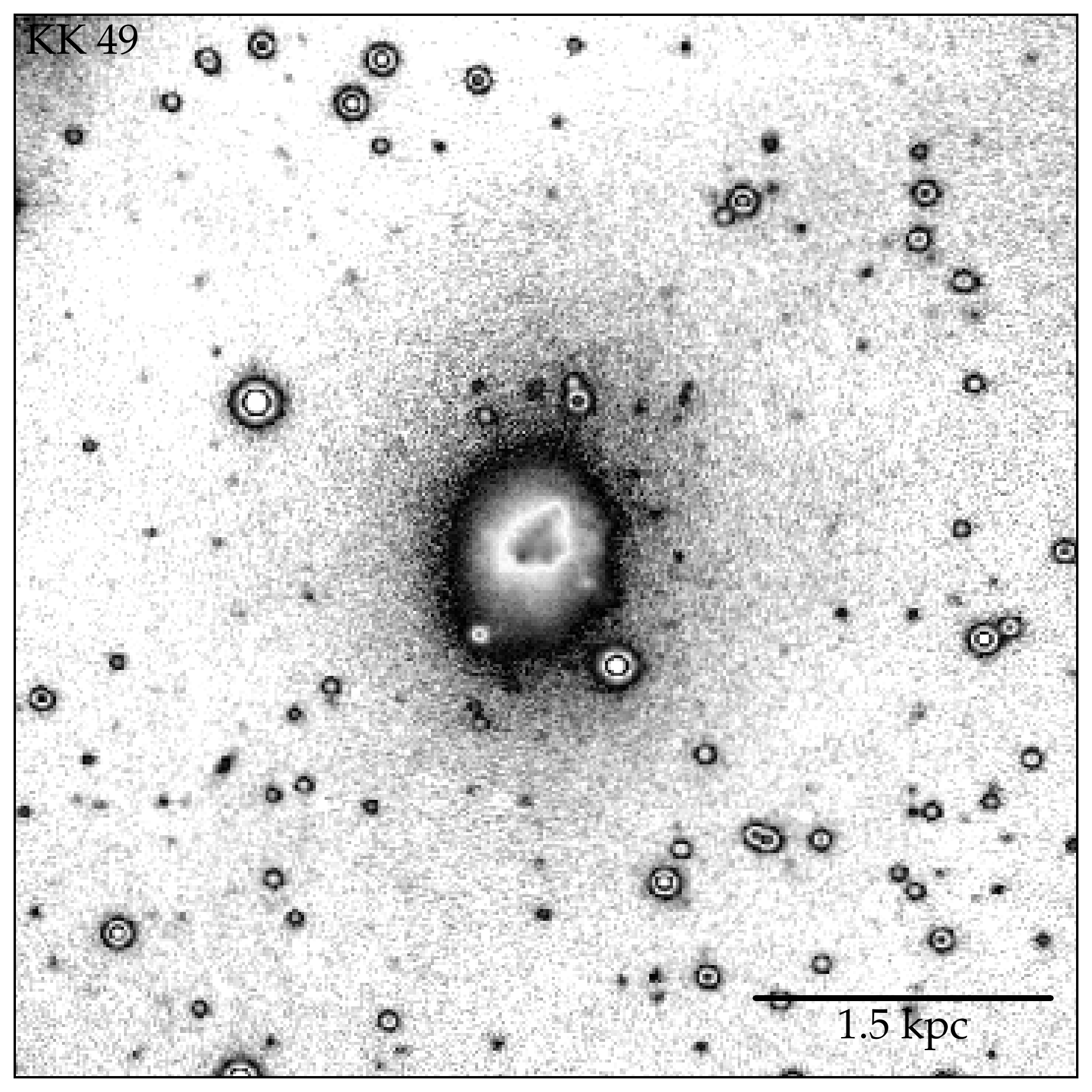}&
\includegraphics[trim = 2mm 2mm 2mm 2mm, clip = true, scale = 0.3]{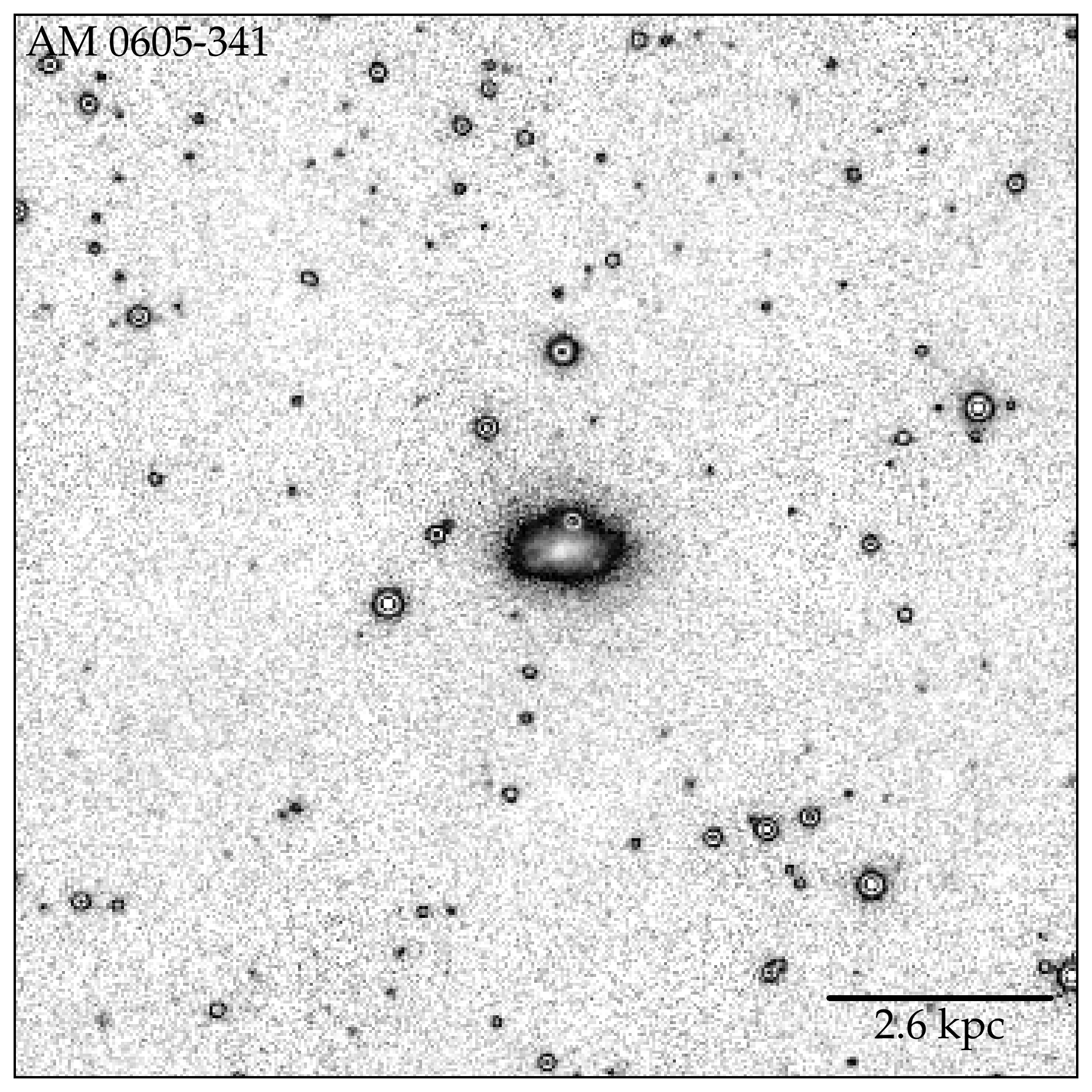}&
\includegraphics[trim = 2mm 2mm 2mm 2mm, clip = true, scale = 0.3]{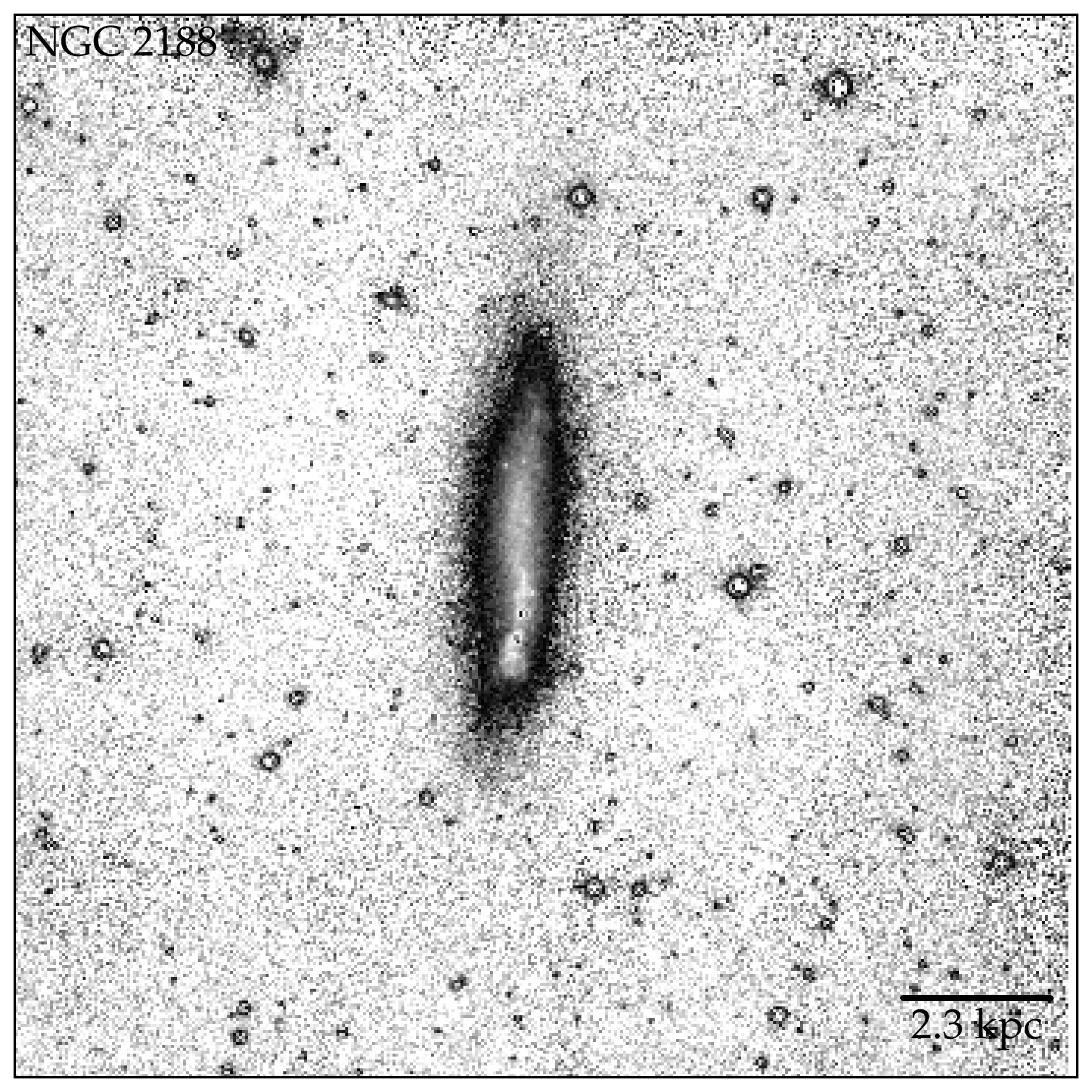}\\

\includegraphics[trim = 2mm 2mm 2mm 2mm, clip = true, scale = 0.3]{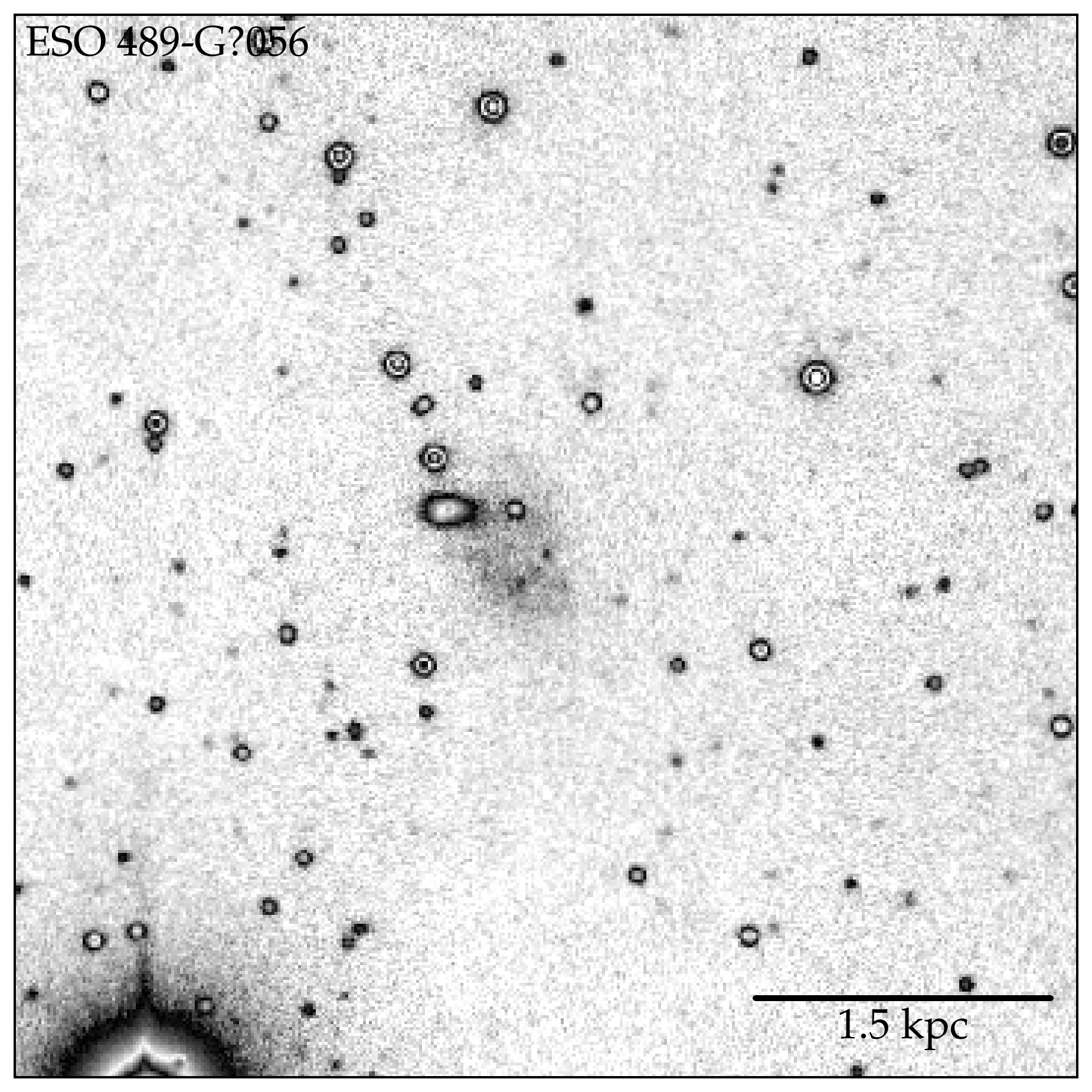}&
\includegraphics[trim = 2mm 2mm 2mm 2mm, clip = true, scale = 0.3]{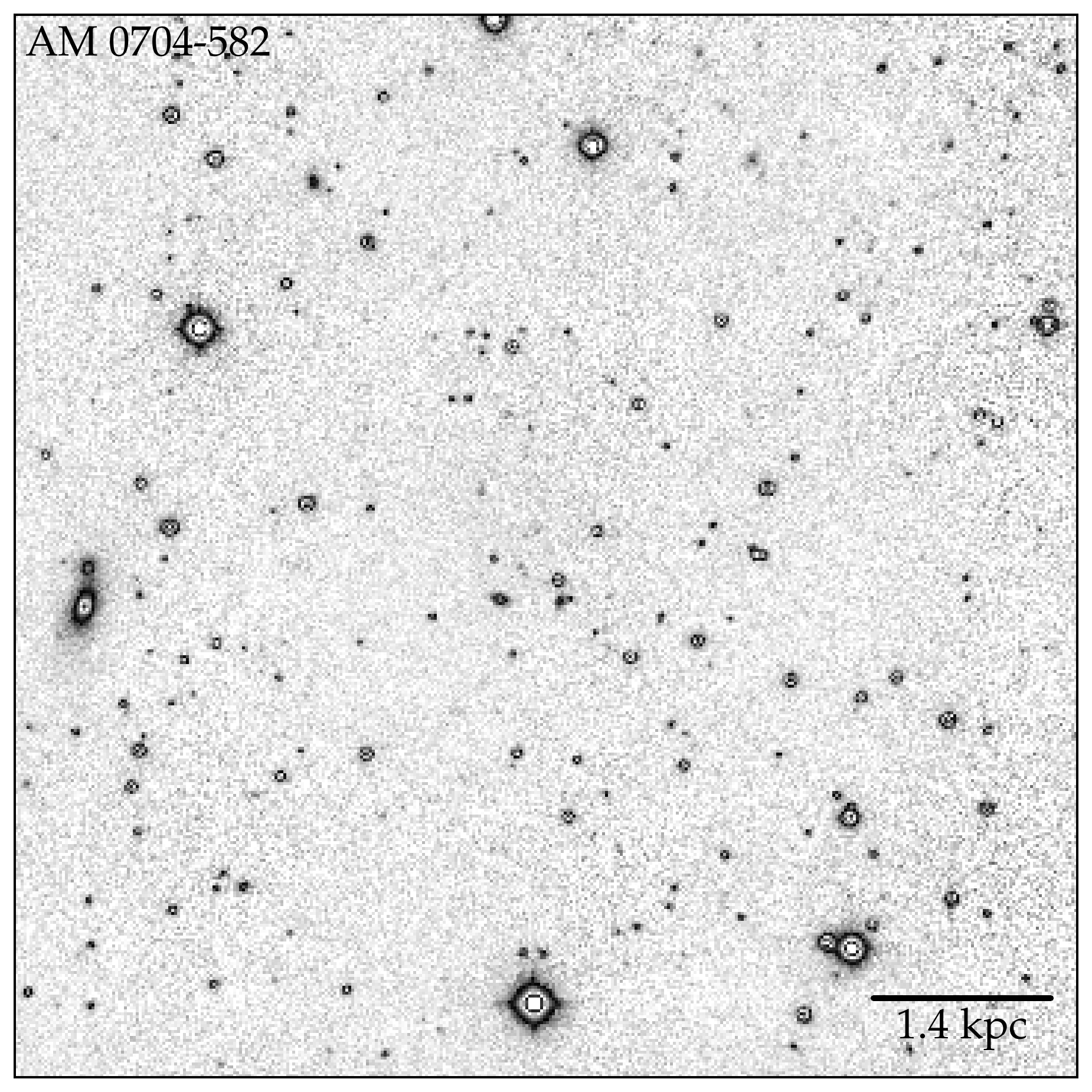}&
\includegraphics[trim = 2mm 2mm 2mm 2mm, clip = true, scale = 0.3]{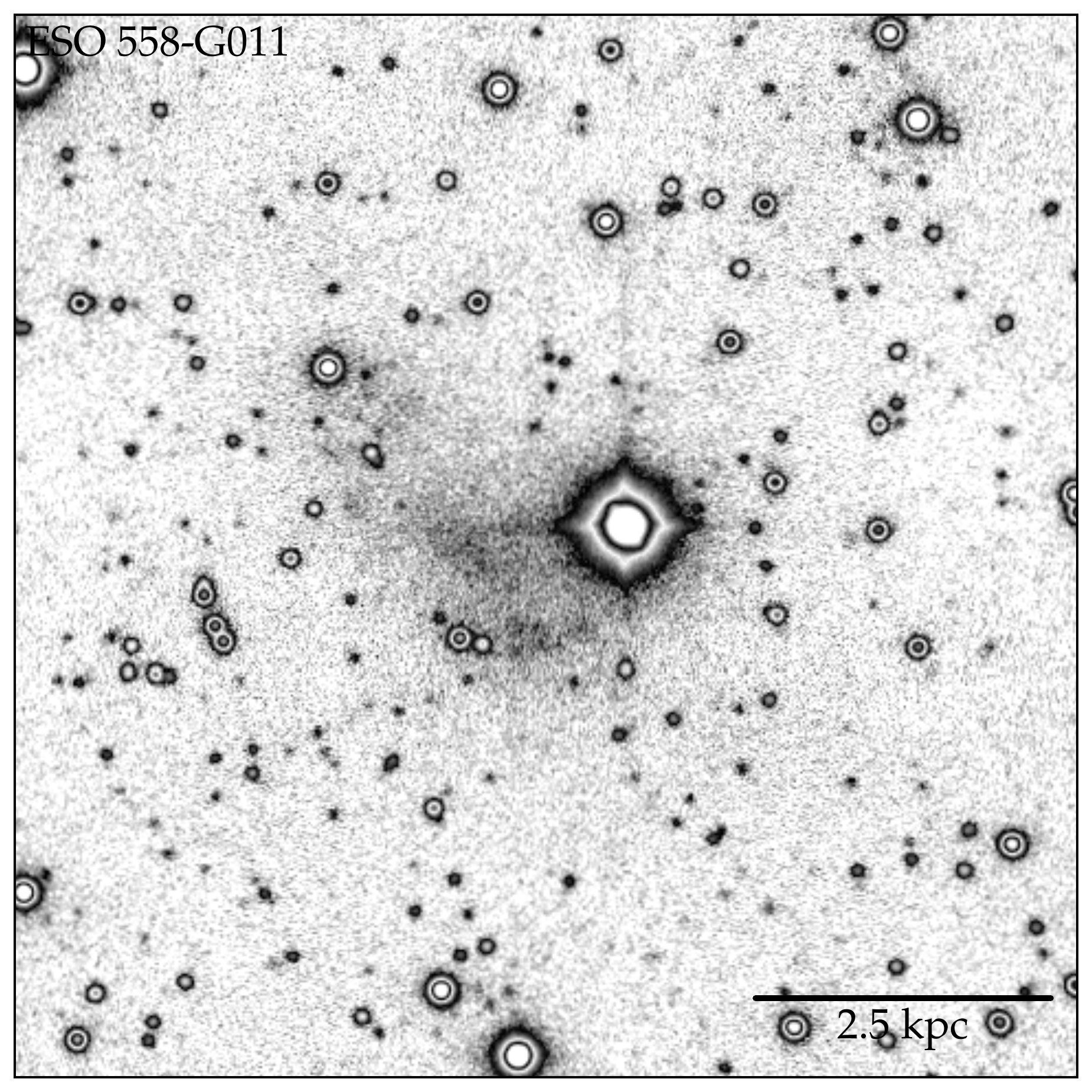}\\

\includegraphics[trim = 2mm 2mm 2mm 2mm, clip = true, scale = 0.3]{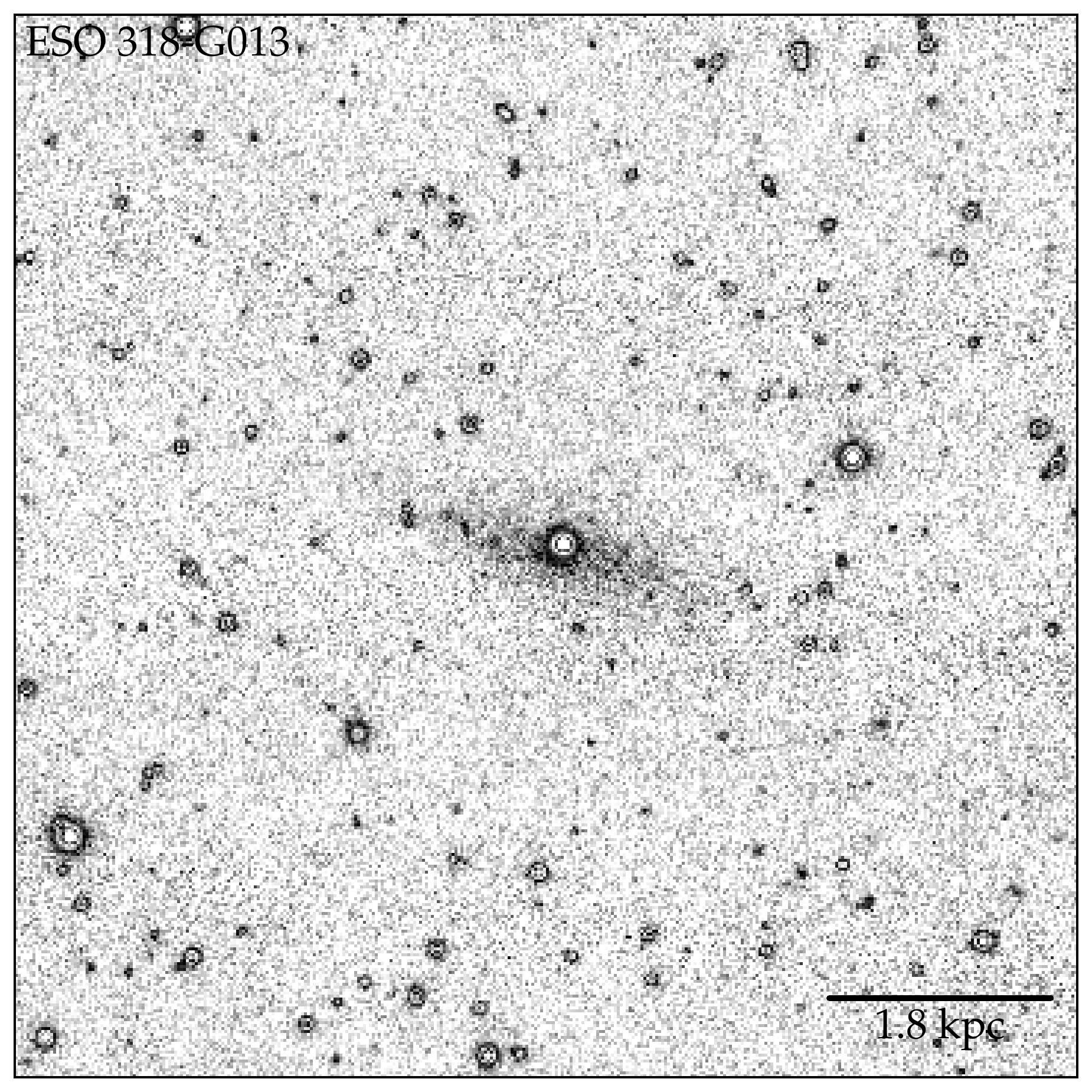}&
\includegraphics[trim = 2mm 2mm 2mm 2mm, clip = true, scale = 0.3]{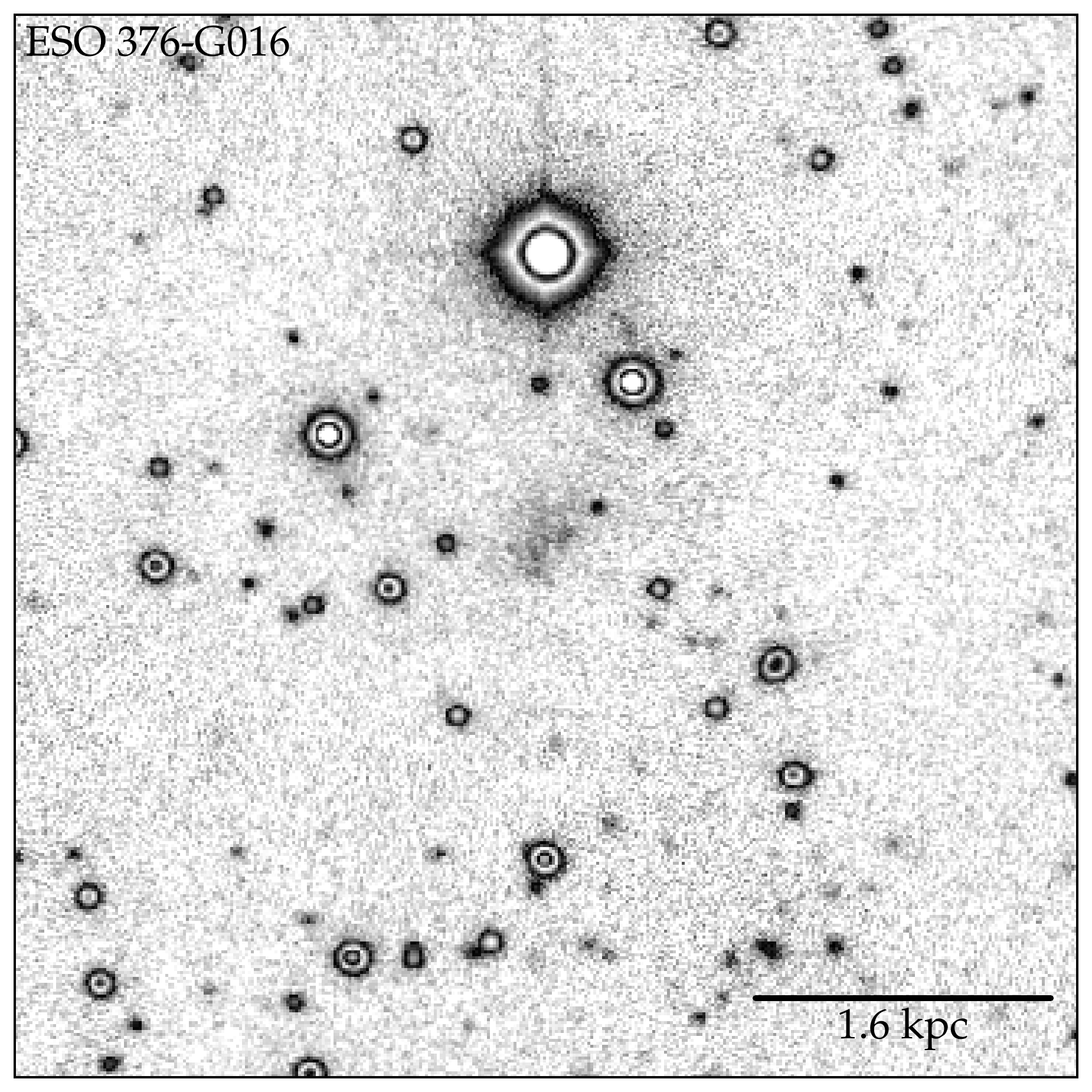}&
\includegraphics[trim = 2mm 2mm 2mm 2mm, clip = true, scale = 0.3]{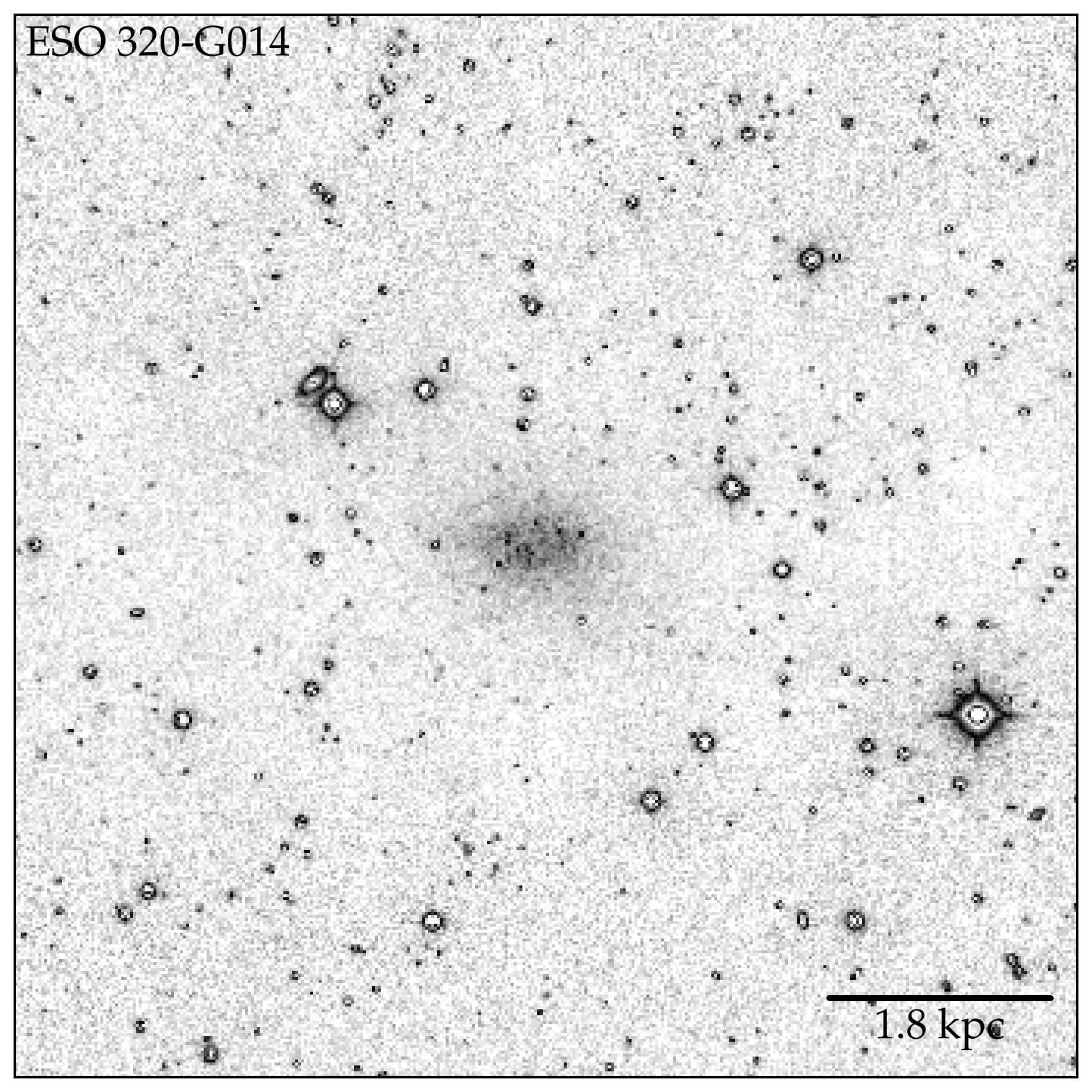}\\

\end{array}
$
\caption{Deep H-band images from the 3.9m AAT using the IRIS2 detector. The scalebar represents 1 arcmin and the corresponding linear distance scale is displayed. The image is oriented such that North is up and East is to the left. The intensity scale begins at the sky level and covers a range of 1000 counts for each galaxy. It is represented by a repeating divergent greyscale, which begins with white (to represent low intensity) through to black (higher intensity) and then back to white (even higher intensity) and so on. }
\label{results1}
\end{figure*}

\addtocounter {figure} {-1}

\begin{figure*}
\setlength{\arraycolsep}{1pt}
\setlength{\parskip}{0pt}
$
\begin{array}{cccc}

\includegraphics[trim = 2mm 2mm 2mm 2mm, clip = true, scale = 0.3]{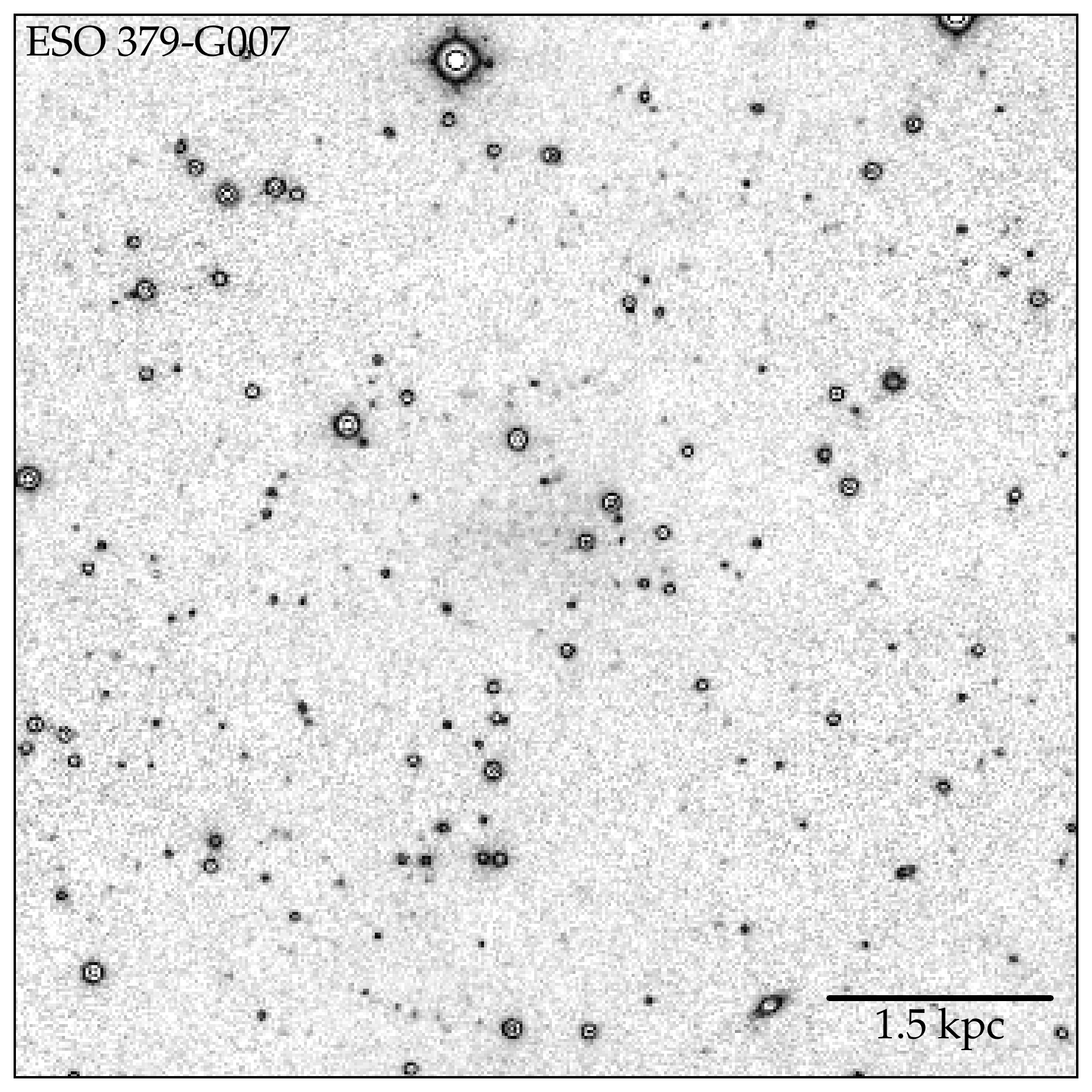}&
\includegraphics[trim = 2mm 2mm 2mm 2mm, clip = true, scale = 0.3]{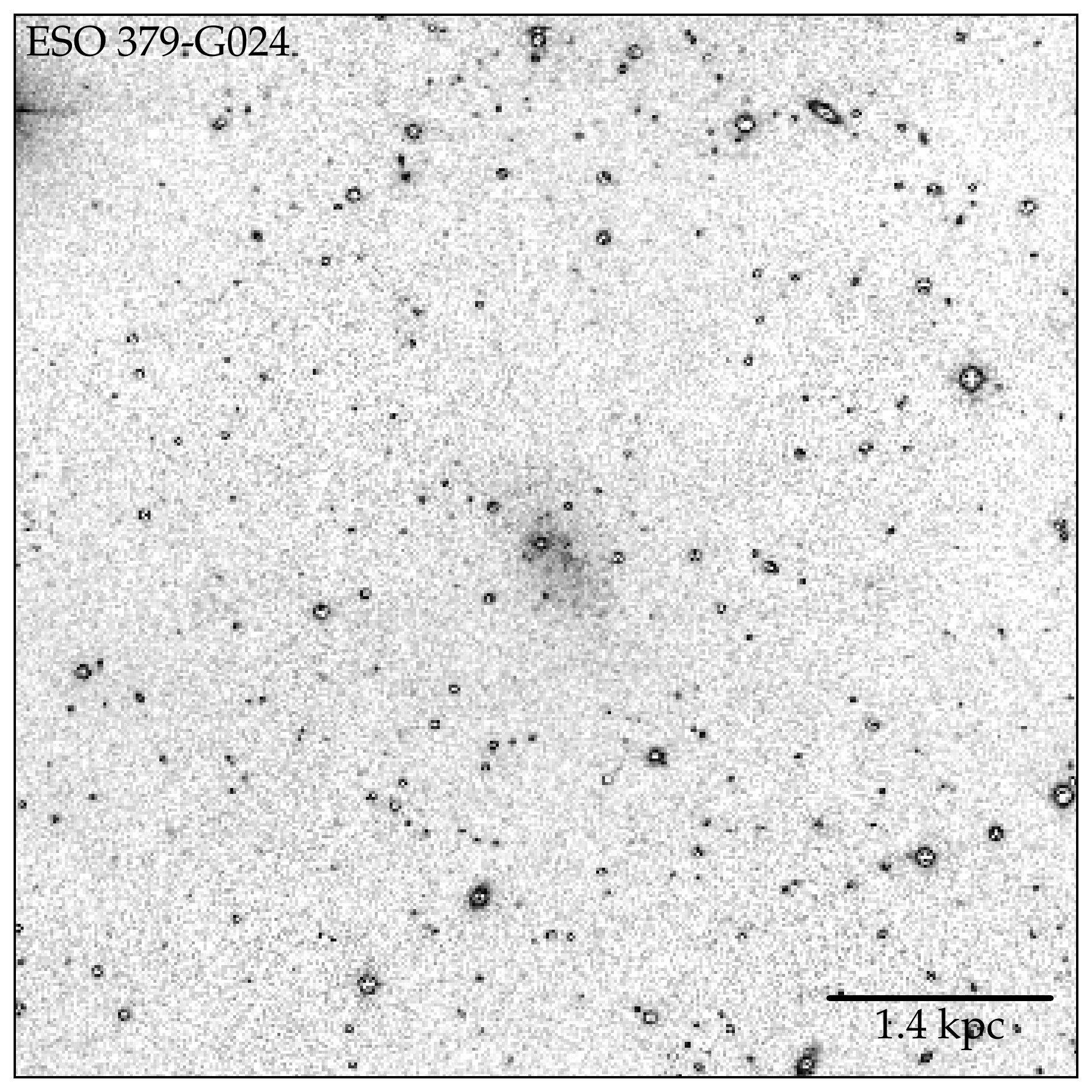}&
\includegraphics[trim = 2mm 2mm 2mm 2mm, clip = true, scale = 0.3]{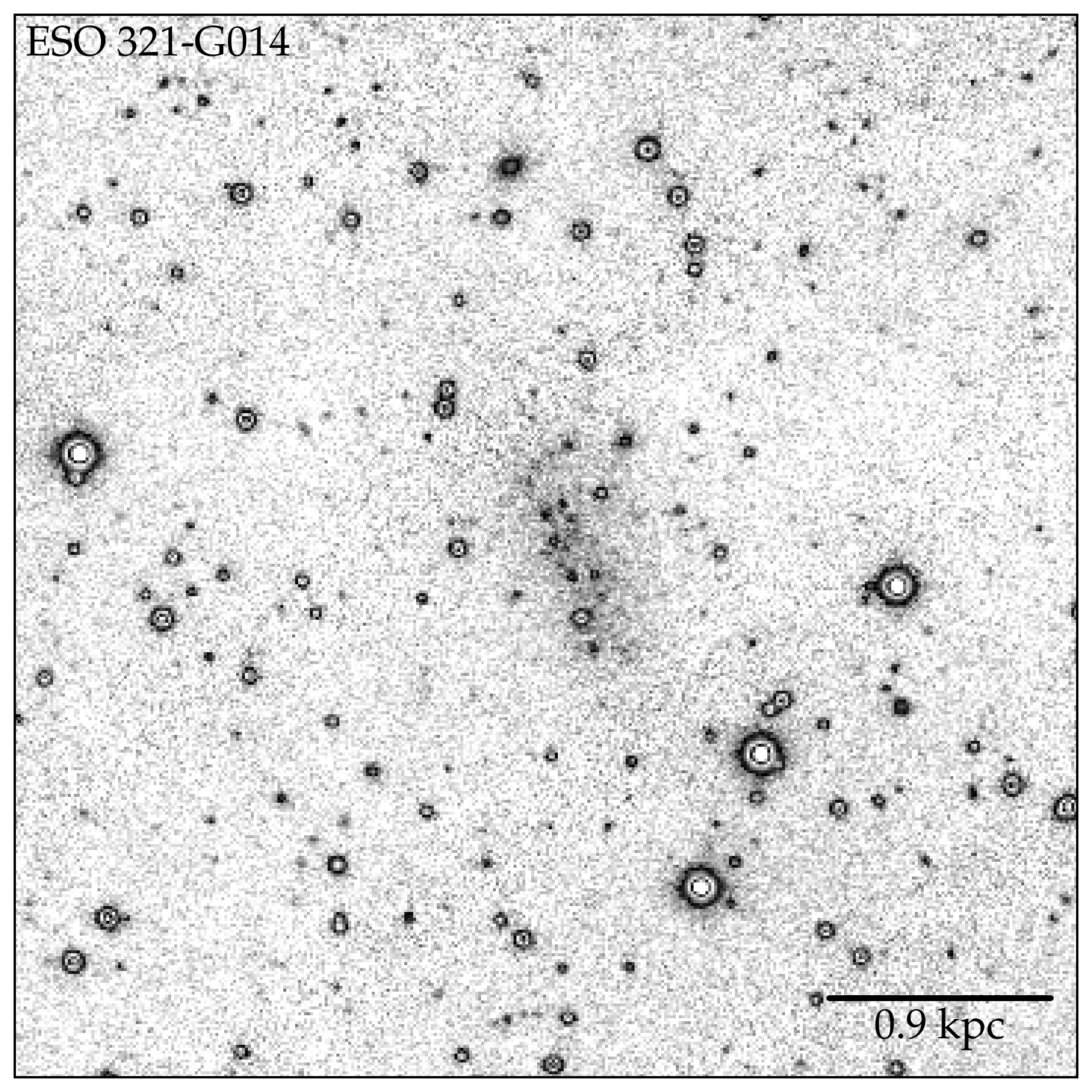}\\

\includegraphics[trim = 2mm 2mm 2mm 2mm, clip = true, scale = 0.3]{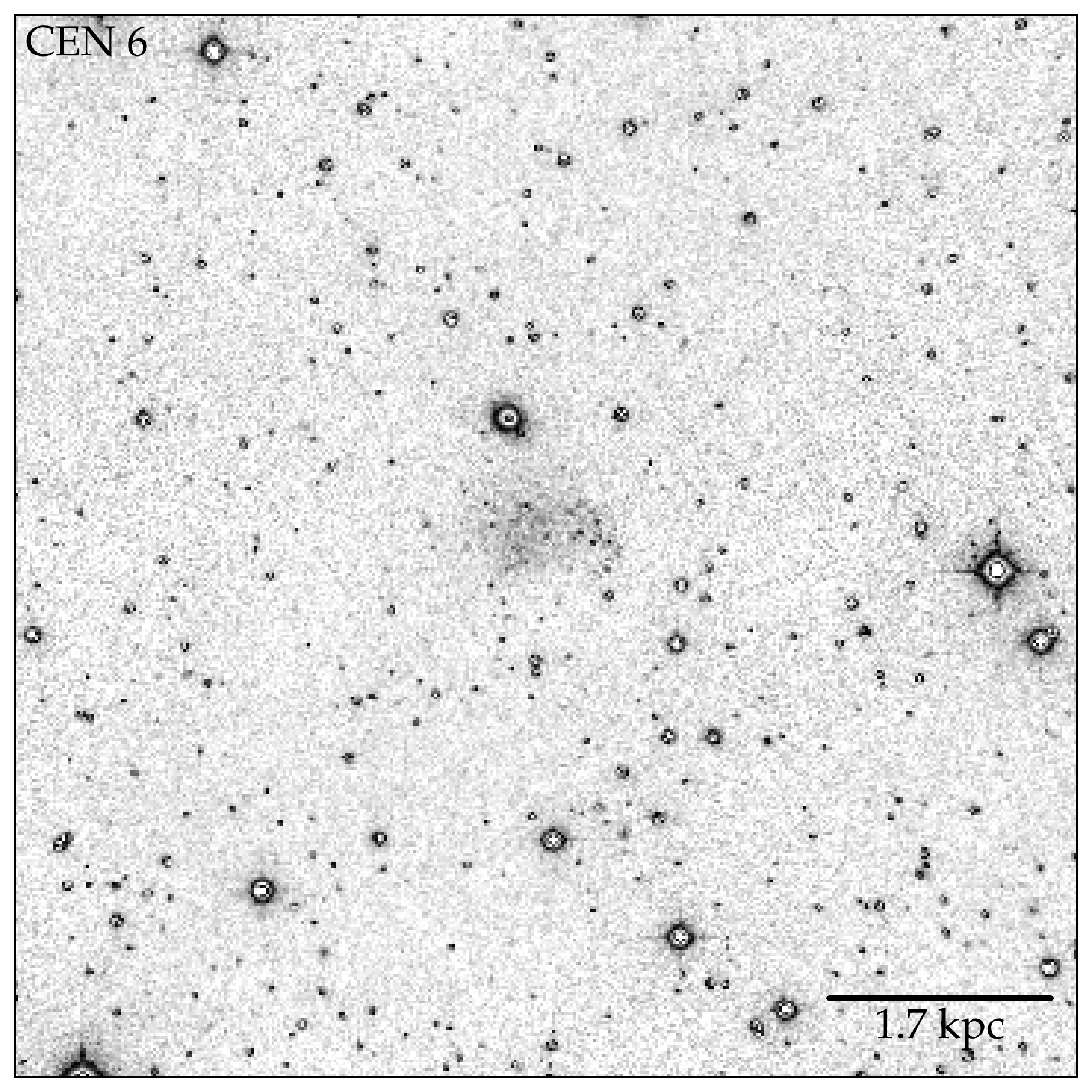}&
\includegraphics[trim = 2mm 2mm 2mm 2mm, clip = true, scale = 0.3]{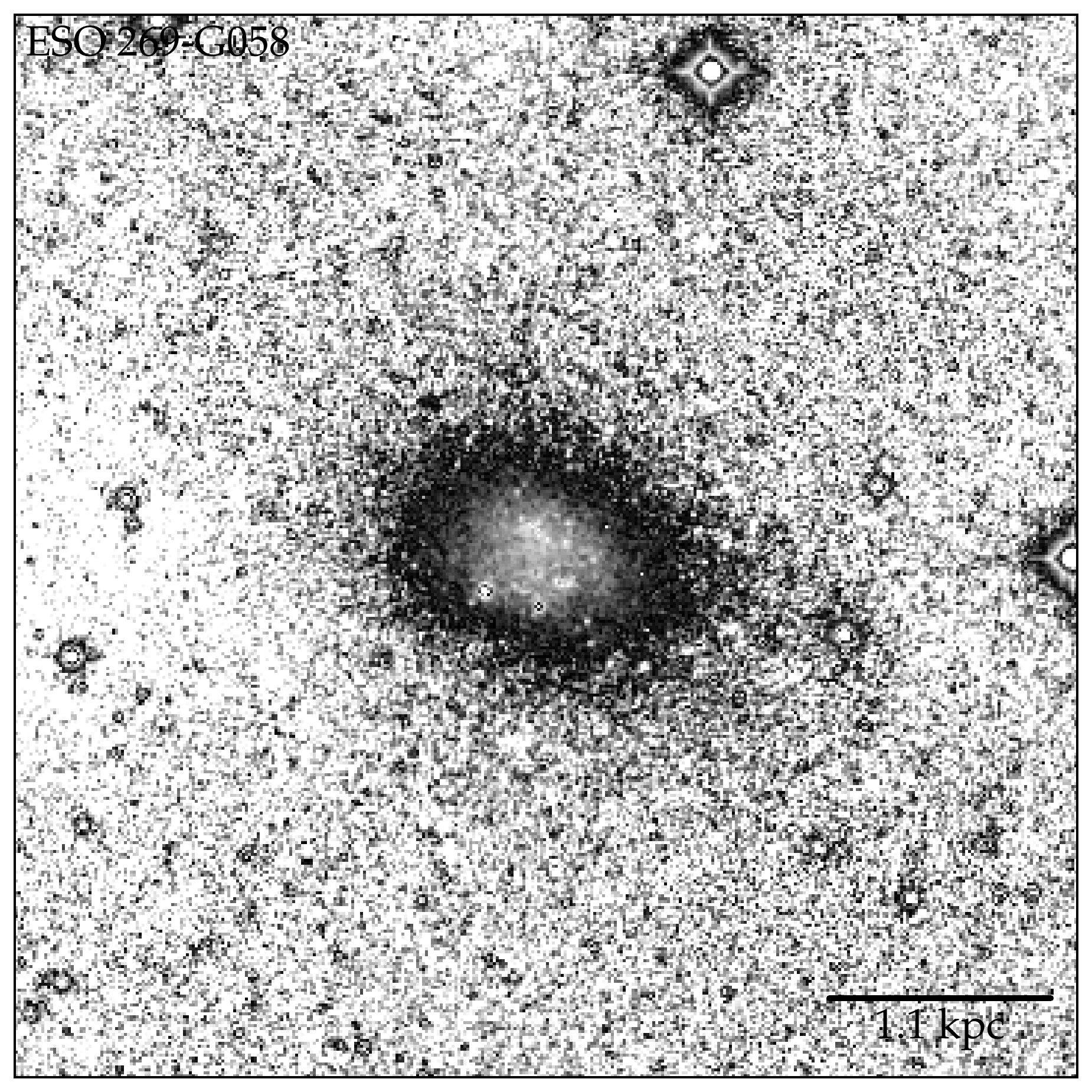}&
\includegraphics[trim = 2mm 2mm 2mm 2mm, clip = true, scale = 0.3]{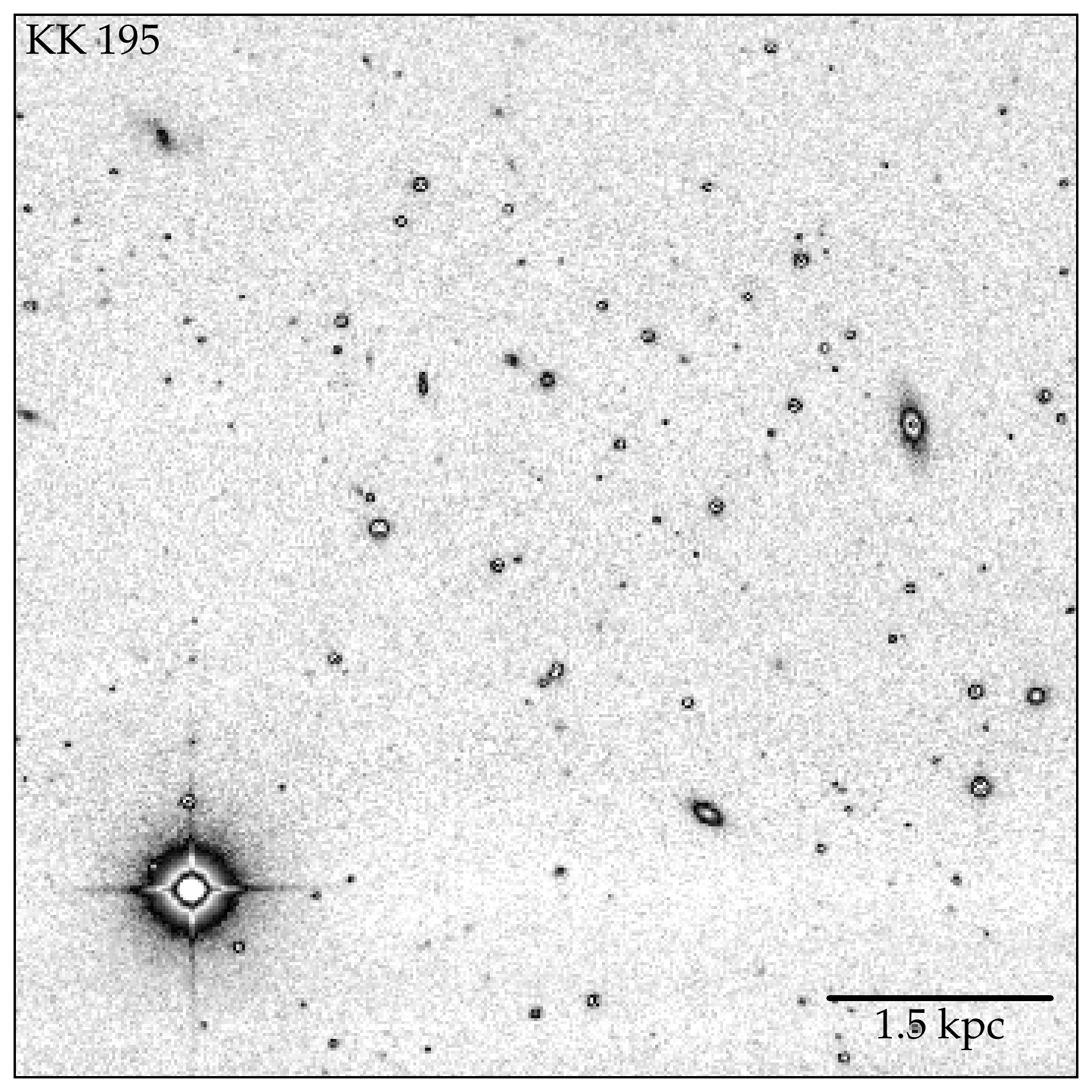}\\

\includegraphics[trim = 2mm 2mm 2mm 2mm, clip = true, scale = 0.3]{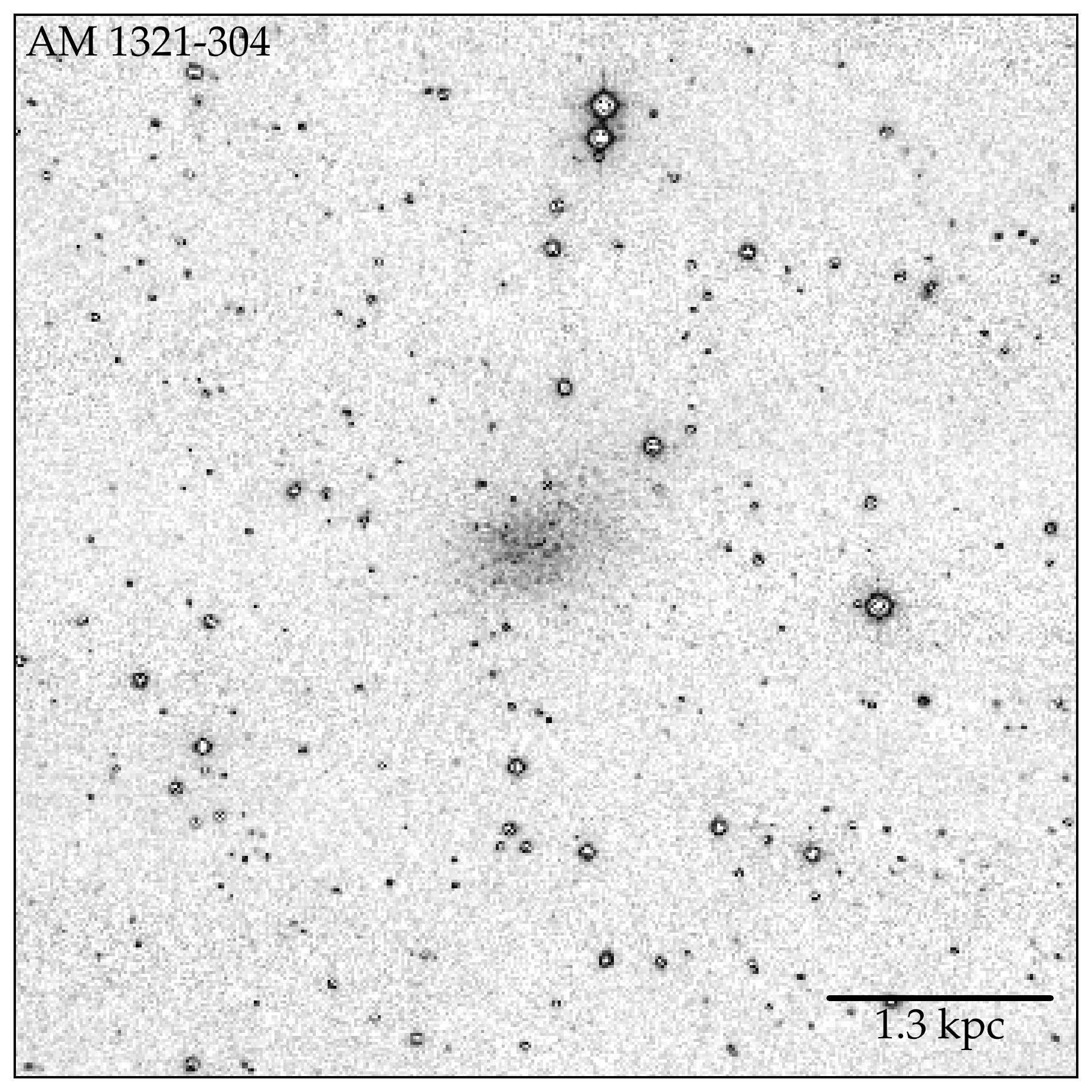}&
\includegraphics[trim = 2mm 2mm 2mm 2mm, clip = true, scale = 0.3]{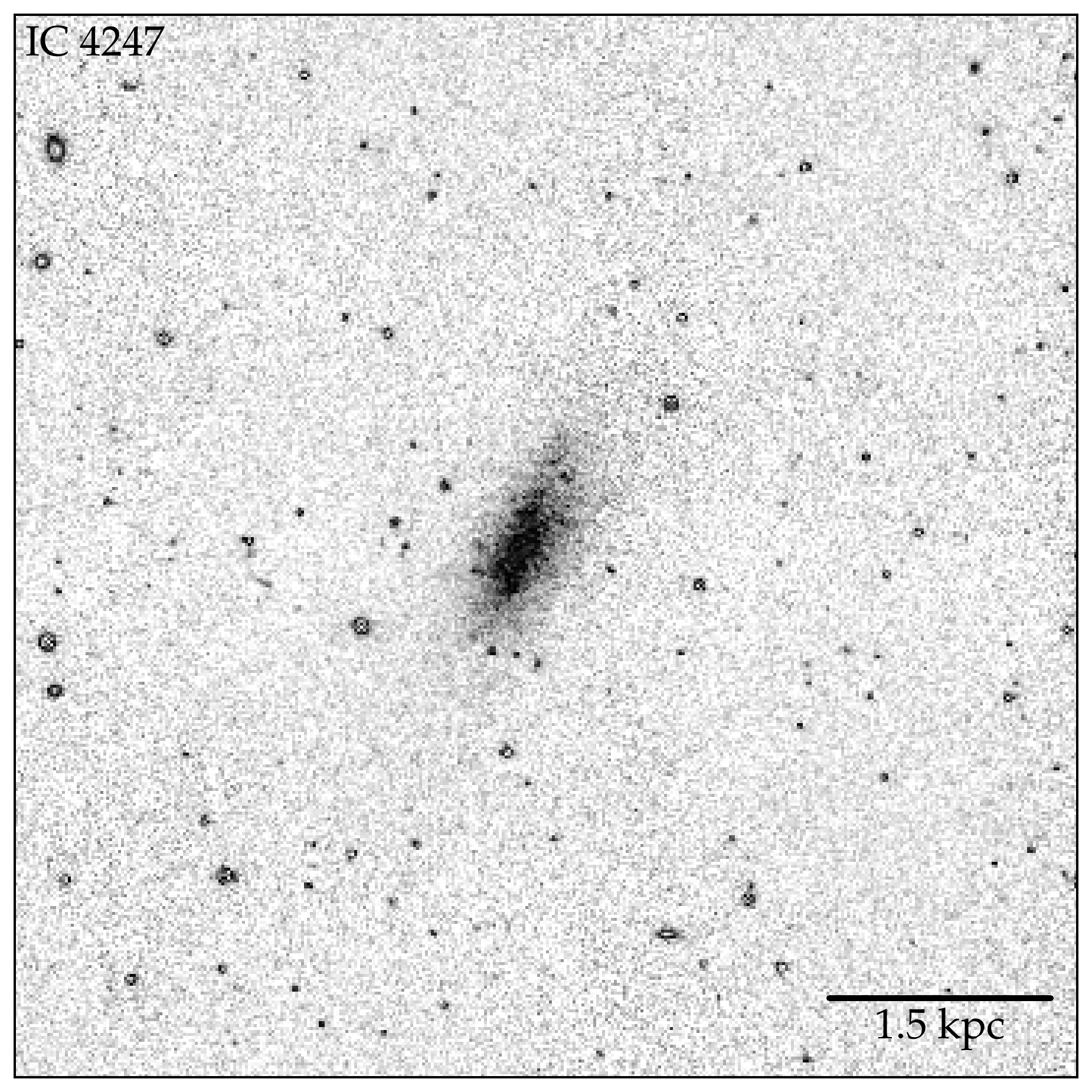}&
\includegraphics[trim = 2mm 2mm 2mm 2mm, clip = true, scale = 0.3]{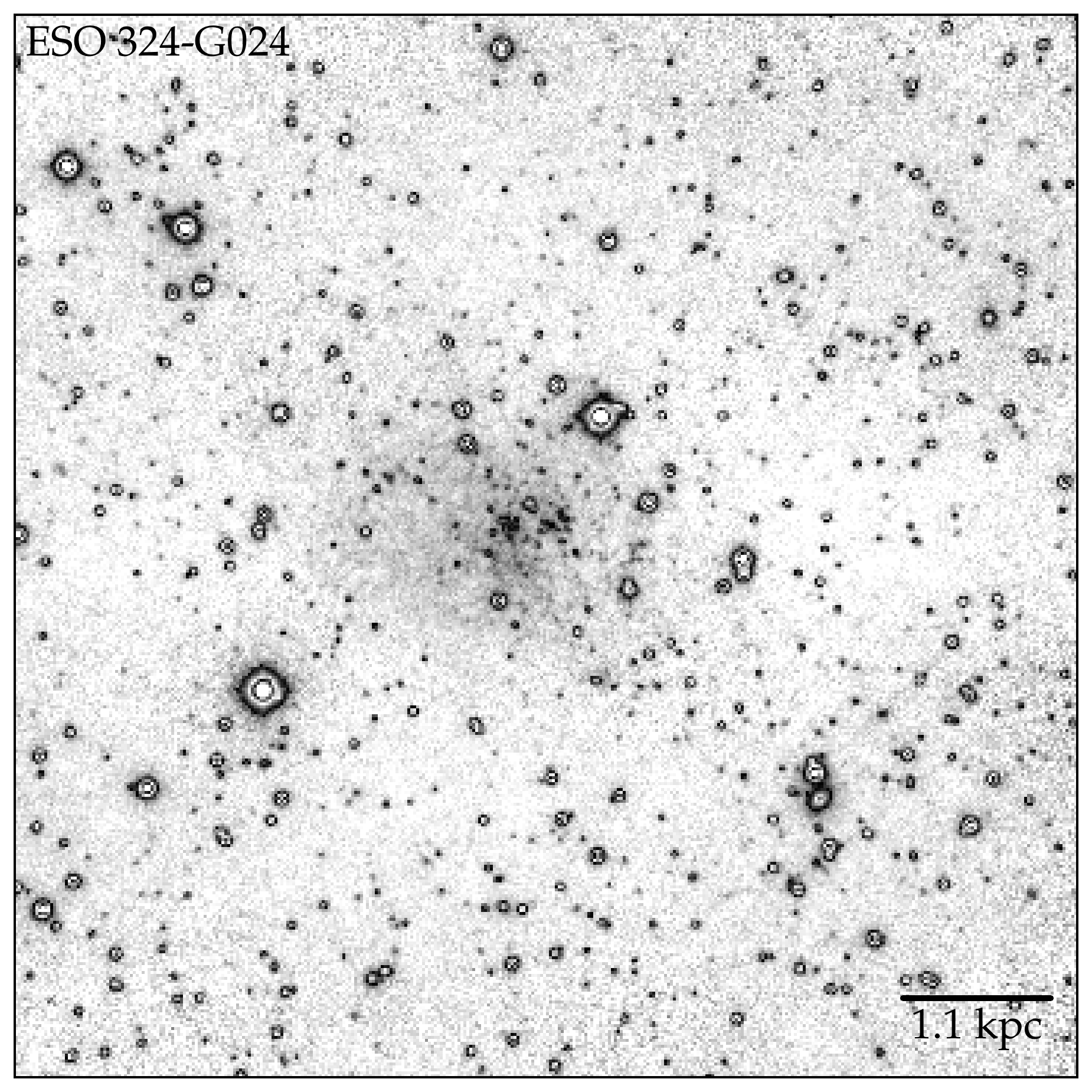}\\

\includegraphics[trim = 2mm 2mm 2mm 2mm, clip = true, scale = 0.3]{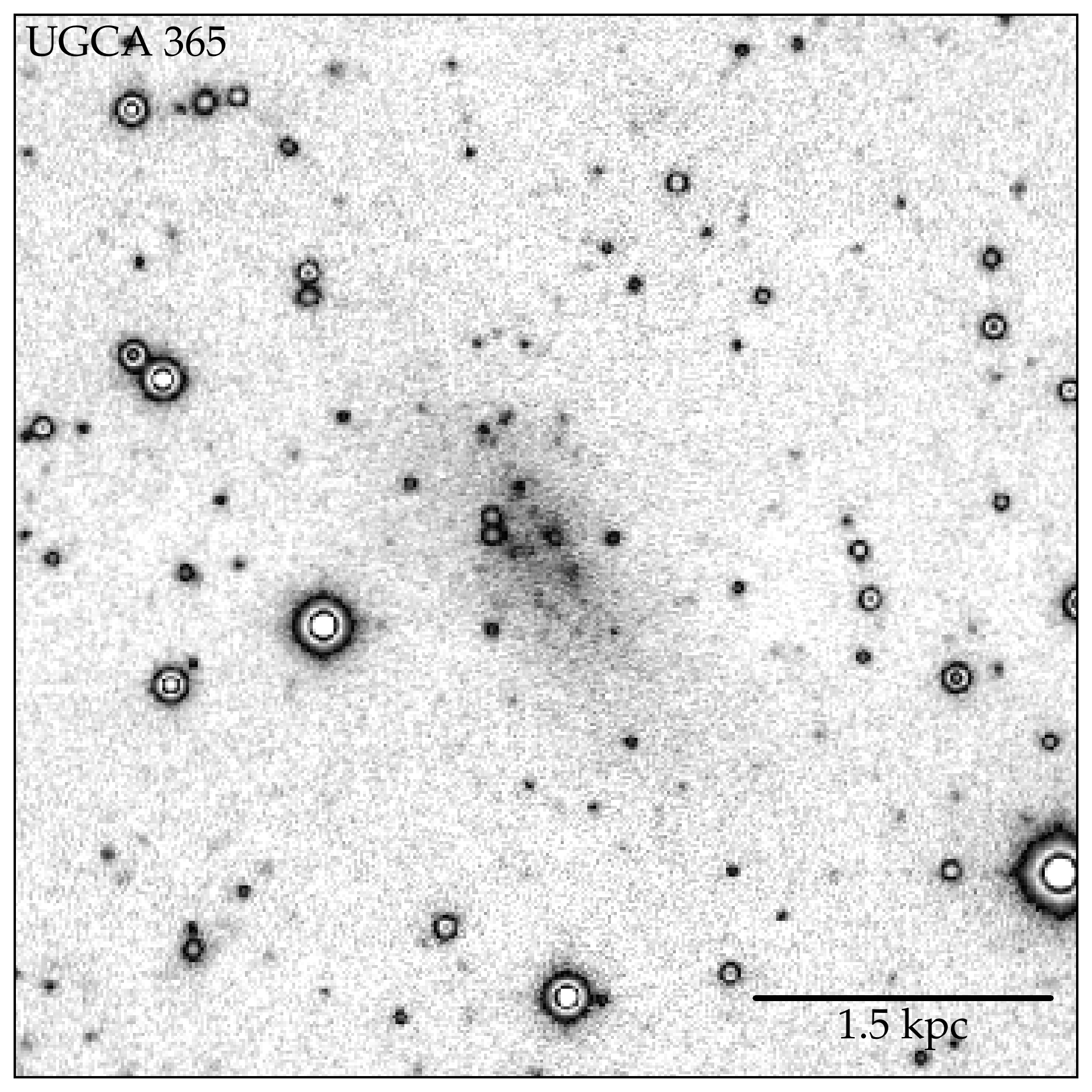}&
\includegraphics[trim = 2mm 2mm 2mm 2mm, clip = true, scale = 0.3]{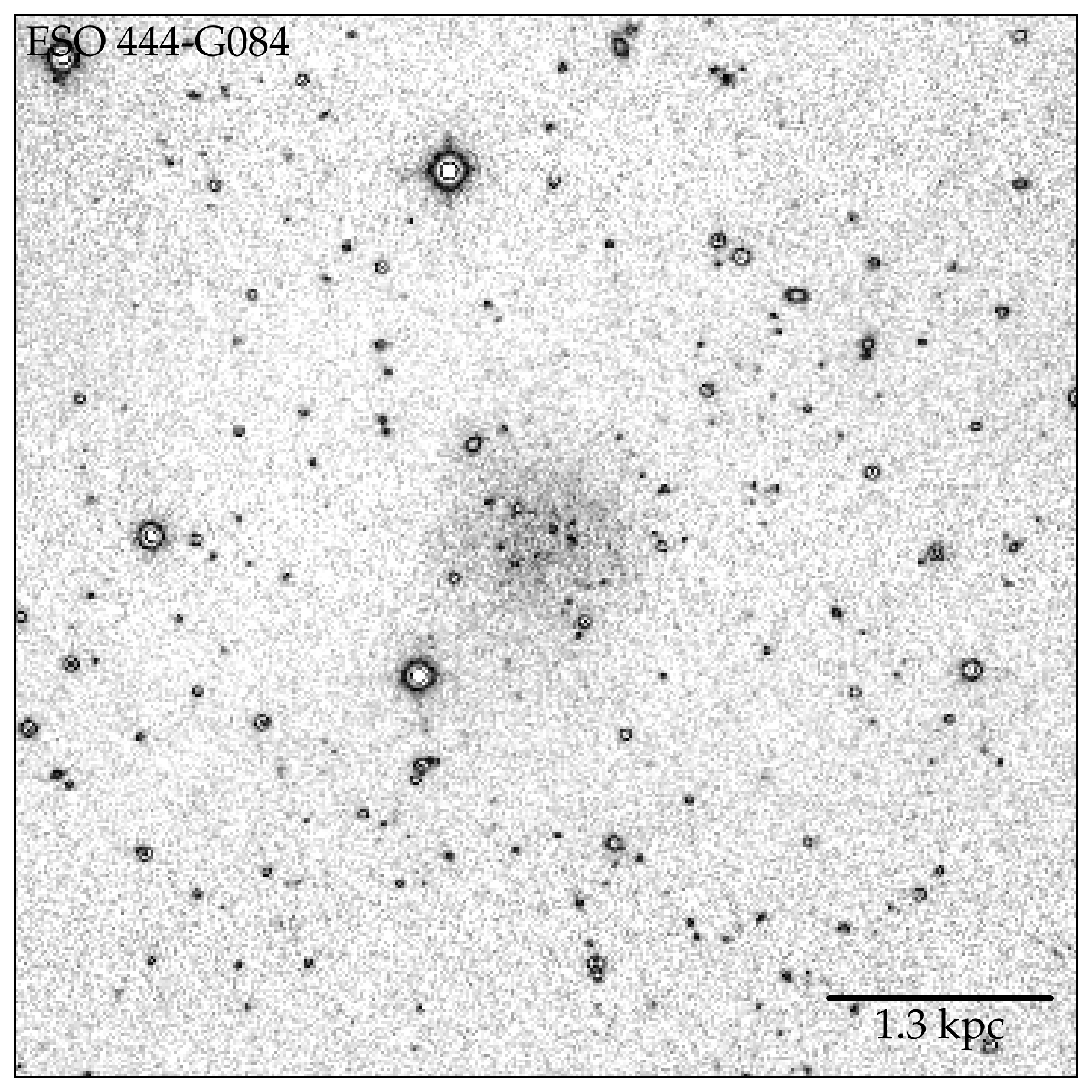}&
\includegraphics[trim = 2mm 2mm 2mm 2mm, clip = true, scale = 0.3]{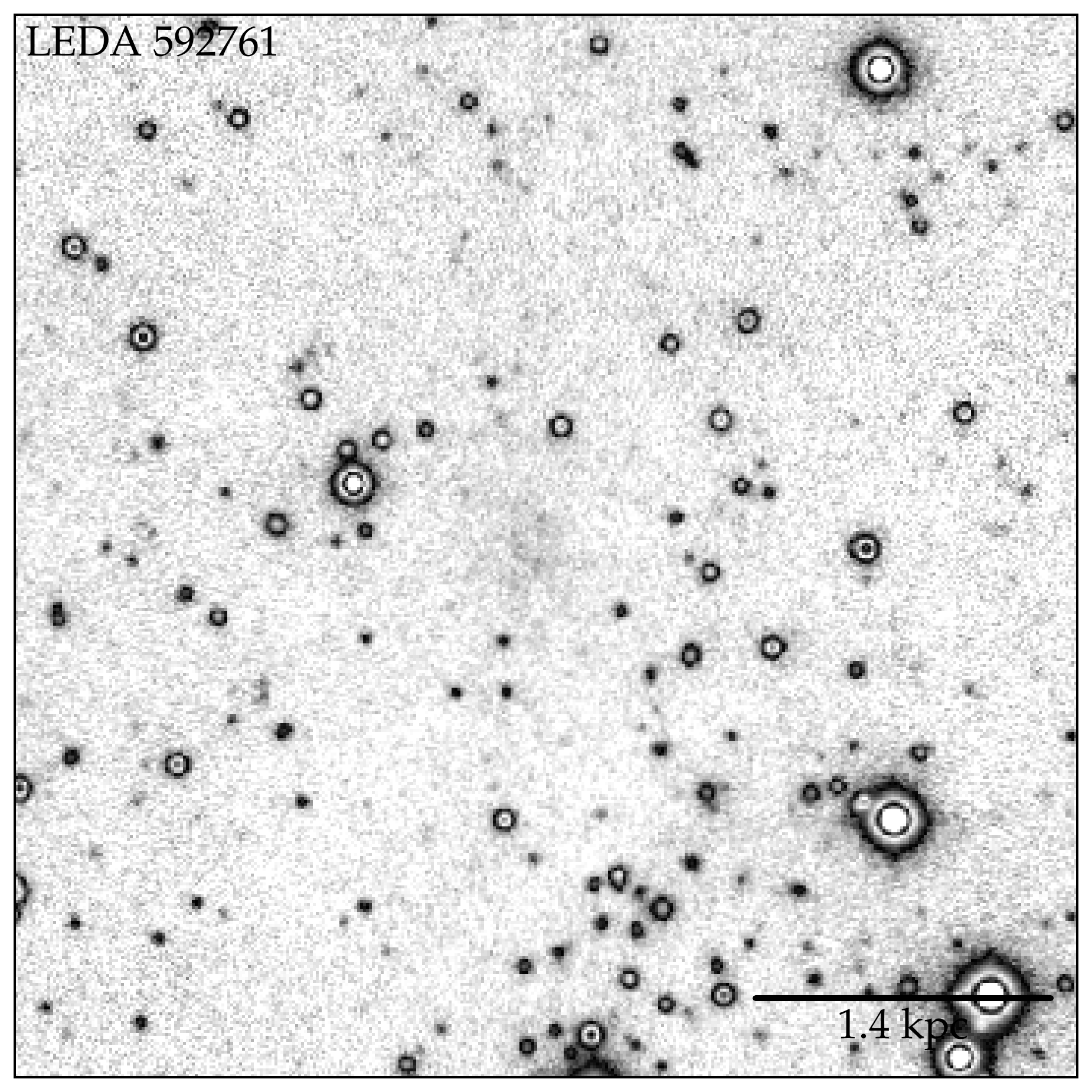}\\
\end{array}
$
\caption{-- continued.}
\label{results2}
\end{figure*}

\addtocounter {figure} {-1}

\begin{figure*}
\setlength{\arraycolsep}{1pt}
\setlength{\parskip}{0pt}
$
\begin{array}{cccc}

\includegraphics[trim = 2mm 2mm 2mm 2mm, clip = true, scale = 0.3]{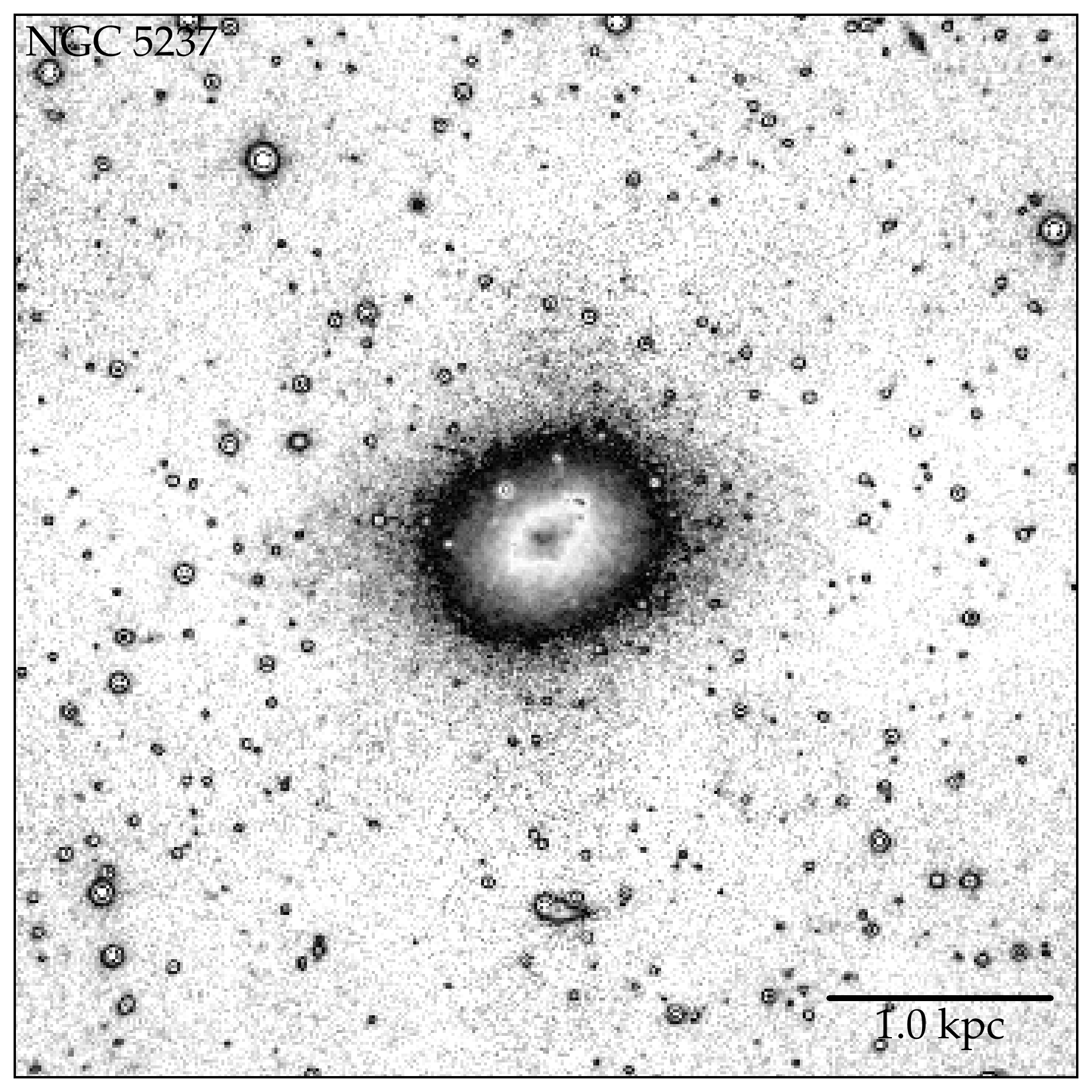}&
\includegraphics[trim = 2mm 2mm 2mm 2mm, clip = true, scale = 0.3]{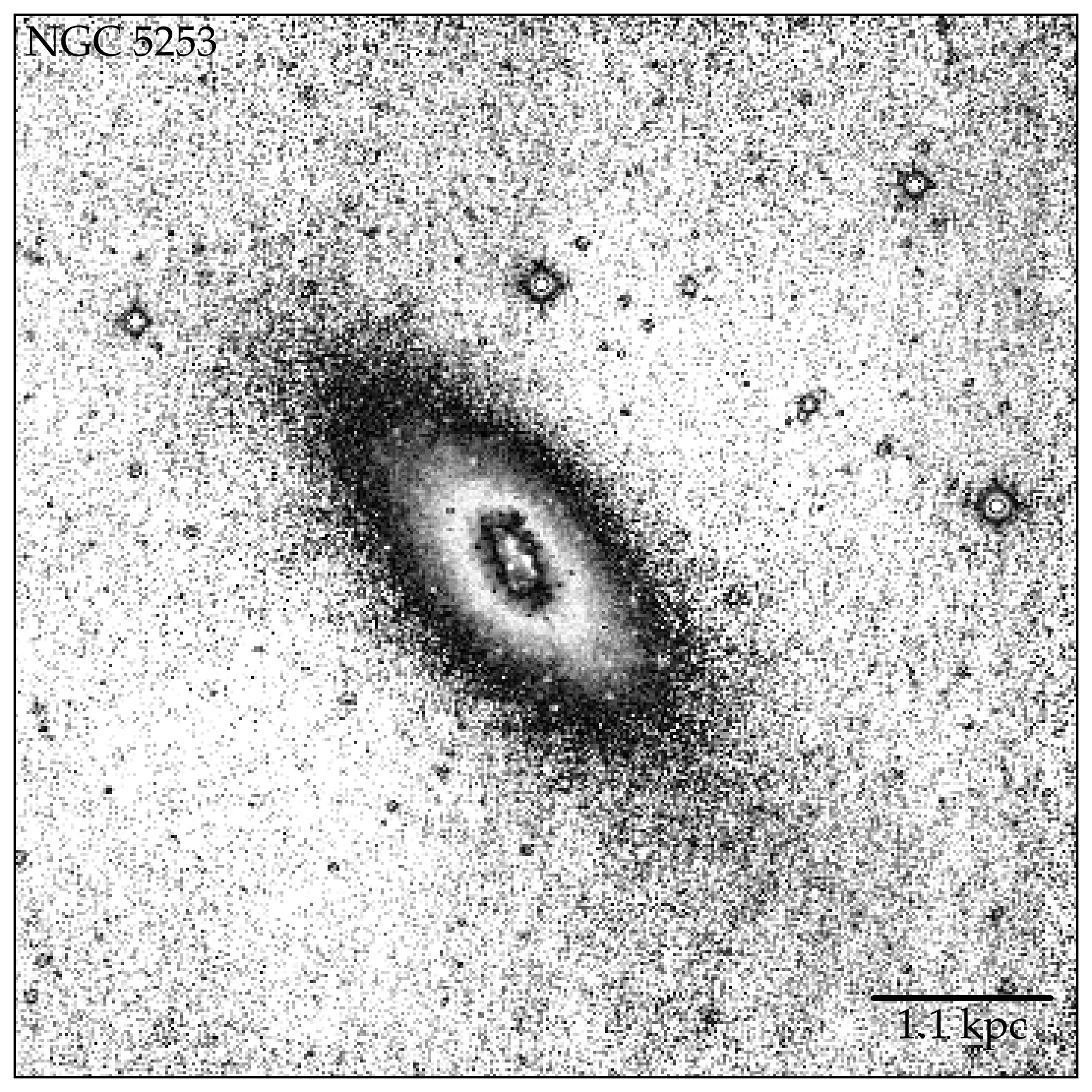}&
\includegraphics[trim = 2mm 2mm 2mm 2mm, clip = true, scale = 0.3]{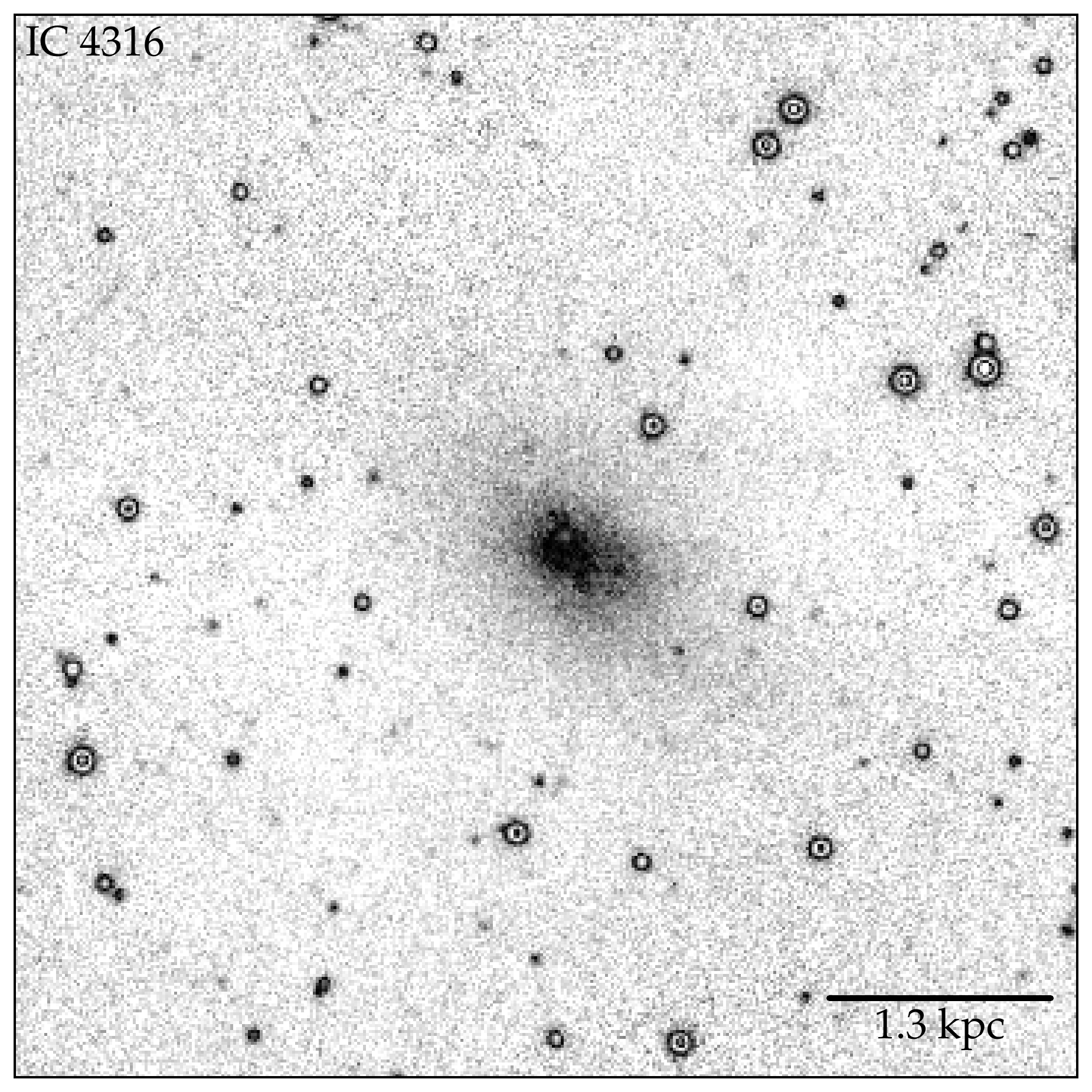}\\

\includegraphics[trim = 2mm 2mm 2mm 2mm, clip = true, scale = 0.3]{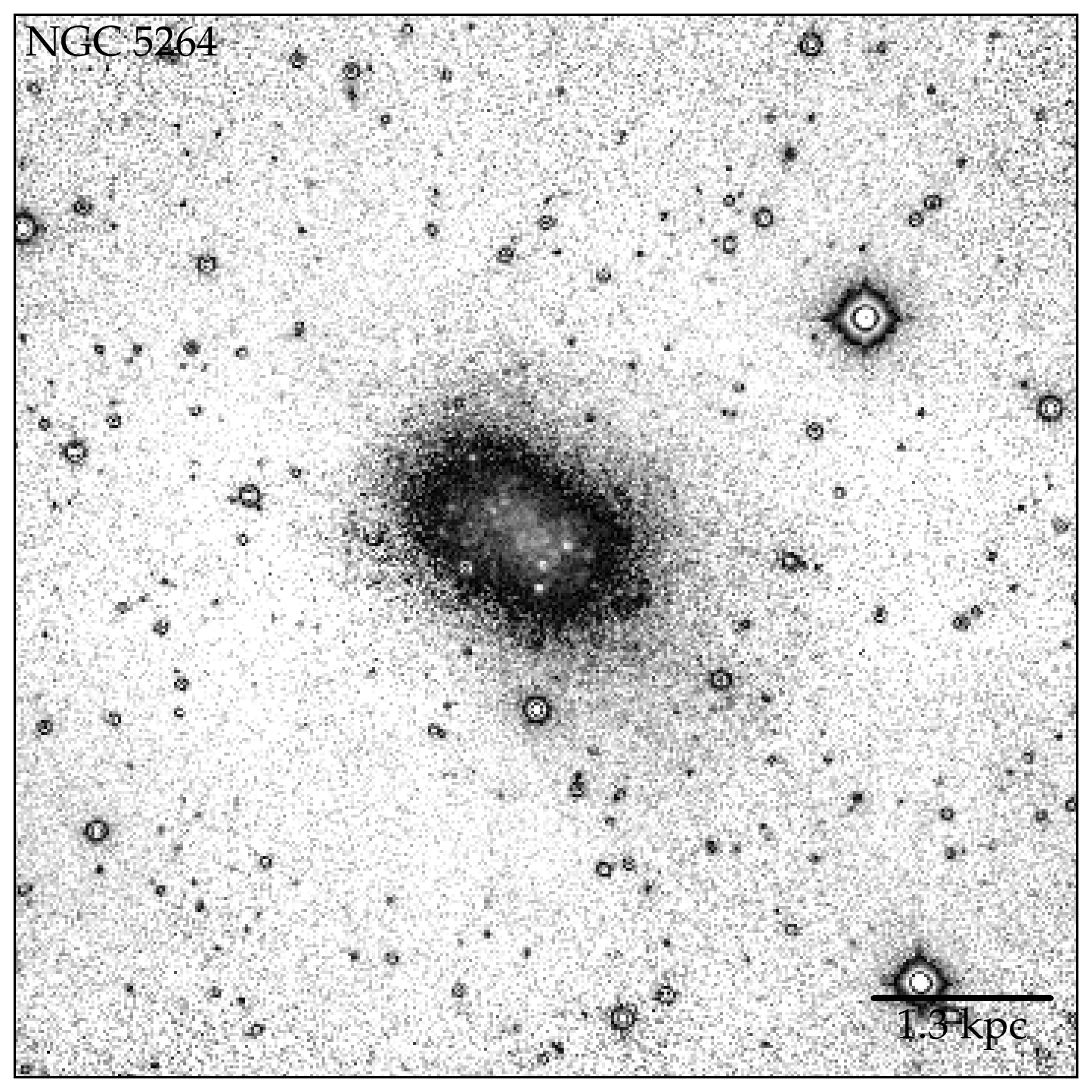}&
\includegraphics[trim = 2mm 2mm 2mm 2mm, clip = true, scale = 0.3]{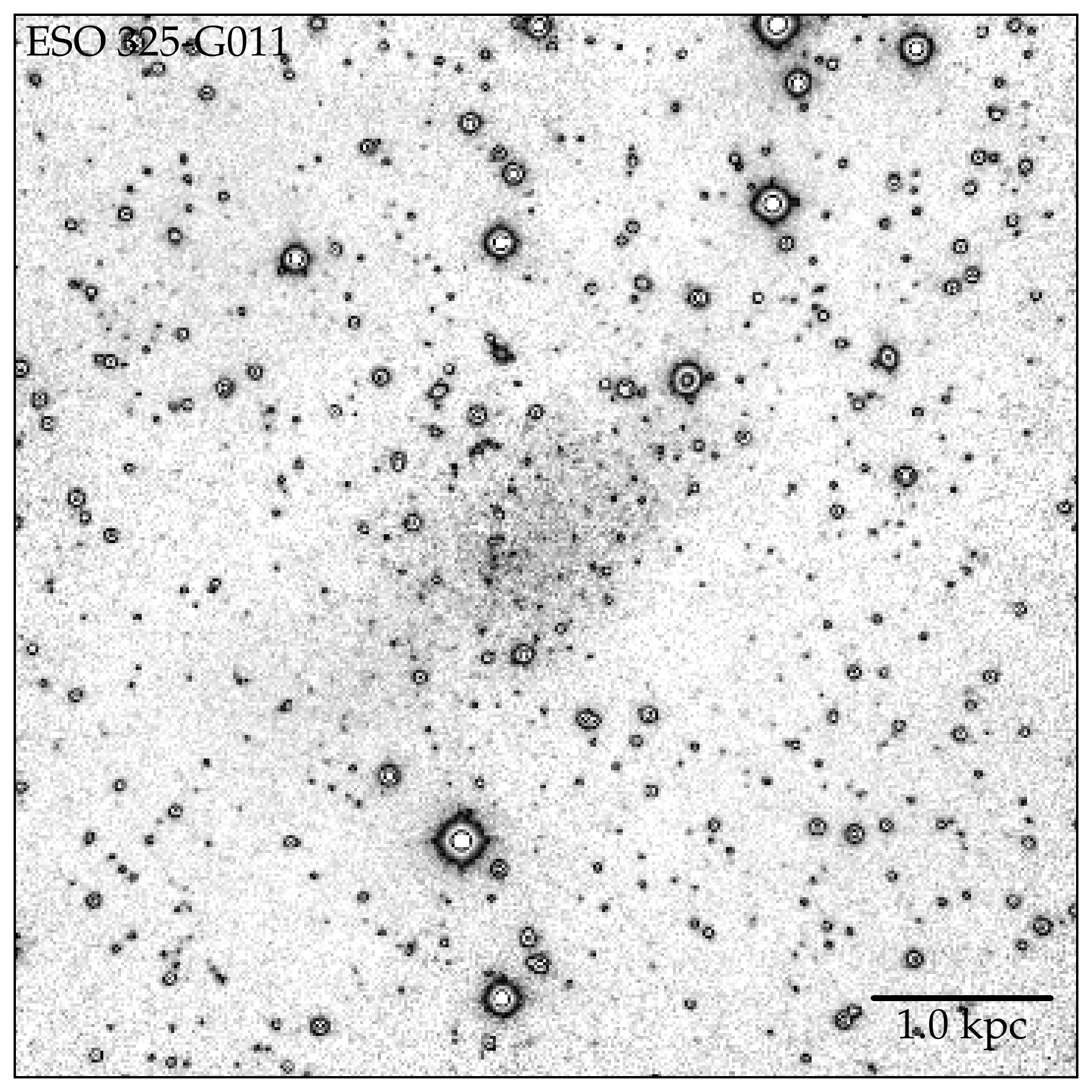}&
\includegraphics[trim = 2mm 2mm 2mm 2mm, clip = true, scale = 0.3]{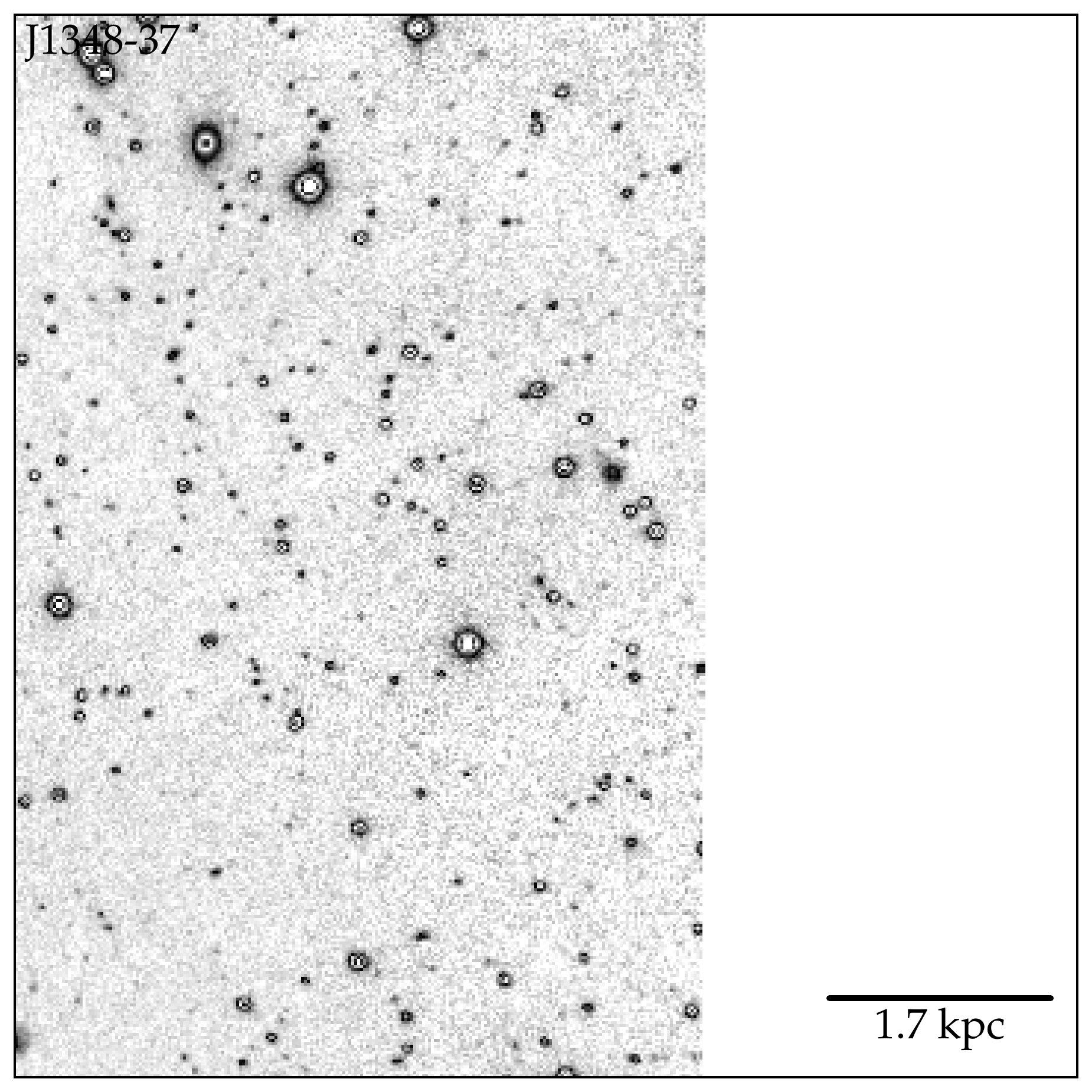}\\

\includegraphics[trim = 2mm 2mm 2mm 2mm, clip = true, scale = 0.3]{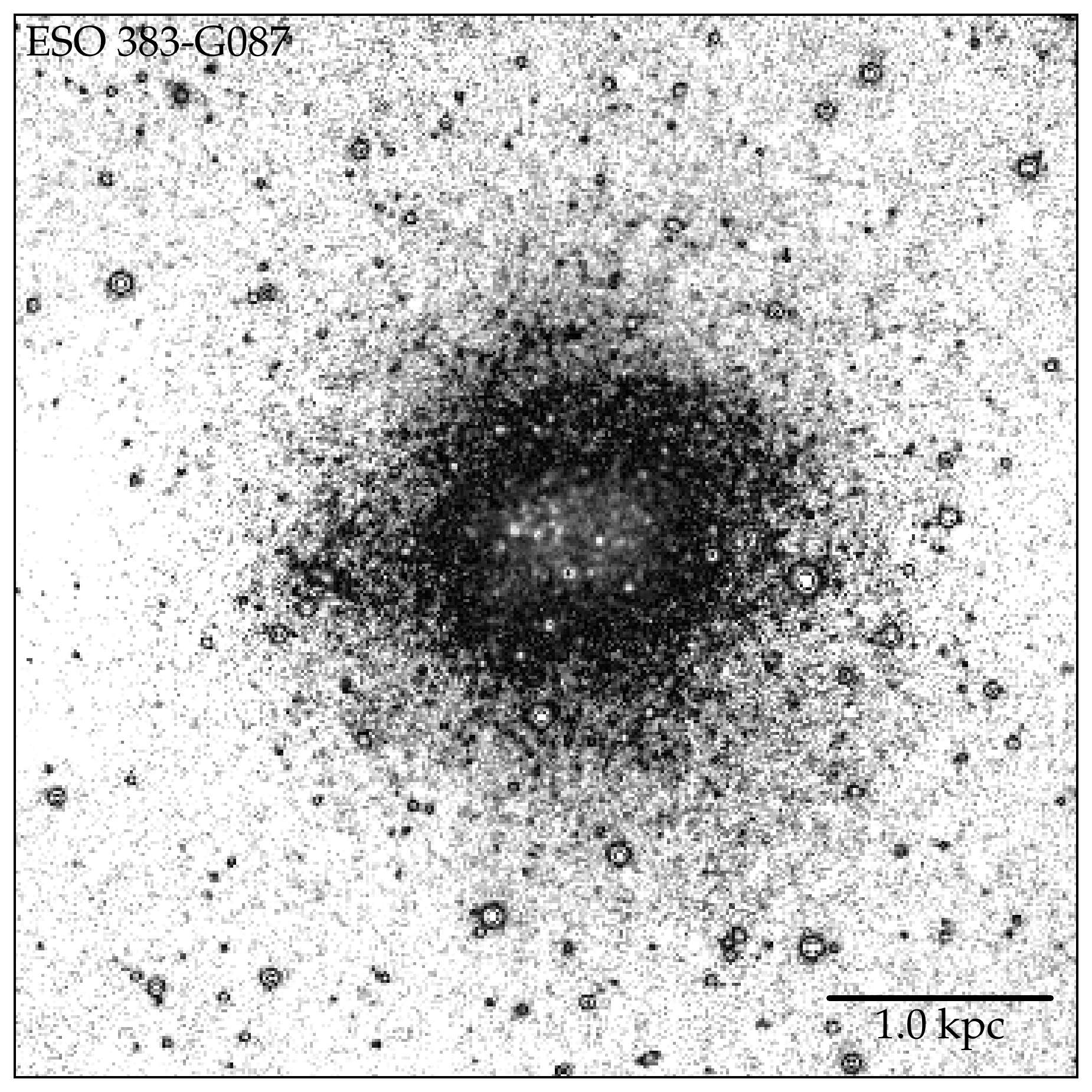}&
\includegraphics[trim = 2mm 2mm 2mm 2mm, clip = true, scale = 0.3]{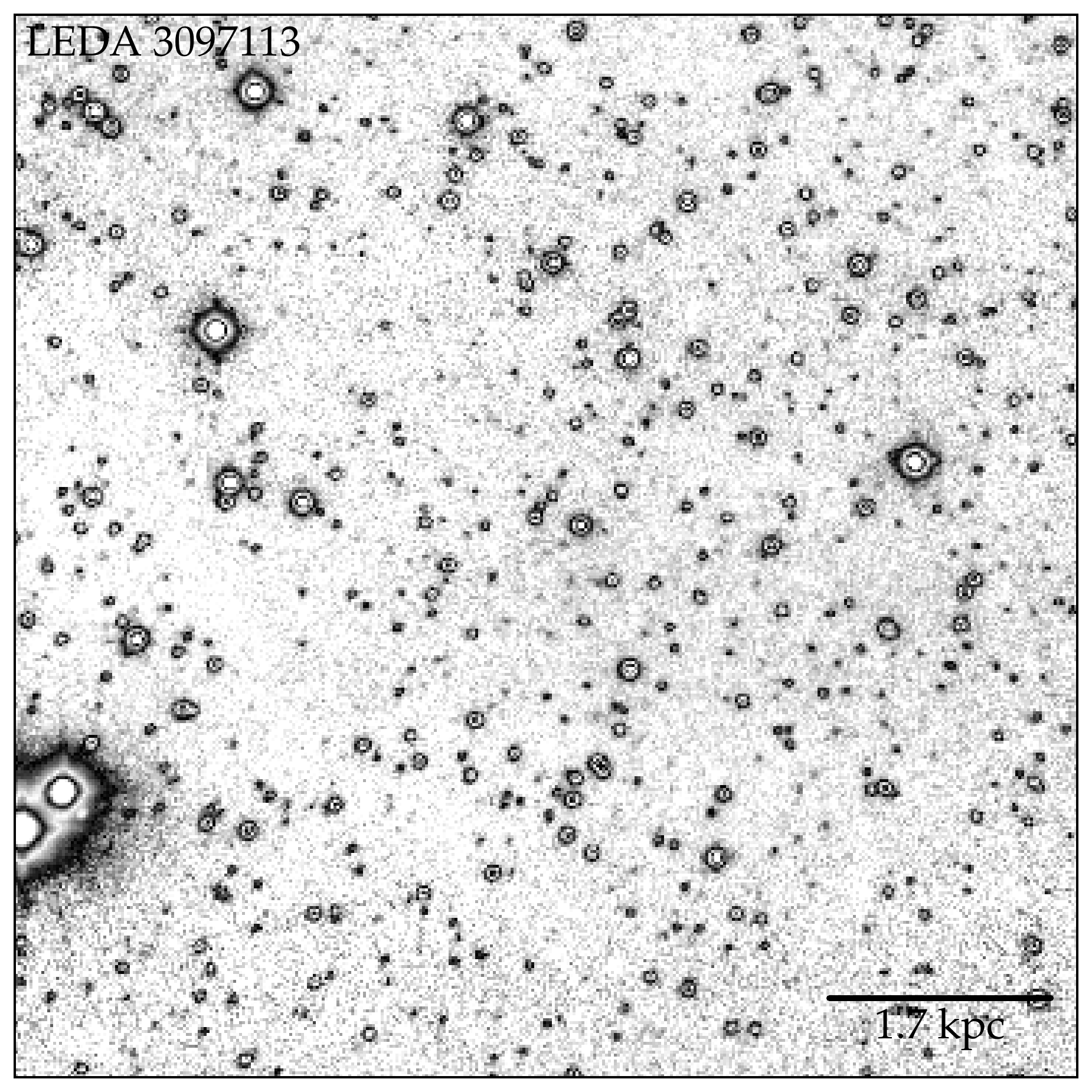}&
\includegraphics[trim = 2mm 2mm 2mm 2mm, clip = true, scale = 0.3]{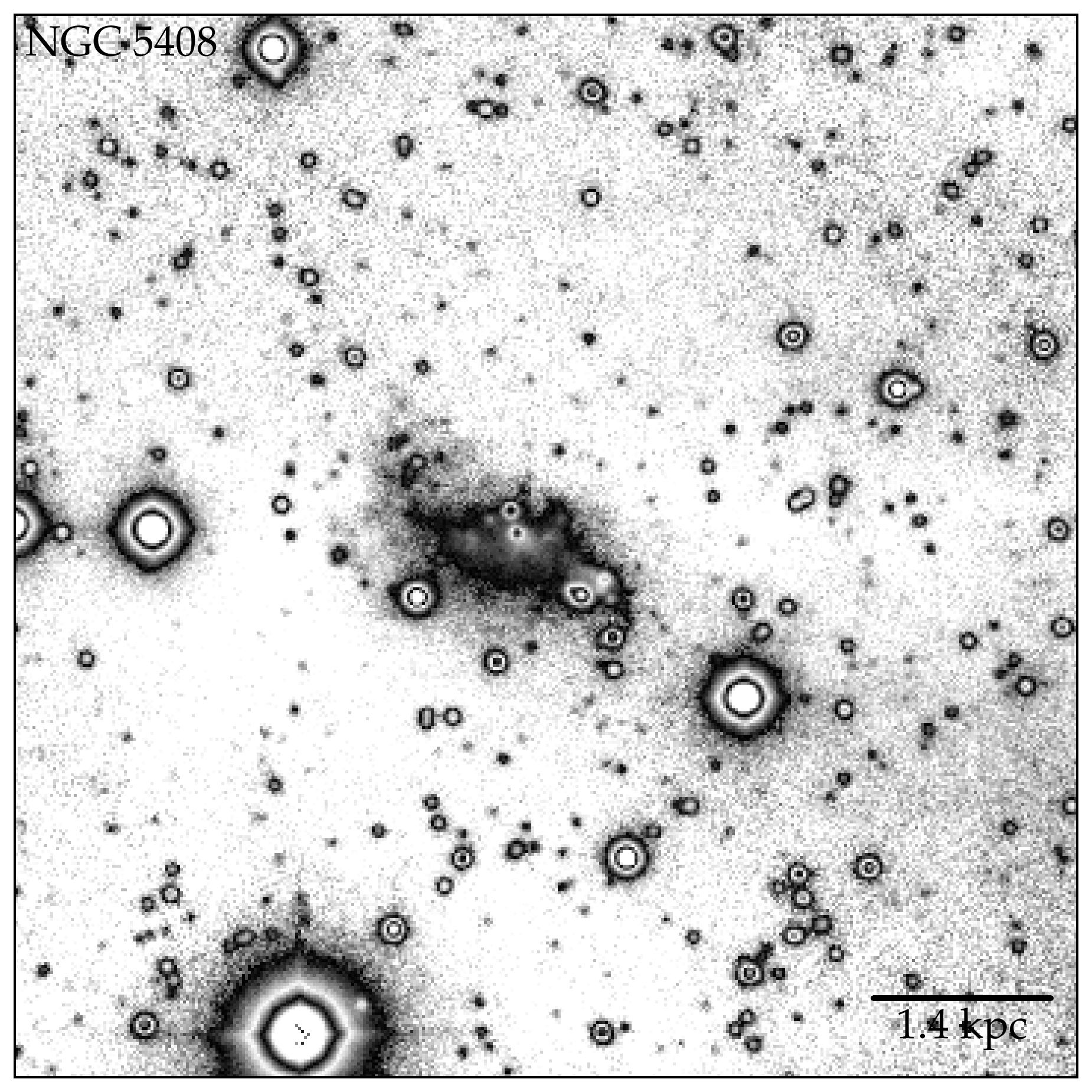}\\

\includegraphics[trim = 2mm 2mm 2mm 2mm, clip = true, scale = 0.3]{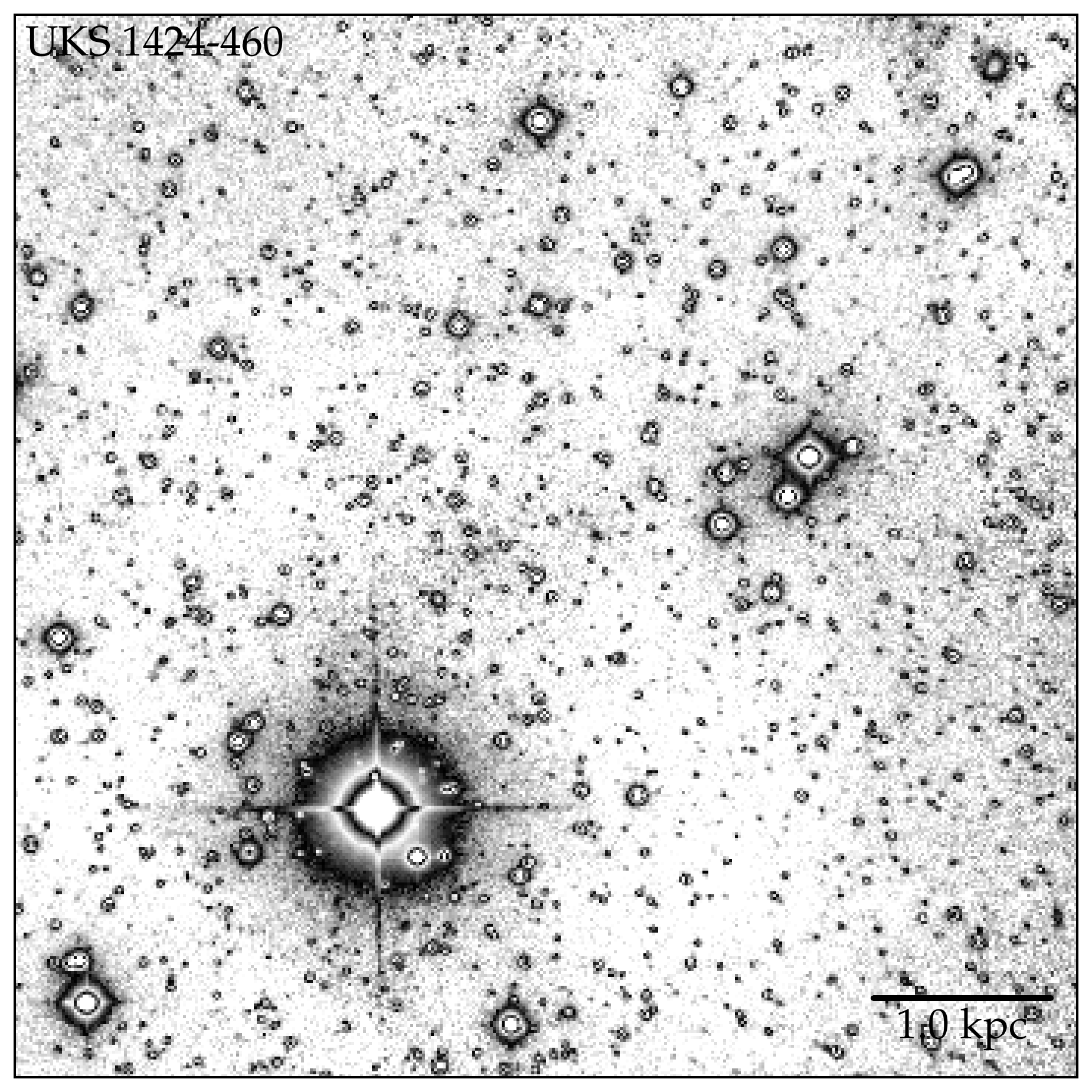}&
\includegraphics[trim = 2mm 2mm 2mm 2mm, clip = true, scale = 0.3]{images/figs/J1434-49_ac.pdf}&
\includegraphics[trim = 2mm 2mm 2mm 2mm, clip = true, scale = 0.3]{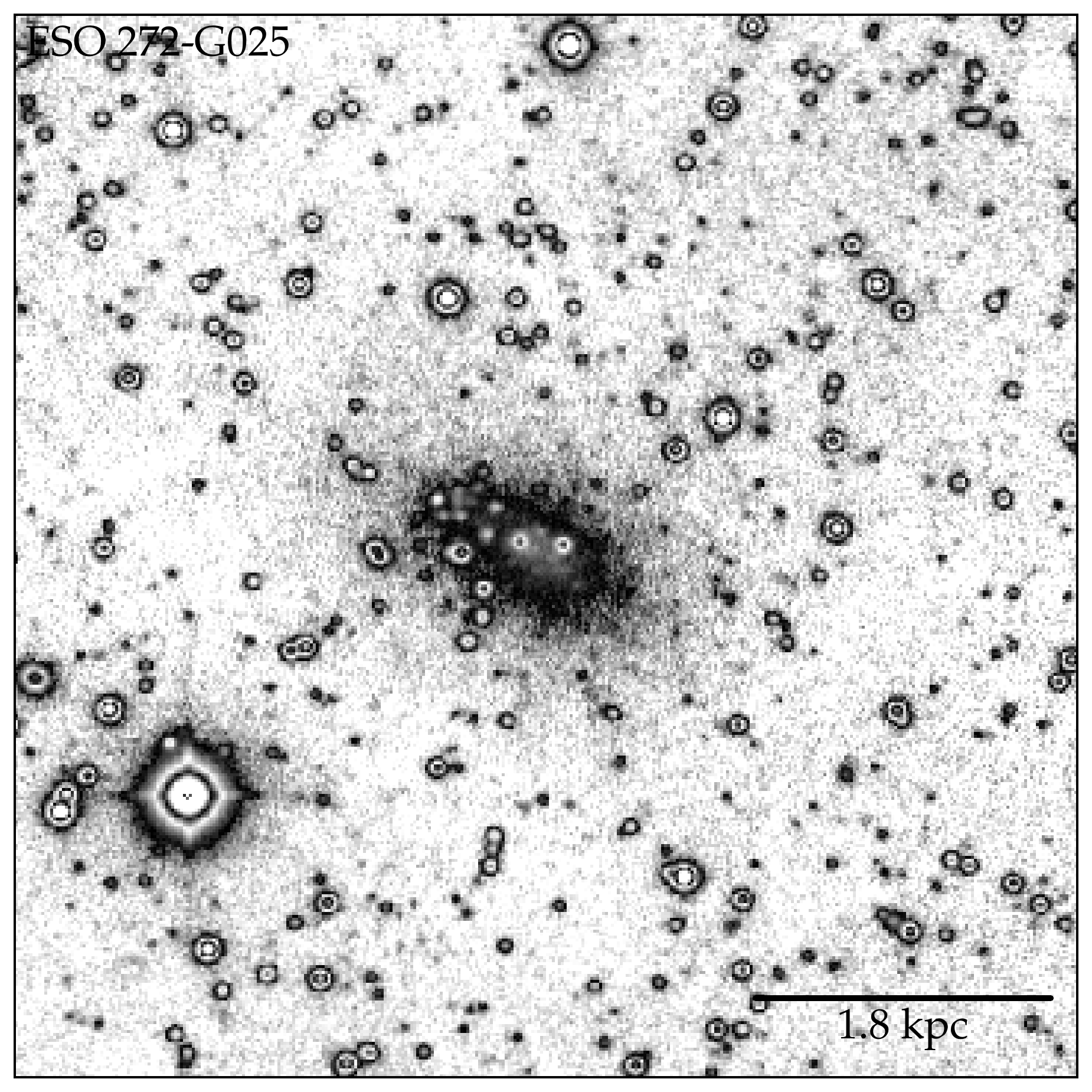}\\

\end{array}
$
\caption{-- continued.}
\label{results3}
\end{figure*}

\addtocounter {figure} {-1}

\begin{figure*}
\setlength{\arraycolsep}{1pt}
\setlength{\parskip}{0pt}
$
\begin{array}{cccc}

\includegraphics[trim = 2mm 2mm 2mm 2mm, clip = true, scale = 0.3]{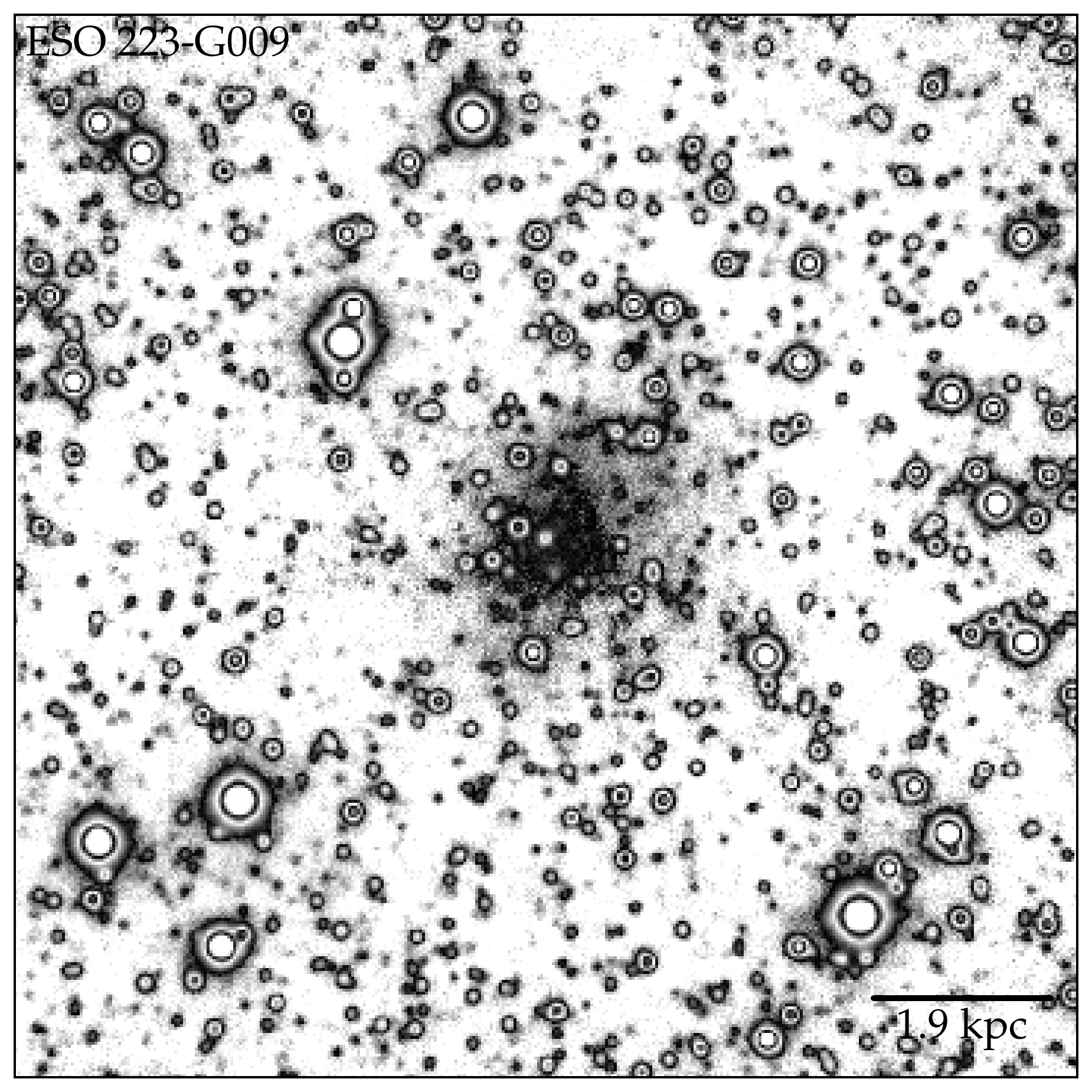}&
\includegraphics[trim = 2mm 2mm 2mm 2mm, clip = true, scale = 0.3]{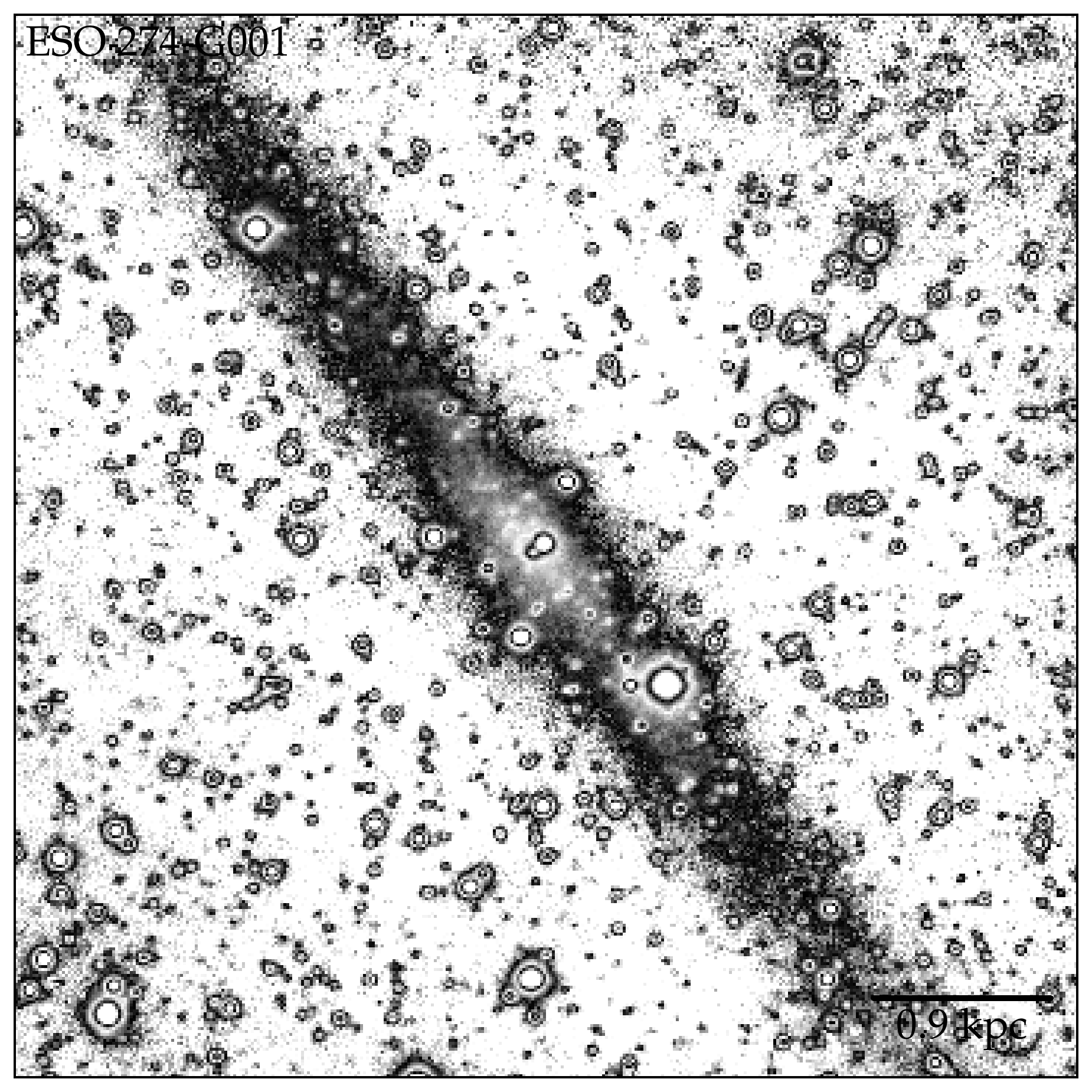}&
\includegraphics[trim = 2mm 2mm 2mm 2mm, clip = true, scale = 0.3]{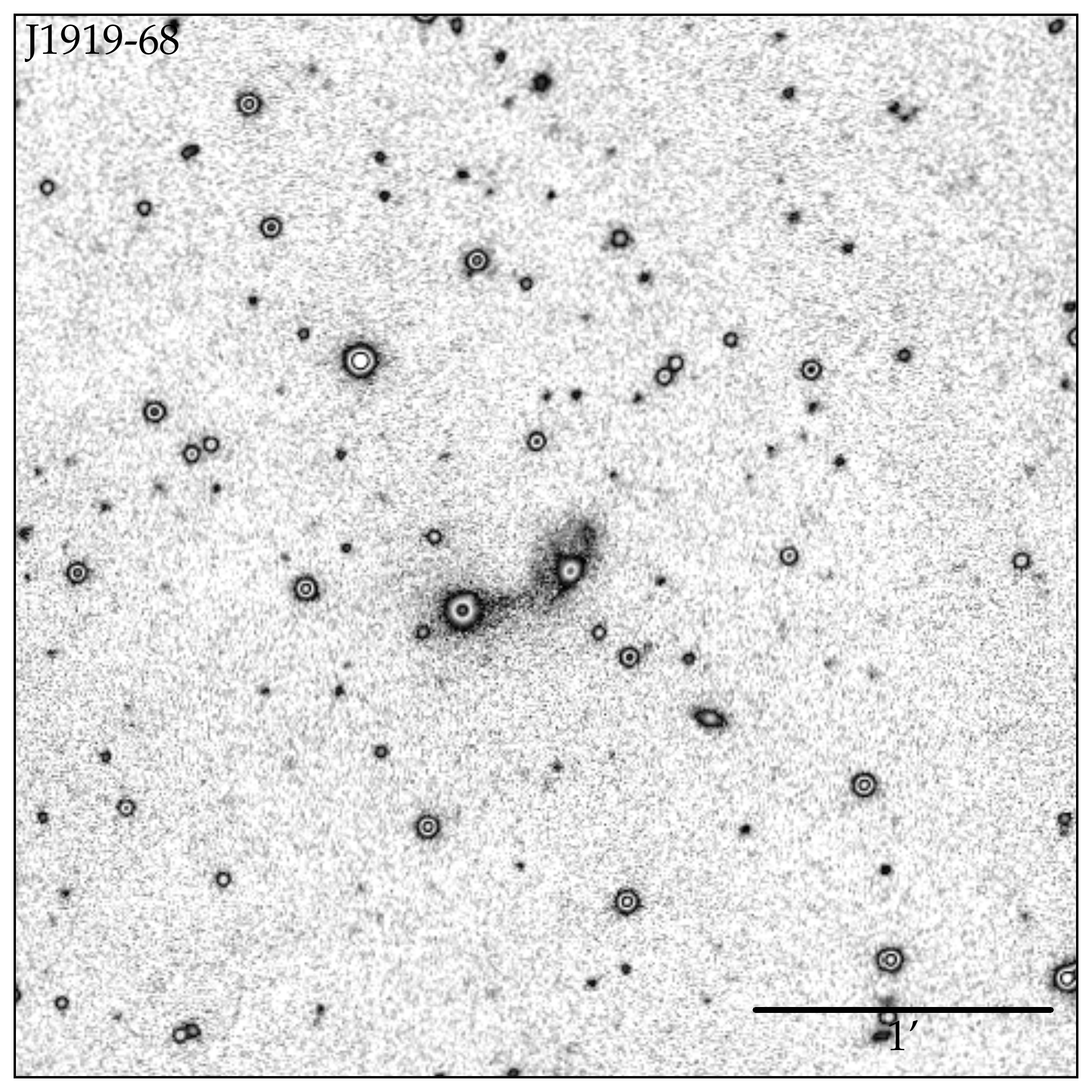}\\

\includegraphics[trim = 2mm 2mm 2mm 2mm, clip = true, scale = 0.3]{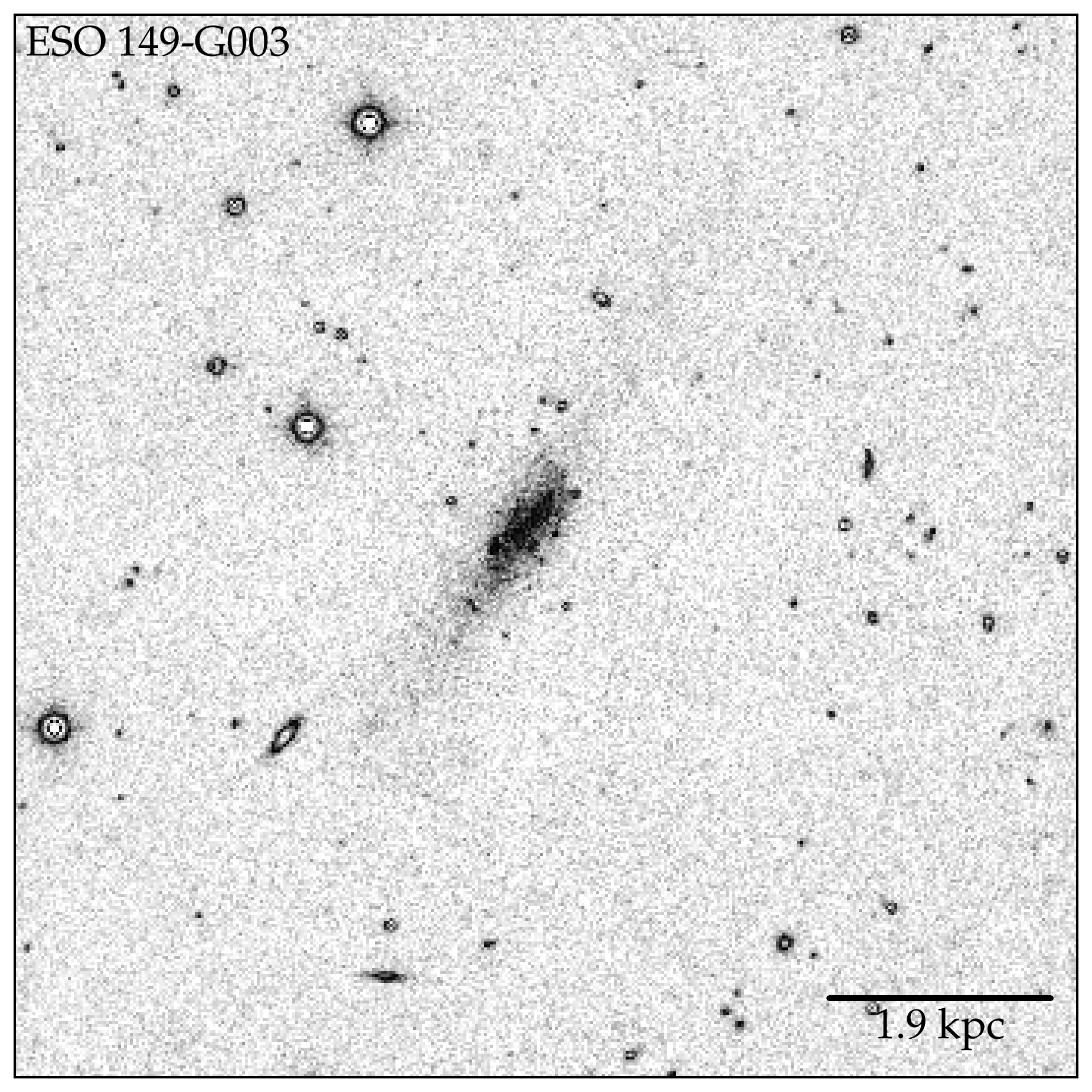}&

\end{array}
$
\caption{-- continued.}
\label{results4}
\end{figure*}

\begin{figure*}
\setlength{\arraycolsep}{1pt}
\setlength{\parskip}{1pt}
$
\begin{array}{cccc}

\includegraphics[trim = 0mm 0mm 6mm 2mm, clip = true, scale = 0.47]{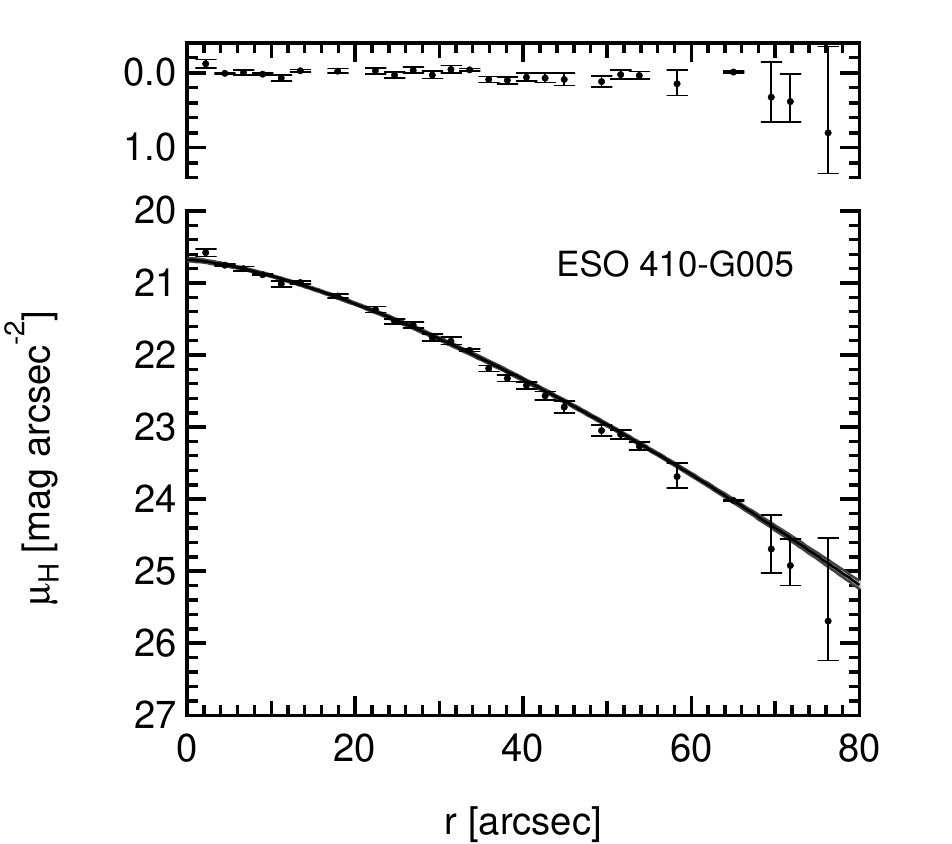}&
\includegraphics[trim = 0mm 0mm 6mm 2mm, clip = true, scale = 0.47]{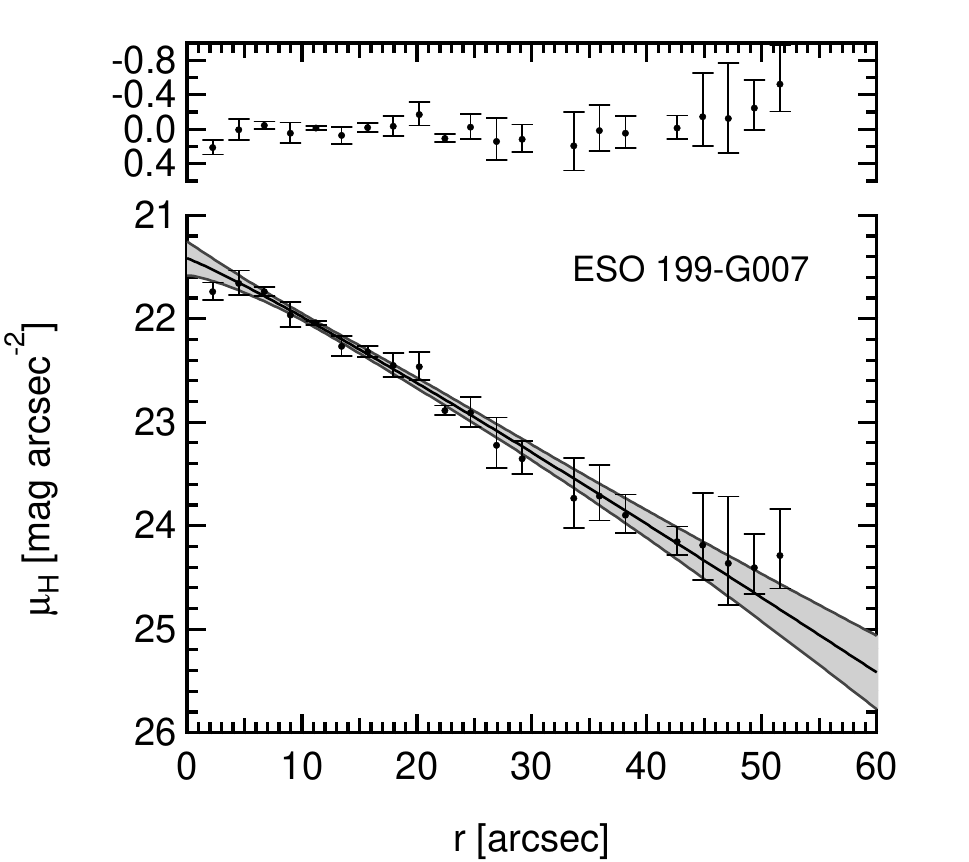}&
\includegraphics[trim = 0mm 0mm 6mm 2mm, clip = true, scale = 0.47]{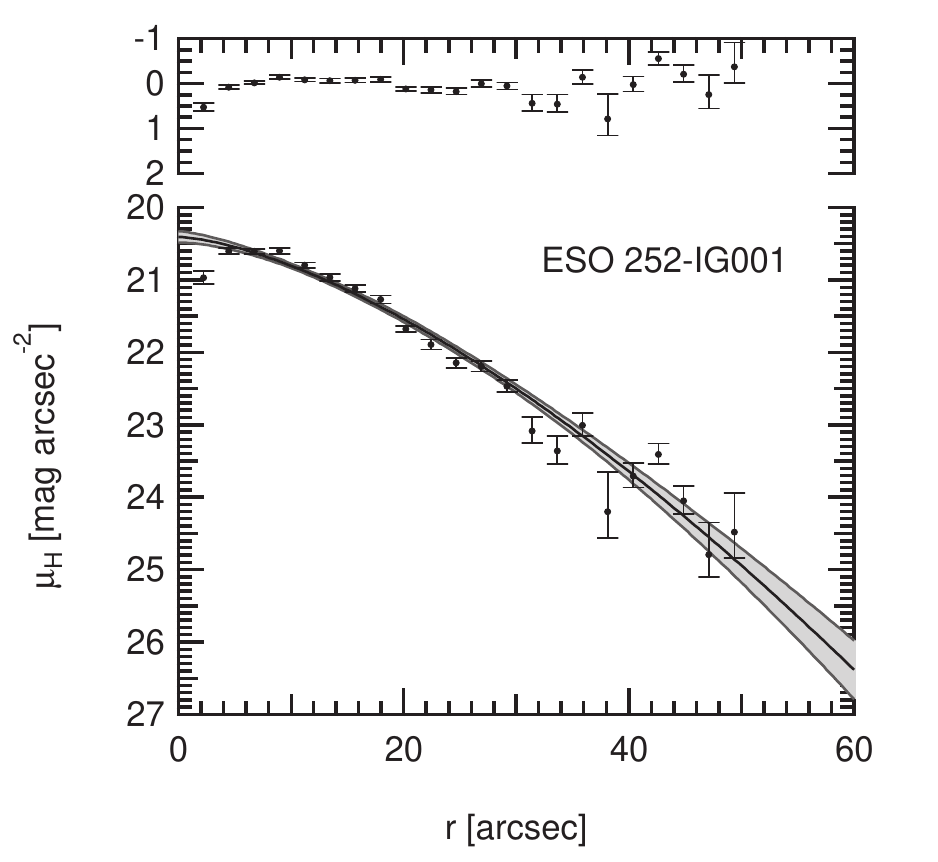}&
\includegraphics[trim = 0mm 0mm 6mm 2mm, clip = true, scale = 0.47]{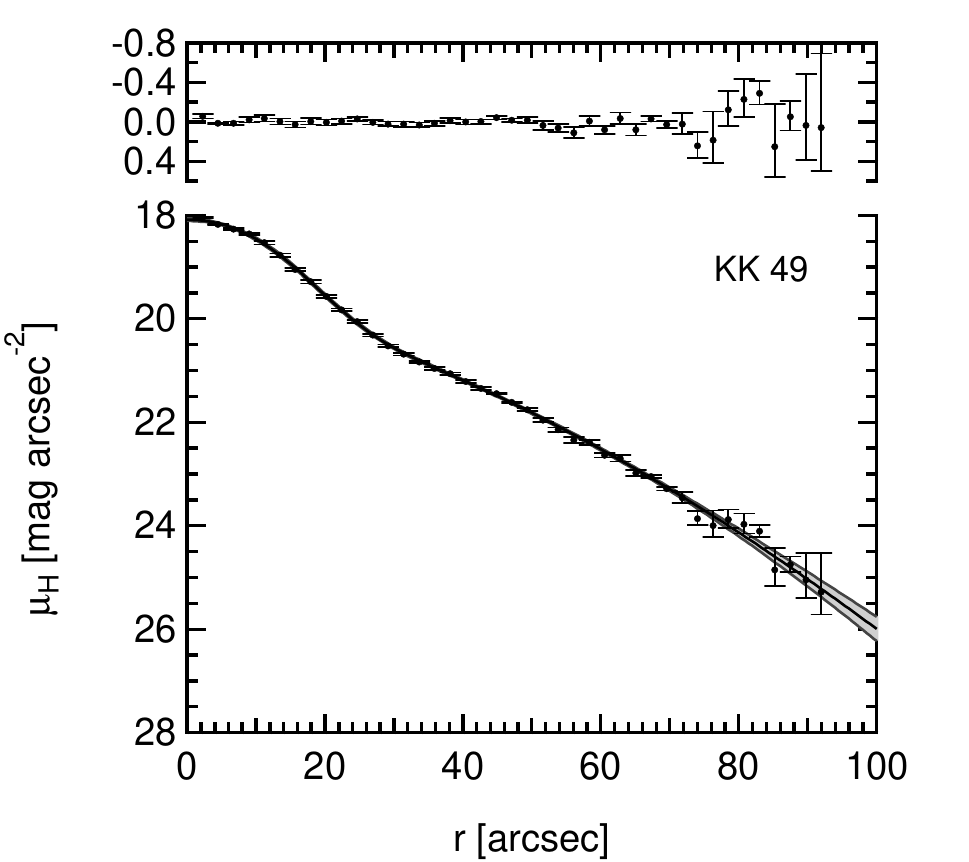}\\

\includegraphics[trim = 0mm 0mm 6mm 2mm, clip = true, scale = 0.47]{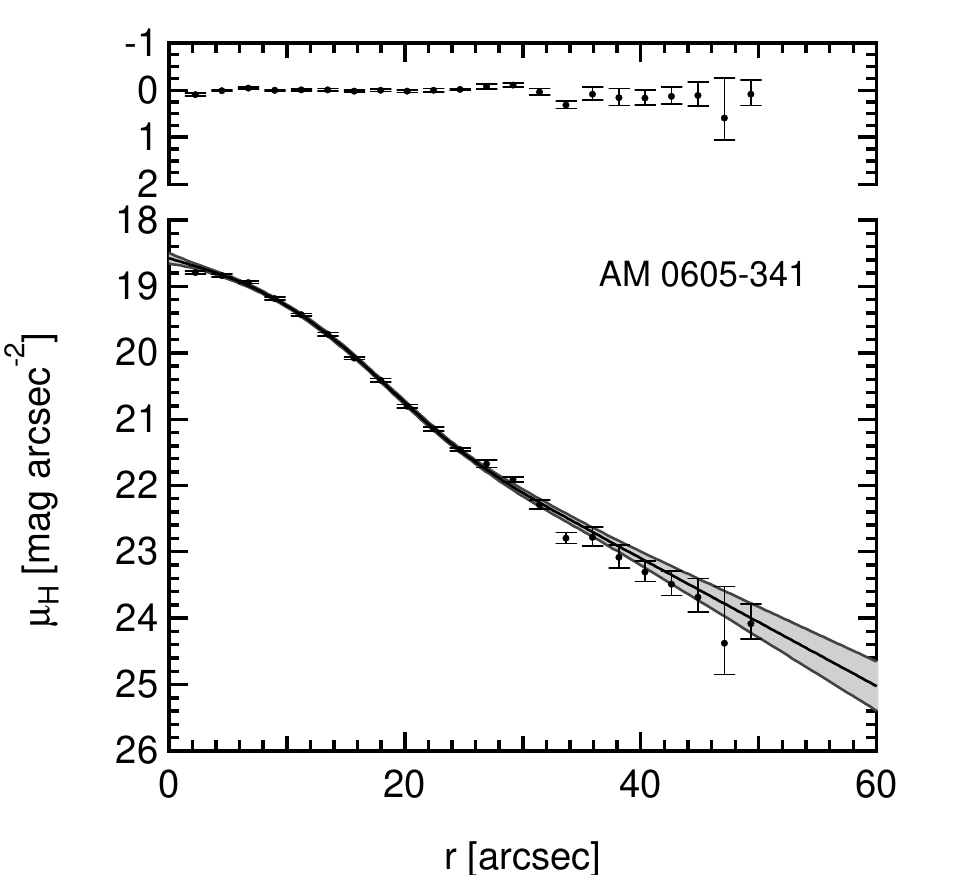}&
\includegraphics[trim = 0mm 0mm 6mm 2mm, clip = true, scale = 0.47]{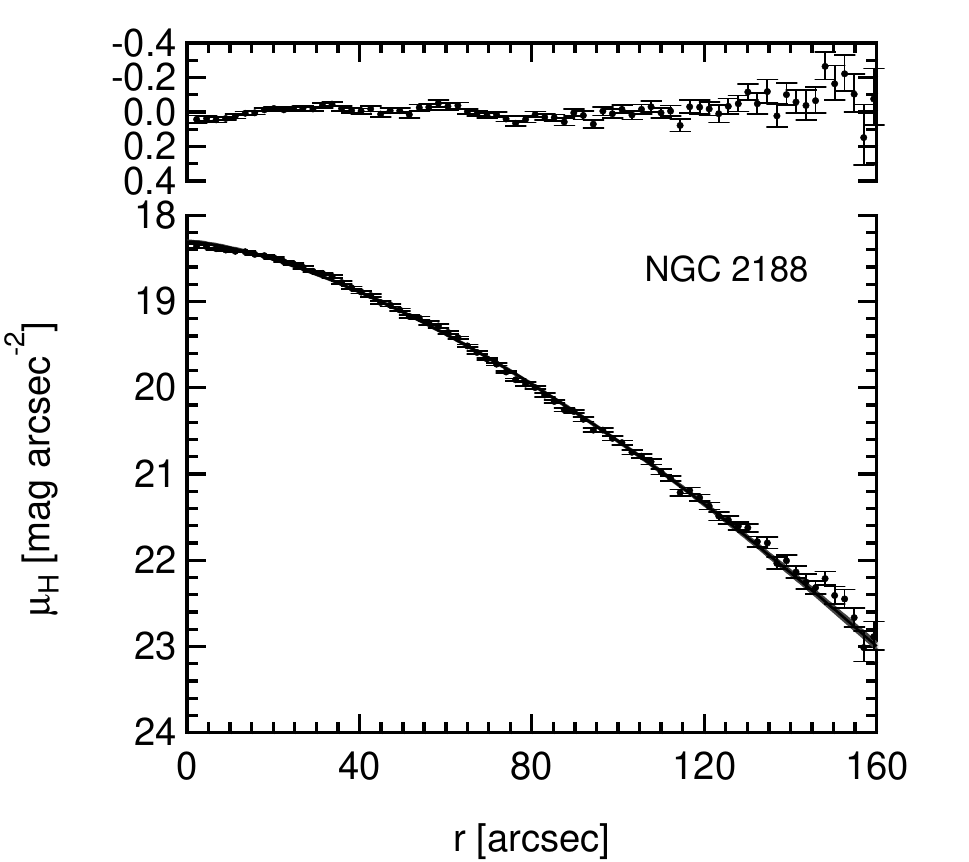}&
\includegraphics[trim = 0mm 0mm 6mm 2mm, clip = true, scale = 0.47]{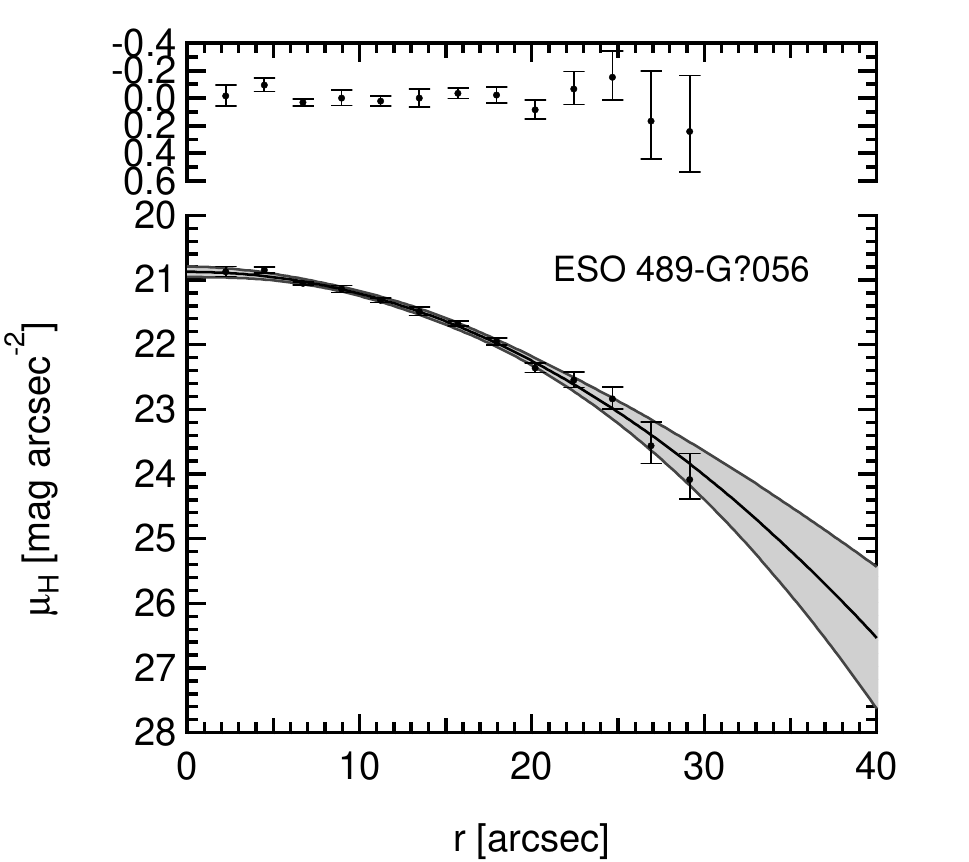}&
\includegraphics[trim = 0mm 0mm 6mm 2mm, clip = true, scale = 0.47]{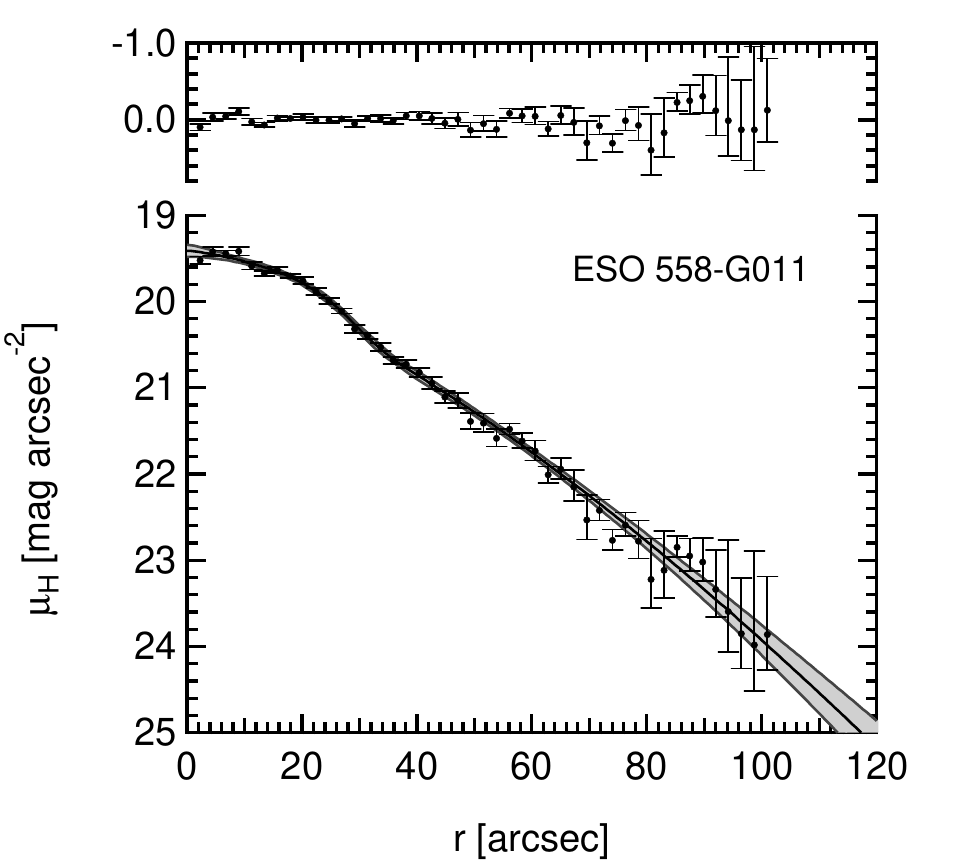}\\

\includegraphics[trim = 0mm 0mm 6mm 2mm, clip = true, scale = 0.47]{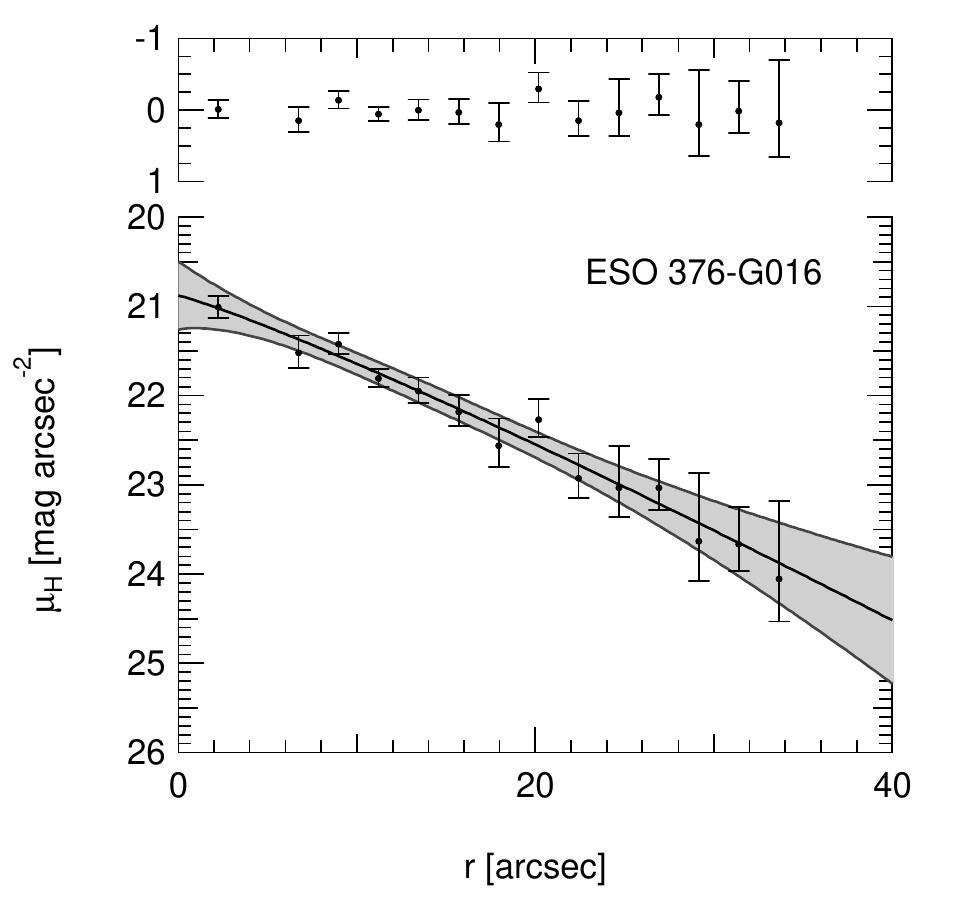}&
\includegraphics[trim = 0mm 0mm 6mm 2mm, clip = true, scale = 0.47]{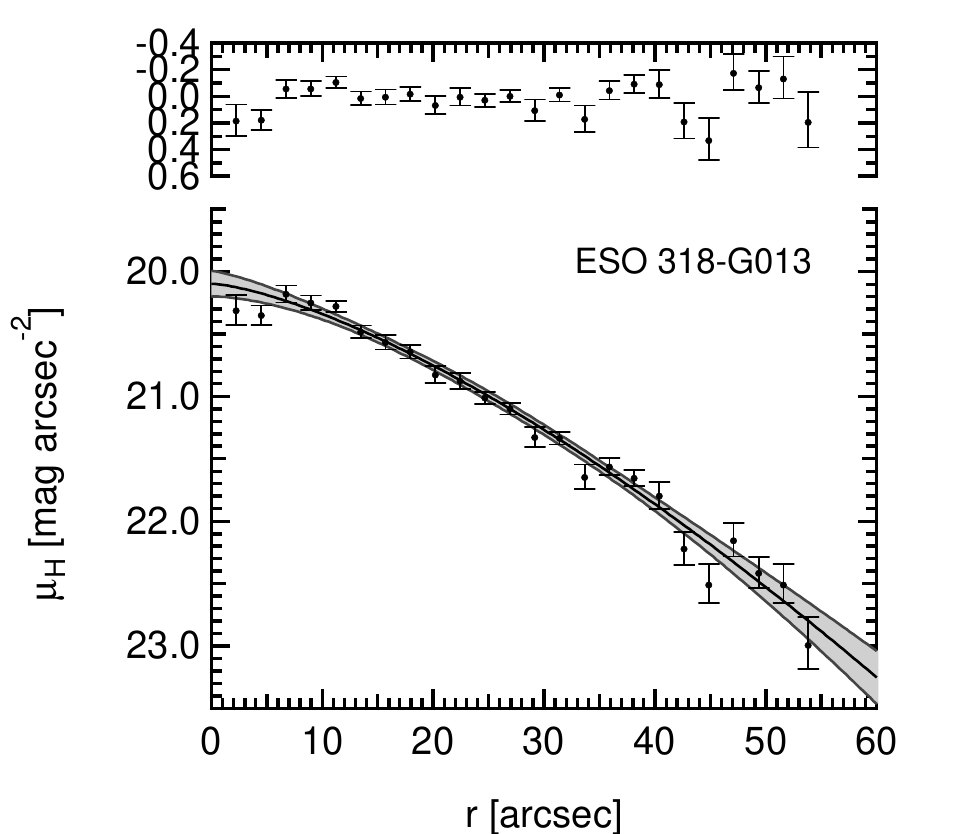}&
\includegraphics[trim = 0mm 0mm 6mm 2mm, clip = true, scale = 0.47]{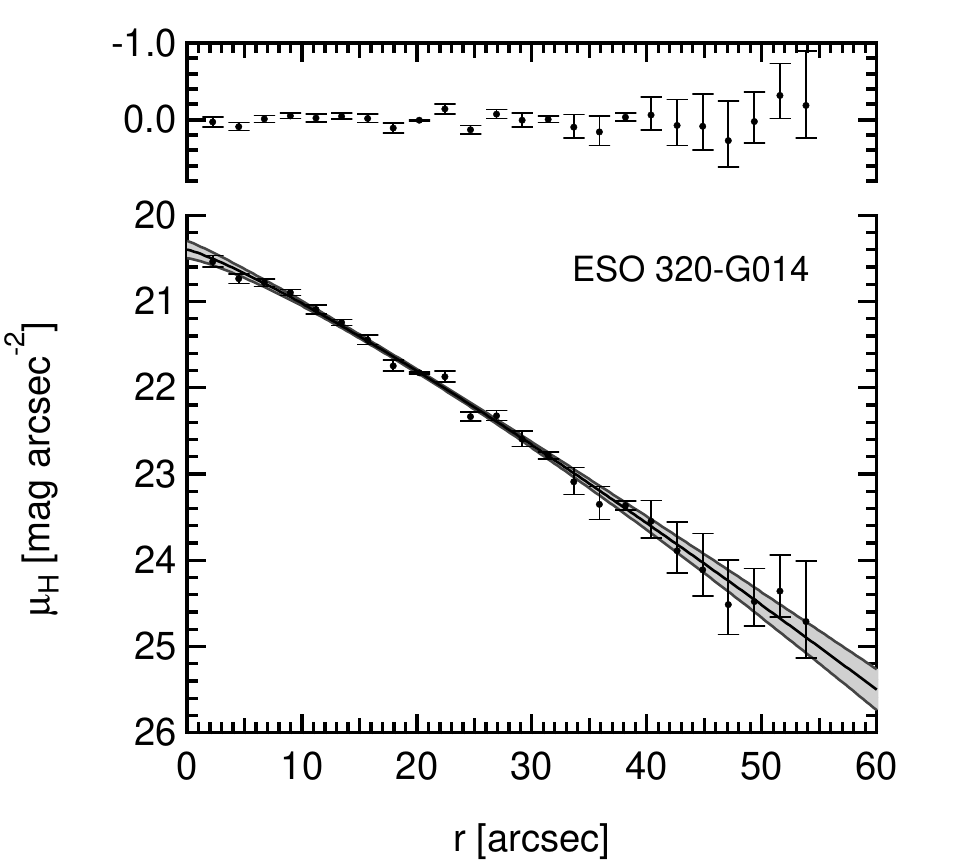}&
\includegraphics[trim = 0mm 0mm 6mm 2mm, clip = true, scale = 0.47]{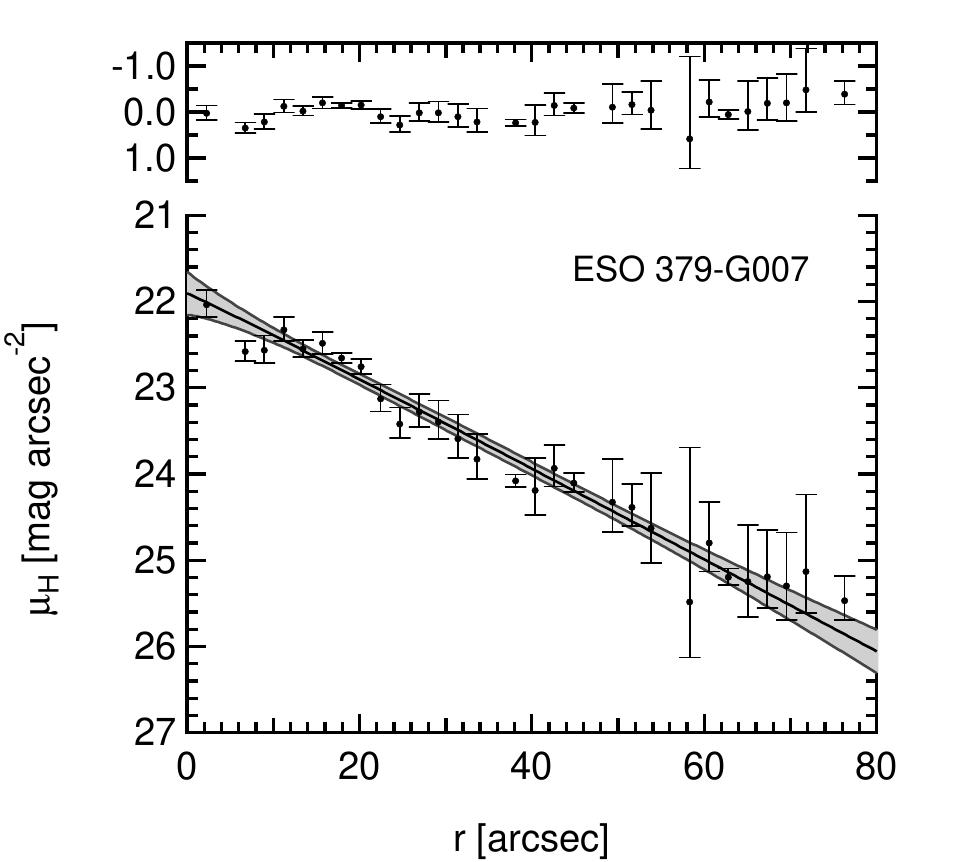}\\

\includegraphics[trim = 0mm 0mm 6mm 2mm, clip = true, scale = 0.47]{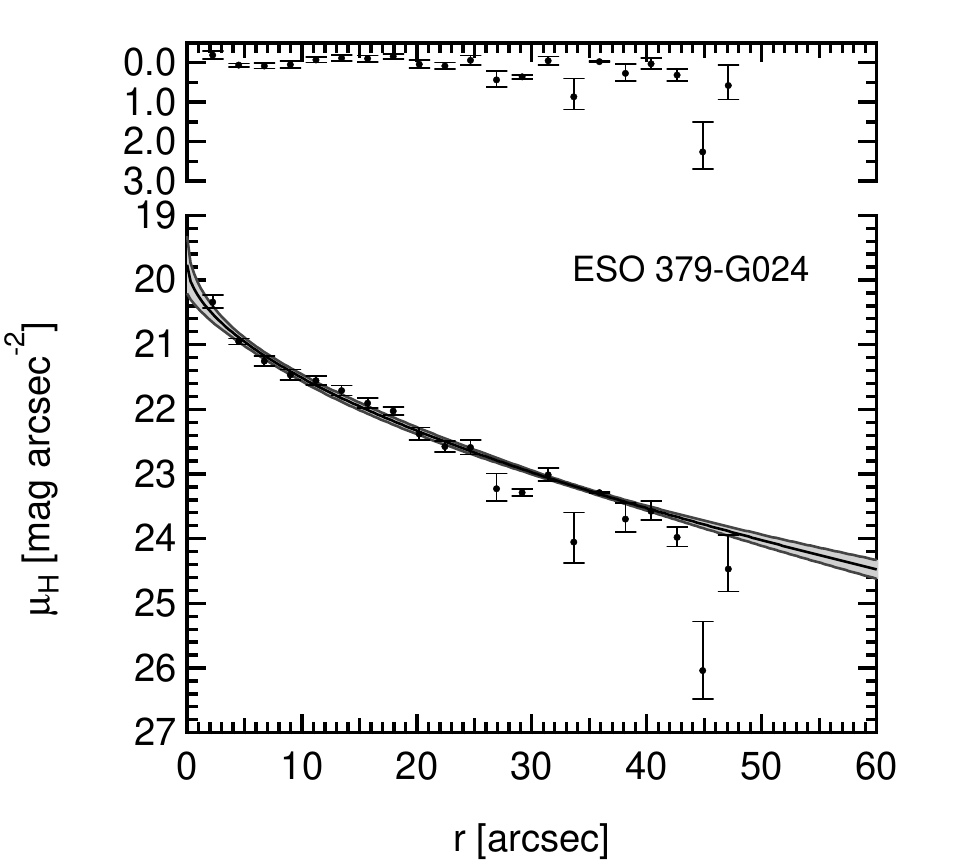}&
\includegraphics[trim = 0mm 0mm 6mm 2mm, clip = true, scale = 0.47]{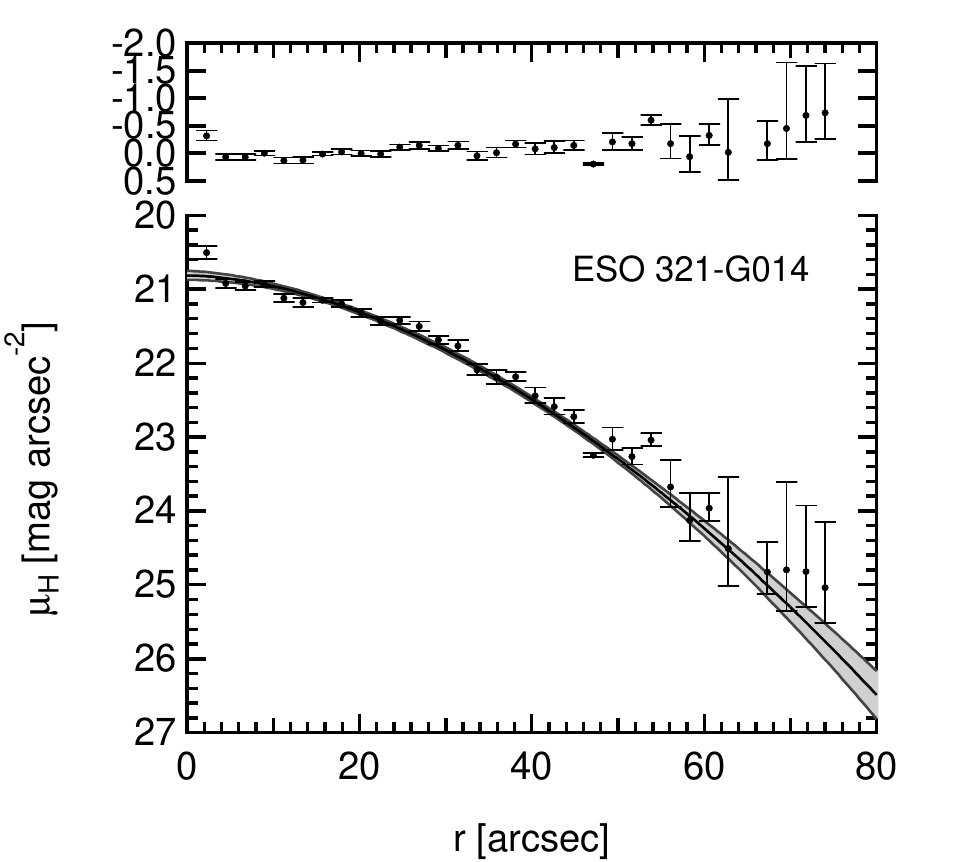}&
\includegraphics[trim = 0mm 0mm 6mm 2mm, clip = true, scale = 0.47]{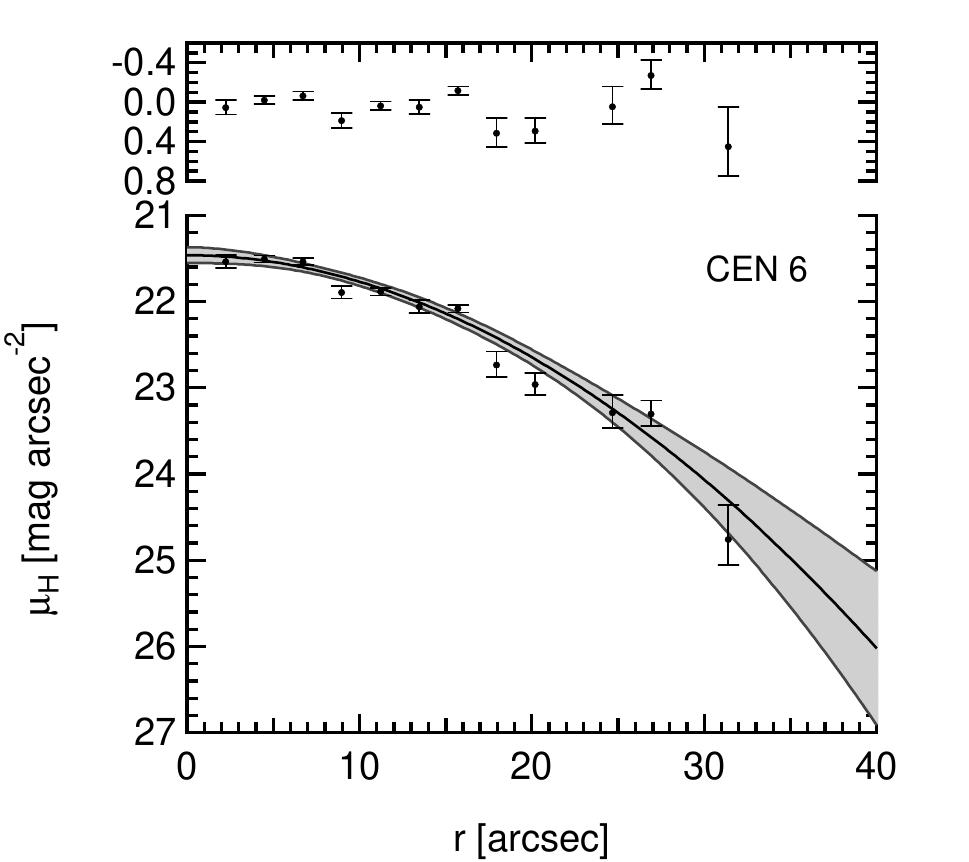}&
\includegraphics[trim = 0mm 0mm 6mm 2mm, clip = true, scale = 0.47]{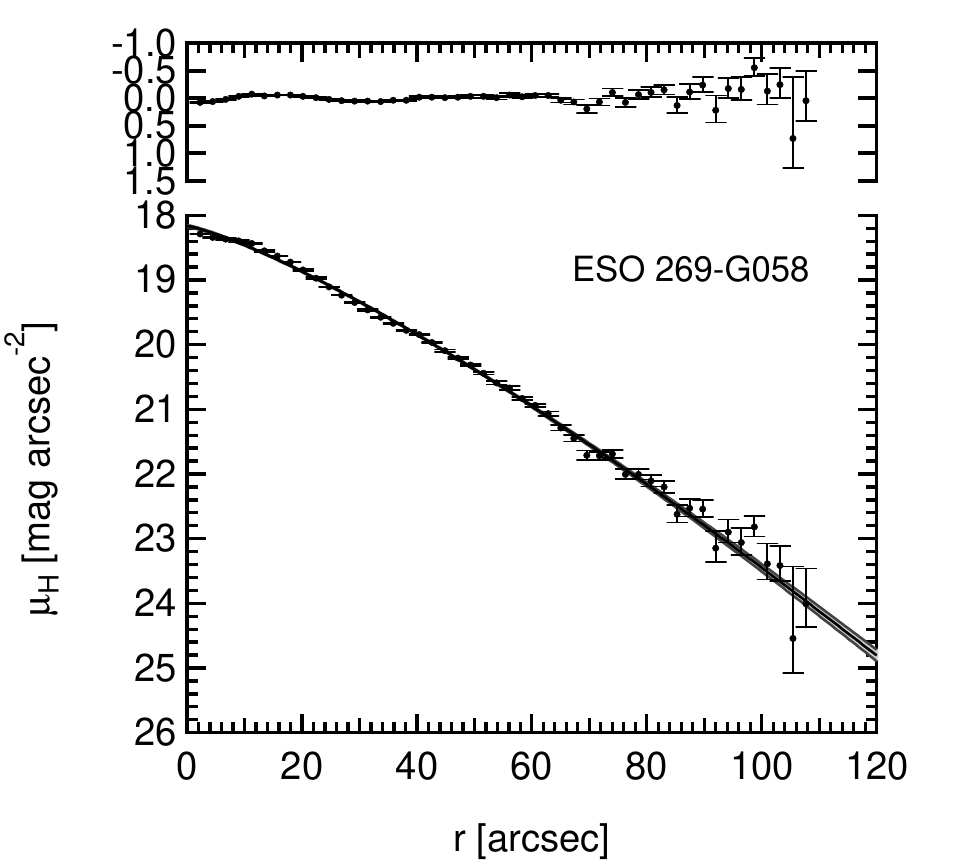}\\

\includegraphics[trim = 0mm 0mm 6mm 2mm, clip = true, scale = 0.47]{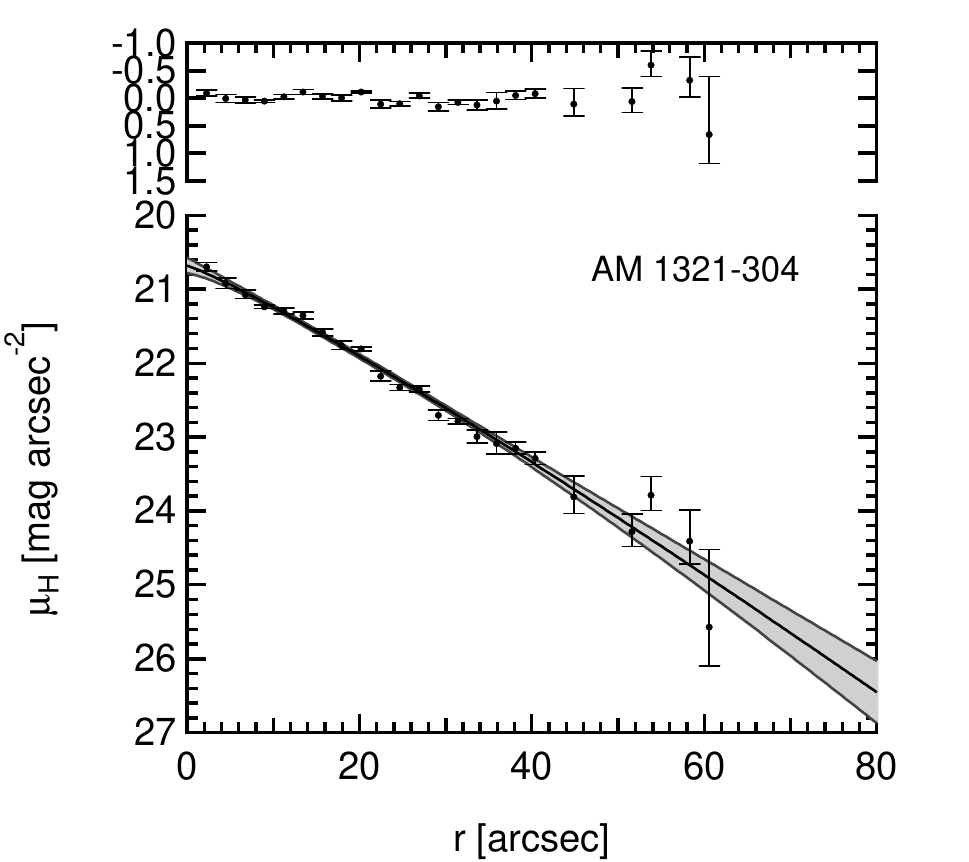}&
\includegraphics[trim = 0mm 0mm 6mm 2mm, clip = true, scale = 0.47]{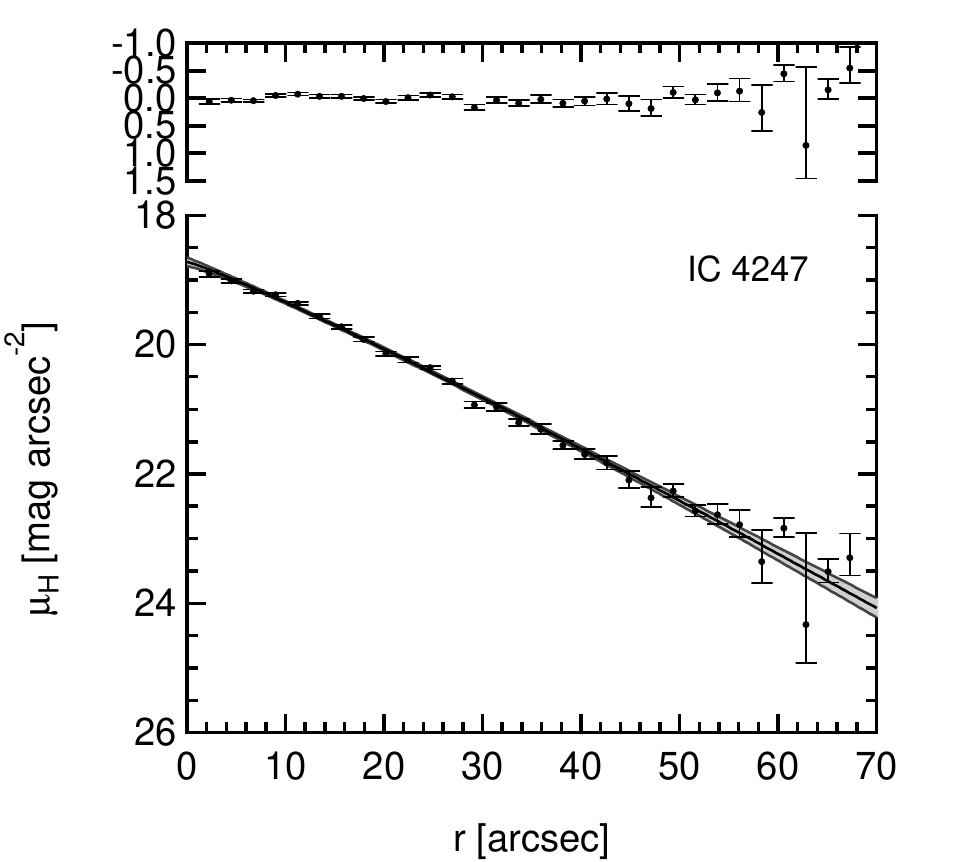}&
\includegraphics[trim = 0mm 0mm 6mm 2mm, clip = true, scale = 0.47]{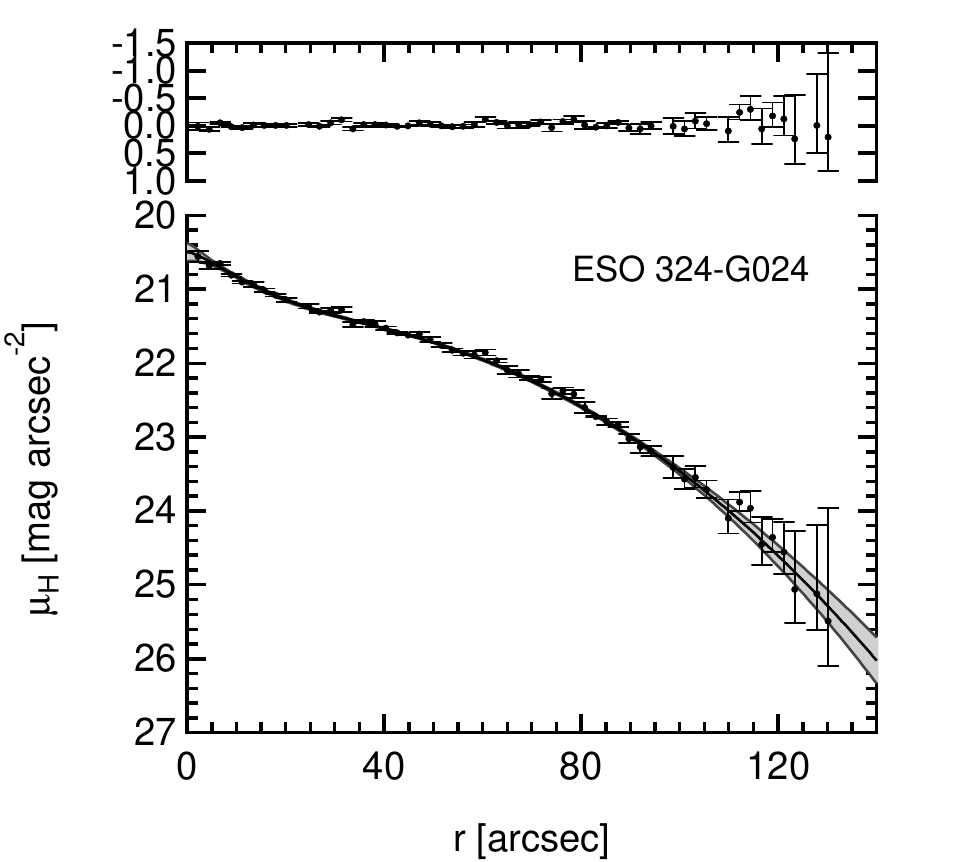}&
\includegraphics[trim = 0mm 0mm 6mm 2mm, clip = true, scale = 0.47]{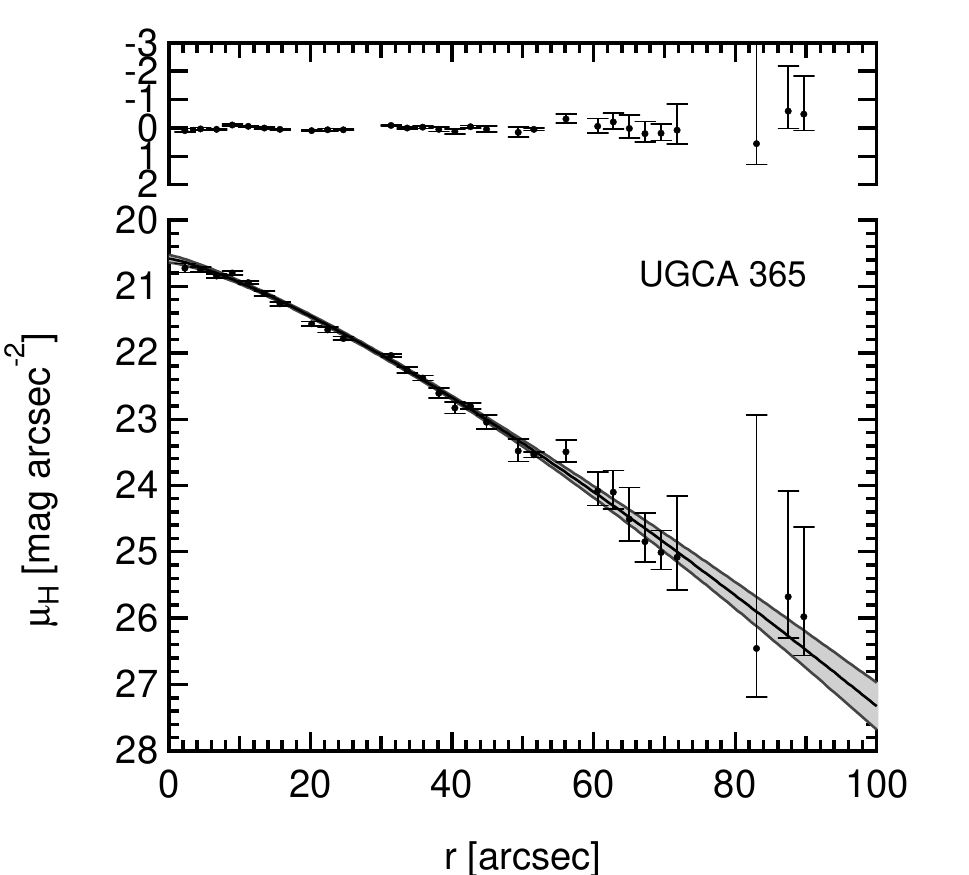}\\

\end{array}
$

\caption{The derived H-band surface brightness profiles for detected sample galaxies except J1919-68 (see text). The best-fitting S\'{e}rsic profile is shown as a solid line with three sigma confidence bands show in grey.}
\label{SBresults1}
\end{figure*}

\addtocounter {figure} {-1}

\begin{figure*}
\setlength{\arraycolsep}{1pt}
\setlength{\parskip}{1pt}
$
\begin{array}{cccc}
\includegraphics[trim = 0mm 0mm 6mm 2mm, clip = true, scale = 0.47]{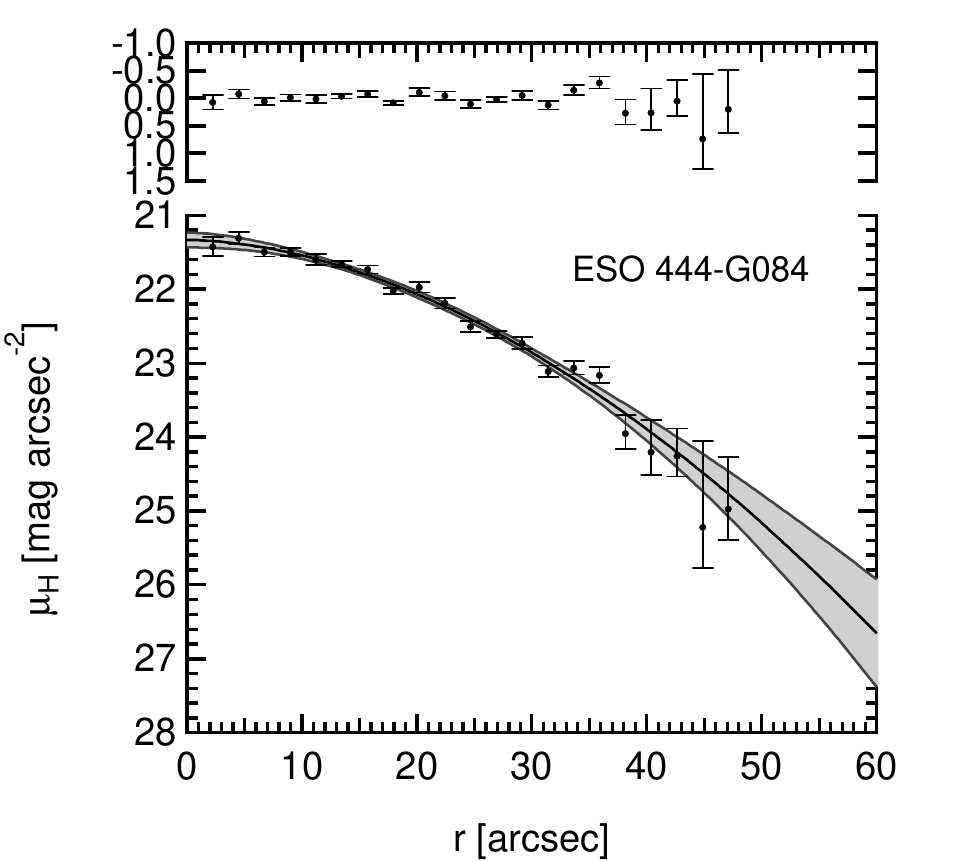}&
\includegraphics[trim = 0mm 0mm 6mm 2mm, clip = true, scale = 0.47]{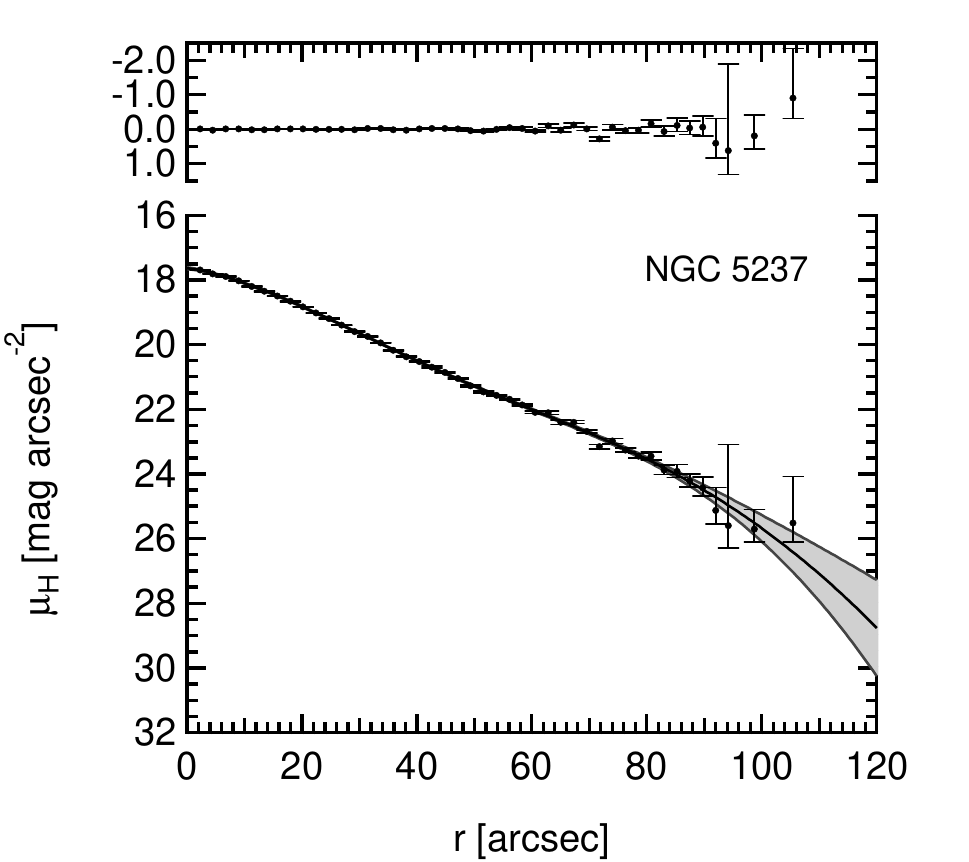}&
\includegraphics[trim = 0mm 0mm 6mm 2mm, clip = true, scale = 0.47]{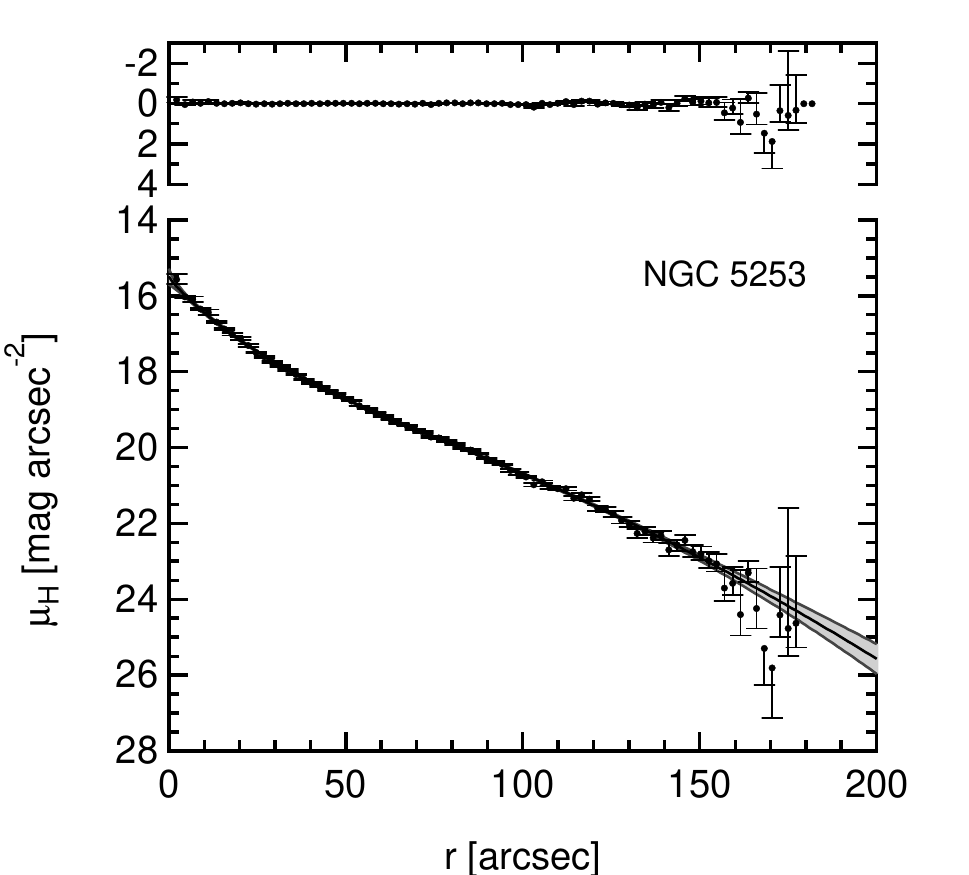}&
\includegraphics[trim = 0mm 0mm 6mm 2mm, clip = true, scale = 0.47]{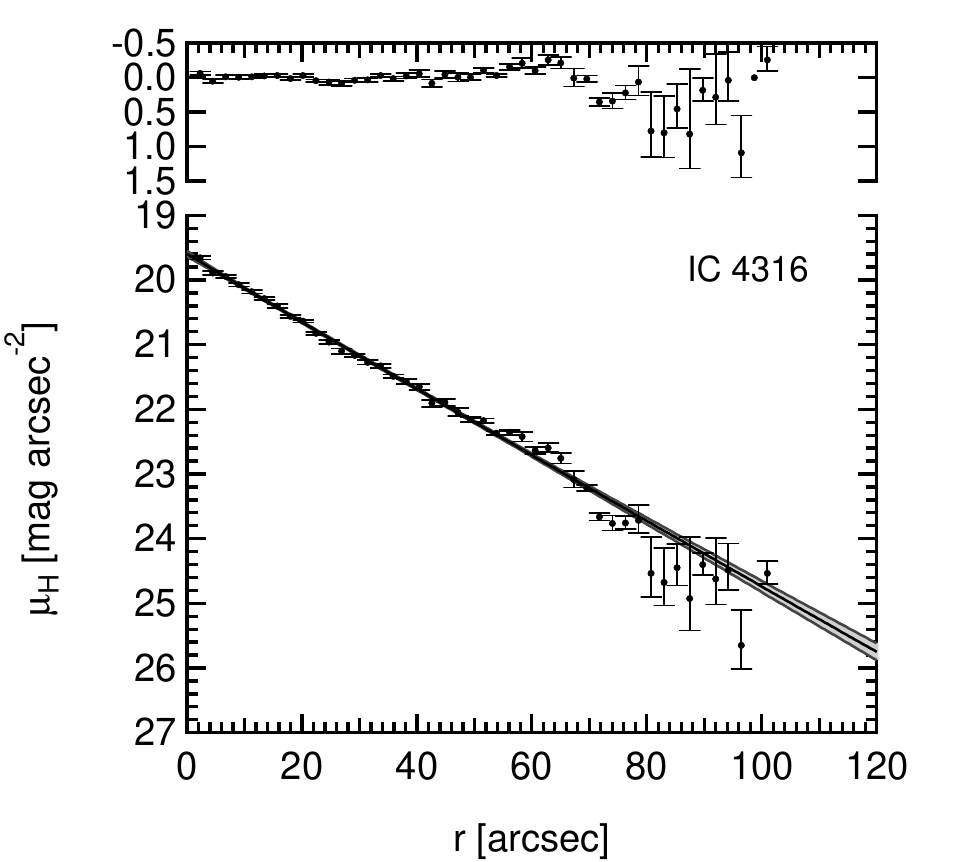}\\

\includegraphics[trim = 0mm 0mm 6mm 2mm, clip = true, scale = 0.47]{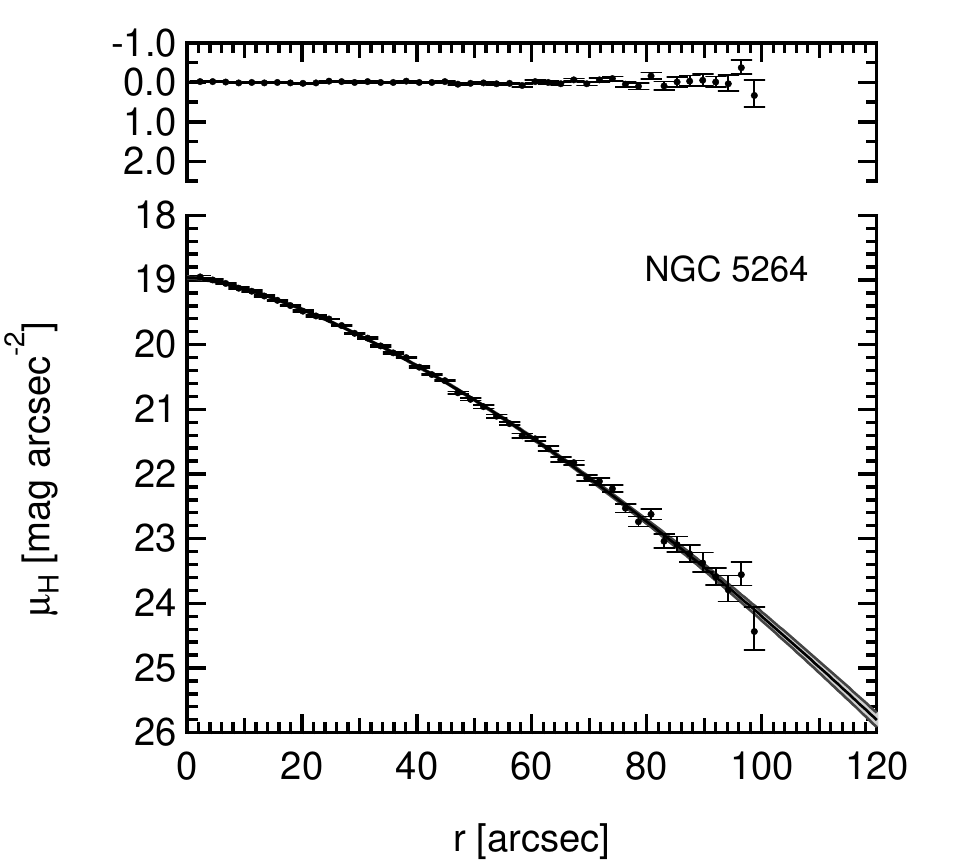}&
\includegraphics[trim = 0mm 0mm 6mm 2mm, clip = true, scale = 0.47]{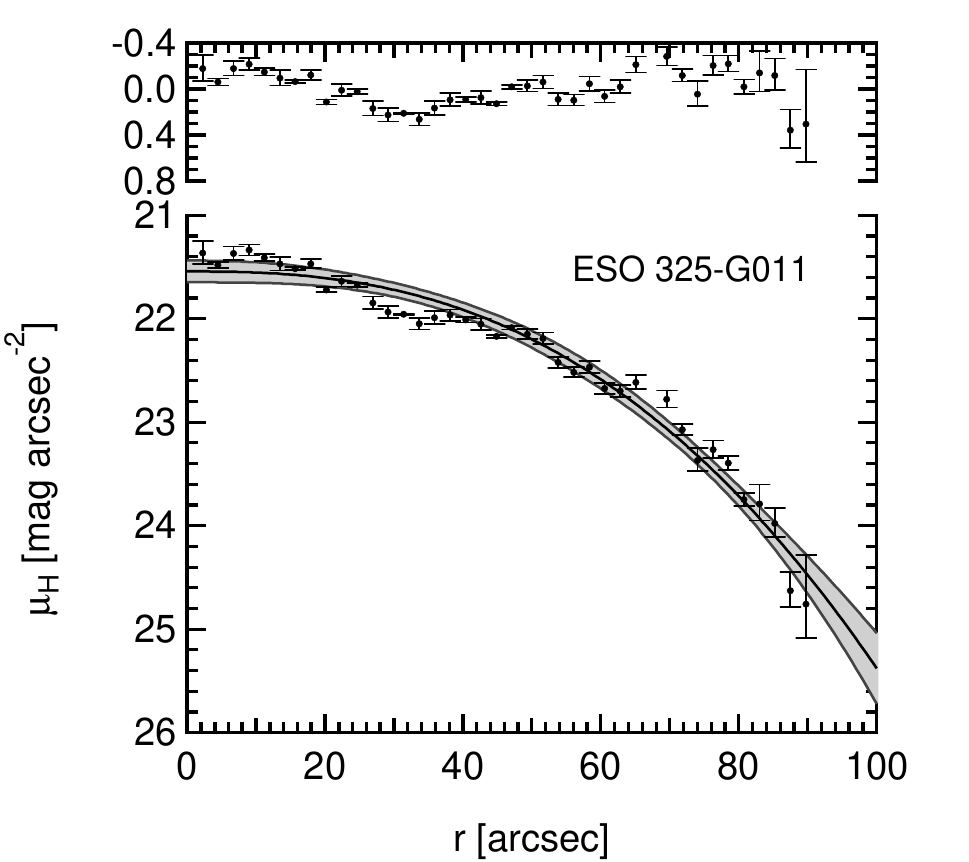}&
\includegraphics[trim = 0mm 0mm 6mm 2mm, clip = true, scale = 0.47]{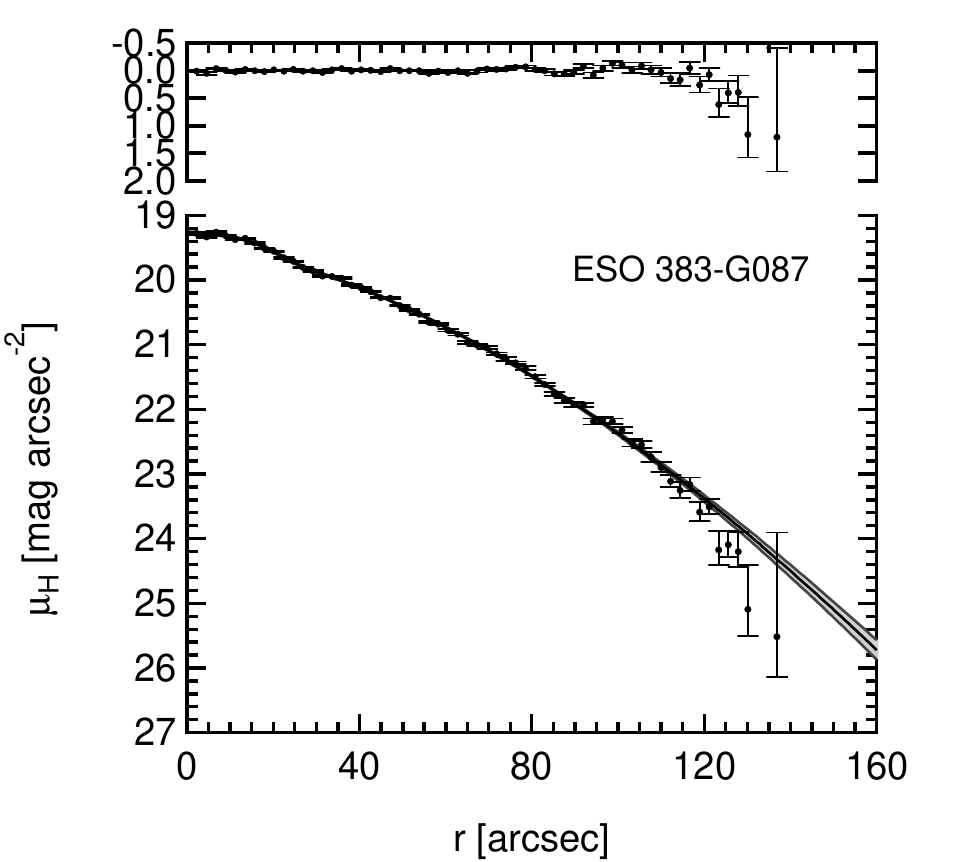}&
\includegraphics[trim = 0mm 0mm 6mm 2mm, clip = true, scale = 0.47]{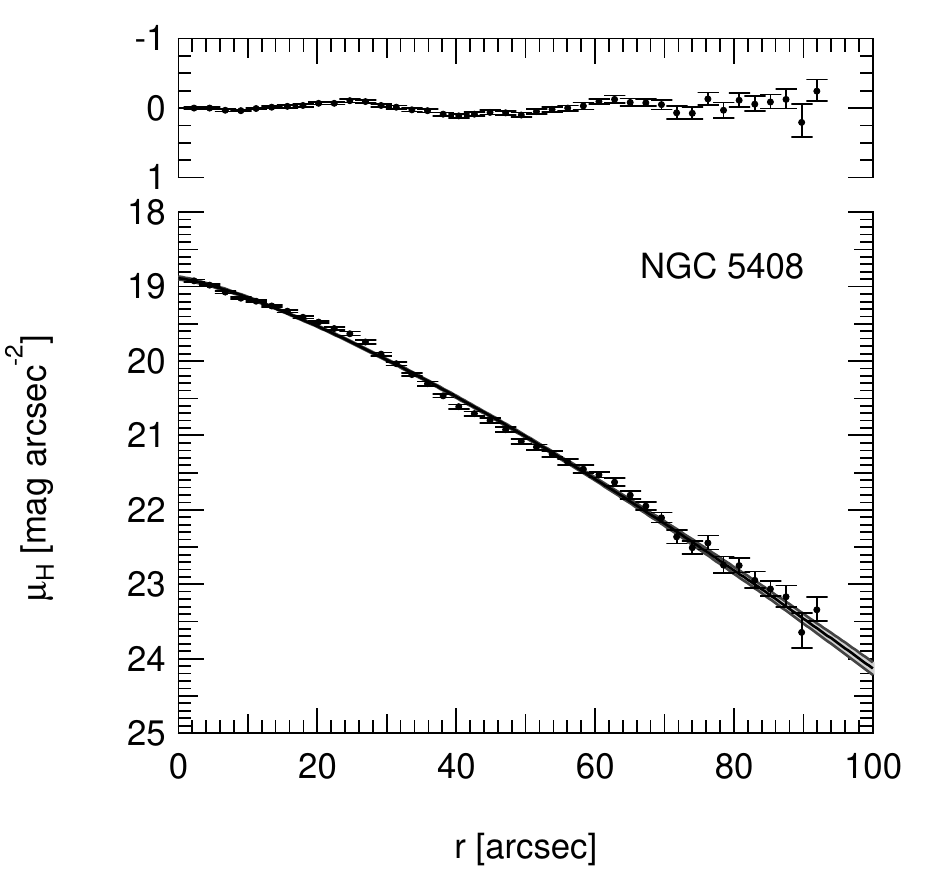}\\

\includegraphics[trim = 0mm 0mm 6mm 2mm, clip = true, scale = 0.47]{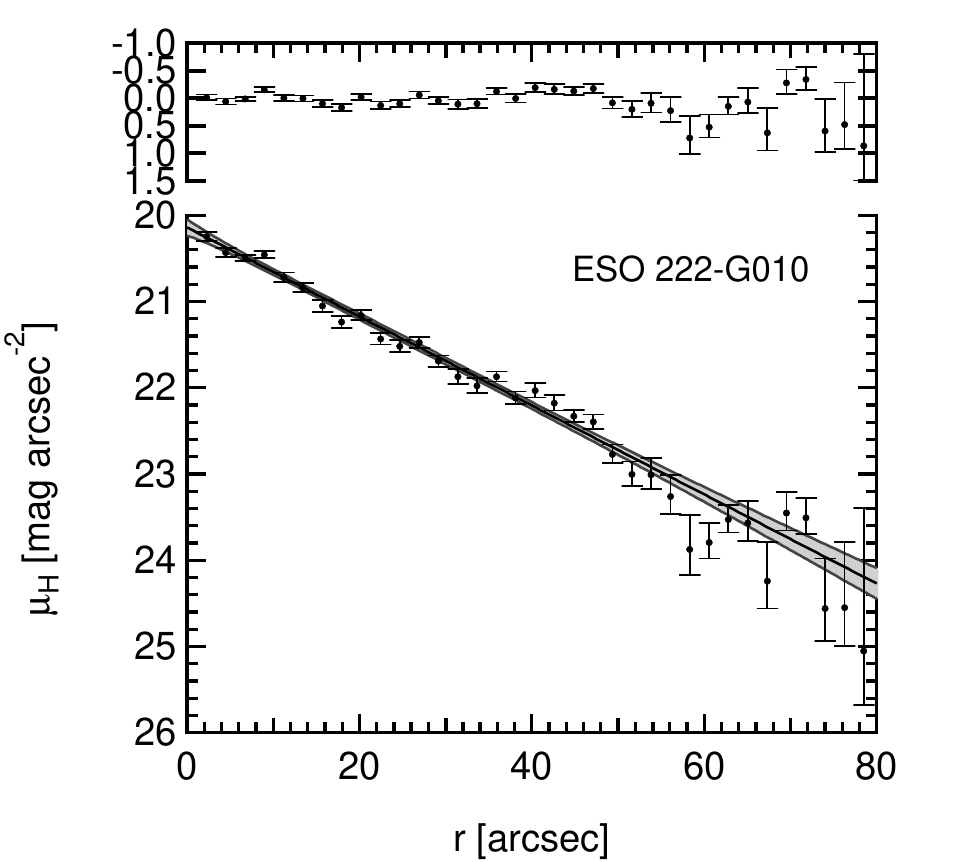}&
\includegraphics[trim = 0mm 0mm 6mm 2mm, clip = true, scale = 0.47]{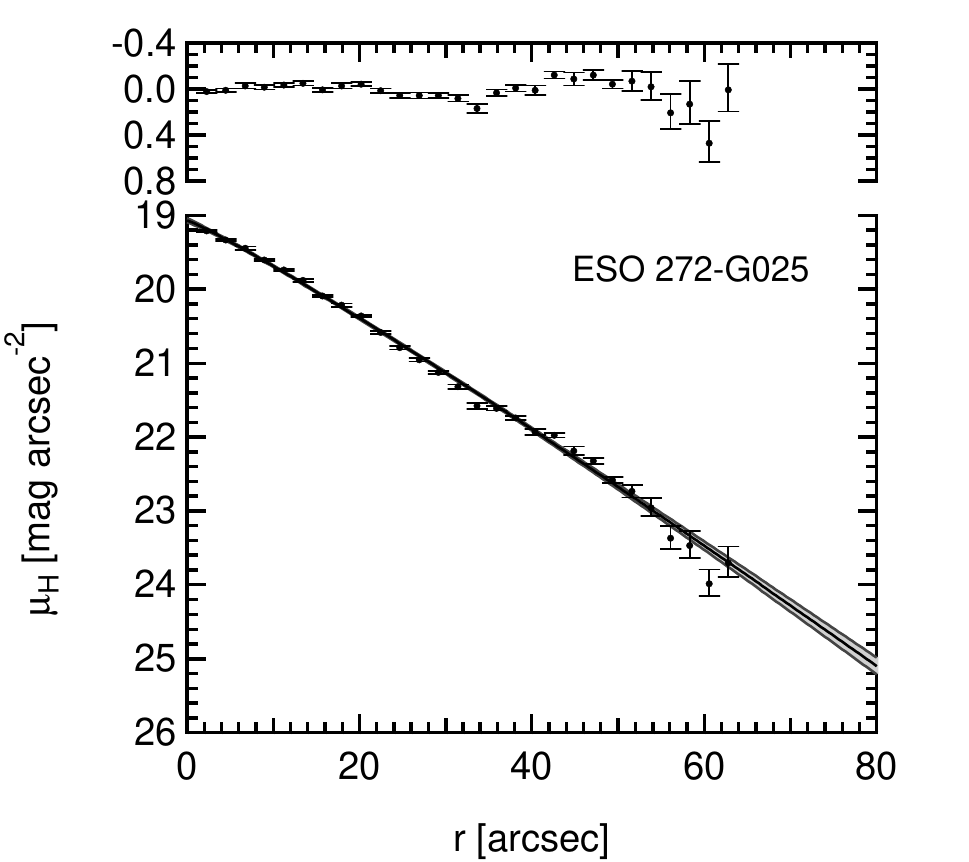}&
\includegraphics[trim = 0mm 0mm 6mm 2mm, clip = true, scale = 0.47]{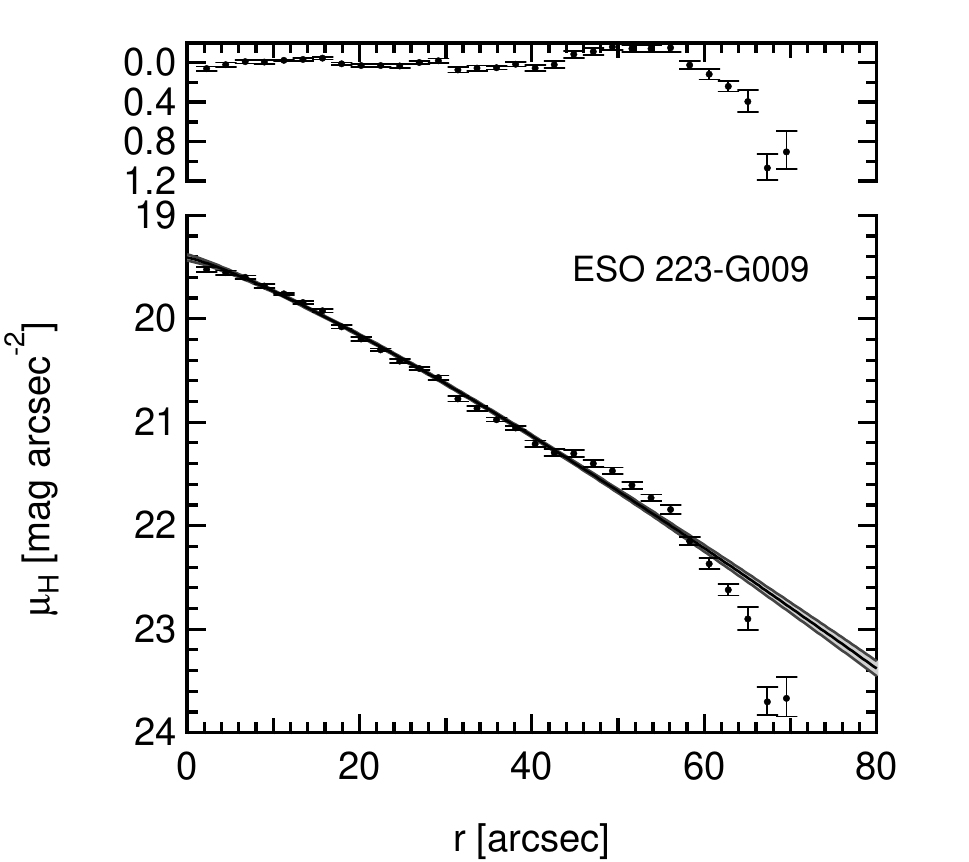}&
\includegraphics[trim = 0mm 0mm 6mm 2mm, clip = true, scale = 0.47]{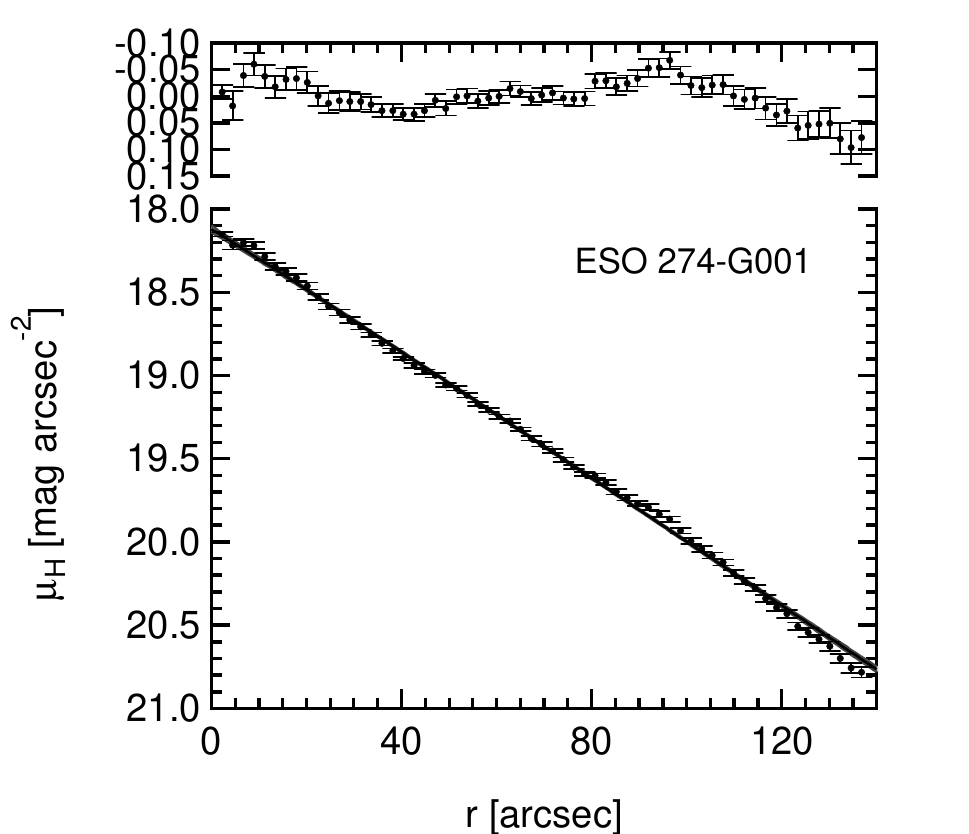}\\

\includegraphics[trim = 0mm 0mm 6mm 2mm, clip = true, scale = 0.47]{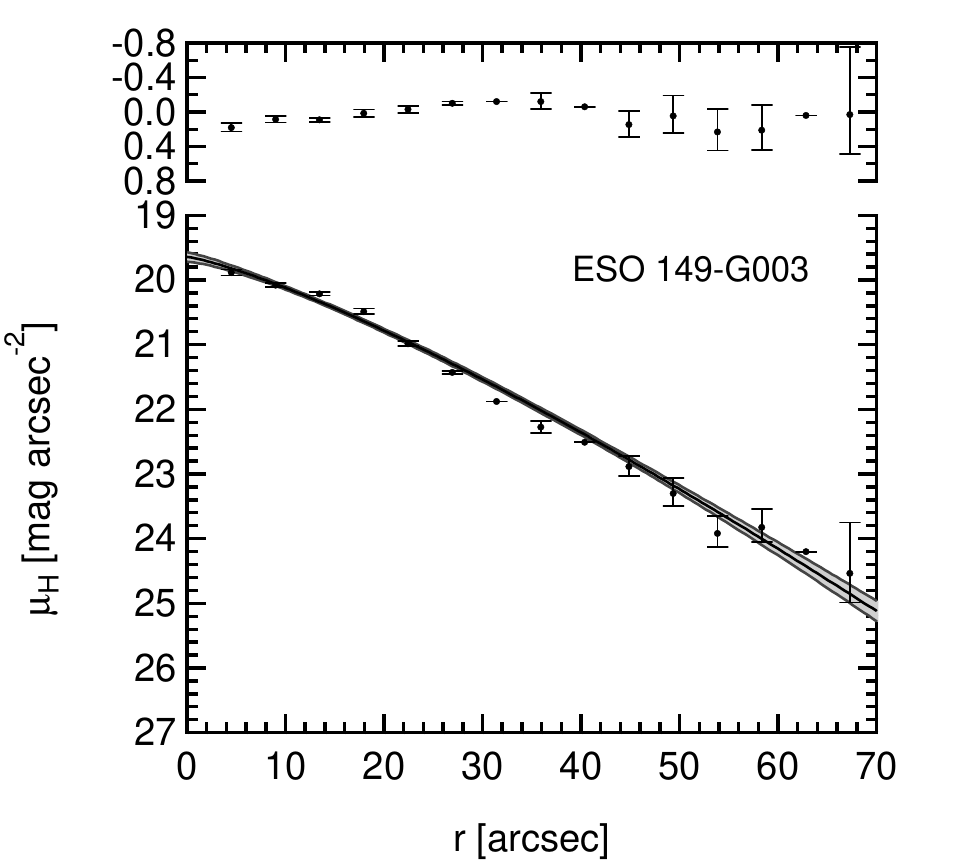}

\end{array}
$
\caption{-- continued.}
\label{SBresults2}
\end{figure*}

\end{document}